%% file: sigma_ridge.tex
\documentclass{article}
\usepackage[utf8]{inputenc}

\usepackage{amsmath}
\usepackage[shortlabels]{enumitem}
\usepackage{amssymb}
\usepackage{amsfonts}
\usepackage{amsthm}
\usepackage{bm}
\usepackage{color}
\usepackage{natbib}
\usepackage{graphicx}
\usepackage{mathtools}
\usepackage{hyperref}
\usepackage{float}

\usepackage{caption}
\usepackage{subcaption}
\DeclareCaptionLabelFormat{r-parens}{#2)} 
\usepackage{adjustbox}
\usepackage{pgfplots}
\usepackage{changes}
\usepackage{pgf}
\usepackage{tikz}
\usetikzlibrary{backgrounds}
\usetikzlibrary{positioning}
\usepgfplotslibrary{fillbetween}
\usepackage{tikzscale}
\usetikzlibrary{arrows.meta}
\usetikzlibrary{backgrounds}
\usepgfplotslibrary{patchplots}

\pgfplotsset{compat=newest}

\pgfplotsset{%
layers/standard/.define layer set={%
    background,axis background,axis grid,axis ticks,axis lines,axis tick labels,pre main,main,axis descriptions,axis foreground%
}{grid style= {/pgfplots/on layer=axis grid},%
    tick style= {/pgfplots/on layer=axis ticks},%
    axis line style= {/pgfplots/on layer=axis lines},%
    label style= {/pgfplots/on layer=axis descriptions},%
    legend style= {/pgfplots/on layer=axis descriptions},%
    title style= {/pgfplots/on layer=axis descriptions},%
    colorbar style= {/pgfplots/on layer=axis descriptions},%
    ticklabel style= {/pgfplots/on layer=axis tick labels},%
    axis background@ style={/pgfplots/on layer=axis background},%
    3d box foreground style={/pgfplots/on layer=axis foreground},%
    },
}

\hypersetup{colorlinks,citecolor=blue,urlcolor=blue,linkcolor=blue}
\usepackage[bbgreekl]{mathbbol}

\usepackage{soul}

\DeclareSymbolFontAlphabet{\mathbb}{AMSb}
\DeclareSymbolFontAlphabet{\mathbbl}{bbold}

\newcommand{\beginsupplement}{%
        \setcounter{table}{0}
        \renewcommand{\thetable}{S\arabic{table}}%
        \setcounter{figure}{0}
        \renewcommand{\thefigure}{S\arabic{figure}}%
     }

\newtheorem{definition}{Definition}

\newtheorem{theorem} {Theorem} 
\newtheorem{lemma}{Lemma}

\newtheorem{proposition} {Proposition} 
\newtheorem{corol} {Corollary} 
\newtheorem*{theorem-non}{Theorem}

\theoremstyle{remark}

\newtheoremstyle{named}{}{}{\itshape}{}{}{.}{.5em}{#1 #3}
\theoremstyle{named}

\newcommand{\RR}{\mathbb{R}}

\newcommand{\nn}{\mathcal{N}}

\newcommand{\risk}[1]{\bm{R}(#1)}
\newcommand{\loobasic}{\operatorname{CV}}
\newcommand{\loo}{\operatorname{CV}^*}
\newcommand{\Ln}{\bm{L}}
\newcommand{\Rn}{\bm{R}}
\newcommand{\Fn}{\bm{F}}
\newcommand{\Rloo}{\loo} 

\DeclareMathOperator{\Tr}{Tr}

\newcommand{\group}[1]{\mathcal{G}_{#1}}
\newcommand{\p}[1]{\left(#1\right)}
\newcommand{\sqb}[1]{\left[#1\right]}
\newcommand{\cb}[1]{\left\{#1\right\}}

\newcommand{\blambda}{\bm{\lambda}}
\newcommand{\bLambda}{\bm{\Lambda}}
\newcommand{\lambdainit}{\widetilde{\lambda}_{\text{init}}}
\newcommand{\blambdainit}{\widetilde{\blambda}_{\text{init}}}

\newcommand{\bw}{\bm{w}}
\newcommand{\balpha}{\bm{\alpha}}
\newcommand{\bX}{\bm{X}}
\newcommand{\bM}{\bm{M}}
\newcommand{\bH}{\bm{H}}
\newcommand{\bu}{\bm{u}}
\newcommand{\bv}{\bm{v}}
\newcommand{\bN}{\bm{N}}
\newcommand{\bA}{\bm{A}}
\newcommand{\ba}{\bm{a}}
\newcommand{\bb}{\bm{b}}
\newcommand{\bB}{\bm{B}}
\newcommand{\bC}{\bm{C}}
\newcommand{\bD}{\bm{D}}
\newcommand{\bS}{\bm{S}}
\newcommand{\bR}{\bm{P}}
\newcommand{\bY}{\bm{Y}}
\newcommand{\bZ}{\bm{Z}}
\newcommand{\bzeta}{\bm{\zeta}}
\newcommand{\boldeta}{\bm{\eta}}
\newcommand{\bd}{\bm{d}}
\newcommand{\bI}{\bm{I}}
\newcommand{\bSigma}{\bm{\Sigma}}

\newcommand{\bvarepsilon}{\bm{\varepsilon}}
\newcommand{\yresponse}{Y}

\newcommand{\Img}{\operatorname{Im}}

\newcommand{\EE}[2][]{\mathbb{E}_{#1}\left[#2\right]}
\newcommand{\EEInline}[2][]{\mathbb{E}_{#1}[#2]}
\newcommand{\Var}[2][]{\operatorname{Var}_{#1}\left[#2\right]}

\newcommand{\abs}[1]{\left\lvert#1\right\rvert}
\newcommand{\Norm}[1]{\left\lVert#1\right\rVert}

\DeclareMathOperator*{\argmin}{argmin}   

 \newcommand\omicron{o}

\DeclarePairedDelimiter{\norm}{\lVert}{\rVert}

\newcommand{\sigmacv}{\bbsigma}

\newcommand{\mhatlambda}{\widehat{\blambda}(\bbsigma)}
\newcommand{\groupg}[1]{\mathcal{G}_{#1}}

\newif\ifdoubleblind
\doubleblindfalse

\newcommand{\citepsupplement}{\citep}
\newcommand{\citetsupplement}{\citet}
\usepackage[margin=1.5in]{geometry}

\date{Draft manuscript: March 2021}
\author{Nikolaos Ignatiadis\thanks{Department of Statistics, Stanford University (ignat@stanford.edu)} \;\; \and \;\; Panagiotis Lolas\thanks{Department of Mathematics, Stanford University (panagd@stanford.edu)}}

\title{$\sigmacv$-Ridge: group regularized ridge regression via \\
empirical Bayes noise level cross-validation}

\begin{document}
\maketitle
\ifdoubleblind \vspace{-2cm} \fi
\begin{abstract}
Features in predictive models are not exchangeable, yet common supervised models treat them as such. Here we study ridge regression when the analyst can partition the features into $K$ groups based on external side-information. For example, in high-throughput biology, features may represent gene expression, protein abundance or clinical data and so each feature group represents a distinct modality. The analyst's goal is to choose optimal regularization parameters $\blambda = (\lambda_1, \dotsc, \lambda_K)$ -- one for each group. In this work, we study the impact of $\blambda$ on the predictive risk of group-regularized ridge regression by deriving limiting risk formulae under a high-dimensional random effects model with $p\asymp n$ as $n \to \infty$. Furthermore, we propose a data-driven method for choosing $\blambda$ that attains the optimal asymptotic risk: The key idea is to interpret the residual noise variance $\sigma^2$, as a regularization parameter to be chosen through cross-validation. An empirical Bayes construction maps the one-dimensional parameter $\sigma$ to the $K$-dimensional vector of regularization parameters, i.e.,  \smash{$\sigma \mapsto \widehat{\blambda}(\sigma)$}. Beyond its theoretical optimality, the proposed method is practical and runs as fast as cross-validated ridge regression without feature groups ($K=1$).\\
\newline
\noindent%
\end{abstract}

\section{Introduction}
The predictive performance of supervised learning methods that predict a response $Y_i$ from high-dimensional features $x_i$ can be improved by using external knowledge about the features, i.e.,  side-information that is not contained in the numerical values of the $x_i$. For example, in high-throughput biology, the features may comprise of distinct modalities, such as gene expression, protein abundance or clinical data.  The gene expression features in turn correspond to different genetic pathways or perhaps to the same genes measured across multiple tissues.~\citet{van2016better} use the term ``co-data'' for such external information, while
\citet{tay2020feature} use the term ``features of features''. How can we use such side-information to improve predictive performance in a principled way?

The conceptual move away from exchangeable features to features with side-information is straight-forward. For $Y_i \in \RR,\;x_i \in \RR^p,\;i=1,\dotsc,n$ consider the regularized regression,
\begin{equation}
\label{eq:general_penalized_reg}
\widehat{\bw} \in \argmin_{\bw} \cb{\frac{1}{2n}\sum_{i=1}^n \p{\yresponse_i - x_i^\intercal \bw}^2 \; + \;\text{Pen}(\bw)}.
\end{equation}
In the exchangeable setting, without a-priori information about the features, $\text{Pen}(\bw)$ is typically chosen as a symmetric regularizer, such as $\text{Pen}(\bw) = \lambda \Norm{\bw}_2^2/2$ ~\citep{hoerl1970ridge, tikhonov1963solution} or $\text{Pen}(\bw) =\lambda \Norm{\bw}_1$ \citep{tibshirani1996regression} with the regularization parameter $\lambda$ tuned, say, through cross-validation. The natural way then of accounting for feature side-information in~\eqref{eq:general_penalized_reg} is to choose a regularizer $\text{Pen}(\bw)$ that is \emph{not} symmetric. However, if one seeks to turn this conceptual extension into a practical method, one is faced with a key difficulty: an explosion in the number of regularization parameters that need to be tuned.

In this work we seek to shed insight into supervised learning with side-information, by theoretically and empirically studying the simplest practically relevant form of~\eqref{eq:general_penalized_reg} with feature co-data. We consider a situation in which the domain scientist can partition the features $\cb{1,\dotsc, p}$ into $K$ disjoint groups $\mathcal{G}_g \subset \cb{1,\dotsc, p}, \;g=1,\dotsc,K$ and seeks to run Ridge regression with one regularization parameter $\lambda_g$ per group. Let $\blambda = (\lambda_1, \dotsc, \lambda_K)$, $\bw = (w_1,\dotsc,w_p)$ and $\bw_{\mathcal{G}_g} = (w_j)_{j \in \mathcal{G}_g}$, then the domain scientist fits the regression,
\begin{equation}
\label{eq:groupridge}
\widehat{\bw} = \widehat{\bw}(\blambda) \in \argmin_{\bw}\cb{ \frac{1}{2n}\sum_{i=1}^n \p{\yresponse_i-x_i^\intercal \bw}^2 + \sum_{g=1}^K \frac{\lambda_g}{2} \norm{\bw_{\mathcal{G}_g}}_2^2}.
\end{equation}
The issue, as already alluded, is the following: How should one choose the $K$-dimensional regularization vector $\blambda =(\lambda_1,\dotsc,\lambda_K)$ needed to solve~\eqref{eq:groupridge} and what is the impact of this choice? Our starting point for answering this question is the following generative model. Let $p_g = \abs{\mathcal{G}_g}$ be the number of features in group $g$ and $\sigma >0$, $\alpha_1^2,\dotsc,\alpha_K^2 > 0$. Then generate (independently)
\begin{equation}
\label{eq:normal_normal}
\begin{aligned}
&w_j \sim \nn(0,\; \alpha_{g}^2\big / p_g)\; &j \in \mathcal{G}_g,\;\; g=1,\dotsc,K\\
&x_i \sim \mathbb P^X,\;\varepsilon_i \sim \nn(0,\;\sigma^2),\; \yresponse_i = x_i^\intercal \bw + \varepsilon_i \;\;\; &i=1,\dotsc,n
\end{aligned}
\end{equation}
Under model~\eqref{eq:normal_normal}, we can precisely characterize the limiting risk of predictions $x \mapsto x^\intercal\widehat{\bw}(\blambda)$ for any value of $\blambda$ by utilizing recent advances in random matrix theory (RMT), cf.~\citet{dobriban2018high} and thus we can study the impact of different choices of $\blambda$. Furthermore, under~\eqref{eq:normal_normal}, we can plausibly choose $\blambda \in [0, \infty)^K$ with a fully \textbf{model-based approach} as we now explain. The solution to~\eqref{eq:groupridge} with the choice of parameters
\begin{equation}
\label{eq:model_based_lambda}
\lambda_g = \frac{p_g}{n} \cdot\frac{\sigma^2}{\alpha_g^2},
\end{equation}
is the posterior mean of $\bw$ under model~\eqref{eq:normal_normal} and so one could fit~\eqref{eq:normal_normal} to estimate $\balpha = (\alpha_1,\dotsc,\alpha_k)$ and $\sigma^2$, and then solve~\eqref{eq:groupridge} with plug-in estimates of~\eqref{eq:model_based_lambda}. This approach, however, comes with caveats. First, estimation typically proceeds by optimization of a non-convex objective, such as restricted maximum likelihood. Second, a data scientist interested in predictive performance may be apprehensive of choosing parameters based on purely model-based criteria. Instead, they may prefer to directly optimize distribution-free measures of \textbf{predictive performance}. For example, they  may choose $\blambda \in [0, \infty)^K$ by minimizing the cross-validated mean squared error. For small $K$, this may be achieved by exhaustive grid search, otherwise, one would  resort to heuristics for the optimization of non-convex objectives.

The methodological contribution of this paper is the development of \textbf{$\sigmacv$-Ridge regression}, a hybrid of the two aforementioned approaches --model-based tuning and cross-validation -- with several favorable properties, which we outline next.

\begin{enumerate}[align=left, wide]
    \item \textbf{Single regularization parameter:} $\sigmacv$-Ridge regression depends on a single, interpretable regularization parameter, which can be chosen by cross-validation. The data scientist can inspect parameter and coefficient paths as a function of the regularization parameter.
    \item \textbf{Computationally tractable:} The method  has the same computational complexity as cross-validated Ridge regression with a single $\lambda$. All underlying computations may be solved to machine precision without any danger of local minima.
    \item \textbf{Asymptotic Optimality in high dimensions:} The method provably matches the predictive performance of the best estimator in the class~\eqref{eq:groupridge} in a high-dimensional nonparametric   random effects model that generalizes~\eqref{eq:normal_normal}. Under the same model, $\sigmacv$-Ridge regression also provably outperforms the Group Lasso~\citep{yuan2006model} with optimal tuning\footnote{$K$ is fixed in our asymptotics and so the Group Lasso may be preferable in settings with many sparse groups.}.
    \item \textbf{Practical:} The method works well in practical situations and datasets, wherein model~\eqref{eq:normal_normal} may not hold.
\end{enumerate}

Throughout this manuscript, we emphasize both the theoretical contributions that are required to study $\sigmacv$-Ridge regression and the practical value of our approach in applications.

\subsection{How does $\sigmacv$-Ridge regression work?}
\label{subsec:sigmaridgeintro}
Our core proposal is a hybrid of cross-validation and model-based hyperparameter tuning. Assume momentarily \textbf{a)} that model~\eqref{eq:normal_normal} holds and \textbf{b)} that $\sigma$ is known to the analyst. Then let $\widehat{\balpha}(\sigma)$ be a model-based estimate (more of which in Section~\ref{sec:method}) of $\balpha$ in~\eqref{eq:normal_normal} with $\sigma$ known. $\widehat{\balpha}(\sigma)$ induces a model-based estimate of $\blambda,$ i.e.,  $\widehat{\lambda}_g(\sigma) = p_g/n \cdot \sigma^2/\widehat{\alpha}_g(\sigma)^2$.

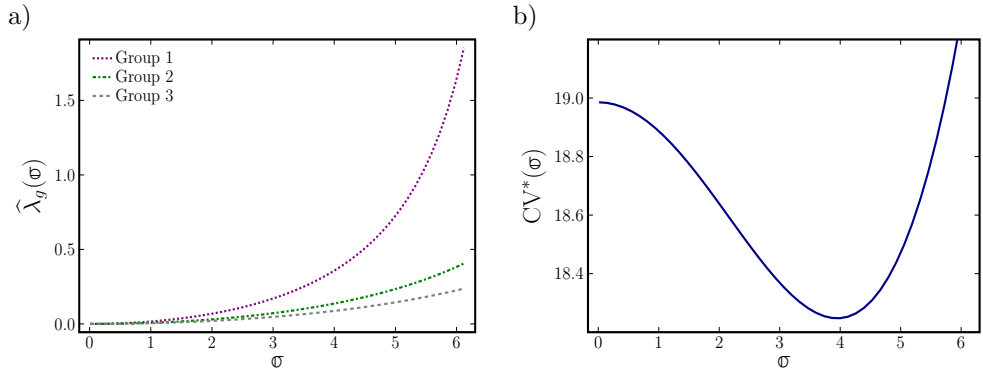
\begin{figure}
\centering
\begin{tabular}{ll}
a) & b)\\
\begin{adjustbox}{width=0.45\linewidth, totalheight=1.8in}\input{figures/intro_figure_left.tikz}\end{adjustbox} &
\begin{adjustbox}{width=0.45\linewidth, totalheight=1.8in}\input{figures/intro_figure_right.tikz}\end{adjustbox} 
\end{tabular}
\caption{\textbf{$\sigmacv$-Ridge regression:} We apply $\sigmacv$-Ridge regression to a single simulation from model~\eqref{eq:normal_normal} with $x_i \sim \mathcal{N}(0,I)$, \smash{$n=400, K=3, p_g=25$}, \smash{$\alpha_g^2 = 4\cdot g$} and $\sigma^2=16$. \textbf{Panel a)} shows the map \smash{$\sigmacv \mapsto \mhatlambda$} from $\sigmacv$ to the per-group regularization parameter. The first group, with the lowest signal, receives the largest penalty. As $\sigmacv$ increases, we regularize more aggressively. \textbf{Panel b)} shows the leave-one-out cross-validation error $\loo(\sigmacv)$ as a function of $\sigmacv$. The minimizer of the curve can be used as a data-driven choice for $\sigmacv$.}
\label{fig:sigma_ridge_intro}
\end{figure}

In practice of course we do not know $\sigma$, nor do we necessarily believe that model~\eqref{eq:normal_normal} holds, so that $\sigma$ may not even be well-defined. Instead we treat $\sigma$ as a one-dimensional tuning parameter that may be chosen by cross-validation. The model-based procedure outlined above is then interpreted merely as a data-driven map from a one-dimensional regularization parameter $\sigma$ to a $K$-dimensional regularization vector $\blambda$. To avoid notational ambiguity, we use the typeface $\sigmacv$ henceforth for our tuning parameter, i.e., 
\begin{equation}
\label{eq:sigma_to_tuning}
\sigmacv \mapsto \mhatlambda, 
\end{equation}
and reserve the letter $\sigma$ only for our theoretical development as the residual noise standard deviation when model~\eqref{eq:normal_normal} is true. We then seek to choose \smash{$\widehat{\sigmacv}$} in a model-agnostic way by cross-validation, so that \smash{$\widehat{\bw}\p{\widehat\blambda\p{\widehat{\sigmacv}}}$} provides close to best out-of-sample predictive performance among estimators of the family $\{\widehat{\bw}(\mhatlambda),\; \sigmacv \in (0,\infty)\}$. We illustrate the idea in Figure~\ref{fig:sigma_ridge_intro}.

\subsection{Related work}
\label{subsec:related_work}

Our theoretical contribution continues a rich line of work \citep{tulino2004random, dicker2016ridge, dobriban2018high, hastie2019surprises, liu2019ridge, xu2019number, dobriban2020wonder, lolas2020regularization} that uses recent advances from Random Matrix Theory to precisely characterize the performance of regression  methods under high-dimensional asymptotics with dense, weak effects. Such an asymptotic perspective is relevant in application domains, e.g., genetics~\citep{boyle2017expanded}, wherein most features are predictive of the response of interest, but the signal of each feature individually is weak. The model of dense and weak effects is to be contrasted with the traditional approach to studying high-dimensional regression through sparsity~\citep{buhlmann2011statistics,donoho2009message}.

From a methodological perspective,  our work     is inspired by~\citet{van2016better}, who introduce the Ridge regression problem with groups and provide an empirical Bayes procedure to learn the optimal penalties for logistic regression and Cox regression. However, in the case of linear regression, the approach of~\citet{van2016better} assumes that the noise level $\sigma^2$ is known or can be well-estimated from the ridge regression residuals. Instead, we provide an end-to-end estimation strategy for linear ridge regression, that furthermore is provably optimal in the high-dimensional regime.

Along the lines of~\citet{van2016better}, there has been a stream of recent empirical work developing practical and reliable methods for supervised learning with feature co-data~\citep{tai2007incorporating, foo2008efficient, bergersen2011weighted, boulesteix2017ipf, velten2018adaptive, munch2018adaptive, perrakis2019scalable,  nabi2020decoupling,  pramanik2020structure, tay2020feature, wiel2020fast}. The goal of all these works is complementary and related to our paper: ``co-data'' is ubiquitous in modern scientific and technological applications; and so it is important to enhance the data analytic toolbox with methods that leverage side-information to improve predictive power. However, all of these previous works do not come with theoretical guarantees\footnote{One exception is the work of~\citet{pramanik2020structure}, who derive limiting risk expression for an approximate message passing algorithm that uses side-information under the strong assumption of Gaussian covariates with identity covariance.}.

The problem of choosing multiple tuning parameters is, of course, not new. For example, when fitting generalized additive models with flexible spline expansions, one may have multiple tuning parameters to control e.g., anisotropic smoothness. The R package \texttt{mgcv}~\citep{wood2000modelling, wood2004stable, wood2017generalized} provides computational routines for efficiently tuning and solving large-scale generalized additive models with many hyperparameters. \texttt{mgcv} is general enough, that it subsumes problem~\eqref{eq:groupridge} and can choose tuning parameters by optimizing the GCV (generalized cross-validation) criterion, or by estimating the parameters in model \eqref{eq:normal_normal} by restricted maximum likelihood and then using the plug-in rule on~\eqref{eq:model_based_lambda}. However, \texttt{mgcv} only works for $p \leq n$ and we are not aware of theoretical guarantees in high-dimensions\footnote{It is however plausible, that the proof techniques in the present paper, along with results of~\citet{jiang2016high} could be used to prove asymptotic optimality of \texttt{mgcv}. More generally, many methods have been developed to estimate the parameters in model~\eqref{eq:normal_normal} in the case of one group ($K=1$), with the motivation of estimating heritability in genetic studies~\citep{dicker2014variance, dicker2016maximum, janson2017eigenprism, veerman2019estimation}. Extensions of heritability methods to $K \geq 2$ would provide alternative model-based approaches towards tuning group-regularized ridge regression.}.

Finally, we note that breaking symmetry in~\eqref{eq:general_penalized_reg} does not necessarily require introducing additional regularization parameters. The main example of an asymmetric $\text{Pen}(\bw)$ with a single regularization parameter is the Group Lasso penalty~\citep{yuan2006model}, $\text{Pen}_{\text{glasso}}(\bw) = \lambda^{\text{glasso}} \cdot \sum_{g=1}^K \sqrt{{p_g}/{p}} \Norm{ \bw_{\mathcal{G}_g}}_2.\;\;$
The group Lasso automatically selects a sparse subset of group features, i.e., most $\widehat{\bw}_{\mathcal{G}_g}$ are set to zero. Section~\ref{subsec:grouplasso} provides more details on the connection of $\sigmacv$-Ridge regression to the Group Lasso.

\subsection{Outline}
In Section~\ref{sec:method} we elaborate on the high-level description from Section~\ref{subsec:sigmaridgeintro} and describe $\sigmacv$-Ridge regression in detail. In Section~\ref{sec:risk_expressions} we introduce the asymptotic framework and provide theoretical results for group-regularized ridge regression; in particular we provide a sharp expression of the limiting predictive risk when the feature covariance matrix is block-diagonal. In Section~\ref{sec:optimality} we build upon the results from Section~\ref{sec:risk_expressions} and prove that $\sigmacv$-Ridge regression asymptotically achieves optimal prediction among all procedures of the form~\eqref{eq:groupridge}. This result holds for arbitrary feature covariance. Section~\ref{sec:numerics} demonstrates promising performance of $\sigmacv$-Ridge regression in simulations and real datasets. In Section~\ref{sec:discussion} we conclude with a discussion.

\section{The proposed method: $\sigmacv$-Ridge regression}
\label{sec:method}

\subsection{Model based tuning with known noise variance $\sigma^2$}

To motivate our proposal, let us assume that model~\eqref{eq:normal_normal} holds with known $\sigma^2$. We write $\bX$ for the $n \times p$ design matrix with rows $x_i^\intercal$ and $\bY = (Y_1, \dotsc, Y_n)$, $\bvarepsilon = (\varepsilon_1,\dotsc,\varepsilon_n)$. We then run ridge regression with deterministic tuning parameter $\lambdainit > 0$\footnote{A data-driven choice for $\lambdainit$ will be provided later.} (i.e., we solve~\eqref{eq:groupridge} with $\blambdainit = (\lambdainit, \dotsc, \lambdainit)$) to get
\begin{equation}
\label{eq:moments_decomp}
\widetilde{\bw}=\underbrace{\p{\frac{\bX^\intercal  \bX}{n}+\lambdainit \bI}^{-1}\frac{\bX^\intercal \bX}{n}}_{=:\bM} \bw \;+\;\underbrace{\frac{1}{\sqrt{n}}\p{\frac{\bX^\intercal  \bX}{n}+\lambdainit \bI}^{-1}\bX^\intercal}_{=:\bN}\frac{\bvarepsilon}{\sqrt{n}},
\end{equation}
where $\bI$ is the $p \times p$ identity matrix. Then under~\eqref{eq:normal_normal}
\begin{equation}\label{eq:system}
\EE{\Norm{\widetilde{\bw}_{\groupg{g}}}_2^2 \mid \bX} =  \sum_{h=1}^K \norm{\bM_{\group{g},\group{h}}}_F^2 \frac{\alpha_h^2}{p_h} + \Norm{\bN_{\group{g},\cdot}}_F^2\frac{\sigma^2}{n},\; g=1,\dotsc,K.
\end{equation}
Here $\norm{\cdot}_F$ is the Frobenius norm of a matrix, $\bN_{\group{g},\cdot}$ is the matrix of the rows of $\bN$ corresponding to the $g$-th group and $\bM_{\group{g},\group{h}}$ is the $p_g \times p_h$ matrix that arises if we keep only the rows that correspond to the $g$-th group from $\bM$ and the columns that correspond to the $h$-th group. 

Under known $\sigma$, the above system of equations directly identifies $\alpha_1,\dotsc,\alpha_K$, and estimation can proceed through the method of moments. Recalling that the optimal model-based regularization parameters take the form $\lambda_g=\lambda_g(\sigma) = p_g/n \cdot \sigma^2/\alpha_g^2$ and writing $\bA$ for the $K\times K$ matrix with entries \smash{$A_{gh}=\norm{\bM_{\group{g},\group{h}}}_F^2/n$} and $\bu, \bv$ for the vectors in $\RR^K$ with entries \smash{$u_g = \EEInline{\norm{\widetilde{\bw}_{\groupg{g}}}_2^2 \mid \bX}$}, \smash{$v_g = \norm{\bN_{\group{g},\cdot}}_F^2/n$}, we may rewrite the above system of equations as
\begin{equation}
    \label{eq:moment_estimator}
    \bd(\sigma) = \bA^{-1}\p{\frac{\bu}{\sigma^2} -\bv},\;\; d_g(\sigma_i) = 1/\lambda_g(\sigma),\; g=1,\dotsc,K.
\end{equation}
The data-driven method of moments estimator plugs in $\widehat{\bu}$ with $\widehat{u}_g = \norm{\widetilde{\bw}_{\groupg{g}}}_2^2$ in place of $\bu$.

\subsection{$\sigmacv^2$ -- reinterpreting $\sigma^2$ as a regularization parameter}\label{sub:reinterpreting_sigmacv}
As we already motivated in the introduction, we treat the variance $\sigma^2$ as a one-dimensional tuning parameter and then use a plug-in rule on~\eqref{eq:moment_estimator}. Making this explicit by using the letter $\sigmacv$ instead of $\sigma$, as in~\eqref{eq:sigma_to_tuning}, we let
\begin{equation}
    \label{eq:mapping}
    \widehat{\lambda}_g(\sigmacv) = 1/\widehat{d}_g(\sigmacv),\;\text{ where }  \widehat{\bd}(\sigmacv) = \argmin_{\bd \in [0,\infty)^K}\cb{ \Norm{\bA\bd - \frac{\widehat{\bu}}{\sigmacv^2} + \bv}_2^2}.
\end{equation}
The above nonnegative least squares problem can be solved in $O(K^3)$ operations. In our setting and applications, $K$ is small compared to $n,p$, so that this cost is negligible.

To provide intuition we first consider the ``trivial'' case of
\textbf{one group} ($K=1$). Then, \smash{$\widehat{\lambda}(\sigmacv)$} in~\eqref{eq:mapping} takes the following form:
\begin{equation}
\widehat{\lambda}(\sigmacv) =  \Norm{\bM}^2_F \bigg/ \p{ \frac{n\Norm{\widetilde{\bw}}^2_2}{\sigmacv^2} - \Norm{\bN}_F^2}_+.
\end{equation}
\begin{figure}[!ht]
\centering
\begin{tabular}{ll}
a) & b)\\
\begin{adjustbox}{width=0.45\linewidth, totalheight=1.8in}\input{figures/christmas_tree_left.tikz}\end{adjustbox} &
\begin{adjustbox}{width=0.45\linewidth, totalheight=1.8in}\input{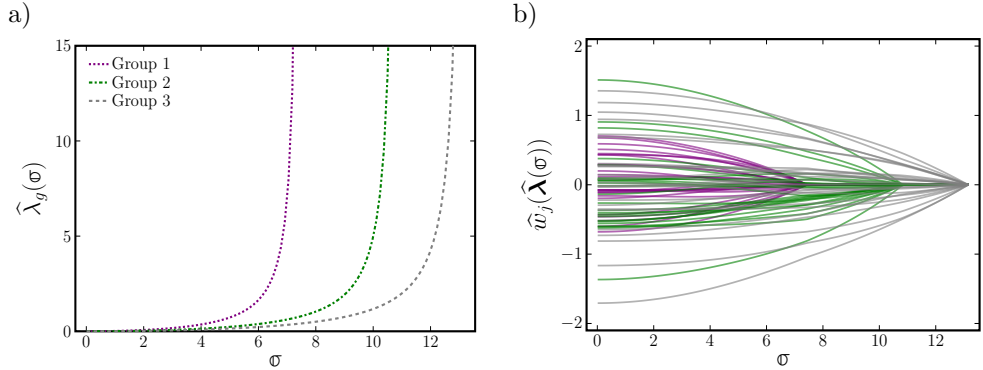}\end{adjustbox}
\end{tabular}
\caption{\textbf{Regularization and coefficient paths in $\sigmacv$-Ridge regression:} The data shown here is from the same simulation as in~Figure~\ref{fig:sigma_ridge_intro}. \textbf{Panel a)} corresponds to panel b) of Fig.~\ref{fig:sigma_ridge_intro}, with the $x$-axis extended to larger values of $\sigmacv$, and shows the map \smash{$\sigmacv \mapsto \mhatlambda$} from $\sigmacv$ to the per-group regularization parameter. For large values of $\sigmacv$ the group-wise regularization parameter blows up to $+\infty$. \textbf{Panel b)} shows the evolution of all coefficients in~\eqref{eq:groupridge}, color-coded by group, as $\sigmacv$ increases. The coefficients shrink more as $\sigmacv$ increases, and for large enough $\sigmacv$, complete groups of features may be set to $0$ (group-sparsity).}
\label{fig:sigma_as_tuning}
\end{figure}
In words, $\widehat{\lambda}(\sigmacv)$ is a data-driven non-decreasing mapping of $\sigmacv \in [0,\infty]$ to $\lambda \in [0,\infty]$, i.e., a reparametrization.  In the case of \textbf{multiple groups} ($K>1)$ we instead interpret $\sigmacv$ as yielding a regularization parameter path \smash{$\sigmacv \mapsto \mhatlambda$} as in~\eqref{eq:sigma_to_tuning}. Figure~\ref{fig:sigma_as_tuning} illustrates this idea and also shows that the path can induce group-level sparsity similar to the Group Lasso, by setting $\widehat{\lambda}_g = \infty$ for some groups. The following proposition lists some properties of the regularization path.

\begin{proposition}[Properties of $\sigmacv$-regularization parameter path]
\label{prop:propo1}
\begin{enumerate}
    Assume the matrix $\bA$ is invertible and $\widehat{u}_g > 0$ for all $g$. Then
    \item For $\sigmacv^2 \geq \max_g \cb{\widehat{u}_g/ v_g} =:\sigmacv_{\text{max}}$,  we have $\widehat{\lambda}_g(\sigmacv) = \infty$ for all $g$, i.e. $\widehat{\bw}\p{\widehat{\blambda}(\sigmacv)} = 0$.
    \item As $\sigmacv \to 0$, we have that $\min_g \widehat{\lambda}_g(\sigmacv) \to 0$, i.e., at least one group is not penalized.
    \item Suppose $\sigmacv_1, \sigmacv_2 >0$ lead to the same active groups, i.e.,  $\mathcal{S}=\mathcal{S}(\sigmacv_1)= \mathcal{S}(\sigmacv_2)$, where $\mathcal{S}(\sigmacv)=\cb{g \in \cb{1,\dotsc,K}: \widehat{\lambda}_g(\sigmacv) < \infty}$. Then, for $g \in \mathcal{S}$ it holds that:
  $$ \widehat{\lambda}_g(\sigmacv_1) = \cb{ \frac{\sigmacv_2^2}{\sigmacv_1^2} \widehat{\lambda}_g(\sigmacv_2)^{-1} + \p{\frac{\sigmacv_2^2}{\sigmacv_1^2} -1}\tilde{v}_{\mathcal{S},g}}^{-1},\;\,\text{ where } \tilde{v}_{\mathcal{S},g} = \p{ \p{\bA_{\cdot, \mathcal{S}}^\intercal \bA_{\cdot, \mathcal{S}}}^{-1}\bA_{\cdot, \mathcal{S}}^\intercal \cdot \bv}_g.$$
\end{enumerate}
\end{proposition}
The above properties are deterministic and do not depend on the validity of~\eqref{eq:normal_normal}.

\subsection{Choosing $\sigmacv$ through accelerated leave-one-out cross-validation}

In light of the interpretation above, for any value of $\sigmacv$ we have a supervised algorithm that proceeds in three steps:
\begin{enumerate}
    \item[1.] Compute $\widehat{\blambda}(\sigmacv)$ as in \eqref{eq:mapping}.
    \item[2.] Let $\widehat{\bw} = \widehat{\bw}\p{\widehat{\blambda}(\sigmacv)}$ the Ridge regression coefficient from~\eqref{eq:groupridge}.
    \item[3.] Predict the response for $x \in \RR^p$ as $x^\intercal \widehat{\bw}$.
\end{enumerate}

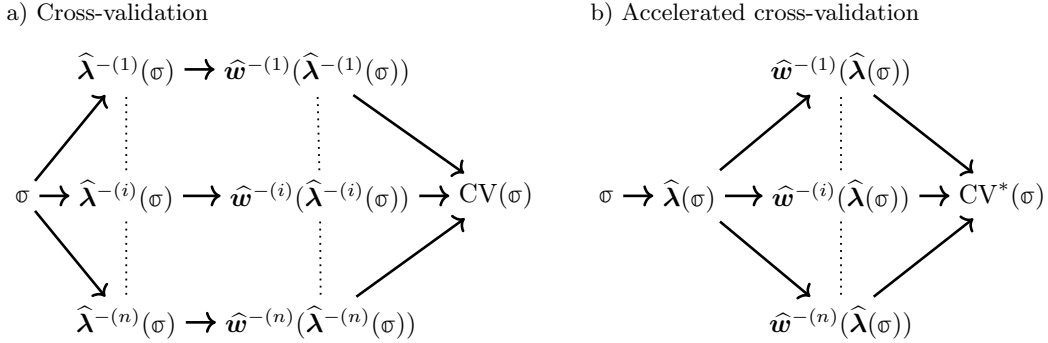
\begin{figure}
\begin{subfigure}{0.55\textwidth}
\caption{Cross-validation}
\begin{tikzpicture}
\node[text centered] (s) {$\sigmacv$};
\node[right = 0.4 of s, text centered] (lambdai) {$\widehat{\blambda}^{-(i)}( \sigmacv)$};
\node[right = 0.5 of lambdai] (betai) {$\widehat{\bw}^{-(i)}(\widehat{\blambda}^{-(i)}( \sigmacv ))$};
\node[above = 1 of lambdai, text centered] (lambda1) {$\widehat{\blambda}^{-(1)}( \sigmacv)$};
\node[right = 0.4 of lambda1, text centered] (beta1) {$\widehat{\bw}^{-(1)}(\widehat{\blambda}^{-(1)}( \sigmacv ))$};
\node[below = 1 of lambdai, text centered] (lambdan) {$\widehat{\blambda}^{-(n)}( \sigmacv)$};
\node[right = 0.4 of lambdan, text centered] (betan) {$\widehat{\bw}^{-(n)}(\widehat{\blambda}^{-(n)}( \sigmacv ))$};
\node[right = 0.4 of betai] (loobasic) {$\loobasic(\sigmacv)$};
\node[below = 0.1 of lambdan] (placeholder) {};
\node[right = 0.7 of loobasic] (placeholder2) {};
\draw[->, line width= 1] (s) --  (lambdai);
\draw[->, line width= 1] (lambdai) --  (betai);
\draw[->, line width= 1] (lambda1) --  (beta1);
\draw[->, line width= 1] (lambdan) --  (betan);
\draw[->, line width= 1] (s) --  (lambda1);
\draw[->, line width= 1] (s) --  (lambdan);
\draw[dotted, line width=0.7] (lambdai) --  (lambda1);
\draw[dotted, line width=0.7] (lambdai) --  (lambdan);
\draw[dotted, line width=0.7] (betai) --  (beta1);
\draw[dotted, line width=0.7] (betai) --  (betan);
\draw[->, line width= 1] (betai) --  (loobasic);
\draw[->, line width= 1] (beta1) --  (loobasic);
\draw[->, line width= 1] (betan) --  (loobasic);
\end{tikzpicture}
\end{subfigure}
\begin{subfigure}{0.45\textwidth}
\caption{Accelerated cross-validation}
\begin{tikzpicture}
\node[text centered] (s) {$\sigmacv$};
\node[right = 0.4 of s, text centered] (lambda) {$\widehat{\blambda}( \sigmacv)$};
\node[right = 0.5 of lambda] (betai) {$\widehat{\bw}^{-(i)}(\widehat{\blambda}( \sigmacv ))$};
\node[above = 1 of betai, text centered] (beta1) {$\widehat{\bw}^{-(1)}(\widehat{\blambda}( \sigmacv ))$};
\node[below = 1 of betai, text centered] (betan) {$\widehat{\bw}^{-(n)}(\widehat{\blambda}( \sigmacv ))$};
\node[right = 0.4 of betai] (loo) {$\loo(\sigmacv)$};
\node[below = 0.1 of betan] (placeholder) {};
\draw[->, line width= 1] (s) --  (lambda);
\draw[->, line width= 1] (lambda) --  (betai);
\draw[->, line width= 1] (lambda) --  (beta1);
\draw[->, line width= 1] (lambda) --  (betan);
\draw[dotted, line width=0.7] (betai) --  (beta1);
\draw[dotted, line width=0.7] (betai) --  (betan);
\draw[->, line width= 1] (betai) --  (loo);
\draw[->, line width= 1] (beta1) --  (loo);
\draw[->, line width= 1] (betan) --  (loo);
\end{tikzpicture}
\end{subfigure}
\caption{\textbf{Leave-one-out cross-validation (LOOCV)  and accelerated LOOCV for $\sigmacv$-Ridge regression.} The accelerated method computes the map $\sigmacv \mapsto \widehat{\blambda}(\sigmacv)$ based on the full dataset and treats it as fixed when computing the leave-one-out error.}
\label{fig:cv_diagram}
\end{figure}

As such, we can now use any method of tuning hyperparameters to choose $\sigmacv$. Here we consider leave-one-out cross-validation (LOOCV). In a direct application of LOOCV, we would calculate $\widehat{\blambda}^{(i)}(\sigmacv)$ and $\widehat{\bw}^{-(i)}(\widehat{\blambda}^{-(i)})$ for each $i$, where the ``$-(i)$'' notation means that the supervised algorithm is trained based on all observations except the $i$-th. The LOOCV error is then defined as,
\begin{equation}
    \label{eq:loo_basic}
\loobasic(\sigmacv) := \frac{1}{n}\sum_{i=1}^n\p{y_i - \widehat{\bw}^{-(i)}(\widehat{\blambda}^{-(i)}( \sigmacv ))^\intercal x_i} ^2.
\end{equation}

See Fig.~\ref{fig:cv_diagram}a for a schematic of the above procedure. Finally we could choose $\sigmacv$ as the minimizer of $\loobasic(\sigmacv)$. A downside however of this approach is that it will be computationally expensive to refit the whole model $n$ times. Instead we propose to omit the first-step of the leave-one-out procedure and keep the map $\sigmacv \mapsto \widehat{\blambda}(\sigmacv)$ fixed throughout, even though it also depends on the full training set. As our ``leave-one-out'' prediction for observation $i$ we use $ \widehat{\bw}^{-(i)}(\widehat{\blambda}( \sigmacv ))^\intercal  x_i$. We define the accelerated leave-one-out objective (also see Fig.~\ref{fig:cv_diagram}b) as
\begin{equation}
    \label{eq:loo}
\loo(\sigmacv) := \frac{1}{n}\sum_{i=1}^n\p{Y_i - \widehat{\bw}^{-(i)}(\widehat{\blambda}( \sigmacv ))^\intercal x_i}^2.
\end{equation}
The upshot is that now we may directly apply the well-known shortcut formula, cf.~\citet{meijer2013efficient} and references therein. That is, letting \smash{$\widehat{\bLambda}(\sigmacv)$} the $p\times p$ diagonal matrix with $j$-th entry equal to $\widehat{\lambda}_g(\sigmacv)$, when $j \in \groupg{g}$, and
\begin{equation}
\label{eq:bH_def}
\bH(\sigmacv) = \bX(\bX^\intercal \bX + \widehat{\bLambda}(\sigmacv))^{-1}\bX^\intercal,\;\; \widehat{\bY}(\sigmacv) = \bH(\sigmacv) \bY,
\end{equation}
then it holds that
\begin{equation}
    \label{eq:loo_shortcut}
\loo(\sigmacv) = \frac{1}{n}\sum_{i=1}^n\p{ \frac{Y_i - \hat{Y}_i(\sigmacv)}{1-H_{ii}(\sigmacv)}}^2.
\end{equation}
Henceforth we propose to choose $\sigmacv$ as $\widehat{\sigmacv} \in \argmin_{ \sigmacv } \loo(\sigmacv)$
\footnote{The idea of using hybrid empirical Bayes/cross-validation approaches for regularization parameter tuning is not new: in the context of group-regularized ridge logistic and Cox regression,~\citet{van2016better} use empirical Bayes to learn a regularization parameter vector \smash{$\widehat{\blambda}^{\text{GR}}$} with entries \smash{$\widehat{\lambda}_g^{\text{GR}}$}. Then, a further tuning parameter $\lambda >0$ is introduced, and cross-validation is used to pick a regularization vector from the family \smash{$(\lambda \cdot \widehat{\blambda}^{\text{GR}})_{\lambda > 0}$}. For $\sigmacv$-Ridge,  $\sigmacv$ is the regularization parameter to be chosen by cross-validation and we theoretically show that it achieves optimal predictive performance~(Theorem~\ref{theo:optimality}).}.
Our theoretical analysis in Section~\ref{sec:optimality} pertains to the choice of $\sigmacv$ through accelerated leave-one-out cross-validation and  demonstrates that it leads to asymptotically optimal predictions.

There is one missing step required to implement the full procedure; the choice of $\lambdainit$ in~\eqref{eq:moments_decomp}. We let $\lambdainit$ be the optimal one-dimensional ridge regression parameter, i.e., the minimizer of $\loo((\lambda,\dotsc,\lambda))$, where in analogy to~\eqref{eq:loo}, we define (with some abuse of notation):
\begin{equation}
    \label{eq:loo_blambda}
\loo(\blambda) := \frac{1}{n}\sum_{i=1}^n\p{Y_i - \widehat{\bw}^{-(i)}(\blambda)^\intercal x_i}^2.
\end{equation}
Finally, we note that to solve~\eqref{eq:groupridge}, we need to factorize $\bX^\intercal \bX/n + \bLambda$, where $\bLambda$ is the diagonal matrix with $j$-th entry $\lambda_g$, when $j \in \group{g}$. We use the Cholesky decomposition when $p \leq 4n$ and the Woodbury matrix identity \citep[Section 1.7]{van2015lecture} otherwise.
\section{Asymptotics of group ridge regression} 
\label{sec:risk_expressions}
Before turning to study $\sigmacv$-Ridge regression, we first study the performance of group-regularized ridge regression~\eqref{eq:groupridge} for a  general choice of $\blambda = (\lambda_1,\dotsc,\lambda_K)$. The core setting for our asymptotic results is that of ridge regression with random design~\citep{hsu2012random} and random effects, that generalizes the Gaussian-Gaussian model~\eqref{eq:normal_normal}:
\begin{equation}
\label{eq:random_effects}
\begin{aligned}
&w_j \sim (0,\; \alpha_{g}^2\big / p_g)\; &j \in \mathcal{G}_g,\;\; g=1,\dotsc,K\\
&x_i \sim \mathbb P^X,\;\varepsilon_i \sim (0,\;\sigma^2),\; \yresponse_i = x_i^\intercal \bw + \varepsilon_i \;\;\; &i=1,\dotsc,n
\end{aligned}
\end{equation}
The notation $Z \sim (\mu, \tau^2)$ denotes a random variable with $\EE{Z}=\mu$ and $\Var{Z}=\tau^2$. For our high-dimensional (HD) asymptotics we make the following assumptions on model~\eqref{eq:random_effects}:

\begin{enumerate}
\item [(HD1)] The number of groups $K$ is fixed and $p_g/n \to \gamma_g>0$ as $n\to \infty$ for all groups $g$. We also write \smash{$\gamma=\sum_{g=1}^K \gamma_g$} for the asymptotic aspect ratio $p/n$.
\item [(HD2)] Let $\bSigma$ be the covariance matrix of $x_1$.  There exist $h_1,h_2>0$ fixed such that all the eigenvalues of $\bSigma$ lie in $[h_1,h_2].$
\item [(HD3)] $x_i \sim \mathbb P^X$ may be written as $\bSigma^{1/2}z_i$ where $z_i \sim \mathbb P^Z$ has i.i.d. entries with mean zero, variance one and uniformly bounded ($8+\eta$)-th moments for some $\eta>0$.
\item [(HD4)] The $(4+\eta)$-th moments of $\sqrt{p}w_j, j=1,\dotsc,p$ and $\varepsilon_i, i=1,\dotsc,n$ are uniformly bounded.
\end{enumerate}
According to (HD1), all feature groups grow at the same rate as $n, p \to \infty$.
The remaining assumptions (HD2-HD4) are common for the high-dimensional analysis of ridge regression (without grouping information), see for example \cite{dobriban2018high}, \cite{hastie2019surprises} and \cite{ledoit_peche}, and are typically considered to be mild. In the latter works, (HD1) is replaced by the assumption that $p/n \to \gamma >0$ as $n \to \infty$.

The key object of our asymptotic study is the out-of-sample prediction risk of an estimator $\widehat{\bw}$ of $\bw$ conditionally on the training set $(\bX,\bY)$ and true coefficient vector $\bw$,
\begin{equation}
    \label{eq:risk_def}
    \risk{\widehat{\bw}} = \EE{ \p{\yresponse_{\text{test}} - x_{\text{test}}^\intercal \widehat{\bw}}^2 \mid \bX,\bY,\bw}. 
\end{equation}
where $(x_{\text{test}},\yresponse_{\text{test}})$ is a fresh draw from~\eqref{eq:random_effects} (with $\bw$ fixed). With some abuse of notation, we also write $\risk{\blambda}$ for the risk of $\widehat{\bw}(\blambda)$ from~\eqref{eq:groupridge}, i.e. $\risk{\blambda} = \risk{\widehat{\bw}(\blambda)}$.

Our first asymptotic result is that the out-of-sample prediction risk $\risk{\blambda}$ concentrates around its marginalization with respect to $\bvarepsilon,\bw$,
\begin{equation}
    \label{eq:Ln_def}
    \Ln(\widehat{\bw}) = \EE{\risk{\widehat{\bw}} \mid \bX}, \;\; \Ln(\blambda) = \Ln(\widehat{\bw}(\blambda)).
\end{equation}
\begin{lemma}\label{lem:risk_close_to_L}
Consider model~\eqref{eq:random_effects} under assumptions $\text{(HD1-4)}$. It almost surely holds that $\abs{\risk{\blambda}-\Ln(\blambda)}\xrightarrow{}0$ as $n\to \infty$
uniformly over $\blambda$ in compact subsets of $(0,\infty)^K$. $\Ln(\blambda)$ is equal to 
\begin{equation*}
\begin{aligned}
\Ln(\blambda)=\sigma^2\;&+\;\frac{\sigma ^2}{n}\Tr\left(\frac{\bX^\intercal  \bX}{n}\left(\frac{\bX^\intercal \bX}{n}+\bLambda\right)^{-1}\bSigma \left(\frac{\bX^\intercal \bX}{n}+\bLambda\right)^{-1}\right)\\ &+\;\frac{1}{p}\Tr\left(\left(\frac{\bX^\intercal \bX}{n}+\bLambda\right)^{-1}
\bSigma \left(\frac{\bX^\intercal \bX}{n}+\bLambda\right)^{-1}
\bLambda \bar{\bD}\bLambda\right).
\end{aligned}
\end{equation*}
Here, $\bar{\bD}$ is the diagonal matrix whose $j$-th diagonal entry is $\alpha_g^2 p / p_g,$ when $j \in \group{g}$.
\end{lemma}
The proof relies on an application of the Marcinkiewicz-Zygmund interpolation \citep[Chapter 7]{erdos2017dynamical} leveraging the boundedness of the $(4+\eta)$-th moments of $w_j$ and $\varepsilon_i$. The upshot of Lemma~\ref{lem:risk_close_to_L} is that the only source of randomness in $\Ln(\blambda)$ is through trace functionals of the sample covariance matrix $\bX^\intercal \bX/ n$, which can be characterized precisely using techniques from Random Matrix Theory~\citep{bai_book}.

\subsection{Sharp risk predictions under block-diagonal covariance}
\label{subsec:risk_expressions}
As a first application of Lemma~\ref{lem:risk_close_to_L}, we seek to provide exact and deterministic expressions for the limiting predictive risk $\risk{\blambda}$ of group-regularized ridge regression for any choice of $\blambda \in (0,\infty)^K$. Such results have previously been derived in the setting without groups ($K=1$)~\citep{dicker2016ridge, dobriban2018high, hastie2019surprises}. The key assumption in these works is that the empirical distribution of the eigenvalues of the feature covariance matrix $\bSigma$ converges to a limiting spectral distribution $H$; the limiting risk formulae then are functions of only $H$, the regularization parameter $\lambda > 0$ and the asymptotic aspect ratio \smash{$\gamma= \lim_{n \to \infty} p/n$}. In the setting with groups, we assume that such convergence holds within each subgroup and assume in addition to (HD1-4) that
\begin{enumerate}
\item [(A1)] Each group of features $g \in \cb{1,\dotsc,K}$ has a covariance matrix $\bSigma_g$ with limiting spectral distribution $H_g$ as $n \to \infty$.
\end{enumerate}
In the grouped setting, in  contrast to the setting without groups, assumption (A1) does not suffice (as we explain below). We thus also assume that:
\begin{enumerate}
    \item [(A2)] The features are uncorrelated across groups up to finite rank perturbations of the covariance matrix $\bSigma$. Concretely, suppose without loss generality that the feature groups have been arranged in consecutive order, i.e., $\group{1}=\cb{1,\dotsc,p_1}$, $\group{2}=\cb{p_1+1,\dotsc,p_2}$ and so forth. We then assume that for each $g\in \cb{1,\dotsc,K}$ there exists a $p_g \times p_g$ symmetric, positive definite matrix $\Tilde{\bSigma}_g$ with all eigenvalues in $[h_1,h_2]$ (with $h_1,h_2>0$) such that $\text{Rank}(\bSigma^{1/2} - \text{diag}(\Tilde{\bSigma}_1^{1/2}, \dotsc, \Tilde{\bSigma}_K^{1/2})) \leq r$ for a fixed $r \in \mathbb N$\footnote{ Under assumptions (HD2) and (A1-2), $\Tilde{\bSigma}_g$ has the same limiting spectral distribution as $\bSigma_g$, i.e., $H_g$.}.
\end{enumerate}
Assumption (A2) is strong. However, an assumption of such kind is necessary. Otherwise the predictive risk may not converge to an asymptotic limit. Furthermore, even when the asymptotic limit exists, the limiting expression will typically not be a function of only the group-wise spectral distributions $H_1,\dotsc,H_K$. For example, if the eigenvectors of $\bSigma$ are sufficiently delocalized (such as uniformly distributed with respect to the Haar measure), then the limit that arises is going to be different than the limit arising from the Block-Diagonal structure in (A2) with the same $H_1,\dotsc,H_K$\footnote{Our proof techniques can be used to derive limiting expressions in such situations too, but we do not carry out this analysis here.}.

Among assumptions under which the predictive risk converges to a limit, we consider (A2) to provide a realistic approximation for some practical settings. For instance, factor models in finance as in \citet{ait2017using, tao2017inverse} assume that financial returns lie close to a low dimensional space of principal components with residuals that tend to have a block-diagonal covariance structure with blocks corresponding to different market sectors. 

The key result of this section is the following Theorem:
\begin{theorem}[Asymptotic risk of group-regularized ridge regression]
\label{thm:main_risk}
Consider model~\eqref{eq:random_effects} under assumptions (HD1-4), (A1-2) and also assume that $x_i \sim \mathcal{N}(0,\bSigma)$ and $\sigma^2 = \Var{\varepsilon_i} =1$\footnote{These two additional assumptions are not important. The assumption $\sigma^2=1$ is merely aesthetic and simplifies the formulae. The assumption $x_i \sim \mathcal{N}(0,\bSigma)$ simplifies our technical arguments and could be replaced by assumption (HD3).
}. We perform group-regularized ridge regression~\eqref{eq:groupridge} with deterministic parameters $\blambda = (\lambda_1,\cdots,\lambda_K) \in (0,\infty)^K$ to estimate $\bw$. 
The out-of-sample prediction risk $\risk{\blambda}=\risk{\widehat{\bw}(\blambda)}$ converges  almost surely to 
\begin{equation}
\label{eq:outofsample_risk}
1+\gamma f(\lambda_1,\cdots,\lambda_K)+\sum_{j=1}^K\frac{\gamma}{\gamma_j}(\gamma_j \lambda_j-\alpha_j^2\lambda_j^2) \frac{\partial f (\lambda_1,\cdots,\lambda_K)}{\partial \lambda_{j}},
\end{equation}
where $f =  f(\lambda_1,\cdots,\lambda_k) \geq 0$  is the unique solution of the equation\footnote{Supplement~\ref{subs:main_eq_analysis} studies equation~\eqref{eq:asympt_root} and explains how to solve it numerically.}
\begin{equation}
\label{eq:asympt_root}
f=\sum_{j=1}^K\frac{\gamma_j}{\gamma}\int \p{\frac{\lambda_j}{t}+\frac{1}{1+\gamma f}}^{-1}{dH_{j}(t)}.
\end{equation}
\end{theorem}
The limiting risk formula of Theorem~\ref{thm:main_risk} depends only on the limiting group-wise spectra $H_1, \dotsc, H_K$, aspect ratios $\gamma_1, \dotsc, \gamma_K$ and regularization parameters $\lambda_1,\dotsc,\lambda_K$  and so the result directly generalizes existing results in the case $K=1$ ~\citep{dicker2016ridge, dobriban2018high, hastie2019surprises}. The extension to the grouped setting leads to technical complications; the arguments of aforementioned papers rely on symmetry properties (say, invariance to rotations) which no longer hold in the presence of grouping information\footnote{While we were finishing this work, we became aware of parallel work by~\citet{wu2020optimal} who derive asymptotic risk formulae for ridge regression with general quadratic penalties. When specialized to our setting, their asymptotic risk formulae are less natural than ours as they are not phrased in terms of $\lambda_1,\dotsc, \lambda_K$, but instead in terms of a single penalty parameter, say $\lambda_1$, and a function $h$ that depends on the value of the ratios $\lambda_2/\lambda_1, \dotsc, \lambda_K/\lambda_1$ in an implicit way. Furthermore,~\citet{wu2020optimal} do not address the key issue of data-driven choice of the optimal regularization parameters.}.

As a first corollary of Theorem~\ref{thm:main_risk}, we compute the optimal predictive risk attainable by group-regularized ridge regression.

\begin{corol}\label{corol:opt_risk}
Under the assumptions of Theorem~\ref{thm:main_risk}, the (asymptotically) optimal choice of regularization parameters is $\lambda_g^* = \gamma_g/\alpha_g^2$. The optimal limiting risk is $1+\gamma f(\lambda_1^*,\cdots,\lambda_K^*)$ with $f$ given by~\eqref{eq:asympt_root}.
\end{corol}

\subsection{Using a single regularization parameter}

Theorem~\ref{thm:main_risk} enables us to theoretically answer and provide quantitative insights into questions as follows. Consider two groups of features, i.e., $K=2$. Analyst 1 has access only to features $\bX_{\cdot,\group{1}}$ and optimally tunes ridge regression. Analyst 2 also has access to the second group of features, i.e., to both $\bX_{\cdot,\group{1}}$ and $\bX_{\cdot,\group{2}}$. Analyst 2, however, is not aware of the grouping and runs ridge regression with a single (optimal) regularization parameter. When is Analyst 1 better off than Analyst 2? This tradeoff, will depend on the size of $\group{2}$ and the strength of its signal. Intuitively, if the signal in $\group{2}$ is low,
then Analyst 1 is better off, since the additional set of features swamps the regression of Analyst 2 with noise\footnote{If both analysts tune their methods suboptimally, say with very light regularization, then Analyst 2 may have an advantage due to the implicit regularization of noise features, cf. the double descent phenomenon described for Ridge regression by~\citet{hastie2019surprises}. 
Our results also allow the study of this phenomenon.}. 
On the other hand, if signal is strong, then Analyst 1 misses out on informative features.

Our first result describes the asymptotically optimal parameter when the limiting spectral distributions are the same for the covariance matrices of each group.  

\begin{corol}\label{corol:single_lam_opt}
Consider the case of $K$ groups such that $H_g=H_{h}$ for all groups $g,h$. Then, the asymptotically optimal choice of a single regularization parameter\footnote{That is, the $\lambda^* \geq 0$ that minimizes the asymptotic limit of $\risk{(\lambda, \dotsc,\lambda)}$ over all $\lambda \geq 0$.} is  equal to $\lambda ^* = \gamma /(\sum_{g=1}^K\alpha_g^2).$
\end{corol}

If $H_1,\cdots,H_K$, were not all the same, then the above statement would no longer be true, since groups with higher predictor variability would affect $\lambda^*$ in different proportions. For the rest of this section we assume that $\bSigma = \bI$ to provide more explicit formulae.

\begin{corol}
\label{corol:risk_singla_lambda}
The prediction risk for $\blambda=(\lambda, \dotsc, \lambda)$ and $\bSigma=\bI$ converges almost surely to $$\frac{1}{u}-\frac{\gamma \lambda -(\sum_{g=1}^K\alpha_g^2)\lambda^2}{(\lambda+u)^2-\gamma u^2}\;\; \text{, where }u=\frac{1-\gamma-\lambda+\sqrt{(\lambda+\gamma-1)^2+4\lambda}}{2}.$$
The optimal asymptotic risk in this case is \smash{$(\gamma+\lambda^*-1+\sqrt{(\gamma+\lambda^*-1)^2+4\lambda^*})/(2\lambda^*),$} with $\lambda^*$ as in Corollary \ref{corol:single_lam_opt}. 
\end{corol}

As a concrete example, when $K=2$, $\alpha_1 >0$ and $\alpha_2=0$, the optimal prediction risk for ridge regression with a single regularization parameter is asymptotically equal to
\ifdoubleblind 
$$(\gamma+\gamma/\alpha_1^2+\sqrt{(\gamma+\gamma/\alpha_1^2-1)^2+4\gamma/\alpha_1^2})/(2\gamma/\alpha_1^2).$$
\else 
\smash{$(\gamma+\gamma/\alpha_1^2+\sqrt{(\gamma+\gamma/\alpha_1^2-1)^2+4\gamma/\alpha_1^2})/(2\gamma/\alpha_1^2)$}.
\fi 
The last expression is increasing in $\gamma_2$ (holding $\gamma_1$ fixed) and converges to $1+\alpha_1^2$ as $\gamma_2\rightarrow\infty$. The existence of pure-noise features, as expected, hurts the performance of optimally-tuned ridge regression with a single regularization parameter.

We next proceed to answer the motivating question asked in the beginning of this section, i.e., the case $K=2$ with two analysts, wherein Analyst 1 has access only to the first feature group, while Analyst 2 has access to both but is not aware of the grouping information.

\begin{corol}\label{corol:more_features}
Suppose $K=2$ and $\bSigma = \bI$. Analyst 1 only has access to the first group of features $\bX_{\cdot,\group{1}}$ and regresses $\bY\sim \bX_{\cdot,\group{1}}$ using ridge with a single parameter $\tilde{\lambda}$. The asymptotically optimal prediction risk (and corresponding optimal $\tilde{\lambda}$) is equal to:
$$(\alpha_2^2+1)\p{\gamma_1+\tilde{\lambda}-1+\sqrt{(\gamma_1+\tilde{\lambda}-1)^2+4\tilde{\lambda}}}\Big/(2\tilde{\lambda}), \text{ with } \tilde{\lambda}=\gamma_1(\alpha_2^2 + 1)/\alpha_1^2.$$ 
Analyst 2 has access to both feature groups and regresses $\bY\sim [\bX_{\cdot,\group{1}} \bX_{\cdot,\group{2}}]$ using ridge with a single parameter $\lambda^*$. The asymptotically optimal prediction risk (and corresponding optimal $\lambda^*$) is equal to:
\begin{equation*}
    \p{\gamma+\lambda^*-1+\sqrt{(\gamma+\lambda^*-1)^2+4\lambda^*}}\Big/(2\lambda^*),  \text{ with }\lambda^*=\gamma/(\alpha_1^2+\alpha_2^2).
\end{equation*}
\end{corol}

Corollary~\ref{corol:more_features} allows us to find for any fixed choices of $\alpha_1,\gamma_1,\gamma_2$ the threshold below which $\alpha_2$ makes the presence of the second group harmful. Some consequences, are as follows. Consider the regime of strong signal in group 1 ($\alpha_1 \to \infty$ and $\alpha_2 \geq 0$ fixed). If $\gamma>1$, then including the second group of predictor variables and using a single regularization parameter hurts predictive performance. On the other hand, if $\gamma<1$ and $\alpha_2^2/\gamma_2 > 1/(1-\gamma)$,  where $\alpha_2^2/\gamma_2$ is the ``signal-to-noise'' ratio in group 2 and $1/(1-\gamma)$ is the out-of-sample risk of unregularized linear regression, then the presence of the second group improves the prediction risk for any value of $\gamma_1$.

\subsection{Numerical illustration of asymptotic risk predictions}
 
In this subsection we illustrate the theoretical risk curves derived in Theorem~\ref{thm:main_risk}. We consider the case $K=2$, $n=1000$ and $\bSigma=\bI$ and show risk curves as a function of $\gamma_1,\gamma_2$ and the signal strengths $\alpha_1^2, \alpha_2^2$ with $\blambda$ chosen in the following 3 ways: \textbf{1)} $\blambda \in (0,\infty)^2$ is the optimal regularization parameter vector defined in Corollary~\ref{corol:opt_risk}, \textbf{2)} $\blambda$ is the optimal regularization parameter among parameters of the form $\blambda = (\lambda, \lambda)$ (i.e., we include both features but use a single regularization parameter) and \textbf{3)} the optimal parameter among parameters of the form $\blambda = (\lambda, \infty)$, (i.e., we omit the second group of features).

\begin{figure}[!ht]
\centering
\begin{tabular}{ccc}
  \begin{adjustbox}{width=0.32\linewidth}\input{figures/oracle_risks_identity/oracle_risk1.tikz}\end{adjustbox} 
& \begin{adjustbox}{width=0.32\linewidth}\input{figures/oracle_risks_identity/oracle_risk2.tikz}\end{adjustbox} 
& \begin{adjustbox}{width=0.32\linewidth}\input{figures/oracle_risks_identity/oracle_risk3.tikz}\end{adjustbox} \\ 
  \begin{adjustbox}{width=0.32\linewidth}\input{figures/oracle_risks_identity/oracle_risk4.tikz}\end{adjustbox} 
& \begin{adjustbox}{width=0.32\linewidth}\input{figures/oracle_risks_identity/oracle_risk5.tikz}\end{adjustbox} 
& \begin{adjustbox}{width=0.32\linewidth}\input{figures/oracle_risks_identity/oracle_risk6.tikz}\end{adjustbox}
\end{tabular}
\caption{\textbf{Asymptotic predictions for asymptotic risk of group-regularized ridge regression}: The lines are theoretical risk curves as per Theorem~\ref{thm:main_risk} for three different choices of $\blambda$ and the triangles correspond to the finite-sample risk evaluated by simulations. The aspect ratios $\gamma_1,\gamma_2$ and the total signal strength $\alpha_1^2 + \alpha_2^2$ vary across panels, while the relative signal strength in the first group ($\alpha_1^2/(\alpha_1^2+ \alpha_2^2)$) varies along the $x$-axis of the panels. } 
\label{fig:theoretical_risks}
\end{figure}
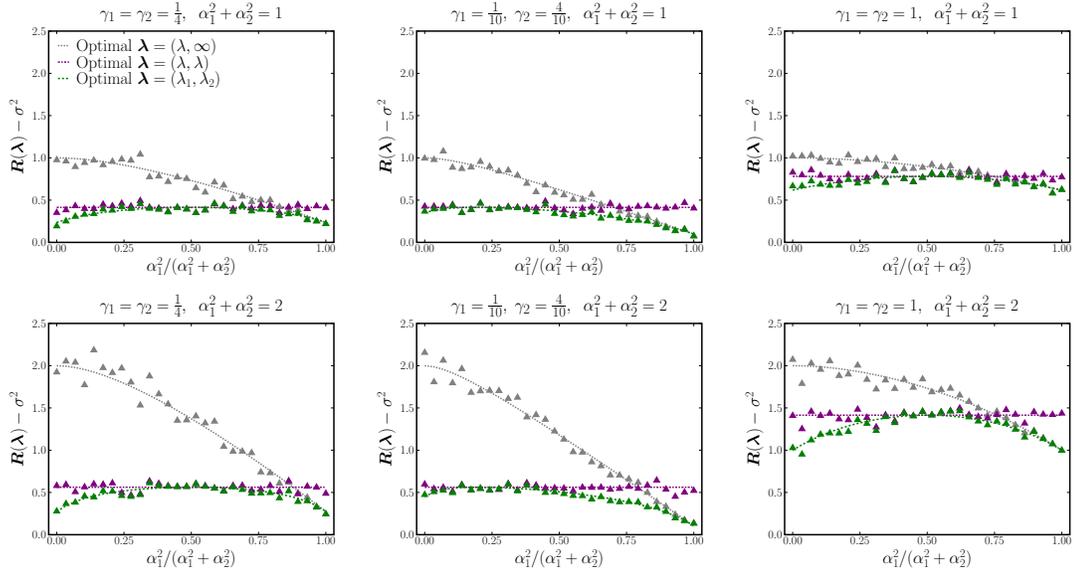

Figure~\ref{fig:theoretical_risks} shows the theoretical risk curves along with an empirical estimate of the test error of the method (computed on $20,000$ test samples) based on a single realization of the simulation; i.e., the triangles correspond to $\risk{\blambda}$~\eqref{eq:risk_def}. We observe the excellent agreement between theoretical and finite-sample risks. As expected, in all panels of Figure~\ref{fig:theoretical_risks}, we observe that group-regularized ridge regression decreases prediction risk the most under strong heterogeneity across groups ($\alpha_1 \neq \alpha_2$). When $\alpha_2 \approx 0$, ridge regression on the first group of features has about the same risk as group-regularized ridge regression. The exact details of the risk curves depend on the corresponding data generating mechanism (through the spectra $H_1,H_2$, aspect ratios $\gamma_1, \gamma_2$ and signal strengths $\alpha_1, \alpha_2$). In Supplementary Figure~\ref{fig:theoretical_risks_exponential} we demonstrate risk curves under a more complicated covariance structure (following \citet{dobriban2018high}); we let $\bSigma_1=\bSigma_2$, each with eigenvalues corresponding to
evenly-spaced quantiles of the Exponential distribution with rate $0.5$. The conclusions are similar.
  
\section{High-dimensional optimality of $\sigmacv$-Ridge}
\label{sec:optimality}
In this section we build upon the asymptotic setting from Section~\ref{sec:risk_expressions} and prove that asymptotically $\sigmacv$-Ridge regression matches the prediction risk of the best possible predictor from the class~\eqref{eq:groupridge}:

\begin{theorem}[Optimality of $\sigmacv$-Ridge Regression]
\label{theo:optimality}
Consider model~\eqref{eq:random_effects} under assumptions $\text{(HD1-4)}$.  Let $V_1 \subset (0,\infty)$ a compact set with $\sigma \in V_1$  and $V_2$ a subset of $(0,\infty]^K$ bounded away from 0.  We choose $\widehat{\sigmacv} \in \argmin_{V_1}\loo{(\sigmacv)}$. Then:
$$ \limsup_{n \to \infty} \p{\risk{\widehat{\blambda}(\widehat{\sigmacv})} - \inf_{\blambda \in V_2}\risk{\blambda}} \leq  0 \text{ almost surely.}$$
This result is true for any (potentially data-driven) choice of $\lambdainit \in V_3$ in~\eqref{eq:moments_decomp}, where $V_3 \subset (0,\infty)$ is also a compact set.
\end{theorem}
For example, under the conditions of Theorem~\ref{prop:risk_prop}, i.e., when $\bSigma$ is block-diagonal, the above Theorem shows that the asymptotic risk of $\sigmacv$-Ridge is equal to the expression in Corollary~\ref{corol:opt_risk}. The result of Theorem~\ref{theo:optimality} is applicable to any feature covariance matrix $\bSigma$ (with eigenvalues bounded away from $0$ and $\infty$).

\subsection{$\sigmacv$-Ridge regression and the Group Lasso}
\label{subsec:grouplasso}
Before proceeding with the proof of Theorem~\ref{theo:optimality}, we first apply it to elucidate the connection of $\sigma$-Ridge regression and the Group Lasso~\citep{yuan2006model}\footnote{The results here were motivated by Section 6 of~\citet{tay2020feature}.}. As already mentioned in Section~\ref{subsec:related_work}, the Group Lasso is a regularized regression method as in~\eqref{eq:general_penalized_reg} that penalizes the Euclidean Norm of group-wise coefficients, 
\begin{equation}
\label{eq:glasso_regression} 
\widehat{\bw}^{\text{glasso}}(\lambda^{\text{glasso}}) \in \argmin_{\bw}  \cb{\frac{1}{2n}\sum_{i=1}^n \p{\yresponse_i - x_i^\intercal \bw}^2 \; + \lambda^{\text{glasso}} \cdot \sum_{g=1}^K \sqrt{\frac{p_g}{p}} \Norm{ \bw_{\mathcal{G}_g}}_2}.
\end{equation}
A beautiful aspect of the Group Lasso is that it enforces group-wise sparsity, i.e., for many groups it holds that $\widehat{\bw}^{\text{glasso}}_{\group{g}} = 0$ and it has strong recovery guarantees of the non-zero groups (similar to the Lasso in the setting without groups)\footnote{We note, that $\sigmacv$-Ridge regression can also select groups of features (cf. Figure~\ref{fig:sigma_as_tuning}). Our focus on this paper, however, is on predictive performance.}.

At first sight, the Group Lasso method~\eqref{eq:glasso_regression} and group-regularized ridge regression~\eqref{eq:groupridge} may seem unrelated. However, it turns out that~\eqref{eq:groupridge} can recover~\eqref{eq:glasso_regression} by appropriate choice of $\blambda$: if we set the components of $\blambda$ as $\lambda_g= \lambda^{\text{glasso}}\cdot \sqrt{p_g/p} \,/\,\norm{\widehat{\bw}^{\text{glasso}}_{\group{g}}}_2$ (with the convention that $\lambda_g=\infty$ if the denominator is $0$), then \smash{$\widehat{\bw}(\blambda) =   \widehat{\bw}^{\text{glasso}}(\lambda^{\text{glasso}})$}.

The intuition behind this result is that we can write $\norm{\bw_{\group{g}}}_2 = \norm{\bw_{\group{g}}}^2_2/\norm{\bw_{\group{g}}}_2$; a formal verification proceeds by checking the Karush-Kuhn-Tucker conditions. In other words, we may think of the Group-Lasso as providing a map from the 1D regularization parameter $\lambda^{\text{glasso}}$ to the $K$-dimensional $\blambda$, that is then used along with the group-regularized ridge objective~\eqref{eq:groupridge}. The construction of the path is motivated by sparsity considerations, while the $\sigmacv$-Ridge path is motivated by models~\eqref{eq:normal_normal} and~\eqref{eq:random_effects}. In view of Theorem~\ref{theo:optimality}, we can prove that:
\begin{corol}[$\sigmacv$-Ridge regression is at least as powerful as the Group Lasso]
\label{corol:glasso} 
Under the assumptions of Theorem \ref{theo:optimality} and for any fixed $\delta > 0$, it holds that
$$\limsup_{n \to \infty}\p{\risk{\widehat{\bw}(\widehat{\blambda}(\widehat{\sigmacv})} -  \inf_{\lambda^{\text{glasso}} \geq \delta} \risk{\widehat{\bw}^{\text{glasso}}(\lambda^{\text{glasso}})} }\leq 0\;\text{ almost surely.}$$
\end{corol}
In interpreting this result, we caution, however, that the Group Lasso has been developed in the context of group-wise selection of features, when most feature groups are assumed to have no signal and the number of groups is potentially large. Consequently, the setting of Corollary~\ref{corol:glasso} favors $\sigmacv$-Ridge regression. Nevertheless, the result does provide some guidance to practitioners about the types of datasets in which $\sigmacv$-Ridge regression would be preferable over the Group Lasso. We return to this comparison, from an empirical perspective, in Section~\ref{sec:numerics}.

\subsection{Proof of Theorem~\ref{theo:optimality}}
The proof of Theorem~\ref{theo:optimality} hinges on Lemma~\ref{lem:risk_close_to_L}, as well as two additional Lemmata that are of independent interest, and that we now describe. Our first lemma justifies the use of the method of moments in~\eqref{eq:system}.
\begin{lemma}\label{lem:system}
Under the assumptions of Theorem \ref{theo:optimality} we have for each $g=1,\cdots,K:$
\begin{equation}
    \label{eq:consistency_system_rhs}
    {\Norm{\widetilde{\bw}_{\groupg{g}}}_2^2 } -\left(  \sum_{h=1}^K \norm{\bM_{\group{g},\group{h}}}_F^2 \frac{\alpha_h^2}{p_h} + \Norm{\bN_{\group{g},\cdot}}_F^2\frac{\sigma^2}{n}\right)\xrightarrow{a.s.}0 \text{ as } n \to \infty.
\end{equation}
In addition, the matrix $\bA$ in~\eqref{eq:mapping} (cf. system ~\eqref{eq:system}) is invertible almost surely for all large $n$ and the inverse has bounded operator norm. It follows that, if we solve~\eqref{eq:mapping} at $\sigmacv = \sigma$, then $\widehat{\lambda}_g(\sigma) - {p_g}{n^{-1}}\cdot{\sigma^2}{\alpha_g^{-2}} \to 0$ almost surely.
\end{lemma}
The first result, i.e.,~\eqref{eq:consistency_system_rhs}, follows very similarly to the proof of Lemma~\ref{lem:risk_close_to_L} and uses Marcinkiewicz-Zygmund interpolation along with the uniform moments bounds we have assumed. The second result, i.e., the study of the matrix $\bA$, is more  challenging, as we need to lower bound the eigenvalues of a matrix that is formed by taking a larger random matrix, squaring its entries and then summing the squares in a block-wise fashion.

We next show that the leave-one-out objective $\Rloo{(\blambda)}$~\eqref{eq:loo_blambda} is uniformly close to $\Ln(\blambda)$.

\begin{lemma}\label{lem:RCV_L}
Under the assumptions of Theorem \ref{theo:optimality} we have $\abs{\Rloo{(\blambda)}-\Ln(\blambda)}\xrightarrow{a.s.}0$ uniformly for $\blambda$ in compact subsets of $(0,\infty)^K$.
\end{lemma}
Along with Lemma~\ref{lem:risk_close_to_L}, it follows that the leave-one-out estimate of the error is close to the true out-of-sample error. Similar results in the case of a single regularization parameter (e.g., ridge regression with a single group) have a long tradition in the statistics literature, see for example~\citep{li1987asymptotic}. More recently, such results have resurfaced under the lens of high-dimensional asymptotics and random matrix theoretic results. For example, in the setting without side-information, \citet{xu2019consistent} study leave-one-out cross-validation for many penalties, but with restrictive assumptions on the distribution of the features $x_i$. \citet{hastie2019surprises} prove a result analogous to Lemma~\ref{lem:RCV_L} in the case of a single group (i.e., without side information) with feature covariance $\bSigma$ that has a limiting spectral distribution. Their proof relies on the fact that the two estimates of the risk have explicit formulae. In the case of group-regularized ridge regression and without assumptions of convergence of the spectral distribution of the population covariance matrix, such explicit formulae are not available, and so the proof (in Supplement~\ref{sec:section4_proofs}) is more involved.

With the key Lemmata~\ref{lem:risk_close_to_L}, \ref{lem:system} and \ref{lem:RCV_L} in hand, we are ready to prove Theorem \ref{theo:optimality}.

\begin{proof}[Proof of  Theorem \ref{theo:optimality}]
We assume that $V_2$ is a compact set; and provide the extension to noncompact $V_2$ in the supplement. Since $\Ln(\blambda)-\risk{\blambda}\xrightarrow{}0$ uniformly over $V_2$ by Lemma~\ref{lem:risk_close_to_L}, it follows that $\inf_{V_2}\Ln(\blambda)-\inf _{V_2}\risk{\blambda}\xrightarrow{a.s.}0.$
Since the formula for $\Ln(\cdot)$ is universal among all distributions on $w_j,\varepsilon_i$ with the moment assumptions that we made, it is, in particular, the expected risk for a model with Gaussian priors on the coefficients and Gaussian errors. In that case, the optimal coefficient vector corresponds to the posterior mean, which is achieved for \smash{$\blambda^* = (\sigma^2 p_gn^{-1}\alpha_g^{-2})_{1\leq g\leq K}$}. We conclude that this choice minimizes $\Ln(\blambda)$,  and so $\limsup_{n \to \infty} \p{\Ln(\blambda^*)-\inf_{V_2}\risk{\blambda}} \leq 0$ almost surely.

It remains to show that $\limsup_{n \to \infty} \p{ \risk{\widehat{\blambda}(\widehat{\sigmacv})}-\Ln(\blambda^*)}\leq 0$ almost surely. One can verify that the functions $\Ln$ are almost surely equicontinuous at $\blambda^*$. By Lemma~\ref{lem:system}, we get for $\sigmacv=\sigma$ that $\widehat{\blambda}(\sigma) - \blambda^* \to 0$ a.s. and consequently $\Ln(\widehat{\blambda}(\sigma))-\Ln(\blambda^*)\xrightarrow{a.s.}0.$ Finally, by Lemmata~\ref{lem:risk_close_to_L}, \ref{lem:RCV_L} and the definition of $\widehat{\sigmacv}$, $\risk{\widehat{\blambda}(\widehat{\sigmacv})}-\Ln(\widehat{\blambda}(\sigma))=\Rloo{(\widehat{\blambda}(\widehat{\sigmacv}))}-\Rloo{(\widehat{\blambda}({\sigma}))}+\omicron{(1)} \leq \omicron{(1)}.$

\end{proof}

\section{Numerical results}
\label{sec:numerics}
We now demonstrate that $\sigmacv$-Ridge regression is practical.  In Section~\ref{subsec:simulation} we conduct a simulation study when the data-generating mechanism is specified by model~\eqref{eq:random_effects}. In Sections~\ref{subsec:data_cll},~\ref{subsec:data_songs}, we apply $\sigmacv$-Ridge regression to real datasets from two distinct domains and show its low out-of-sample prediction error. In these datasets, model~\eqref{eq:random_effects} is unlikely to hold, and so our theoretical results are not applicable, but nevertheless $\sigmacv$-Ridge regression performs favourably. 

Throughout our numerical results, we compare the following four methods. 
\begin{enumerate}[nosep,align=left, wide]
    \item $\sigmacv$-\textbf{Ridge}, i.e., the regression method introduced in this work, tuned via accelerated leave-one-out cross-validation~\eqref{eq:loo}. $\sigmacv$ is chosen from an equidistant grid of 100 points between $10^{-3}\cdot\sigmacv_{\text{max}}$ to $\sigmacv_{\text{max}}$, where $\sigmacv_{\text{max}}$ is defined in the first part of Proposition~\ref{prop:propo1}.
    \item \textbf{Single Ridge}, i.e., Ridge regression with a single regularization parameter $\lambda$ (the same across all feature groups), tuned via LOOCV~\eqref{eq:loo_blambda}. We choose $\lambda \in \mathcal{L}_{\text{grid}}$, where $\mathcal{L}_{\text{grid}}$ is a logarithmically equidistant grid of 100 points from $10^{-6} \cdot \lambda_{\text{max}}$ to $\lambda_{\text{max}}$ and \smash{$\lambda_{\text{max}}=10^3\cdot\Norm{\bX^\intercal \bY/n}_{\infty}$} (the default choice of $\lambda_{\text{max}}$ in the Glmnet package~\citep{ friedman2010regularization}).
    \item \textbf{Multi Ridge}, i.e., Group Ridge regression~\eqref{eq:groupridge} with one regularization parameter per group. $\blambda \in (0,\infty)^K$ is chosen via LOOCV~\eqref{eq:loo_blambda} among 5000 randomly selected points from the product grid $\mathcal{L}_{\text{grid}}^K$, where $\mathcal{L}_{\text{grid}}$ is the grid used for Single Ridge regression.
    \item \textbf{Group Lasso}~\citep{yuan2006model}, as in~\eqref{eq:glasso_regression} using the implementation in the Seagull R package~\citep{klosa2020seagull}. We tune $\lambda^{\text{gl}} = \lambda^{\text{glasso}}$ by monitoring the mean squared error on a holdout set with $30\%$ of the observations. $\lambda^{\text{gl}}$ is chosen from a logarithmically equidistant grid (100 points) that ranges from $10^{-6}\cdot \lambda^{\text{gl}}_{\text{max}}$ to  \smash{$\lambda^{\text{gl}}_{\text{max}} = n^{-1}\max_g\{{\lVert \p{\bX^\intercal \bY}_{\group{g}}\rVert_{\infty}/\sqrt{p_g/p}}\}$}, i.e., the smallest \smash{$\lambda^{\text{gl}}$} so that $\widehat{\bw}^{\text{glasso}}(\lambda^{\text{gl}}) = \bm{0}$. 
\end{enumerate}

\subsection{Simulation study}
\label{subsec:simulation}
In this section we simulate from Model~\eqref{eq:normal_normal}. In addition to the four methods described above, we also compare against the Bayes estimator, i.e., group-ridge regression with the oracle choice of regularization parameters~\eqref{eq:model_based_lambda}. We set $p=800$ and partition the features into $K=32$ groups (each of size $p_g=25$). The $g$-th group of features is generated as in~\eqref{eq:normal_normal} with $\alpha_g=(g-1)/31$, i.e., $\alpha_1=0$, $\alpha_{32}=10$ and the other $\alpha_g$s are linearly spaced between $0$ and $10$. The features are simulated as $x_i \sim \mathcal{N}(0,\bSigma)$ with $\bSigma$ chosen first as the covariance of an autoregressive process of order 1 (AR) with autocorrelation equal to $0.8$\footnote{The correlation persists across blocks, i.e., the covariance is not block-diagonal.} and second as the identity matrix. We set $\sigma=5$ and vary $n \in \cb{p/2, p, 2p}$. We also generate $10000$ test samples to evaluate the mean squared error (MSE). For each setting, we report the squared error averaged over 400 simulation runs.

In each simulation run, we also coarsen the grouping information as follows: We merge  consecutive groups to reduce the total number of groups to \smash{$K \in \cb{2^1,2^2,\dotsc,2^5}$}. The coarsened grouping information is passed on to the regression methods used. Such coarsening impacts $\sigmacv$-Ridge, Multi Ridge and the Group Lasso, but not Single Ridge, nor the oracle Bayes estimator.

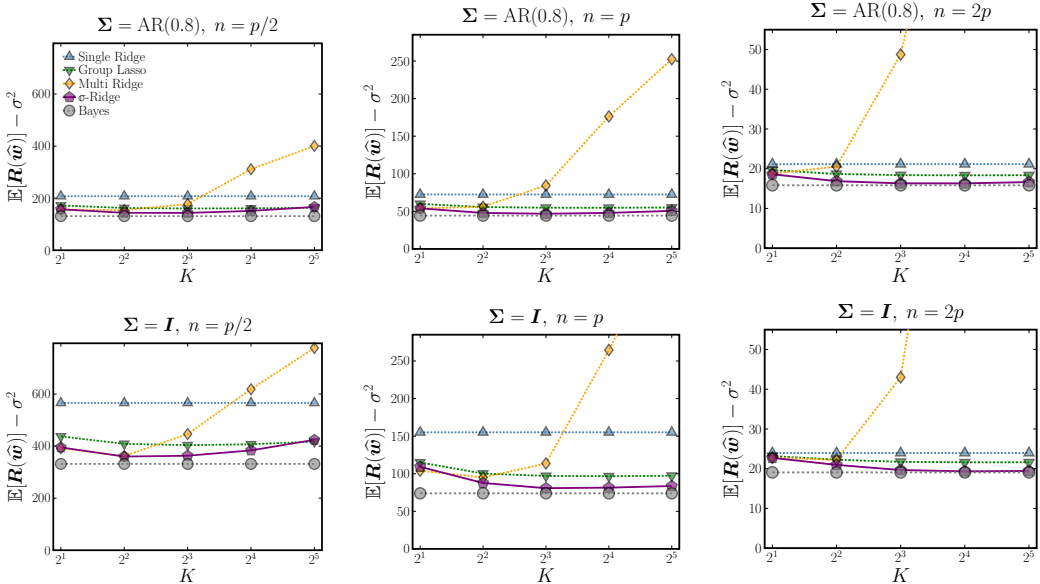
\begin{figure}[!ht]
\centering
\begin{tabular}{ccc}
  \begin{adjustbox}{width=0.31\linewidth}\input{figures/simulations/simulations_ar_phalf.tikz}\end{adjustbox}  &  \begin{adjustbox}{width=0.31\linewidth}\input{figures/simulations/simulations_ar_p.tikz}\end{adjustbox} &   \begin{adjustbox}{width=0.31\linewidth}\input{figures/simulations/simulations_ar_ptwice.tikz}\end{adjustbox} \\
  \begin{adjustbox}{width=0.31\linewidth}\input{figures/simulations/simulations_id_phalf.tikz}\end{adjustbox}  &  \begin{adjustbox}{width=0.31\linewidth}\input{figures/simulations/simulations_id_p.tikz}\end{adjustbox} &   \begin{adjustbox}{width=0.31\linewidth}\input{figures/simulations/simulations_id_ptwice.tikz}\end{adjustbox}
\end{tabular}
\caption{\textbf{Simulation study:} We compare the mean squared error (minus the response noise variance $\sigma^2$) of four methods and the Bayes estimator for two different choices of feature covariance and three different choices of aspect ratio $\gamma \in \cb{2, 1, 0.5}$. The $x$-axis shows the number of feature groups $K$ provided to the methods; the side-information is coarser for smaller $K$.}
\label{fig:simulation}
\end{figure}

The results are shown in Figure~\ref{fig:simulation}. We make the following observations: Throughout all settings, $\sigmacv$-Ridge performs best and often gets close to matching the Bayes risk. When $K$ is small, i.e., when the the side-information made available to $\sigmacv$-Ridge and the Group Lasso is weak and does not fully capture the heterogeneity of the data-generating mechanism, then these methods cannot match the Bayes risk. When the sample size is sufficiently large ($n=p$ or $n=2p$), the gap to the Bayes risk of $\sigmacv$-Ridge strictly decreases as $K$ grows. On the other hand, when the sample size is small ($n=p/2$),  the MSE of $\sigmacv$-Ridge first decreases with increasing $K$, but then decreases. The reason is that for large $K$ the map $\sigmacv \mapsto \widehat{\blambda}(\sigmacv) \in (0,\infty]^K$~\eqref{eq:mapping} becomes more unstable due to larger estimation error. Even in this regime, however, the risk of $\sigmacv$-Ridge is the same as  the best of the other data-driven methods (Group Lasso). Multi Ridge is competitive for small $K$, but its performance quickly deteriorates as $K$ increases and even becomes worse than Single Ridge.

\subsection{Drug response in chronic lymphocytic leukemia} 
\label{subsec:data_cll}
Our first empirical application comes from high-throughput biology. \citet{dietrich2018drug} collected data from different blood cancer patients on the ex-vivo viability of cells after exposure to different drugs, as well as molecular profiling measurements of DNA methylation and RNA-Seq expression. The response of interest $Y_i$ is the ex-vivo viability of the cells of patient $i$ after treatment with Ibrutinib (a drug used to treat chronic lymphocytic leukemia). There are $n=121$ samples with $p=9553$ features that may be partitioned into three groups: $\group{\text{Drugs}}$, the response (ex-vivo viability) to 61 drugs (different from Ibrutinib) measured at 5 different concentrations ($p_{\text{Drugs}}=305$), and $\group{\text{Methyl}}$,  $\group{\text{RNA}}$ corresponding to the $p_{\text{Methyl}} = 4248$,  $p_{\text{RNA}} = 5000$ most variable methylation, resp. RNA-Seq expression measurements. We refer to \citet{dietrich2018drug} for more details, as well as \citet{velten2018adaptive, pramanik2020structure} for further analyses of this dataset. 

\begin{table}
\centering
\input{tables/cll_analysis.tex}
\caption{\textbf{Results on chronic lymphocytic leukemia dataset:} We report data-driven tuning parameters selected by each method, as well as the implied choice of $\widehat{\blambda} = (\widehat{\lambda}_{\text{Drugs}}, \widehat{\lambda}_{\text{Methyl}}, \widehat{\lambda}_{\text{RNA}})$. We note that Single/Multi Ridge directly tune $\blambda$. We also report the time required for running the whole procedure and the RMSE (root mean squared error) estimated by 10-fold cross-validation.}
\label{tab:cll_analysis}
\end{table}

After standardizing the response and the features (to sample mean 0 and sample variance 1), we apply the four different regression methods on the full dataset $(n=121)$. The results are shown in Table~\ref{tab:cll_analysis}, where we show the data-driven choice of tuning parameters and the implied values of \smash{$\widehat{\blambda}$} (for the Group Lasso we define \smash{$\widehat{\blambda}$} as explained in Section~\ref{subsec:grouplasso}). We also report the time it takes to fit the full regression models\footnote{Algorithm run-times were evaluated on a single core of a Macbook Pro with a 2.6 GHz 6-Core Intel Core i7 processor and capture the whole (accelerated) leave-one-out procedure. The goal of the timings is to demonstrate that $\sigmacv$-Ridge regression is practical. Precise timings for all these methods will vary substantially depending on computing device and algorithmic/implementation choices (matrix decompositions used, convex optimization routines, numerical tolerances and so forth).}, including the time required for data-driven tuning. Furthermore, we split the dataset into $10$-folds, and then use cross-validation to evaluate the root mean squared error (RMSE) of the four methods\footnote{Standardization of the features and the response is part of the cross-validation, i.e., it is repeated in each iteration of cross-validation using only the training folds. The error is evaluated at the original response scale.}.  In terms of RMSE, $\sigmacv$-Ridge and the Group Lasso perform best. Both only mildly regularize $\group{\text{Drugs}}$, while they completely discard the groups $\group{\text{Methyl}}$,  $\group{\text{RNA}}$ by setting $\widehat{\lambda}_g = \infty$. This makes sense from a biological perspective: the drug measurements of ex-vivo viability are phenotypically close to the response of interest, i.e., the ex-vivo viability to another drug (Ibrutinib). In contrast, Single Ridge and Multi Ridge apply mild regularization to the latter groups, and their error is larger.

\begin{table}
\centering
\input{tables/cll_analysis_noise.tex}
\caption{\textbf{Results on chronic lymphocytic leukemia dataset:} This Table is analogous to Table~\ref{tab:cll_analysis}; the main difference is that we repeat the analysis of the dataset after adding two additional groups of features that correspond to noise.}
\label{tab:cll_analysis_noise}
\end{table}

We then repeat the same evaluation after adding two groups of noise features to $\bX$. The first noise group consists of permuted drug measurements ($p_{\text{Noise}_1} = 305$),  while the second noise group consists of $p_{\text{Noise}_2} = 100$ i.i.d. Standard Gaussian measurements that are independent of everything else. Results are shown in Table~\ref{tab:cll_analysis_noise} and are qualitatively similar to the results from Table~\ref{tab:cll_analysis}. $\sigmacv$-Ridge and the Group Lasso automatically discard the two noise groups as well, by setting their $\widehat{\lambda}_g = \infty$.

\subsection{Release year in the one million songs dataset}
\label{subsec:data_songs}

As our second empirical example, we seek to predict the release year of different songs based on timbre features. The dataset we use is a subset of the Million Song Dataset~\citep{Bertin-Mahieux2011} that is made available through the UCI Machine Learning repository~\citep{UCI}. We refer the reader to \citet{dobriban2020wonder} for another analysis of this dataset using distributed ridge regression that does not account for the group structure of the features.

The dataset consists of $n=515,345$ samples that have been split into training ($463,715$) and test subsets ($51,630$ samples).
Each sample $i$ corresponds to a song: the response $Y_i$ is the year of release. The raw data for each song consists of segments and 12 timbre attributes per segment. These are converted into features $x_i \in \RR^{156}$ as follows: First, each timbre attribute is averaged across all song segments; this yields a group of 12 features $\group{\text{mean}}$. The next 12 features $\group{\text{std}}$ are computed as the standard deviation of the raw timbre attributes. The features in $\group{\text{cov}}$ consist of the \smash{$66 = \binom{12}{2}$} pairwise covariances of the raw timbre attributes, and similarly $\group{\text{cor}}$ of the 66 pairwise correlations. In total we thus have $K=4$ groups of features with $p_1=p_2=12$ and $p_3=p_4=66$.

\begin{figure}[!ht]
\centering
\begin{tabular}{l}
a)\\
\includegraphics[width=0.8\linewidth]{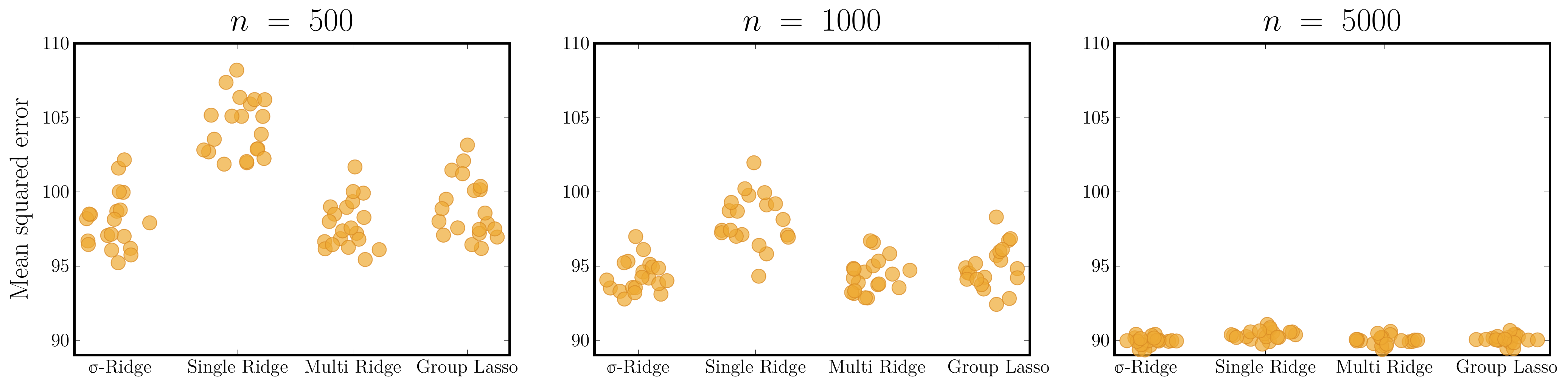} \\
b)\\
\includegraphics[width=0.8\linewidth]{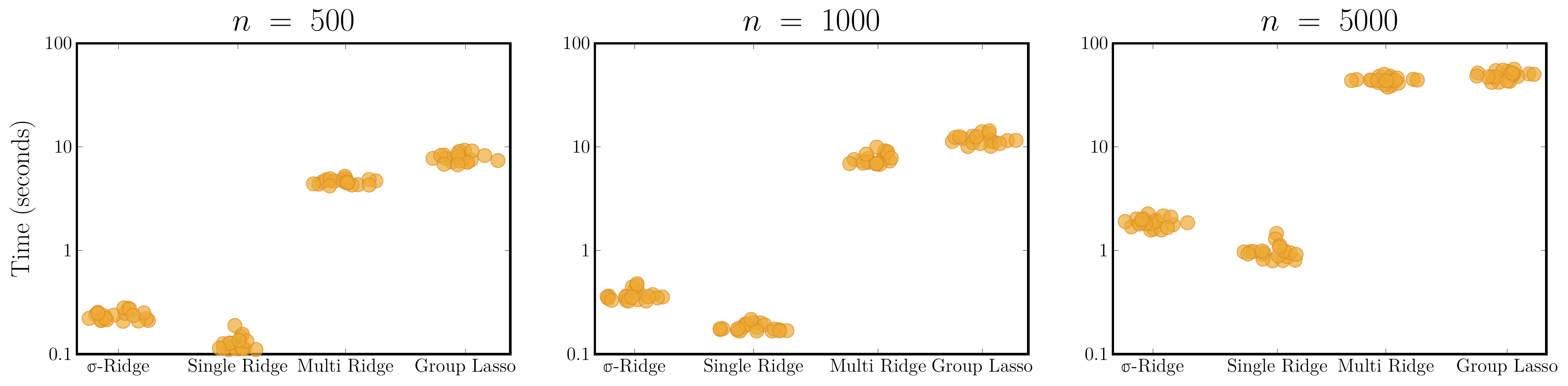}
\end{tabular}
\caption{\textbf{Empirical comparison using the One Million Song dataset~\citep{Bertin-Mahieux2011}:} We compare four methods on subsamples of the full dataset; each column corresponds to a different  subsample size. We repeat the subsampling 20 times. Each point corresponds to a combination of subsample and regression method. In \textbf{a)} we show the mean squared error evaluated on a larger test set, while in \textbf{b)} we compare the timing.}
\label{fig:million_songs_mse_time}
\end{figure}

For $n \in \cb{500,1000,5000}$ and 20 Monte Carlo replications, we randomly subsample the training set to $n$ so as to increase the difficulty of the prediction task. We then standardize (to sample mean 0 and sample variance 1) the response and features and apply the 4 regression methods with data-driven tuning. The mean squared error (at the original response scale) is evaluated based on all the test samples. Figure~\ref{fig:million_songs_mse_time} shows the mean squared error (MSE) and time required to apply each regression method for each Monte Carlo replicate and each subsample size. We observe that $\sigmacv$-Ridge Regression outperforms the other methods in terms of MSE and has running time comparable to Single Ridge.  Multi Ridge and Group Lasso are slower. The advantage of $\sigmacv$-Ridge regression is most pronounced for small sample sizes $(n \in \cb{500,1000})$; for $n=5000$ even Single Ridge performs well.

\begin{figure}[!ht]
\centering
\begin{tabular}{l}
a)$\;n\;=\;500$\\
\includegraphics[width=\linewidth]{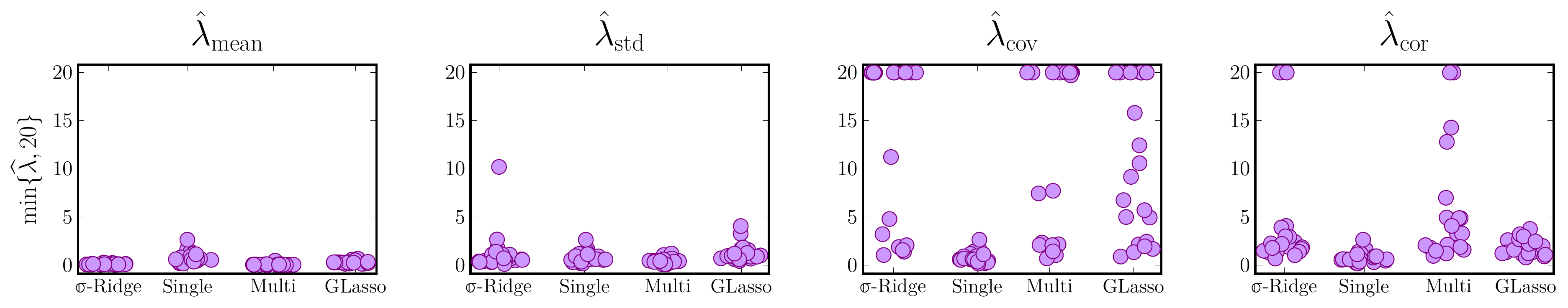} \\
b)$\;n\;=\;1000$\\
\includegraphics[width=\linewidth]{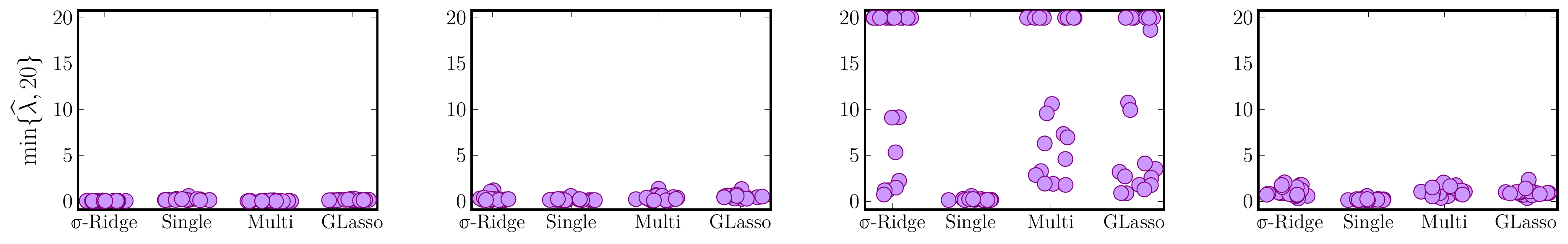}\\
c)$\;n\;=\;5000$\\
\includegraphics[width=\linewidth]{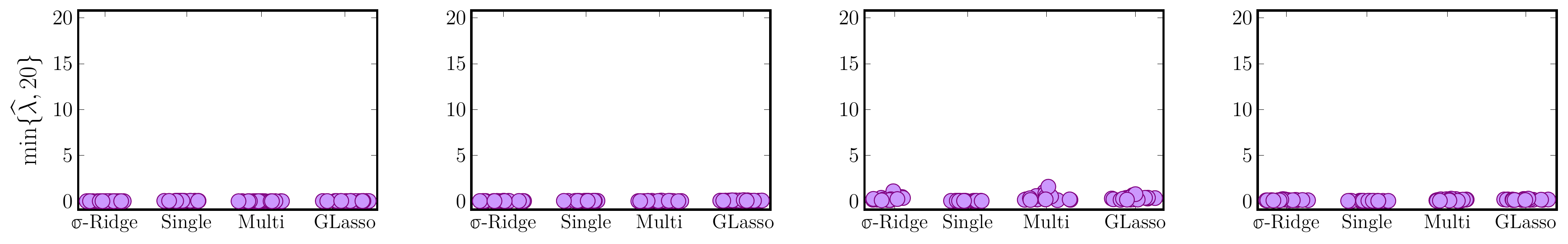}
\end{tabular}
\caption{\textbf{Data-driven penalization in the One Million Song dataset~\citep{Bertin-Mahieux2011}:} This is a companion to Figure~\ref{fig:million_songs_mse_time}. The rows correspond to the columns of Fig.~\ref{fig:million_songs_mse_time}, i.e., to the number of samples in the training data. The columns  correspond to the $K=4$ different feature groups and each panel shows the $\widehat{\lambda}_g$ applied by each method to each feature group and for each subsample.}
\label{fig:million_songs_lambdas}
\end{figure}

In Figure~\ref{fig:million_songs_lambdas} we show the data-driven regularization parameters $\widehat{\blambda}$ assigned by each of the four methods (for each training subsample) to the four feature groups. We first discuss the regularization parameters learned by $\sigmacv$-Ridge regression. We observe that across all subsamples, $\widehat{\lambda}_{\text{mean}}$ is almost $0$; and so the feature group $\group{\text{mean}}$ appears to be important for prediction. $\widehat{\lambda}_{\text{std}}$ on the other hand varies across subsample runs. It is typically larger for smaller training sets: the smaller $n$ is, the stronger the regularization. A similar trend is observed for $\group{\text{cor}}$. The features in $\group{\text{cov}}$ appear to be less important and are regularized substantially also for $n=1000$. The trend for the data-driven choices of $\widehat{\blambda}$ is similar for Multi Ridge and Group Lasso. Single Ridge struggles at $n \in \cb{500,1000}$ as it is forced to penalize the informative features $\group{\text{mean}}$ so as to control overfitting on features in the other groups. In contrast, all other methods leverage the grouping side-information and so do not face this difficulty.

\section{Discussion}
\label{sec:discussion}
In this paper we have presented an end-to-end approach for implementing group-regularized ridge regression in high dimensions, that is both practical and supported theoretically. Side-information in regression settings has become ubiquitous in modern applications with large-scale datasets, and its importance is only going to grow. We hope that our work will spur further methodological and theoretical developments beyond model~\eqref{eq:groupridge}, for example to classification settings, to other penalties and to more general forms of side-information. Recent theoretical advances  in understanding high-dimensional regression and classification~\citep{montanari2019generalization, celentano2019fundamental, liang2020precise, taheri2020sharp} could be instrumental in such an effort. Another avenue of research is the development of ``wrapper'' methods that enable the utilization of side-information by black-box supervised learning methods. For example,~\citet{ren2020knockoffs} show that  feature selection based on a large class of feature importance statistics can be enhanced by accounting for side-information.

From a broader methodological perspective, our conceptual approach is the following: we use a model-based/empirical Bayes approach to capture key aspects of the data generating mechanism in a flexible way (the system \eqref{eq:mapping}); but then we calibrate the result by tuning a 1-dimensional parameter based on a frequentist criterion; in this case the leave-one-out cross-validation error. This paradigm --flexible modeling plus calibration of a 1-dimensional parameter based on frequentist criteria-- has proven fruitful for statistical applications beyond the group-regularized ridge regression problem considered here, e.g., feature selection in regression settings via knockoffs~\citep{candes2018panning, ren2020knockoffs},  multiple testing with side-information \citep{ignatiadis2016data, lei2018adapt, ignatiadis2018covariate}, empirical Bayes shrinkage with side-information~\citep{tan2016steinized, ignatiadis2019ebayes} and conformal prediction~\citep{vovk2005algorithmic, gupta2019nested}.

\section*{Software}
A software package implementing the method is available on Github under the link \url{https://github.com/nignatiadis/SigmaRidgeRegression.jl}. The package has been implemented in the Julia programming language~\citep{bezanson2017julia} and uses the MLJ~\citep{blaom2020mlj} interface for supervised learning. The Github repository also provides code to reproduce all numerical results and plots in this manuscript.

\section*{Acknowledgements}
We thank Michael Celentano, Vaggos Chatziafratis, Iain Johnstone, Kenneth Tay, Stefan Wager and Lexing Ying for enlightening discussions and critical comments on the manuscript. We thank Emmanuel Candès for pointing us to literature on estimation of heritability.

\bibliographystyle{plainnat}
\bibliography{gridge}

\newpage

\appendix
\beginsupplement
\setcounter{page}{1}
\renewcommand{\thepage}{S\arabic{page}}

\section{Proofs for Section~\ref{sec:method}}

\begin{proof}[Proof of Proposition~\ref{prop:propo1}]
$ $\newline

\emph{1.} Note that we are minimizing $\Norm{\bA\bd(\sigmacv) - \bb(\sigmacv)}_2^2$, where $b_g(\sigmacv) = \widehat{u}_g/\sigmacv^2 - v_g$. In particular, when $\sigmacv^2 \geq \widehat{u}_g/v_g$ for all $g$, then $b_g(\sigmacv) \leq 0$ for all $g$. But since all entries of $\bA$ are $\geq 0$ and we optimize over $\bd(\sigmacv) \in [0,\infty)^K$, the lowest objective value must be attained when $\bd(\sigmacv) = \bm{0}$ and so all $\widehat{\lambda}_g(\sigmacv)=\infty$.

\emph{2.} Let us introduce dual variables $\bzeta(\sigmacv)\geq 0$, then the Lagrangian takes the form:
$$ \mathcal{L}(\bd(\sigmacv);\bzeta(\sigmacv)) = \Norm{\bA \bd(\sigmacv) - \bb(\sigmacv)}_2^2 - \bd(\sigmacv)^\intercal \bzeta(\sigmacv). $$ 
Minimizing with respect to $\bd(\sigmacv)$ we find that:
\begin{equation}
\label{eq:kktsystem}
\bA^\intercal \bA \widehat{\bd}(\sigmacv) = \bA^\intercal \bb(\sigmacv) + \bzeta(\sigmacv)/2 =: \boldeta(\sigmacv).
\end{equation}
Note that $b_g(\sigmacv) \to \infty$ as $\sigmacv \to 0$, and since all entries of $\bA$ are $\geq 0$ and $\bA$ is invertible, it also follows that $\eta_g(\sigmacv) \to \infty$ as $\sigmacv \to 0$. Now let $a^* >0$ be the largest entry of $\bA^\intercal \bA$, then it follows that (for all $k \in \cb{1,\dotsc,K}$):
$$ \max_{g}\cb{\widehat{d}_g(\sigmacv)} \cdot a^* \cdot K \geq \eta_k(\sigmacv).$$
Thus $\max_{g}\cb{\widehat{d}_g(\sigmacv)} \to \infty$ as $\sigmacv \to 0$, i.e., $\min_{g}\cb{\widehat{\lambda}_g(\sigmacv)} \to 0$. 

\emph{3.} Let $\mathcal{S} \subset \cb{1,\dotsc,K}$ be the subset of coordinates $g$ so that $\widehat{d}_g(\sigmacv) >0$ (and so $\widehat{\lambda}_g(\sigmacv) < \infty)$. Then, by the Karush-Kuhn-Tucker (KKT) conditions,~\eqref{eq:kktsystem} holds and furthermore (by complementary slackness) $\zeta_g(\sigmacv) = 0$ for all $g \in \mathcal{S}$. Hence, subsetting ~\eqref{eq:kktsystem} to $\mathcal{S}$ and letting $\bM_{\mathcal{S}} = \bA_{\cdot, \mathcal{S}}^\intercal \bA_{\cdot, \mathcal{S}}$ we get
$$ \bM_{\mathcal{S}}\widehat{\bd}_{\mathcal{S}}(\sigmacv) =  \bA_{\cdot, \mathcal{S}}^\intercal \bb(\sigmacv).$$
Now let $\sigmacv_1, \sigmacv_2>0$ have active set $\mathcal{S}$, then:
$$
\begin{aligned}
\bM_{\mathcal{S}}\widehat{\bd}_{\mathcal{S}}(\sigmacv_1) &= \bA_{\cdot, \mathcal{S}}^\intercal \p{\widehat{\bu}/\sigmacv_1^2 - \bv} \\
&= \frac{\sigmacv_2^2}{\sigmacv_1^2}\bA_{\cdot, \mathcal{S}}^\intercal \p{\widehat{\bu}/\sigmacv_2^2 - \bv} + \p{\frac{\sigmacv_2^2}{\sigmacv_1^2} -1} \bA_{\cdot, \mathcal{S}}^\intercal \bv \\
&= \frac{\sigmacv_2^2}{\sigmacv_1^2}\bM_{\mathcal{S}}\widehat{\bd}_{\mathcal{S}}(\sigmacv_2)  + \p{\frac{\sigmacv_2^2}{\sigmacv_1^2} -1} \bA_{\cdot, \mathcal{S}}^\intercal \bv.
\end{aligned}
$$
By multiplying with $\bM_{\mathcal{S}}^{-1}$ we conclude.

\end{proof}

\section{Proofs for Section~\ref{sec:risk_expressions}, Lemma~\ref{lem:risk_close_to_L}}
\label{sec:optimality_proofs}

We will need the following lemma which is adapted from Lemma 7.8, Lemma 7.9 and Lemma 7.10 from \citet{erdos2017dynamical}. This will be an essential ingredient in controlling the concentration of quadratic forms that we will encounter.

\begin{lemma}\label{lem1}
Let $q\geq 2$ and $X_1,\cdots,X_N,Y_1,\cdots,Y_N$ be independent random variables with mean 0, variance 1 and $2q$-th moment bounded by $c_0$. Then, for any deterministic $(b_i)_{1\leq i\leq N}, (a_{ij})_{1\leq i,j\leq N}$ we have for some positive constant $C_q=C_q(c_0)$:

\begin{equation}\label{eq:lem11}
\Norm{\sum_i b_i (X_i^2-1)}_q\leq C_q \p{\sum_i \abs{b_i}^2}^{\frac{1}{2}}\;,
\end{equation}
\begin{equation}\label{eq:lem12}
\Norm{\sum_{i,j}a_{ij}X_iY_j}_q \leq C_q \p{\sum_{i,j}a_{ij}^2}^{\frac{1}{2}}\;,
\end{equation}
\begin{equation}\label{eq:lem13}
\Norm{\sum_{i\neq j}a_{ij}X_iX_j}_q\leq C_q \p{\sum_{i\neq j}a_{ij}^2}^{\frac{1}{2}}\;,
\end{equation}
where $\Norm{\cdot}_q$ is the $L_q$ norm, i.e., $\Norm{U}_q = \EE{\abs{U}^q}^{1/q}$ for a random variable $U$. 
\end{lemma}

\begin{proof}[Proof of Lemma \ref{lem:risk_close_to_L}]

Assume that pointwise convergence holds (which we will prove later). We show that it can be extended to uniform convergence on a fixed compact set $C \subset (0,\infty)^K$. First,
\begin{equation}
\label{eq:w_decomp}
\begin{aligned}
\widehat{\bw} &=\left(\frac{\bX^\intercal \bX}{n}+\bLambda\right)^{-1}\frac{\bX^\intercal \bY}{n}\\
&=\bw-\left(\frac{\bX^\intercal \bX}{n}+\bLambda\right)^{-1}\bLambda \bw+\left(\frac{\bX^\intercal \bX}{n}+\bLambda\right)^{-1} \frac{\bX^\intercal \bvarepsilon}{n}\\
&=\bA\frac{\bvarepsilon}{\sqrt{n}}+\bw-\bB\bw, \text{ with }\bA=\left(\frac{\bX^\intercal \bX}{n}+\bLambda\right)^{-1}\frac{\bX^\intercal}{\sqrt{n}}, \;\; \bB=\left(\frac{\bX^\intercal \bX}{n}+\bLambda\right)^{-1}\bLambda.
\end{aligned}
\end{equation}
By the strong law of large numbers, we have that $\norm{\bw}^2\xrightarrow{a.s.}\sum_{i=1}^K\alpha_i^2<\infty.$ and $\frac{\norm{\bvarepsilon}^2}{n}\xrightarrow{a.s.}\sigma^2.$ In addition,  for $\blambda\in C$ it holds that
\begin{equation}\label{norm_B}\Norm{\bB}\leq 1,\end{equation}
where $\Norm{\cdot}$ is the operator norm and also that
$$\Norm{\bA}\leq \Norm{\frac{\bX}{\sqrt{n}}}\times \Norm{\p{\frac{\bX^\intercal \bX}{n}+\bLambda}^{-1}}\leq \Norm{\bSigma}^{\frac{1}{2}}\times \Norm{\frac{\bZ}{\sqrt{n}}}{\Norm{\bLambda^{-1}}}.$$
Using $\lim_{n \to \infty} \Norm{\bZ/\sqrt{n}}=1+\sqrt{\gamma}$~\citepsupplement{silver_bai_no_outside} we get that almost surely 
\begin{equation}\label{norm_A}
\limsup \max_{\blambda \in C}{\Norm{\bA}}<\infty.
\end{equation}
Combining~\eqref{eq:w_decomp},~\eqref{norm_B} and~\eqref{norm_A} we conclude that almost surely $$\limsup \max_{\blambda\in C}\norm{\widehat{\bw}(\blambda)-\bw}<\infty.$$ By a very similar argument we get that almost surely 
$$\limsup \max_{\blambda\in C}\norm{\nabla_{\blambda}\widehat{\bw}(\blambda)}<\infty.$$
Since $\Rn_n=\sigma^2+\p{\widehat{\bw}-\bw}^\intercal \bSigma \p{\widehat{\bw}-\bw},$ we have by the previous observations that 
$$\limsup \max_{\blambda\in C}\Norm{\nabla_{\blambda} \Rn_n(\blambda)}<\infty.$$ Hence, the sequence of functions $\{\Rn_n\}_{n\geq 1}$ is almost surely uniformly bounded and equicontinuous on $C$. Similarly for the sequence \smash{$\{\Ln_n\}_{n\geq 1},$} hence the difference $\Fn_n=\Rn_n-\Ln_n$ almost surely consists of bounded equicontinuous functions. Since \smash{$\Fn_n\xrightarrow{a.s.} 0$} on a countable dense subset, the only uniform subsequential limit of $\Fn_n$ can be the function 0. Due to the fact that almost surely any subsequence of $\{\Fn_n\}_{n\geq 1}$ has a uniformly convergent subsequence by the Arzela-Ascoli theorem, we conclude that it must almost surely converge uniformly to 0.

It remains to show pointwise convergence. With the same notation as above 
$$\p{\widehat{\bw}-\bw}^\intercal \bSigma \p{\widehat{\bw}-\bw}=\frac{\bvarepsilon^\intercal \bA^\intercal \bSigma \bA\bvarepsilon}{n}+\bw^\intercal \bB^\intercal \bSigma \bB \bw-2\frac{\bvarepsilon^\intercal \bA^\intercal \bSigma \bB\bw}{\sqrt{n}}.$$
Using Lemma~\ref{lem1} for $q=(4+\eta) / 2$ for the three quadratic forms in the expression above we see (details below) that:
$$\frac{\bvarepsilon^\intercal \bA^\intercal \bSigma \bA\bvarepsilon}{n}-\sigma^2 \frac{1}{n}\Tr\left(\frac{\bX^\intercal \bX}{n}\left(\frac{\bX^\intercal \bX}{n}+\bLambda\right)^{-1}\bSigma \left(\frac{\bX^\intercal \bX}{n}+\bLambda\right)^{-1}\right)\xrightarrow{a.s.}0$$
$$\frac{\bvarepsilon^\intercal \bA^\intercal \bSigma  \bB\bw}{\sqrt{n}}\xrightarrow{a.s.}0$$
$$\bw^\intercal \bB^\intercal \bB \bw-\frac{1}{p}\Tr\left(\left(\frac{\bX^\intercal \bX}{n}+\bLambda\right)^{-1}\bSigma \left(\frac{\bX^\intercal \bX}{n}+\bLambda\right)^{-1}\bLambda \bar{\bD} \bLambda \right)\xrightarrow{a.s.}0.$$
From the last three convergence results pointwise convergence of $\Rn_n-\Ln_n$ follows. The proof is completed, once we justify the three almost sure limits above. We provide the details for the second limit; the argument for the other two cases is almost identical. Using Lemma~\ref{lem1} after taking into account the variances of $\bvarepsilon, \bw$ we have $$\frac{\Norm{\bvarepsilon ^\intercal \bA^\intercal \bSigma \bB \bw}_q}{\sqrt{n}}=\mathcal{O}\left((pn)^{-\frac{1}{2}}\Norm{  \bA^\intercal \bSigma \bB }_F\right)=\mathcal{O}\left((pn)^{-\frac{1}{2}} \Norm{\bSigma}\Norm{\bB}\Norm{\bA}\sqrt{p}\right)=\mathcal{O}(n^{-\frac{1}{2}}).$$
It follows by Markov's inequality that, for any $\epsilon>0,$ $$\mathbb{P}\left(\abs{\frac{\bvarepsilon ^\intercal \bA^\intercal \bSigma \bB \bw}{\sqrt{n}}}>\epsilon\right)\leq \epsilon^{-q}\frac{\Norm{\bvarepsilon ^\intercal \bA^\intercal \bSigma \bB \bw}_q^q}{n^{\frac{q}{2}}}=\mathcal{O}\left(n^{-\frac{q}{2}}\right).$$ Since $q/2=1+\eta/4>1,$ by Borel-Cantelli we know that with probability 1 eventually $$\abs{\frac{\bvarepsilon ^\intercal \bA^\intercal \bSigma \bB \bw}{\sqrt{n}}} < \epsilon.$$ This proves the desired convergence, since $\epsilon$ was arbitrary.
\end{proof}

\section{Proofs for Section~\ref{sec:risk_expressions}: Risk formulae}

\subsection{Random Matrix Theory and Free Probability preliminaries}\label{sub:free_prob}

We provide a short review of some tools from free-probability that we are going to use in the proofs in the next subsection. For a comprehensive introduction to free-probability and proofs for the results that are mentioned here, the reader can refer to \citetsupplement{mingo_speicher} and \citetsupplement{free2}. In the last decade the emergence of random matrix theory in statistics has made free-probability methods very fruitful; see for example \citetsupplement{fan_sun} and \citetsupplement{fan}.

For a probability measure $\mu$ on the real line, we define the Stieltjes transform $m_{\mu}(z)=\int \frac{\mu(dx)}{x-z}$ for $z\in \mathbb{C}$ away from the support of $\mu$. This is a holomorphic function and $m_{\mu}(z)\in \mathbb{C}^+$ if and only if $z\in\mathbb{C}^+.$

Before we start with the free-probability tools that we need, we refer to the famous generalized Marcenko-Pastur ditribution. For a real symmetric matrix with eigenvalues $\sigma_1,...,\sigma_p$ (including multiplicity), the empirical spectral distribution is the probability measure on the real line defined as $\frac{1}{p}\sum_{j=1}^p \delta_{\sigma_j}$. For the empirical covariance matrix of i.i.d. random variables $x_1,\cdots,x_n$ with \smash{$x_i=\bSigma^{\frac{1}{2}}z_i,$} where $z_i$ has mean 0-variance 1 i.i.d. entries, $p,n\rightarrow\infty$ and $\frac{p}{n}\rightarrow\gamma>0$, it is well-known that the empirical spectral distribution has a weak limit if $\bSigma$ is either nonrandom or independent of $z_i$'s and has itself a limiting spectral distribution $H$. In particular, we have the following famous result:

\begin{theorem-non}[\citetsupplement{mp} and \citetsupplement{Silverstein}]
The empirical spectral distribution of the empirical covariance matrix converges almost surely to a deterministic measure with Stieltjes transform $m_{\gamma,H}$ that satisfies $$m_{\gamma,H}(z)=\int \frac{dH(t)}{t(1-\gamma-\gamma z m_{\gamma,H})-z}.$$
\end{theorem-non}
\noindent For the null case $H=\delta_1$ we get the standard Marcenko-Pastur distribution with parameter $\gamma$. This is a probability measure that has density 
$$ p_{\gamma}(x)=\frac{\sqrt{(b_{\gamma}-x)(x-a_{\gamma})}}{2\pi\gamma x}, a_{\gamma}<x<b_{\gamma}.$$
The support is given by $a_{\gamma}=(1-\sqrt{\gamma})^2,b_{\gamma}=(1+\sqrt{\gamma})^2.$ In the case of $\gamma\geq 1$ we get an extra mass of size $1-\frac{1}{\gamma}$ at 0.

If we know the Stieltjes transform of the limiting spectral distribution, the density (assuming it exists) can be recovered by the Stieltjes inversion formula: 
$$p(x)=\frac{1}{\pi}\lim_{u\downarrow 0}Im( m_{\gamma,H}(x+iu)).$$
Asymptotic freeness plays a major role for our proofs and we define it here following \citetsupplement{mingo_speicher} and \citetsupplement{tao}. We will use $\tau(W)=\frac{1}{N}\Tr(W)$ to denote the normalized trace of an $N\times N$ matrix $W$, $\overline{\tau}(W)=\mathbb{E}[\tau(W)]$.

\begin{definition}[Definition 2.5.18 in \citetsupplement{tao}]
Consider two sequences of $N\times N$ random matrices $(A_N)_{N\geq 1},(B_N)_{N\geq 1}$.  We call them asymptotically freely independent when for each $m\in \mathbb{N}$ and any polynomials $P_1,\cdots,P_m$ we have $$\overline{\tau}(\Pi_{i=1}^m(P_i(C_i)-\overline{\tau}(P_i(C_i))))\xrightarrow{N\rightarrow\infty} 0,$$ where $C_1,\cdots,C_m$ is any alternating choice of $A_N,B_N$. If the same convergence holds with $\tau$ instead of $\overline{\tau}$, i.e., if,
$$\tau(\Pi_{i=1}^m(P_i(C_i)-\overline{\tau}(P_i(C_i))))\xrightarrow{N\rightarrow\infty} 0, \text{ almost surely},$$ 
then we say that the sequences of matrices are almost surely asymptotically free.
\end{definition}

A well-known result from \citetsupplement{mingo_speicher} is the following. 

\begin{theorem-non} [Page 111 in \citetsupplement{mingo_speicher}]
Let $A_N$ and $B_N$ be two sequences of independent $N\times N$ matrices that converge almost surely in moments to some probability measures. Let $U_N$ be a matrix drawn independently from $A_N,B_N$ with respect to the Haar measure. Then, $A_N,U_N^{-1}B_NU_N$ are almost surely asymptotically free.
\end{theorem-non}

 If $\mu $ has compact support, it is well-known that there exists a function $R_{\mu}$, called the $R$-transform, such that $R_{\mu}(z)+z^{-1}$ is holomorphic in some open set containing $0$ for which $m_{\mu}(R_{\mu}(z)+\frac{1}{z})=-z$. We will use the notation $B_{\mu}(z)=R_{\mu}(z)+\frac{1}{z}.$ We have the following result from \citetsupplement{mingo_speicher}.

\begin{theorem-non} [Page 51 in \citetsupplement{mingo_speicher}]
Let $A,B$ be independent symmetric $N\times N$ random matrices and $U$ an independent random matrix which is uniformly distributed with respect to the Haar measure on $\mathbb{R}^{N\times N}$. If the empirical spectral distributions of $A,B$ converge to deterministic probability measures $\mu,\nu$ respectively, then the empirical spectral distribution of the matrix $A+U^\intercal BU$ converges almost surely to a measure with $R$-transform $R(z)=R_{\mu}(z)+R_{\nu}(z)$. This is known as the free additive convolution of $\mu,\nu$.
\end{theorem-non}

\subsection{Proofs of the risk formulae}
Throughout this section, we assume without loss of generality that the features are indexed so that $\group{1} = \cb{1,\dotsc,p_1}$, $\group{2} = \cb{p_1 +1 \dotsc,\; p_1 + p_2}$ and so forth. Second, given a $\blambda = (\lambda_1, \dotsc, \lambda_K) \in [0,\infty]^K$ we will write $\bLambda$ for the diagonal $p \times p$ matrix with diagonal 
$$(\underbrace{\lambda_1, \dotsc,\lambda_1}_{p_1 \text{ times}},  \lambda_2, \dotsc, \lambda_{K-1}, \underbrace{\lambda_K, \dotsc, \lambda_K}_{p_K \text{ times}}).$$
Using Lemma~\ref{lem:risk_close_to_L}, it suffices to study the asymptotics of $\Ln(\blambda)$ and so it suffices to study:
\begin{equation}
\label{eq:ln_def_supplement}
\begin{aligned}
1&+\frac{1}{n}\Tr\left(\frac{\bX^\intercal  \bX}{n}\left(\frac{\bX^\intercal \bX}{n}+\bLambda\right)^{-1}\bSigma \left(\frac{\bX^\intercal \bX}{n}+\bLambda\right)^{-1}\right)\\
&+\frac{1}{p}\Tr\left(\left(\frac{\bX^\intercal \bX}{n}+\bLambda\right)^{-1}
\bSigma \left(\frac{\bX^\intercal \bX}{n}+\bLambda\right)^{-1}
\bLambda \bar{\bD}\bLambda\right).
\end{aligned}
\end{equation}
It is convenient to define the following quantities. First, we provide explicit notation for the block-diagonal and low-rank components in the decomposition of $\bSigma$:
$$\bSigma_B= \text{diag}(\Tilde{\bSigma}_1, \dotsc, \Tilde{\bSigma}_K),\;\;\; \bSigma_{L} = \bSigma-\bSigma_B.$$
We also define:
$$\bM=\p{\frac{\bZ^\intercal \bZ}{n}+\bSigma^{-\frac{1}{2}}\bLambda \bSigma^{-\frac{1}{2}}}^{-1},\,\,\bM_B=\p{\frac{\bZ^\intercal \bZ}{n}+\bSigma_B^{-\frac{1}{2}}\bLambda \bSigma_B^{-\frac{1}{2}}}^{-1}.$$
We study the two non-constant terms in~\eqref{eq:ln_def_supplement} separately. The first term can be rewritten as
\begin{equation*}
\begin{aligned}
&\frac{1}{n}\Tr\p{\p{\frac{\bX^\intercal \bX}{n}+\bLambda}^{-1}\bSigma}- \frac{1}{n}\Tr\p{\p{\frac{\bX^\intercal \bX}{n}+\bLambda}^{-1}\bSigma\p{\frac{\bX^\intercal \bX}{n}+\bLambda}^{-1}\bLambda}\\
=\;&\frac{1}{n}\Tr\p{\bSigma^{-\frac{1}{2}}{\bM}\bSigma^{-\frac{1}{2}}\bSigma}-\frac{1}{n}\Tr\p{\bSigma^{-\frac{1}{2}}{\bM}\bSigma^{-\frac{1}{2}}\bSigma\bSigma^{-\frac{1}{2}}{\bM}\bSigma^{-\frac{1}{2}}\bLambda}\\
=\;&\frac{1}{n}\Tr({\bM})- \frac{1}{n}\Tr\p{{\bM}^2\bSigma^{-\frac{1}{2}}\bLambda\bSigma^{-\frac{1}{2}}}.
\end{aligned}
\end{equation*}
$\bSigma_B^{-\frac{1}{2}}$ is a finite rank perturbation of $\bSigma^{-\frac{1}{2}}$ and similarly,  $\bM_B$ is a finite rank perturbation of $\bM$ (since it is true for their inverses). Furthermore, $\bSigma_B^{-\frac{1}{2}},\bSigma^{-\frac{1}{2}}$ and $\bM_B, \bM$ have (almost surely) $\mathcal{O}(1)$ operator norm. Thus, for a fixed $\bLambda$ we have
\begin{equation}
\label{eq:part_1_low_rank}
\begin{aligned}
&\frac{1}{n}\Tr(\bM)-\frac{1}{n}\Tr\p{\bM^2\bSigma^{-\frac{1}{2}}\bLambda\bSigma^{-\frac{1}{2}}}\\
=\;& \frac{1}{n}\Tr(\bM_B)- \frac{1}{n}\Tr\p{\bM_B^2\bSigma^{-\frac{1}{2}}\bLambda\bSigma^{-\frac{1}{2}}}+\mathcal{O}(n^{-1})\\
=\;& \frac{1}{n}\Tr(\bM_B)- \frac{1}{n}\Tr\p{\bM_B^2\bSigma_B^{-\frac{1}{2}}\bLambda\bSigma_B^{-\frac{1}{2}}}+\mathcal{O}(n^{-1}).
\end{aligned}
\end{equation}
We turn to second non-constant term in~\eqref{eq:ln_def_supplement}. It can be rewritten as 
\begin{equation*}
\frac{1}{p}\Tr\p{\bSigma^{-\frac{1}{2}}\bM \bSigma^{-\frac{1}{2}} \bSigma \bSigma^{-\frac{1}{2}} \bM \bSigma^{-\frac{1}{2}}\bLambda \bar{\bD} \bLambda}=\frac{1}{p}\Tr\p{\bSigma^{-\frac{1}{2}}\bM^2\bSigma^{-\frac{1}{2}}\bLambda  \bar{\bD} \bLambda}.
\end{equation*}
Arguing by the low rank approximations, we find that:
\begin{equation}
\label{eq:part_2_low_rank}
\frac{1}{p}\Tr\p{\bSigma^{-\frac{1}{2}}\bM^2\bSigma^{-\frac{1}{2}}\bLambda  \bar{\bD} \bLambda} = \frac{1}{p}\Tr\p{\bSigma_B^{-\frac{1}{2}}\bM_B^2\bSigma_B^{-\frac{1}{2}}\bLambda\bar{\bD} \bLambda}+\mathcal{O}(p^{-1}).
\end{equation}
The upshot of~\eqref{eq:part_1_low_rank} and~\eqref{eq:part_2_low_rank} is that the  risk asymptotics are independent of the exact form of the low-rank difference $\bSigma_L$. Hence as we move on with our proof, we are going to assume without loss of generality that $\bSigma$ is block-diagonal, i.e., that $\tilde{\bSigma}_g = \bSigma_g$ for all groups $g$ and $\bSigma = \text{diag}(\bSigma_1, \dotsc, \bSigma_K)$. We will also drop the subscript $B$ in $\bM_B$ and $\bSigma_B$ for the rest of the proof (since $\bSigma_B =\bSigma$ and $\bM = \bM_B$ under our new assumptions).

Hence, we now continue the study of the two non-constant terms in~\eqref{eq:ln_def_supplement} making use of the aforementioned simplification. It will be useful to first note that the matrices $\bar{\bD},\bLambda,\bSigma$ commute with each other. In addition, we may verify that,
\begin{equation}
\partial_j \bM=-\bM \begin{pmatrix}
0_{p_1}\\
&\cdots\\
&&\!\! \bSigma_j^{-1}\\
&&&0_{p_K}
\end{pmatrix} \bM.
\end{equation}
By using the above we can further simplify the first non-constant term in~\eqref{eq:ln_def_supplement} as:
$$\Tr\p{\bM^2\bSigma^{-1}\bLambda}=\Tr\p{\bM\bSigma^{-1}\bLambda\bM}=-\sum_{j=1}^K\lambda_j \Tr\p{\partial_j\bM}.$$
Noting that replacing $n^{-1}$ by $\gamma/p$ only changes an $\omicron(1)$ term, we finally see that the term can be rewritten as
$$\gamma \frac{1}{p}\Tr\p{\bM+\sum_{j=1}^K \lambda_j \partial_j\bM} + \omicron(1).$$
We now turn to the second non-constant term in~\eqref{eq:ln_def_supplement}, which is equal to:
$$\frac{1}{p}\Tr\p{\bM^2 \bSigma^{-1} \bLambda^2\bar{\bD}}=\frac{1}{p}\Tr\p{\bM\bSigma^{-1}\bLambda^2\bar{\bD} \bM}=-\frac{1}{p}\Tr\p{\sum_{j=1}^K\frac{\gamma}{\gamma_j}\alpha_j^2\lambda_j^2\partial_{j}\bM} + \omicron(1).$$
We conclude that the risk can be approximated (up to $\omicron(1)$ terms) by
\begin{equation}
\label{eq:risk_before_free_prob}
1+\frac{1}{p}\Tr\p{\gamma \bM+\sum_{j=1}^K \frac{\gamma}{\gamma_j}(\gamma_j\lambda_j-\alpha_j^2\lambda_j^2)\partial_j \bM}.
\end{equation}
To complete the proof we will characterize the limit of~\eqref{eq:risk_before_free_prob} more precisely using the following proposition as an intermediate step:
\begin{proposition}\label{prop:risk_prop}
We have for $z\in \mathbb{C}-\mathbb{R}$ 
$$\frac{1}{p}\Tr\p{\frac{\bZ^\intercal \bZ}{n}+\bSigma^{-\frac{1}{2}}\bLambda \bSigma^{-\frac{1}{2}}-z\bI}^{-1}\rightarrow m(z),$$
where $$m(z)=\sum_{j=1}^K\frac{\gamma_j}{\gamma}\int \p{\frac{\lambda_j}{t}-z+\frac{1}{1+\gamma m(z)}}^{-1}dH_{j}(t)$$ and $\Img(z)\Img(m(z))>0.$ In other words, the empirical spectral distribution of $$\frac{\bZ^\intercal \bZ}{n}+\bSigma^{-\frac{1}{2}}\bLambda \bSigma^{-\frac{1}{2}}$$ converges to a probability measure $\mu$ with Stieltjes transform $m(z)=\int\frac{1}{x-z} \mu(dx)$ described by the equation above.

\end{proposition}

\begin{proof}

We know that the empirical spectral distribution of $\bZ^\intercal \bZ/n$ converges almost surely to the Marcenko-Pastur distribution with parameter $\gamma$, which has a Stieltjes transform $m_{\gamma}(z)$ that satisfies $$m_{\gamma}(z)=\frac{1}{1-\gamma-z-\gamma zm_{\gamma}(z)}.$$ 
Solving $m_{\gamma}(B_1(z))=-z$ gives $B_1(z)=\frac{1}{z}+\frac{1}{1-\gamma z}$, so the R-transform of the Marcenko-Pastur distribution is $R_1(z)=1/(1-\gamma z)$.
From our assumptions, it also follows that the empirical spectral distribution of $\bSigma^{-\frac{1}{2}}\bLambda\bSigma^{-\frac{1}{2}}$ also converges to a deterministic measure which we call $\nu$. If $U$ is uniformly distributed with respect to the Haar measure, then $\frac{\bZ^\intercal \bZ}{n}$ has the same distribution as $U\frac{\bZ^\intercal \bZ}{n}U^\intercal$. We see that the empirical spectral distribution of $$\frac{\bZ^\intercal \bZ}{n}+\bSigma^{-\frac{1}{2}}\bLambda \bSigma^{-\frac{1}{2}}$$ converges almost surely to a measure with R-transform 
$$R(z)=R_1(z)+R_{\nu}(z)=\frac{1}{1-\gamma z}+R_{\nu}(z)\; \Longrightarrow \; B(z)=\frac{1}{1-\gamma z}+B_{\nu}(z).$$
If $m$ is the Stieltjes transform of the limit, we get that $z= 1/(1+\gamma m(z))+B_{\nu}(-m(z))$ and so by change of variables, we get:
$$m(z)=m_{\nu}\p{z-\frac{1}{1+\gamma m(z)}}=\sum_{j=1}^K\frac{\gamma_j}{\gamma}\int \p{\frac{\lambda_j}{t}-z+\frac{1}{1+\gamma m(z)}}^{-1}dH_{j}(t).$$ Detailed derivations of the Stieltjes transform equations for additive free  convolutions can be found in \citetsupplement{capitaine2016spectrum} (see, for example, page 17 for the case of additive free convolution with a Marcenko-Pastur distribution).

\end{proof}

\begin{proof}[Proof of Theorem \ref{thm:main_risk}]

As a first consequence of Proposition \ref{prop:risk_prop}, we observe that by taking $z \to 0$, it follows that
\begin{equation}
\label{eq:trBm_at_0}
\frac{1}{p}\Tr(\bM)\xrightarrow{a.s.} m(0).
\end{equation}
Here we used the fact that the spectrum of $\frac{\bZ^\intercal \bZ}{n}+\bSigma^{-\frac{1}{2}}\bLambda \bSigma^{-\frac{1}{2}}$ is bounded away from zero. In the statement of our theorem, $m(0)$ will play the role of $f = f(\blambda) = m(0)$. Writing $\bM$ also as a function of $\blambda$, i.e., $\bM = \bM(\blambda) = \bM(\lambda_1,\dotsc,\lambda_K)$, we thus have proven convergence
\begin{equation}
\label{eq:trBmlambda_at_0}
\frac{1}{p}\Tr(\bM(\blambda))\xrightarrow{a.s.} f(\blambda).
\end{equation}
To conclude, it suffices to show that
\begin{equation}
\label{eq:trBmlambda_deriv_at_0}
\frac{1}{p}\Tr(\partial_{1}\bM(\blambda))\xrightarrow{a.s.} \partial_{1}f(\blambda),
\end{equation}
and similarly for the other groups.~\eqref{eq:trBmlambda_at_0}, however does not directly imply~\eqref{eq:trBmlambda_deriv_at_0}, since a sequence of functions can converge almost surely to a deterministic function, but the sequence of derivatives may not converge. We now work to circumvent this problem. We prove the convergence for the partial derivative with respect to $\lambda_1$ and similarly one can prove the convergence of the other partial derivatives.

We claim the following slight extension of~\eqref{eq:trBmlambda_at_0}, which we will verify later.

\paragraph{Claim:} For $z\in\mathbb{C}$ in a neighborhood of the fixed $\lambda_1>0$, it holds that almost surely $h_{p}(z)=\frac{1}{p}\Tr(\bM(z,\lambda_2,\cdots,\lambda_k))$ converges to a function $h_{\infty}(z)$. $h_{\infty}(z)$ depends on $\lambda_2,\cdots,\lambda_k$, but for notational simplicity we suppress this dependency.

\bigskip

Now, let $\Gamma$ be a circle contained in that neighborhood of $\lambda_1$. Considering a countable dense subset $\Tilde{\Gamma}$ of $\Gamma$, then we know that almost surely $h_{p}(z)\rightarrow h_{\infty}(z)$ for all $z\in\Tilde{\Gamma}$.

Next, observe that 
$$
\begin{aligned}
\abs{h_p'(z)}&\leq \Norm{\bM(z,\lambda_2,\cdots,\lambda_K)\begin{pmatrix}
\bSigma_1^{-1}\\
&0_{p-p_1}\\
\end{pmatrix}\bM(z,\lambda_2,\cdots,\lambda_K)}\\
&\leq \norm{\bSigma_1^{-1}}\norm{\bM(z,\lambda_2,\cdots,\lambda_K)}^2.
\end{aligned}
$$
In words, $\abs{h_p'(\cdot)}$ is uniformly bounded for $z\in\mathbb{C}$ close to $\lambda_1$, say by some constant $\Tilde{C}$ independent of $p$. We conclude that $$\abs{h_p(z_2)-h_p(z_1)}\leq \int_0^1 \abs{h_p'(tz_2+(1-t)z_1)} d t\abs{z_2-z_1}\leq \Tilde{C}\abs{z_2-z_1}$$
for $z_1,z_2$ in a neighborhood of $\lambda_1.$ As a result, the functions $h_p(z)$ are uniformly bounded and equicontinuous. Thus, $h_{p}$ has a subsequence that converges uniformly by the Arzela-Ascoli theorem. Since $h_p$ are holomorphic, the limit is also holomorphic and agrees with $h_{\infty}(z)$ on $\Tilde{\Gamma},$ and so the limit has to be same almost surely for any subsequence. Hence, almost surely $h_{p}\rightarrow h_{\infty}$ uniformly on $\Gamma$ and $h_{\infty}$ is holomorphic.

We know by the Cauchy integral formula that for all $z$ in the interior of $\Gamma$ $$h_p'(z)=\frac{1}{2\pi i}\oint_{\Gamma}\frac{h_p(w)}{(w-z)^2}dw.$$ As a consequence, we almost surely have in some smaller open neighborhood $U$ of $\lambda_1>0$ that simultaneously $h_{p}(z)\rightarrow h_{\infty}(z),h_p'(z)\rightarrow h'_{\infty}(z)$ for all $z\in U$. Thus, on the real line where \smash{$h_{\infty}(\lambda_1)=f(\lambda_1,\dotsc,\lambda_K)$}, we have \smash{$\partial_{\lambda_1}\frac{1}{p}\Tr(\bM(\lambda_1,\cdots,\lambda_K))\xrightarrow{a.s.}\partial_{\lambda_1}f(\lambda_1,\dotsc,\lambda_K).$}

It remains to verify our claim, namely that 
\smash{$\frac{1}{p}\Tr(\bM(z,\lambda_2,\cdots,\lambda_K))$} converges almost surely if $z$ is close to $\lambda_1$.  To this end, we define $$\bLambda_z=
\begin{pmatrix}
z \bI_{p_1}\\
&\!\!\lambda_2\bI_{p_2}\\
&&\cdots\\
&&&\lambda_K \bI_{p_K}
\end{pmatrix}=\begin{pmatrix}
z \bI_{p_1}\\
&\underline{\bLambda}\\
\end{pmatrix}.$$

We next claim that for $z$ close to $\lambda_1>0$ the eigenvalues of \smash{$\bZ^\intercal \bZ/n+\bSigma^{-\frac{1}{2}}\bLambda_z\bSigma^{-\frac{1}{2}}$} all lie eventually in a fixed compact subset of $\{w\in \mathbb{C}:Re(w)>0\}$. To see why, notice that the operator norm is bounded from above by \smash{$\norm{\frac{\bZ^\intercal \bZ}{n}}+\norm{\bSigma^{-1}}\norm{\bLambda_z}$} and \smash{$\limsup{\norm{\frac{\bZ^\intercal \bZ}{n}}}=(1+\sqrt{\gamma})^2$} almost surely by a well-known result of \citetsupplement{silver_choi}. The lower bound is straightforward for $z$ close to $\lambda_1.$ As a result, we can find for any $\delta>0$ a polynomial with complex coefficients such that \smash{$\abs{P_{\delta}(\sigma)-\frac{1}{\sigma}}\leq \delta$} for any eigenvalue $\sigma$ of \smash{$\bZ^\intercal \bZ/n+\bSigma^{-\frac{1}{2}}\bLambda_z\bSigma^{-\frac{1}{2}}$}. Furthermore, if we put the matrix \smash{$\bZ^\intercal \bZ/n+\bSigma^{-\frac{1}{2}}\bLambda_z\bSigma^{-\frac{1}{2}}$} in upper triangular form we see that almost surely for all large $p$ we have $$\frac{1}{p}\abs{\Tr\p{\frac{\bZ^\intercal \bZ}{n}+\bSigma^{-\frac{1}{2}}\bLambda_z\bSigma^{-\frac{1}{2}}}^{-1} -\Tr\p{P_{\delta}\p{\frac{\bZ^\intercal \bZ}{n}+\bSigma^{-\frac{1}{2}}\bLambda_z\bSigma^{-\frac{1}{2}}}}}\leq \delta.$$ Since $\delta>0$ was arbitrary, it is enough to show that $\Tr(P_{\delta}(\bZ^\intercal \bZ/n+\bSigma^{-\frac{1}{2}}\bLambda_z\bSigma^{-\frac{1}{2}}))/p$ converges almost surely. Thus, it is enough to show that $\Tr((\bZ^\intercal \bZ/n+\bSigma^{-\frac{1}{2}}\bLambda_z\bSigma^{-\frac{1}{2}})^j)/p$ converges almost surely for any $j\in\mathbb{N}$. This follows by free independence, since the matrix $n^{-1}{\bZ^\intercal \bZ}$ is invariant in law under conjugation by orthogonal matrices.
\end{proof}

\begin{proof}[Proof of Corollary~\ref{corol:opt_risk}]
We notice the universality of our results, namely that the limits do not depend on the distribution of $w_j$ and $\varepsilon_i$ (as long as the moment bounds hold). As a consequence, this implies that the Bayes optimal parameters $p_g/(n\alpha_g^2),$ which give the $\widehat{\bw}$ that is the posterior mean of $\bw$ in the case of jointly Gaussian $w_j$ and $\varepsilon_i$, is optimal asymptotically more generally. By equicontinuity of $\Ln(\blambda)$ in $\blambda$ we may replace $p_g/(n\alpha_g^2)$ by $\lambda_g^* = \gamma_g/\alpha_g^2$ to arrive at the asymptotically optimal choice of regularization parameters. For this choice of optimal parameters the risk is asymptotically equal to \smash{$1+\gamma f(\frac{\gamma_1}{\alpha_1^2},\dotsc,\frac{\gamma_k}{\alpha_k^2}).$} This is because for the optimal choice of parameters, the sum in~\eqref{eq:outofsample_risk} cancels out.
\end{proof}

\begin{proof}[Proof of Corollary \ref{corol:single_lam_opt}]

Let $g(\lambda)=f(\lambda,\lambda,\dotsc,\lambda)$ and call $H$ the limiting spectral distribution of all groups. Then, $$g'(\lambda)=-\int \frac{tdH(t)}{(\lambda+\frac{t}{1+\gamma g})^2}+\gamma g'(\lambda)\int \frac{dH(t)}{(1+\lambda \frac{1+\gamma g}{t})^2}.$$
We set 
$$T_1(\lambda)=\int \frac{tdH(t)}{(\lambda+\frac{t}{1+\gamma g})^2},\;\;\; T_2(\lambda)=-1+\gamma \int \frac{dH(t)}{(1+\lambda \frac{1+\gamma g}{t})^2}.$$
Then, $g'(\lambda)=T_1(\lambda)T_2(\lambda)^{-1}.$ For the partial derivatives $\partial_{i}f(\lambda,\lambda,\dotsc,\lambda)$ we find $$\partial_i f(\lambda,\lambda,\cdots,\lambda)=-\frac{\gamma_i}{\gamma}\int \frac{tdH(t)}{(\lambda+\frac{t}{1+\gamma g})^2}+\partial_i f(\lambda,\lambda,\dotsc,\lambda)\sum_{j=1}^K\gamma_j \int \frac{ dH(t)}{(1+\lambda\frac{1+\gamma g}{t})^2},$$
and so $$\partial_i f(\lambda,\lambda,\dotsc,\lambda)=\frac{\gamma_i}{\gamma}T_1(\lambda)T_2^{-1}(\lambda)=\frac{\gamma_i}{\gamma}g'(\lambda).$$ This implies that the risk is asymptotically $$1+\gamma g(\lambda)+\sum_{i=1}^K(\gamma_i\lambda-\alpha_i^2\lambda^2)g'(\lambda)=1+\gamma g(\lambda)+(\gamma \lambda-\alpha^2\lambda^2)g'(\lambda),$$ where $\alpha^2=\alpha_1^2+\dotsc+\alpha_K^2.$ As a result, the risk in this case is equal to the risk of ridge regression in a model with only 1 group and variance of each individual weight equal to $\alpha^2p^{-1}.$ The value of $\lambda$ that minimizes this is $\lambda^*=\frac{\gamma}{\alpha^2}$ by Corollary~\ref{corol:opt_risk}. 
\end{proof}

\begin{proof}[Proof of Corollary \ref{corol:risk_singla_lambda}]

Let $u=(1+\gamma f)^{-1}.$ Then, $$\frac{1}{u}=1+\sum_{j=1}^K\gamma_j\frac{1}{\lambda_j+u}\;\Longrightarrow\; \frac{\partial_{\lambda_1}u}{u^2}=\sum_{j=1}^K\frac{\gamma_j\partial_{\lambda_1}u}{(\lambda_j+u)^2}+\frac{\gamma_1}{(\lambda_1+u)^2}.$$ For $\lambda_1=\dotsc=\lambda_K=\lambda$ we get (suppressing the dependency of $u$ on $\lambda$ in  the equations below)
$$\partial_{\lambda_1}u(\lambda,\cdots,\lambda)=\frac{\gamma_1 u^2}{(\lambda+u)^2-\gamma u^2}$$
$$\Longrightarrow \; \partial_{\lambda_1}f(\lambda,\cdots,\lambda)=-\frac{1}{\gamma}\frac{\partial_{\lambda_1}u(\lambda,\cdots,\lambda)}{u^2}=\frac{\gamma_1}{\gamma}\frac{1}{\gamma u^2-(\lambda+u)^2}.$$
Upon repeating this argument for all groups $g$, Theorem~\ref{thm:main_risk} implies that the asymptotic risk is $$\frac{1}{u}-\frac{\gamma \lambda -(\sum_{j=1}^K\alpha_j^2)\lambda^2}{(\lambda+u)^2-\gamma u^2}.$$ 
We next solve $$\frac{1}{u}=1+\frac{\gamma}{\lambda+u}\; \Longrightarrow \; u=\frac{1-\gamma-\lambda+\sqrt{(\lambda+\gamma-1)^2+4\lambda}}{2}.$$ 
Finally, applying Corollaries~\ref{corol:opt_risk} and~\ref{corol:single_lam_opt} we find that the optimal risk (based on a single $\lambda$) is $$\frac{\gamma+\lambda^*-1+\sqrt{(\gamma+\lambda^*-1)^2+4\lambda^*}}{2\lambda^*}.$$ 
\end{proof}

\begin{proof}[Proof of Corollary \ref{corol:more_features}]
Since most of the work has already been done, we sketch the proof only. The problem if we ignore the second group becomes equivalent to our original problem, but with only one group and residual variance $1+\alpha_2^2.$ Thus, the Bayes optimal parameter in this scenario becomes $\Tilde{\lambda}=\gamma_1(\alpha_2^2+1)/\alpha_1^2$ which gives asymptotically the posterior mean for any $n$. In other words, the problem is the same as in the case $K=1$, but the error has to be rescaled by $1+\alpha_2^2$ and the variance of the weights has to be divided by that number exactly to compensate for that. The result then follows by Corollary~\ref{corol:risk_singla_lambda}.
\end{proof}

We now also prove the two remarks following the statement of Corollary~\ref{corol:risk_singla_lambda}. For both remarks, $\alpha_1\rightarrow\infty$, so $\tilde{\lambda},\lambda^*\rightarrow 0$.

We consider the first remark now, i.e., $\gamma > 1$. We also assume that $\gamma_1>1$ (the cases $\gamma_1=1$ and $\gamma_1 < 1$ being similar). The optimal ridge risk with both groups and a single parameters grows as $(\gamma-1)/\lambda^*,$ while the optimal risk of ridge with a single group grows as $(\alpha_2^2+1)(\gamma_1-1)/\tilde{\lambda}.$ We find $$\frac{\frac{\gamma-1}{\lambda^*}}{(\alpha_2^2+1)\frac{\gamma_1-1}{\tilde{\lambda}}}\rightarrow \frac{1-\gamma^{-1}}{1-\gamma_1^{-1}}> 1.$$ 
Now consider the case $\gamma < 1$. If we use both groups, then the optimal risk converges to $1/(1-\gamma)$, while with only the first group, the optimal risk converges to $(\alpha_2^2+1)(1-\gamma_1)$. Now we solve
$$ \frac{\alpha_2^2+1}{1-\gamma_1}\leq \frac{1}{1-\gamma}\;\Longrightarrow\; \alpha_2^2\leq \frac{\gamma_2}{1-\gamma}.$$
If $\alpha_2^2$ exceeds the threshold \smash{$\gamma_2/(1-\gamma),$} then the performance in the case that we know the values of the predictors in the second group is enhanced by including them, even if we use a single regularization parameter.

\subsection{Analysis of the Main Equation}\label{subs:main_eq_analysis}
In this section we study the main equation of Theorem~\ref{thm:main_risk}. Let,
$$P(f)=\sum_{j=1}^K\frac{\gamma_j}{\gamma}\int \p{\frac{\lambda_j}{t}+\frac{1}{1+\gamma f}}^{-1}d H_j(t),$$
then we are trying to solve $P(f)=f$. Numerically, this problem can be solved by bisection. Alternatively, a fixed point algorithm can be used to find $f$, as we now explain. These algorithms also prove constructively that $P(f)=f$ indeed has a root.

We first consider the case $\gamma<1$ separately. Then we initialize $f$ arbitrarily, say $f_0=0$, and iteratively set $f_{m+1}=P(f_m)$ until convergence. Observe that for $f\geq 0$ $$P'(f)=\sum_{j=1}^K\frac{\gamma_j}{\gamma}\int \gamma \p{\frac{\lambda_j}{t}+\frac{1}{1+\gamma f}}^{-2}(1+\gamma f)^{-2}dH_j(t)\leq \gamma \sum_{j=1}^K\frac{\gamma_j}{\gamma} \int dH_j(t)=\gamma<1.$$ We conclude that for $\gamma<1$ the function $P$ is a contraction. As a result, the fixed point algorithm converges to the unique solution in $[0,\infty)$ and in fact $\abs{f_m-f}=\mathcal{O}(\gamma^m)$. 
  
Next, we consider a general $\gamma$. We define $u=\frac{1}{1+\gamma f}$. Then, $$\frac{1}{u}=1+\sum_{j=1}^K\gamma_j\int \p{\frac{\lambda_j}{t}+u}^{-1} d H_j(t) \; \Longrightarrow \;  1=u+\sum_{j=1}^K\gamma_j \int \frac{u}{\frac{\lambda_j}{t}+u} d H_j(t).$$
The function on the right hand side of the equation is strictly increasing in $u\geq 0$, starting at 0 and going to $\infty$ as $u\rightarrow\infty.$ Thus, there exists a unique such $u^*$, hence a unique solution $f\geq 0$. In addition, we have $$u=\frac{1}{1+\sum_{j=1}^K\gamma_j\int (\frac{\lambda_j}{t}+u)^{-1} d H_j(t)}=G(u).$$
Observe that $G$ is strictly increasing in $u\geq 0$. Thus, if we initialize $u_0$ arbitrarily and iteratively define $u_{m+1}=G(u_m)$. If $u_0<u^*$, then we can prove inductively that $u_m<u_{m+1}<u^*$. To see why, we compute $$\frac{u_m}{u_{m+1}}=u_m+\sum_{j=1}^K\gamma_j \int \frac{u_m}{\frac{\lambda_j}{t}+u_m} d H_j(t)<1.$$ In addition, $u_{m+1}=G(u_{m})<G(u^*)=u^*$. Similarly, if $u_0>u^*,$ then inductively we have $u_m>u_{m+1}>u^*.$ As a consequence, the sequence of iterates $u_m$ converges in both cases, and by the uniqueness of the fixed point $u^*$ we have $u_m\rightarrow u^*.$

\section{Proofs for Section~\ref{sec:optimality}}
\label{sec:section4_proofs}

\subsection{Lemma~\ref{lem:system}}
\begin{proof}[Proof of Lemma~\ref{lem:system}]
We first prove~\eqref{eq:consistency_system_rhs} and then lower bound the smallest eigenvalue of the matrix $\bA$.

\paragraph{Concentration: } Fix $\tilde{\lambda} \in (0,\infty)$. In our argument here we consider $\lambdainit$ as deterministic and equal to $\tilde{\lambda}$. Arguing as in the proof of Lemma \ref{lem:risk_close_to_L} we can show that for $\tilde{\lambda} \in [\underline{\lambda}, \bar{\lambda}]$, a compact subset of $(0,\infty)$, the same result holds uniformly almost surely. This justifies data-driven choices of $\lambdainit$.

Now, let $\bM_{\group{g},\cdot}$ be the $p_g\times p$ matrix that consists of the rows of $\bM$ that correspond to the group $g$. It is enough to prove the following three asymptotic results.
\begin{equation}\label{eq:asympt_1}
    \Norm{\bM_{\group{g},\cdot}\bw}^2-\left(\sum_{h=1}^K \norm{\bM_{\group{g},\group{h}}}_F^2 \frac{\alpha_h^2}{p_h}\right)\xrightarrow{a.s.}0.
\end{equation}
\begin{equation}\label{eq:asympt_2}
 \frac{\Norm{\bN_{\group{g},\cdot}\bvarepsilon}^2}{n}- \Norm{\bN_{\group{g},\cdot}}_F^2\frac{\sigma^2}{n}\xrightarrow{a.s.}0.   
\end{equation}
\begin{equation}\label{eq:asympt_3}
\bw^\intercal \bM_{\group{g}}^\intercal \bN_{\group{g},\cdot}\frac{\bvarepsilon}{\sqrt{n}}\xrightarrow{a.s.}0.
\end{equation}
Let us start with the proof of~\eqref{eq:asympt_1}. First,$$\Norm{\bM_{\group{g},\group{h}}}_F \leq \Norm{\bM}_F \leq \sqrt{p}\Norm{\bM} = \mathcal{O}(\sqrt{p}).$$
Applying Lemma~\ref{lem1} for $q=(4+\eta)/2$ to the quadratic form $\Norm{\bM_{\group{g},\group{h}}\bw_{\group{h}}\sqrt{p_h}}^2 / p_h$ we see that:
$$ \mathbb P\sqb{{ \abs{\Norm{\bM_{\group{g},\group{h}}\bw_{\group{h}}\sqrt{p_h}}^2 / p_h - \EE{\Norm{\bM_{\group{g},\group{h}}\bw_{\group{h}}}^2}} \geq \varepsilon}} \leq \mathcal{O}\p{\frac{p^{q/2}}{p_h^q}} = \mathcal{O}\p{p^{-q/2}}.$$
Thus, an application of the Borel-Cantelli lemma yields that $$\norm{\bM_{\group{g},\group{h}}\bw_{\group{h}}}^2-\norm{\bM_{\group{g},\group{h}}}_F^2 \frac{\alpha_h^2}{p_h}\xrightarrow{a.s.}0.$$ This implies~\eqref{eq:asympt_1}. The derivation of~\eqref{eq:asympt_2} is analogous. Furthermore, also from Lemma~\ref{lem1} for $q=(4+\eta)/2$ there exists a constant $C=C(q,\alpha_1,\cdots,\alpha_K,\sigma)$ $$\Norm{\ \bw^\intercal \bM_{\group{g},\cdot}^\intercal \bN_{\group{g},\cdot}\frac{\bvarepsilon}{\sqrt{n}}}_q\leq C\frac{\Norm{\bM_{\group{g},\cdot}^\intercal \bN_{\group{g},\cdot}}_F}{\sqrt{pn}}\leq C\frac{\Norm{\bM_{\group{g},\cdot}^\intercal \bN_{\group{g},\cdot}
}}{\sqrt{n}}\leq \mathcal{O}\p{\frac{1}{\sqrt{n}}},$$ since $\lVert\bM_{\group{g},\cdot}^\intercal\rVert\leq 1$ and $\lVert\bN_{\group{g},\cdot}\rVert = \mathcal{O}(1) $. Another application of the Borel-Cantelli lemma proves~\eqref{eq:asympt_3}.

\paragraph{Invertibility: } We now turn to prove the more technically challenging result of the Lemma; namely that $\bA$ has an inverse and the operator norm of its inverse is uniformly bounded almost surely.
Let $M_{ij}$ be the $(i,j)-$th entry of $\bM$ and $\bM_s$ the matrix whose $(i,j)-th$ entry is $M_{ij}^2$. Analogously, let $\bC_s$ the matrix given by taking the square of $n^{-1}\bX^\intercal \bX$ entrywise. We first observe that the $K \times K$ matrix $\bA$ is formed by taking the sum of entries in submatrices of $\bM_s$ divided by $n$. By analogy, we also define $\bB$ as the $K\times K$ matrix whose $(k,\ell)-$th entry is the sum of squares of entries in the $p_k \times p_{\ell}$ submatrix of $\bC_s$ divided by $n$.

Now consider the eigendecomposition $n^{-1}\bX^\intercal \bX = \sum_{i=1}^p d_i V_i V_i^\intercal$, where $d_1$ is the largest eigenvalue of $n^{-1}\bX^\intercal \bX$. Then, the same eigenvectors diagonalize $\bM$ and
$$\bM = \sum_{i=1}^p \tilde{d}_i V_iV_i^\intercal, \;\; \tilde{d}_i = d_i/(d_i + \tilde{\lambda}).$$
Let us also write $V_{ki}$ for the $i-$th coordinate of $V_k$ and let $\ba=(a_1,\cdots,a_p)^\intercal$ be an arbitrary vector. It holds that $M_{ij} =\sum _{k=1}^p\tilde{d}_k  V_{ki}V_{kj}$ and so:
 
 \begin{equation}\label{eq:squared_entries_bound}
 \begin{aligned}
 \ba^\intercal \bM_s \ba &= \sum_{i,j=1}^p a_ia_jM_{ij}^2 \\
 &=\sum_{i,j=1}^p a_ia_j\sum_{k,l=1}^p\tilde{d}_k \tilde{d}_l V_{ki}V_{kj}V_{li}V_{lj}\\
 &=\sum_{k,l=1}^p\tilde{d}_k \tilde{d}_l\left(\sum_{i=1}^pa_i V_{ki}V_{li}\right)^2 \\
 &\geq \frac{1}{(d_1+\underline{\lambda})^2} \sum_{k,l=1}^pd_k d_l \left(\sum_{i=1}^pa_i V_{ki}V_{li}\right)^2\\
 &=\frac{\ba^\intercal \bC_s \ba}{(d_1+\underline{\lambda})^2}.
 \end{aligned}
 \end{equation}
 Now given any $K$-dimensional vector $\tilde{\ba}$, we can expand it as $\ba$ with $a_i = a_j = \tilde{a}_g$ for all $i,j \in \group{g}$ and $g=1,\dotsc,K$, which yields:
 \begin{equation}
     \tilde{\ba}^\intercal \bA \tilde{\ba} \geq  \frac{\tilde{\ba}^\intercal \bB \tilde{\ba}}{(d_1+\underline{\lambda})^2}.
 \end{equation}
Since $d_1$ is uniformly bounded almost surely, we conclude that, in order to show that the smallest eigenvalue of $\bA$ is bounded away from zero, it suffices to show that the smallest eigenvalue of $\bB$ is bounded away from $0$.
  
Using Lemma \ref{lem1} we can show  that for any matrices $A_1,A_2$ bounded in operator norm
\begin{equation}
\label{eq:tr_A1_A2}
\frac{\Tr\left(A_1\frac{\bZ^\intercal\bZ}{n}A_2\frac{\bZ^\intercal\bZ}{n}\right)}{n}-\frac{\Tr(A_1A_2)}{n}-\frac{\Tr(A_1)}{n}\frac{\Tr(A_2)}{n}\xrightarrow{a.s.}0.
\end{equation}
We now sketch the argument using a leave-one-out technique. Recall that $z_i^\intercal$ is the $i-$th row of $\bZ$ and let $\bZ_i$ the matrix that we get if we delete that row from $\bZ,$ then
$$
\begin{aligned}
\frac{1}{n}\Tr\left(A_1\frac{\bZ^\intercal\bZ}{n}A_2\frac{\bZ^\intercal\bZ}{n}\right)&=\frac{1}{n^2}\sum_{i=1}^n z_i^\intercal A_1\left(\frac{\bZ_i^\intercal \bZ_i}{n}+\frac{z_i z_i^\intercal}{n}\right)A_2z_i \\ &=\frac{1}{n}\sum_{i=1}^n\frac{z_i^\intercal A_1 z_i}{n} \frac{z_i^\intercal A_2 z_i}{n}\;+\;\frac{1}{ n}\sum_{i=1}^n z_i^\intercal A_1 \frac{\bZ_i^\intercal \bZ_i}{n}A_2 z_i/n.
\end{aligned}
$$
With Lemma~\ref{lem1} and Borel Cantelli we can show that almost surely (and uniformly in $i$)
$$ z_i^\intercal A_1 \frac{\bZ_i^\intercal \bZ_i}{n}A_2 z_i/n = \frac{1}{n}\Tr\left(A_1\frac{\bZ_i^\intercal\bZ_i}{n}A_2\right)+ o(1),$$ 
and that
$$\frac{1}{n}\Tr\left(A_1\frac{\bZ_i^\intercal\bZ_i}{n}A_2\right)=  \frac{\Tr(A_1A_2)}{n} + o(1).$$
Thus:
$$\frac{1}{n}\sum_{i=1}^n z_i^\intercal A_1 \frac{\bZ_i^\intercal \bZ_i}{n}A_2 z_i/n \,=\,  \frac{\Tr(A_1A_2)}{n} + o(1).$$
By the same reasoning
$$\frac{1}{n}\sum_{i=1}^n\frac{z_i^\intercal A_1 z_i}{n} \frac{z_i^\intercal A_2 z_i}{n} =  \frac{\Tr(A_1)}{n}\frac{\Tr(A_2)}{n} + o(1).$$
Now let $\bSigma_{\group{k},\group{\ell}}\in\mathbb{R}^{p_k\times p_{\ell}}$ be the submatrix of $\bSigma$ that corresponds to the covariance of the $k$-th and $\ell$-th groups and also let $R_k \in\mathbb{R}^{p\times p_k}$ the submatrix of $\bSigma^{1/2}$ consisting of the columns that correspond to the $k$-th group. Then, as a result of~\eqref{eq:tr_A1_A2} the $(k,\ell)$-th entry of $\bB$ is
$$
\begin{aligned}
\frac{\Tr\left(R_k^\intercal \frac{\bZ^\intercal\bZ}{n} R_{\ell}R_{\ell}^\intercal \frac{\bZ^\intercal\bZ}{n}R_k\right)}{n}&=\frac{\Tr(R_k^\intercal R_k)}{n}\frac{\Tr(R_{\ell}^\intercal R_{\ell})}{n}+\frac{\Tr(R_k R_k^\intercal R_{\ell} R_{\ell}^\intercal)}{n}+\omicron{(1)}\\   
&=\frac{\Norm{\bSigma_{\group{k},\group{\ell}}}_F^2}{n}+\frac{\Tr(\bSigma_{\group{k},\group{k}})}{n}\frac{\Tr(\bSigma_{\group{\ell},\group{\ell}})}{n}+\omicron{(1)},
\end{aligned}
$$
and so
$$
\begin{aligned}
\bB &=\left(\frac{\Tr(\bSigma_{\group{k},\group{k}})}{n}\frac{\Tr(\bSigma_{\group{\ell},\group{\ell}})}{n}\right)_{1\leq k,\ell\leq K} +\left(n^{-1}\Norm{\bSigma_{\group{k},\group{\ell}}}_F^2\right)_{1\leq k,\ell\leq K}+\omicron{(1)} \\
&\succeq \left(n^{-1}\Norm{\bSigma_{\group{k},\group{\ell}}}_F^2\right)_{1\leq k,\ell\leq K}+\omicron{(1)},
\end{aligned}
$$ which has eigenvalues uniformly bounded away from $0$. To see why, we revisit the key idea of the argument from \eqref{eq:squared_entries_bound}.  Let $\bSigma_s$ be the $p\times p$ matrix with $(i,j)$-th entry equal to $\Sigma_{ij}^2$. Also let $\tilde{\bSigma}$ be the $K \times K$ matrix with $(k, 
\ell)$-entry $\tilde{\Sigma}_{k\ell} = n^{-1}\Norm{\bSigma_{\group{k},\group{\ell}}}_F^2$. Now let $\bSigma = \sum_{i=1}^p \sigma_i U_i U_i^\top$ be the spectral decomposition of $\bSigma$, where $\sigma_p$ is the smallest eigenvalue of $\bSigma$. Next let  $\ba=(a_1,\cdots,a_p) \in\mathbb{R}^p$ arbitrary and write $U_{ki}$ for the $i-$th coordinate of $U_k$, then:
 \begin{equation}\label{eq:squared_entries_bound_revisited}
 \begin{aligned}
 \ba^\intercal \bSigma_s \ba &= \sum_{i,j=1}^p a_ia_j \Sigma_{ij}^2 \\
 &=\sum_{i,j=1}^p a_ia_j\sum_{k,l=1}^p \sigma_k \sigma_l U_{ki}U_{kj}U_{l i}U_{l j}\\
 &=\sum_{k,l=1}^p \sigma_k \sigma_{l}\left(\sum_{i=1}^p a_i U_{ki}U_{l i}\right)^2 \\
 &\geq \sigma_p^2 \sum_{k,l=1}^p \left(\sum_{i=1}^pa_i U_{ki}U_{l i}\right)^2\\
 &=\sigma_p^2 \Norm{\ba}_2^2.
 \end{aligned}
 \end{equation}
Fix $\tilde{\ba} \in \RR^K$ and expand it as $\ba$ with $a_i = a_j = \tilde{a}_g$ for all $i,j \in \group{g}$ and $g=1,\dotsc,K$. Then:
 \begin{equation}
 \begin{aligned}
    \tilde{\ba}^\intercal \tilde{\bSigma} \tilde{\ba} &= n^{-1} \ba^\intercal \bSigma_s \ba \\
    & \geq   n^{-1} \sigma_p^2 \Norm{\ba}_2^2 \\
    & \geq n^{-1} \sigma_p^2 \sum_{k=1}^K p_k \tilde{a}_k^2 \\
    & \geq \min_{k=1,\dotsc,K}\cb{\frac{p_k}{n}} \cdot \sigma_p^2 \cdot \Norm{\tilde{\ba}}_2^2.
\end{aligned}
\end{equation}
 
\end{proof}

\subsection{Lemma~\ref{lem:RCV_L}}

\begin{proof}[Proof of Lemma \ref{lem:RCV_L}]
For the proof of this lemma we assume without loss of generality that $\sigma=1.$

For the leave-one-out cross-validation risk we have the famous shortcut formula (which can be found, for example, in \cite{hastie2019surprises}) $$\loo_n(\blambda)=\frac{1}{n}\sum_{i=1}^n\left(\frac{y_i-x_i^\intercal \widehat{\bw}(\blambda)}{1-(S_{\bLambda})_{ii}}\right)^2,$$ where $S_{\bLambda}=\frac{1}{n}\bX\left(\frac{\bX^\intercal \bX}{n}+\bLambda\right)^{-1}\bX^\intercal$ is the group-regularized ridge regression smoother matrix. We will use the notation 
$$\bM=\left(\frac{\bZ^\intercal \bZ}{n}+\bSigma^{-\frac{1}{2}}\bLambda\bSigma^{-\frac{1}{2}}\right)^{-1}.$$
We divide the proof into three main steps.

\begin{enumerate}
\item We show that for any $\bLambda$ the denominators approach $1/(1+n^{-1}\Tr(\bM))$.  
\item We find asymptotic approximations for the sums of the numerators to complete the proof of pointwise convergence.
\item We control $\nabla_{\blambda}\loo_n(\blambda)$ to prove that convergence is uniform on compact subsets of $(0,\infty)^K.$ 
\end{enumerate}
\bigskip

\noindent\emph{Step 1:} It is convenient to first note that  $(S_{\bLambda})_{ii}=\frac{z_i^\intercal \bM z_i}{n}$. Next let $$\bM_i=\left(\frac{\bZ^\intercal \bZ}{n}+\bSigma^{-\frac{1}{2}}\bLambda \bSigma^{-\frac{1}{2}}-\frac{z_iz_i^\intercal}{n}\right)^{-1}.$$
 We have the resolvent identity 
\begin{equation}\label{resolvent}
\bM^{-1}=\bM_i^{-1}+n^{-1}z_iz_i^\intercal\; \Longrightarrow\; \bM_i=\bM+\bM_i\frac{z_iz_i^\intercal}{n}\bM \; \Longrightarrow\;\frac{z_i^\intercal \bM z_i}{n}=\frac{\frac{z_i^\intercal \bM_i z_i}{n}}{1+\frac{z_i^\intercal \bM_i z_i}{n}}
\end{equation}
Using Lemma \ref{lem1} (upon conditioning on $M_i$), the union bound and Borel-Cantelli we have that \begin{equation}\label{max_goes_0}
\max_{i}\abs{\frac{z_i^\intercal \bM_iz_i}{n}-n^{-1}\Tr(\bM_i)}\xrightarrow{a.s.}0.
\end{equation}
Now we prove that $n^{-1}\Tr(\bM_i)$ is close to $n^{-1}\Tr(\bM).$, i.e. that 
\begin{equation}\label{traces_are_close}
\max_{i}n^{-1}\abs{\Tr(\bM_i)-\Tr(\bM)}\xrightarrow{a.s.}0.
\end{equation}
To see this, first note that $\bM^{-1} \succeq \bM_i^{-1} \succeq \bSigma^{-1/2}\bLambda \bSigma^{1/2}$, i.e., $\bM \preceq \bM_i \preceq \bSigma^{1/2}\bLambda^{-1}\bSigma^{1/2}$, and so $\bM,\bM_i$ have uniformly bounded eigenvalues. Second, $\bM_i-\bM=\bM_i\frac{z_iz_i^\intercal}{n}\bM$ has rank 1. These results together imply that
$$0\leq \Tr(\bM_i)- \Tr(\bM)\leq \norm{\bM_i-\bM}=\mathcal{O}(1)$$ almost surely and uniformly in $i,n$ for any fixed $\bLambda$, hence~\eqref{traces_are_close} follows. Combining ~\eqref{max_goes_0} and ~\eqref{traces_are_close} we see that $$\max_i\abs{\frac{z_i^\intercal \bM_i z_i}{n}-n^{-1}\Tr(\bM)}\xrightarrow{a.s.}0.$$ Replacing the quadratic forms in ~\eqref{resolvent} by the normalized trace of $\bM$ (which is uniformly bounded) we get \begin{equation}\label{quadratic_form}
\max_{i}\abs{(S_{\bLambda})_{ii}-\frac{n^{-1}\Tr(\bM)}{1+n^{-1}\Tr(\bM)}}\xrightarrow{a.s.}0
\end{equation}

\noindent\emph{Step 2:} The average of the numerators in the shortcut formula for the leave-one-out cross-validation error is 
\begin{equation*}
\begin{aligned}
\frac{\Norm{\bY-\bX\widehat{\bw}(\bLambda)}^2}{n}&=\frac{1}{n}\Norm{\bX\bw+\bvarepsilon-\bX\left(\frac{\bX^\intercal \bX}{n}+\bLambda\right)^{-1}\left(\frac{\bX^\intercal \bX}{n}\bw+\frac{\bX^\intercal \bvarepsilon}{n}\right)}^2\\
&=\frac{1}{n}\Norm{\bX\left(\frac{\bX^\intercal \bX}{n}+\bLambda\right)^{-1}\bLambda \bw+\left(I_n-\frac{1}{n}\bX\left(\frac{\bX^\intercal \bX}{n}+\bLambda\right)^{-1}\bX^\intercal\right)\bvarepsilon}^2,
\end{aligned}
\end{equation*}
which, by the same concentration argument as we did for the out-of-sample error in Lemma~\ref{lem:risk_close_to_L}, is asymptotically approximated almost surely by
\begin{equation}\label{CV_num}\begin{split}
\frac{1}{n}\Norm{\left(I_n-\frac{1}{n}\bX\left(\frac{\bX^\intercal \bX}{n}+\bLambda\right)^{-1}\bX^\intercal\right)}^2_F\\+\;\frac{1}{p}\Tr\left(\left(\frac{\bX^\intercal \bX}{n}+\bLambda\right)^{-1}\frac{\bX^\intercal \bX}{n}\left(\frac{\bX^\intercal \bX}{n}+\bLambda\right)^{-1}\bLambda\bar{\bD}\bLambda\right)\end{split}
\end{equation}
Firstly we provide asymptotics for the first term in ~\eqref{CV_num}. It is equal to 
\begin{equation}\label{term1}\begin{aligned}
E_1&=1-\frac{2}{n}\Tr\left(\left(\frac{\bX^\intercal \bX}{n}+\bLambda\right)^{-1}\frac{\bX^\intercal \bX}{n}\right)+\frac{1}{n}\Norm{\bX\left(\frac{\bX^\intercal \bX}{n}+\bLambda\right)^{-1}\frac{\bX^\intercal}{n}}_F^2\\
&=1-\frac{2}{n} \Tr(S_{\bLambda})+\frac{1}{n}\Tr\left(\left(\bM\frac{\bZ^\intercal \bZ}{n}\right)^2\right).\end{aligned}
\end{equation} 
We see that \begin{equation}\label{term_1_term}
\frac{1}{n}\Tr\left(\left(\bM\frac{\bZ^\intercal \bZ}{n}\right)^2\right)=\frac{1}{n}\sum_{1\leq i,j\leq n}\frac{(z_i^\intercal \bM z_j)^2}{n^2},
\end{equation}
where the terms for $i=j$ are just the diagonal entries of $S_{\bLambda}$ squared. Hence, by ~\eqref{quadratic_form} it follows that the average of those terms gives asymptotically $\left(\frac{1}{n} \Tr(S_{\bLambda})\right)^2,$ so from ~\eqref{term1} and ~\eqref{term_1_term} we see that \begin{equation}\label{E_1_approx}
E_1=\p{1-\frac{1}{n}\Tr(S_{\bLambda})}^2+\frac{1}{n}\sum_{i\neq j}\frac{(z_i^\intercal \bM z_j)^2}{n^2}+\omicron{(1)}.
\end{equation}
Using again the resolvent identity $\bM_i=\bM+\bM_i\frac{z_iz_i^\intercal}{n}\bM$ we deduce that \begin{equation}\label{eq:Mz_i}
z_j^\intercal \bM z_i=\frac{z_j^\intercal \bM_i z_i}{1+\frac{z_i^\intercal \bM_i z_i}{n}},
\end{equation}
and so that
\begin{equation}
\begin{aligned}
\frac{1}{n}\sum_{i}\sum_{j\neq i}\frac{(z_i^T \bM z_j)^2}{n^2}&=
\frac{1}{n}\sum_{i}\frac{\left(1+\frac{z_i^\intercal \bM_i z_i}{n}\right)^{-2}}{n}z_i^\intercal \bM _i\left(\sum_{j\neq i}\frac{z_j z_j^\intercal}{n}\right) \bM_i z_i\\
&=(1+\omicron{(1)})\frac{\left(1+n^{-1}\Tr(\bM)\right)^{-2}}{n}\sum_i \frac{1}{n}z_i^\intercal \bM_i\left(\sum_{j\neq i}\frac{z_j z_j^\intercal}{n}\right) \bM_i z_i.
\end{aligned}
\end{equation}
At this point we observe that by Lemma \ref{lem1}, the union bound and Borel-Cantelli we have:
\begin{equation}\label{eq:CV_i_neq_j_terms}\begin{split}
\max_i{\frac{1}{n}\abs{z_i^\intercal \bM_i\left(\sum_{j\neq i}\frac{z_jz_j^\intercal}{n}\right) \bM_iz_i- \Tr\left(\bM_i^2\sum_{j\neq i}\frac{z_jz_j^\intercal}{n}\right)}}\xrightarrow{a.s.}0.\end{split}
\end{equation}
Here we also used the fact that $\sum_{j\neq i}z_jz_j^\intercal/n \preceq \bZ^\intercal \bZ/n$. Next we will need the following lemma, which we are going to prove after the end of this proof.

\begin{lemma}\label{lem:M_i_squared}
With the assumptions we made it holds that: 
\begin{equation}\max _i \frac{1}{n}\abs{\Tr\p{\bM_i^2\sum_{j\neq i}\frac{z_jz_j^\intercal}{n}}-\Tr\p{\bM^2\frac{\bZ^\intercal \bZ}{n}}}\xrightarrow{a.s.}0.
\end{equation}
\end{lemma}
\noindent Combining Lemma \ref{lem:M_i_squared} and ~\eqref{eq:CV_i_neq_j_terms} we conclude that \begin{equation}\label{eq:i_neq_j_terms}\begin{split}
\max_i{\frac{1}{n}\abs{z_i^\intercal \bM_i\left(\sum_{j\neq i}\frac{z_jz_j^\intercal}{n}\right) \bM_iz_i- \Tr\left(\bM^2\frac{\bZ^\intercal \bZ}{n}\right)}}\xrightarrow{a.s.}0.
\end{split}
\end{equation}
Finally, combining ~\eqref{E_1_approx} and ~\eqref{eq:i_neq_j_terms} we have proved that
\begin{equation}
\label{eq:E_1_final}
\begin{aligned}
E_1&=\left(1-\frac{\Tr(S_{\bLambda})}{n}\right)^2+\frac{n^{-1}\Tr\left(\bM^2\frac{\bZ^\intercal \bZ}{n}\right)}{(1+n^{-1}\Tr(\bM))^2}+\omicron{(1)}\\
&=(1+n^{-1}\Tr(\bM))^{-2}\left(1+n^{-1}\Tr\left(\bM^2\frac{\bZ^\intercal \bZ}{n}\right)\right)+\omicron{(1)}.
\end{aligned}
\end{equation}
Now it is time to examine the second term of ~\eqref{CV_num}. Let $\bar{\bSigma}=\bSigma^{-\frac{1}{2}}\bLambda\bar{\bD}\bLambda\bSigma^{-\frac{1}{2}}$. The second term of ~\eqref{CV_num} equals \begin{equation}\label{eq:E_2}
\begin{aligned}
E_2&=\frac{1}{p}\Tr\left(\left(\frac{\bX^\intercal \bX}{n}+\bLambda\right)^{-1}\frac{\bX^\intercal \bX}{n}\left(\frac{\bX^\intercal \bX}{n}+\bLambda\right)^{-1} \bLambda \bar{\bD}\bLambda\right)\\
&=\frac{1}{p}\Tr\left(\bM\frac{\bZ^\intercal \bZ}{n}\bM\bar{\bSigma}\right)\\
&=\frac{1}{p}\sum_{i=1}^n\frac{z_i^\intercal \bM\bar{\bSigma}\bM z_i}{n}.
\end{aligned}
\end{equation}
Using the resolvent identity $\bM_i=\bM+\frac{\bM_iz_iz_i^\intercal \bM}{n}$ we derive
\begin{equation}
\bM z_i=\p{1-\frac{z_i^\intercal \bM z_i}{n}}\bM_i z_i=\frac{\bM_iz_i}{1+\frac{z_i^\intercal \bM_iz_i}{n}}.
\end{equation}
Replacing this in ~\eqref{eq:E_2} we get:

\begin{equation}\label{E_2_new}\begin{split}
E_2=\frac{1}{p}\sum_{i=1}^n\left(1-(S_{\bLambda})_{ii}\right)^{2}\frac{z_i^\intercal \bM_i\bar{\bSigma}\bM_iz_i}{n}.\end{split}
\end{equation}
In the same way as we proved ~\eqref{eq:i_neq_j_terms} we can prove that \begin{equation}\label{eq:E_2_terms}\begin{split}
\max_{i}\abs{\frac{z_i^\intercal \bM_i\bar{\bSigma}\bM_iz_i}{n}-\frac{1}{n}\Tr(\bM^2\bar{\bSigma})}\xrightarrow{a.s.}0.\end{split}
\end{equation}
Combining ~\eqref{quadratic_form} and ~\eqref{eq:E_2_terms} we get:
\begin{equation}\label{eq:E_2_final}
E_2=\frac{(1+n^{-1}\Tr(\bM))^{-2}}{p}\Tr(\bM^2\bar{\bSigma})+\omicron{(1)}.
\end{equation}
Using ~\eqref{quadratic_form},~\eqref{eq:E_1_final} and ~\eqref{eq:E_2_final} we have:
\begin{equation}\label{eq:CV_final}\begin{split}
\loo_n(\blambda)=1+ \frac{1}{n}\Tr\left(\bM^2\frac{\bZ^\intercal \bZ}{n}\right)+\frac{1}{p}\Tr\left(\bM^2\bSigma^{-\frac{1}{2}}\bLambda\bar{\bD}\bLambda \bSigma^{-\frac{1}{2}}\right)+\omicron{(1)}.\end{split}
\end{equation}
Omitting the $\omicron{(1)}$, this is exactly the expression of $\Ln_n(\blambda)$.

\emph{Step 3:} For $\blambda$ taking values in a compact set $C$ of $(0,\infty)^K$ we have by the shortcut formula that the functions $\loo_n(\blambda)$ are almost surely uniformly bounded and uniformly Lipschitz, hence uniformly equicontinuous. The same holds for $\Ln_n,$ so also for the difference $\loo_n-\Ln_n.$ By the pointwise convergence that we showed in the second step we have that almost surely in a countable dense subset $C_{d}$ of $C$ we have $\loo_n-\Ln_n  \xrightarrow{a.s.}0.$ Since by the Arzela-Ascoli theorem almost surely any subsequence of $\{\loo_n-\Ln_n\}_{n\geq 1}$ has a uniformly convergent subsequence and since by pointwise convergence on $C_d$ the only uniform subsequential limit can be $0$, we know that $$\sup_{\blambda\in C}\abs{\loo_n(\blambda)-\Ln_n(\blambda)}\xrightarrow{a.s.}0.$$ The proof is completed.

\end{proof}

\begin{proof}[Proof of Lemma \ref{lem:M_i_squared}]

Let $\bR=\bSigma^{-\frac{1}{2}}\bLambda \bSigma^{-\frac{1}{2}}.$ We have
$$\frac{1}{n}\Tr\left(\bM_i^2\sum_{j\neq i}\frac{z_jz_j^\intercal}{n}\right)=\frac{1}{n}\Tr\left(\bM_i^2(\bM_i^{-1}-\bR)\right)=\frac{\Tr(\bM_i)-\Tr(\bM_i^2\bR)}{n}.$$
and
$$\frac{1}{n}\Tr \left(\bM^2\sum_{j}\frac{z_jz_j^\intercal}{n}\right)=\frac{1}{n}\Tr\left(\bM^2(\bM^{-1}-\bR)\right)=\frac{\Tr(\bM)-\Tr(\bM^2\bR)}{n}.$$
In light of~\eqref{traces_are_close}, it suffices to show that $$\max_{i}\abs{\frac{\Tr((\bM_i^2-\bM^2)\bR)}{n}}\xrightarrow{a.s.}0.$$
We have that $\bM_i^2-\bM^2=\bM_i^2\left(\bM^{-2}-\bM_i^{-2}\right)\bM^2$ has rank at most 2 (by expanding the middle terms), so by the fact that $\bM_i^2,\bM^2, \bR$ have uniformly bounded operator norm, it follows that almost surely 
$$\max_{i}\abs{\frac{\Tr((\bM_i^2-\bM^2)\bR)}{n}}=\mathcal{O}(n^{-1}).$$
This completes the proof of the lemma.

\end{proof}

\subsection{Theorem~\ref{theo:optimality}}

\begin{proof}[Proof of Theorem~\ref{theo:optimality}]
It only remains to extend the argument to noncompact sets. We fix $M>0$ large and consider for each $\blambda\in V_2$ the matrix $\bLambda_M$ that we get by truncating entries of $\bLambda$ that are larger than $M$ to $M.$ Let $\widehat{\bw}(\bLambda)$ be the estimator of $w$ using ridge regression with penalty matrix $\bLambda$ and $\widehat{\bw}(\bLambda_M)$ the estimator using the penalty matrix $\bLambda_M$. We first check that it suffices to show that 
\begin{equation}
\label{eq:wts_equiv_at_inf}
\lim_{M\to \infty}\limsup_{n\to \infty}\sup_{\bLambda \in V_2}\Norm{\widehat{\bw}(\bLambda_M)-\widehat{\bw}(\bLambda )}=0 \text{  almost surely}.
\end{equation}
To see that the above suffices, note that we can write
$$
\begin{aligned}
&\Rn(\widehat{\bw}(\bLambda_M))-\Rn(\widehat{\bw}(\bLambda))\\
=\;&\p{\widehat{\bw}(\bLambda_M) - \bw}^\intercal \bSigma \p{\widehat{\bw}(\bLambda_M) - \bw} -  \p{\widehat{\bw}(\bLambda) - \bw}^\intercal \bSigma \p{\widehat{\bw}(\bLambda) - \bw} \\
=\;& \widehat{\bw}(\bLambda_M)^\intercal \bSigma  \widehat{\bw}(\bLambda_M) - \widehat{\bw}(\bLambda)^\intercal \bSigma  \widehat{\bw}(\bLambda) - 2\bw^\intercal\bSigma\p{\widehat{\bw}(\bLambda_M) - \widehat{\bw}(\bLambda)} \\
=\;& \p{\widehat{\bw}(\bLambda_M) - \widehat{\bw}(\bLambda)}^\intercal \bSigma  \widehat{\bw}(\bLambda_M) + \widehat{\bw}(\bLambda)^\intercal \bSigma \p{\widehat{\bw}(\bLambda_M)- \widehat{\bw}(\bLambda)} - 2\bw^\intercal\bSigma\p{\widehat{\bw}(\bLambda_M) - \widehat{\bw}(\bLambda)}.
\end{aligned}
$$
We can show that $\Norm{\bSigma\widehat{\bw}(\bLambda_M)}_2$, $\Norm{\bSigma\widehat{\bw}(\bLambda)}_2$, $\Norm{\bSigma\bw}_2$ are bounded uniformly in $n$, $M$ and $\bLambda$ almost surely, and so~\eqref{eq:wts_equiv_at_inf} implies that
\begin{equation}
\label{eq:risks_equiv_at_inf}
\lim_{M\to \infty}\limsup_{n\to \infty}\sup_{\bLambda \in V_2}\abs{   \Rn(\widehat{\bw}(\bLambda_M))-\Rn (\widehat{\bw}(\bLambda))}=0 \text{  almost surely}.
\end{equation}
It remains to prove~\eqref{eq:wts_equiv_at_inf}. We temporarily fix $\bLambda$. We also let $\bS_1 = \left(n^{-1}\bX^\intercal \bX +\bLambda \right)^{-1},$ $\bS_2 = \left(n^{-1}\bX^\intercal \bX +\bLambda_M \right)^{-1}.$  Next, assume without loss of generality that only the last $j$ groups have $\lambda_g$ exceeding $M,$ and let $\bLambda_1,\bLambda_2$ be the diagonal submatrices of $\bLambda$ that correspond to the first $K-j$ and last $j$ groups respectively, and write $$\frac{\bX^\intercal \bX}{n}=\begin{pmatrix}\bS_{11}& \bS_{12}\\
\bS_{21}& \bS_{22}\\
\end{pmatrix},$$ where $\bS_{22}$ is the sample covariance of the features in the last $j$ groups. Then, by block-diagonal inversion:
$$\bS_1 = \begin{pmatrix}
\left(\bS_{11}+\bLambda_1 +\bS_{12}\left(\bS_{22}+\bLambda_2\right)^{-1} \bS_{21}\right)^{-1}& -\left(\bS_{11}+\bLambda_1 - \bS_{12}\left(\bS_{22}+\bLambda_2\right)^{-1} \bS_{21}\right)^{-1} \bS_{12}\left(\bS_{22}+\bLambda_2\right)^{-1}\\
* & \left(\bS_{22}+\bLambda_2 -\bS_{21}\left(\bS_{11}+\bLambda _1\right)^{-1} \bS_{12}\right)^{-1}\\
\end{pmatrix},$$ where $*$ is completed to make $\bS_1$ symmetric. We also note that $\bS_{11}, \bS_{12}, \bS_{22}$ are bounded in operator norm due to the fact that:
$$\max\cb{\Norm{\bS_{11}},\Norm{\bS_{12}},\Norm{\bS_{22}}}\leq \Norm{\bS}=n^{-1}\Norm{\bX}^2.$$ 
Thus, uniformly in $\bLambda_1$ we have: $$\bS_1=\begin{pmatrix}
\left(\bS_{11}+\bLambda_1 \right)^{-1}&0\\
0& 0\end{pmatrix}+\mathcal{O}_{op} (\bM^{-1}).$$ The same argument holds for $\bS_2$ and we get $\Norm{\bS_1-\bS_2}=\mathcal{O}(\bM^{-1})$ in operator norm. It follows that $$\Norm{\widehat{\bw}(\bLambda_M)-\widehat{\bw}(\bLambda )}=\Norm{(\bS_1-\bS_2)\frac{\bX^\intercal \bY}{n}}\leq \Norm{\bS_1-\bS_2}\Norm{\frac{\bX}{\sqrt{n}}}\frac{\Norm{\bY}}{\sqrt{n}}=\mathcal{O}(\bM^{-1}),$$
where we have used the fact that $\limsup n^{-1/2}\Norm{\bX} \leq \limsup  \Norm{\bSigma}^{1/2}n^{-1/2}\Norm{\bZ}=(1+\sqrt{\gamma})\Norm{\bSigma}^{1/2}.$

\end{proof}

\subsection{Corollary~\ref{corol:glasso}}

\begin{proof}[Proof of Corollary~\ref{corol:glasso}] 
It is enough to show that any set of coefficients that can be achieved using Group Lasso can also be achieved by suitably choosing the parameters of Group Ridge. Let $\lambda_1'=\lambda^{\text{glasso}} \sqrt{p_i/p},\dotsc,\lambda_K'=\lambda^{\text{glasso}}\sqrt{p_K/p}$ be the parameters of Group Lasso, giving regression weights (coefficients) $\widetilde{\bw}_{\group{g}}\in\mathbb{R}^{p_g},1\leq g \leq K$ for the groups. Let $\lambda_g= \lambda_g'/\Norm{\tilde{\bw}_{\group{g}}}$ be the group ridge parameters for the corresponding ridge regression and write $\blambda =(\lambda_1,\dotsc,\lambda_K)$. We claim that the group-ridge solution $\widehat{\bw}(\blambda)$ from~\eqref{eq:groupridge} is equal to $\widetilde{\bw}$.  First notice that if for a group it holds that $\widetilde{\bw}_{\group{g}} = \bm{0}$, then $\lambda_g = \infty $, and so Ridge will also assign $\widehat{\bw}_{\group{g}}(\blambda) = \bm{0}$. Upon removing the subset of groups $g$ such that $\widetilde{\bw}_{\group{g}} \neq \bm{0} $ from the design matrix $\bX$, we may assume without loss of generality that all $\widetilde{\bw}_{\group{g}} \neq \bm{0} $ and so all $\lambda_g \in (0,\infty)$. Then, \begin{equation}
    \nabla_{\bw} \left(\frac{\Norm{\bY-\bX\bw}^2}{2n}+\sum_{g=1}^K\frac{\lambda_g \Norm{\bw_{\group{g}}}^2}{2}\right)\bigg|_{\bw=\widetilde{\bw}}=\frac{\bX^\intercal (\bX\widetilde{\bw}-\bY)}{n}+\sum_{g=1}^K \frac{\lambda_g'}{\Norm{\widetilde{\bw}_{\group{g}}}}\widetilde{\bw}_{\group{g}}=0,
\end{equation}
since the last expression is exactly the gradient of the Group Lasso loss evaluated at the minimizer $\widetilde{\bw}.$

The conclusion follows by an application of Theorem~\ref{theo:optimality}. To apply the theorem we need to argue that $\lambda_g$ are bounded away from $0$ for large $n$ almost surely. It suffices to show that $\Norm{\widetilde{\bw}_{\group{g}}}_2$ is bounded away from infinity. By comparing the objective value of the Group Lasso at $\widetilde{\bw}$ and at $\bm{0}$, we see that for all $g$,
$$\lambda_g' \cdot \Norm{\widetilde{\bw}_{\group{g}}}_2  \leq \frac{1}{2n}\norm{Y}_2^2.$$
By our assumptions, the RHS is bounded almost surely. The LHS satisfies (deterministically) $ \liminf_{n \to \infty} \lambda_g' \geq \sqrt{\gamma_g}\cdot \delta > 0$ and so we conclude. 
\end{proof}

\newpage
\section{Additional figures}

\begin{figure}[H]
\centering
\begin{tabular}{ccc}
  \begin{adjustbox}{width=0.33\linewidth}\input{figures/oracle_risks_exponential/oracle_risk1.tikz}\end{adjustbox} 
& \begin{adjustbox}{width=0.33\linewidth}\input{figures/oracle_risks_exponential/oracle_risk2.tikz}\end{adjustbox} 
& \begin{adjustbox}{width=0.33\linewidth}\input{figures/oracle_risks_exponential/oracle_risk3.tikz}\end{adjustbox} \\ 
  \begin{adjustbox}{width=0.33\linewidth}\input{figures/oracle_risks_exponential/oracle_risk4.tikz}\end{adjustbox} 
& \begin{adjustbox}{width=0.33\linewidth}\input{figures/oracle_risks_exponential/oracle_risk5.tikz}\end{adjustbox} 
& \begin{adjustbox}{width=0.33\linewidth}\input{figures/oracle_risks_exponential/oracle_risk6.tikz}\end{adjustbox}
\end{tabular}
\caption{\textbf{Asymptotic predictions for asymptotic risk of group-regularized ridge regression}. This figure is analogous to Figure~\ref{fig:theoretical_risks} with a different feature covariance. $\bSigma_1=\bSigma_2$ have eigenvalues equal to the evenly-spaced quantiles of the Exponential distribution with rate $0.5$.}
\label{fig:theoretical_risks_exponential}
\end{figure}
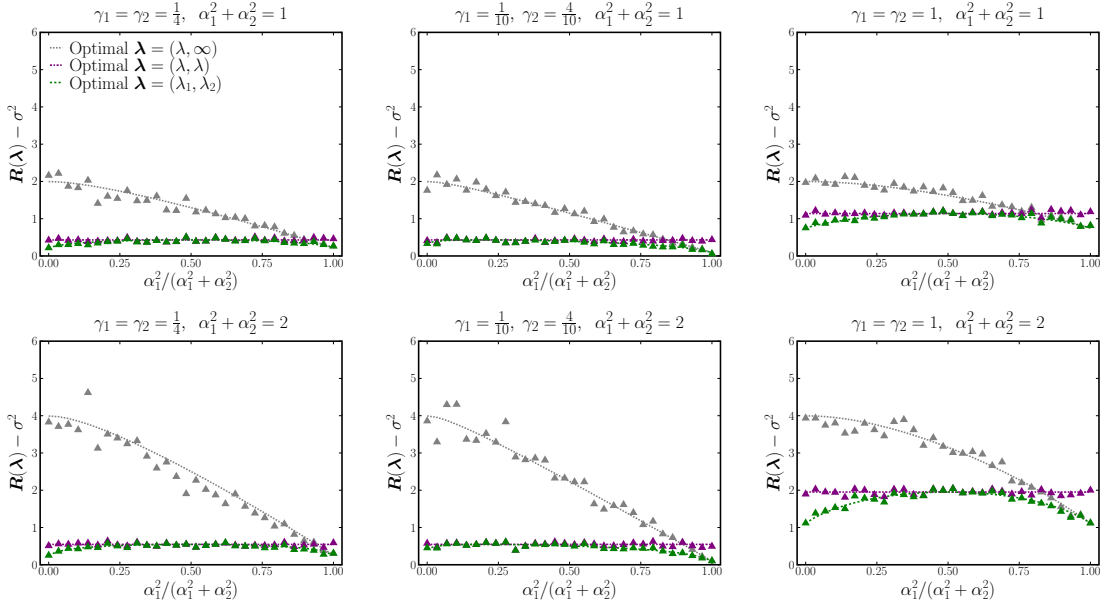

\newpage 
\ifdoubleblind
\onehalfspacing
\bibliographystylesupplement{abbrvnat}
\bibliographysupplement{gridge}
\fi 
\end{document}

%% file: figures/intro_figure_left.tikz
\begin{tikzpicture}[/tikz/background rectangle/.style={fill={rgb,1:red,1.0;green,1.0;blue,1.0}, draw opacity={1.0}}, show background rectangle]
\begin{axis}[point meta max={nan}, point meta min={nan}, legend cell align={left}, title={}, title style={at={{(0.5,1)}}, anchor={south}, font={{\fontsize{25.2 pt}{32.76 pt}\selectfont}}, color={rgb,1:red,0.0;green,0.0;blue,0.0}, draw opacity={1.0}, rotate={0.0}}, legend style={color={rgb,1:red,0.0;green,0.0;blue,0.0}, draw opacity={0.0}, line width={1.8}, solid, fill={rgb,1:red,0.0;green,0.0;blue,0.0}, fill opacity={0.0}, text opacity={1.0}, font={{\fontsize{14.4 pt}{18.720000000000002 pt}\selectfont}}, text={rgb,1:red,0.0;green,0.0;blue,0.0}, at={(0.02, 0.98)}, anchor={north west}}, axis background/.style={fill={rgb,1:red,1.0;green,1.0;blue,1.0}, opacity={1.0}}, anchor={north west}, xshift={1.0mm}, yshift={-1.0mm}, width={137.7mm}, height={99.6mm}, scaled x ticks={false}, xlabel={$\sigmacv$}, x tick style={color={rgb,1:red,0.0;green,0.0;blue,0.0}, opacity={1.0}}, x tick label style={color={rgb,1:red,0.0;green,0.0;blue,0.0}, opacity={1.0}, rotate={0}}, xlabel style={at={(ticklabel cs:0.5)}, anchor=near ticklabel, font={{\fontsize{19.8 pt}{25.740000000000002 pt}\selectfont}}, color={rgb,1:red,0.0;green,0.0;blue,0.0}, draw opacity={1.0}, rotate={0.0}}, xmajorgrids={false}, xmin={-0.1700970721713377}, xmax={6.304980924292732}, xtick={{0.0,1.0,2.0,3.0,4.0,5.0,6.0}}, xticklabels={{$0$,$1$,$2$,$3$,$4$,$5$,$6$}}, xtick align={inside}, xticklabel style={font={{\fontsize{14.4 pt}{18.720000000000002 pt}\selectfont}}, color={rgb,1:red,0.0;green,0.0;blue,0.0}, draw opacity={1.0}, rotate={0.0}}, x grid style={color={rgb,1:red,0.0;green,0.0;blue,0.0}, draw opacity={0.1}, line width={0.9}, solid}, x axis line style={color={rgb,1:red,0.0;green,0.0;blue,0.0}, draw opacity={1.0}, line width={1.8}, solid}, scaled y ticks={false}, ylabel={$\widehat{\lambda}_g(\sigmacv)$}, y tick style={color={rgb,1:red,0.0;green,0.0;blue,0.0}, opacity={1.0}}, y tick label style={color={rgb,1:red,0.0;green,0.0;blue,0.0}, opacity={1.0}, rotate={0}}, ylabel style={at={(ticklabel cs:0.5)}, anchor=near ticklabel, font={{\fontsize{19.8 pt}{25.740000000000002 pt}\selectfont}}, color={rgb,1:red,0.0;green,0.0;blue,0.0}, draw opacity={1.0}, rotate={0.0}}, ymajorgrids={false}, ymin={-0.055600177682499655}, ymax={1.9089697538679482}, ytick={{0.0,0.5,1.0,1.5}}, yticklabels={{$0.0$,$0.5$,$1.0$,$1.5$}}, ytick align={inside}, yticklabel style={font={{\fontsize{14.4 pt}{18.720000000000002 pt}\selectfont}}, color={rgb,1:red,0.0;green,0.0;blue,0.0}, draw opacity={1.0}, rotate={0.0}}, y grid style={color={rgb,1:red,0.0;green,0.0;blue,0.0}, draw opacity={0.1}, line width={0.9}, solid}, y axis line style={color={rgb,1:red,0.0;green,0.0;blue,0.0}, draw opacity={1.0}, line width={1.8}, solid}]
    \addplot[color={rgb,1:red,0.502;green,0.0;blue,0.502}, name path={1ec8013c-5633-4464-8bf7-956d507d4105}, draw opacity={1.0}, line width={1.8}, dotted]
        table[row sep={\\}]
        {
            \\
            0.013159852256890666  2.743639823400004e-6  \\
            0.1459547250309692  0.00033761975251406825  \\
            0.27874959780504777  0.0012327205795672012  \\
            0.4115444705791263  0.0026914934270527184  \\
            0.5443393433532048  0.004719579196816768  \\
            0.6771342161272833  0.007324867157792461  \\
            0.809929088901362  0.010517572208620807  \\
            0.9427239616754405  0.01431033605479869  \\
            1.075518834449519  0.018718354191550633  \\
            1.2083137072235974  0.023759531109061625  \\
            1.341108579997676  0.02945466673718111  \\
            1.4739034527717545  0.03582767784325298  \\
            1.6066983255458331  0.0429058589149553  \\
            1.7394931983199116  0.050720188031533206  \\
            1.8722880710939902  0.05930568439097097  \\
            2.005082943868069  0.06870182556703933  \\
            2.1378778166421473  0.07895303428166843  \\
            2.2706726894162257  0.0901092465751254  \\
            2.4034675621903046  0.1022265758425291  \\
            2.536262434964383  0.11536809041450527  \\
            2.6690573077384614  0.1296047263674751  \\
            2.80185218051254  0.14501636228555628  \\
            2.9346470532866187  0.16169308906649005  \\
            3.067441926060697  0.17973671597569632  \\
            3.2002367988347755  0.199262564553493  \\
            3.333031671608854  0.22040161541417777  \\
            3.465826544382933  0.24330309045973403  \\
            3.5986214171570112  0.26813757597141946  \\
            3.7314162899310896  0.2951008224013575  \\
            3.864211162705168  0.32441839723009647  \\
            3.997006035479247  0.3563514219301928  \\
            4.129800908253325  0.39120369857408094  \\
            4.262595781027404  0.4293306342702289  \\
            4.395390653801482  0.4711505147407071  \\
            4.528185526575561  0.5171588805209781  \\
            4.66098039934964  0.5679470488282781  \\
            4.793775272123718  0.624226245208503  \\
            4.926570144897797  0.686859431511303  \\
            5.059365017671875  0.7569038535765367  \\
            5.192159890445954  0.8356687701791603  \\
            5.324954763220032  0.924795081206935  \\
            5.4577496359941104  1.0263671999379433  \\
            5.590544508768189  1.1430735034260822  \\
            5.723339381542268  1.2784418892793514  \\
            5.856134254316347  1.43719492471534  \\
            5.988929127090425  1.6258019708720648  \\
            6.121723999864503  1.8533687180693506  \\
        }
        ;
    \addlegendentry {Group 1}
    \addplot[color={rgb,1:red,0.0;green,0.502;blue,0.0}, name path={5b86328b-56ea-4964-b608-e561cd273cd8}, draw opacity={1.0}, line width={1.8}, dashdotted]
        table[row sep={\\}]
        {
            \\
            0.013159852256890666  1.2759585810281856e-6  \\
            0.1459547250309692  0.00015698167520514965  \\
            0.27874959780504777  0.0005728615159524226  \\
            0.4115444705791263  0.0012496655671031546  \\
            0.5443393433532048  0.0021886168326235903  \\
            0.6771342161272833  0.003391416770532719  \\
            0.809929088901362  0.004860253026257719  \\
            0.9427239616754405  0.006597809428825099  \\
            1.075518834449519  0.00860727833559284  \\
            1.2083137072235974  0.01089237543227447  \\
            1.341108579997676  0.013457357117546215  \\
            1.4739034527717545  0.016307040625900802  \\
            1.6066983255458331  0.019446827068993297  \\
            1.7394931983199116  0.022882727604938178  \\
            1.8722880710939902  0.026621392977344498  \\
            2.005082943868069  0.030670146701865984  \\
            2.1378778166421473  0.035037022218329605  \\
            2.2706726894162257  0.039730804371819445  \\
            2.4034675621903046  0.044761075637286624  \\
            2.536262434964383  0.05013826756032249  \\
            2.6690573077384614  0.055873717952840796  \\
            2.80185218051254  0.06197973445793407  \\
            2.9346470532866187  0.06846966518472612  \\
            3.067441926060697  0.07535797721355515  \\
            3.2002367988347755  0.08266034388657822  \\
            3.333031671608854  0.09039374193161381  \\
            3.465826544382933  0.09857655962099948  \\
            3.5986214171570112  0.10722871734635517  \\
            3.7314162899310896  0.11637180219913225  \\
            3.864211162705168  0.12602921839140027  \\
            3.997006035479247  0.13622635563840615  \\
            4.129800908253325  0.1469907779624591  \\
            4.262595781027404  0.15835243577695515  \\
            4.395390653801482  0.17034390458248258  \\
            4.528185526575561  0.1830006541695075  \\
            4.66098039934964  0.1963613528933502  \\
            4.793775272123718  0.21046821239090435  \\
            4.926570144897797  0.22536737907459597  \\
            5.059365017671875  0.24110937990470657  \\
            5.192159890445954  0.25774963135334605  \\
            5.324954763220032  0.27534902219139706  \\
            5.4577496359941104  0.293974582829149  \\
            5.590544508768189  0.31370025651855316  \\
            5.723339381542268  0.33460779090406856  \\
            5.856134254316347  0.35678777235026643  \\
            5.988929127090425  0.3803408303863391  \\
            6.121723999864503  0.40537904576360106  \\
        }
        ;
    \addlegendentry {Group 2}
    \addplot[color={rgb,1:red,0.502;green,0.502;blue,0.502}, name path={9ddd8892-15d6-4c28-a826-e6368c3596e6}, draw opacity={1.0}, line width={1.8}, dashed]
        table[row sep={\\}]
        {
            \\
            0.013159852256890666  8.581160979195409e-7  \\
            0.1459547250309692  0.0001055682373995495  \\
            0.27874959780504777  0.0003851826907965093  \\
            0.4115444705791263  0.0008400431309457082  \\
            0.5443393433532048  0.0014707060545453962  \\
            0.6771342161272833  0.002277944504813442  \\
            0.809929088901362  0.0032627504456954597  \\
            0.9427239616754405  0.0044263378194472295  \\
            1.075518834449519  0.005770146305245945  \\
            1.2083137072235974  0.007295845800659396  \\
            1.341108579997676  0.009005341652181474  \\
            1.4739034527717545  0.010900780665669695  \\
            1.6066983255458331  0.012984557932442263  \\
            1.7394931983199116  0.015259324512059337  \\
            1.8722880710939902  0.01772799601848161  \\
            2.005082943868069  0.020393762162430135  \\
            2.1378778166421473  0.0232600973094334  \\
            2.2706726894162257  0.026330772120317047  \\
            2.4034675621903046  0.029609866348853833  \\
            2.536262434964383  0.0331017828800429  \\
            2.6690573077384614  0.036811263102137386  \\
            2.80185218051254  0.040743403716209915  \\
            2.9346470532866187  0.04490367509887877  \\
            3.067441926060697  0.049297941346970686  \\
            3.2002367988347755  0.053932482147555215  \\
            3.333031671608854  0.05881401663315512  \\
            3.465826544382933  0.06394972940026153  \\
            3.5986214171570112  0.06934729888983561  \\
            3.7314162899310896  0.07501492835157847  \\
            3.864211162705168  0.08096137963976636  \\
            3.997006035479247  0.08719601011780714  \\
            4.129800908253325  0.09372881298186721  \\
            4.262595781027404  0.10057046135152567  \\
            4.395390653801482  0.10773235651810247  \\
            4.528185526575561  0.11522668078986567  \\
            4.66098039934964  0.12306645542866929  \\
            4.793775272123718  0.13126560423578357  \\
            4.926570144897797  0.13983902341702875  \\
            5.059365017671875  0.14880265844030408  \\
            5.192159890445954  0.1581735886939952  \\
            5.324954763220032  0.1679701208646437  \\
            5.4577496359941104  0.17821189207917484  \\
            5.590544508768189  0.18891998400388918  \\
            5.723339381542268  0.20011704926289417  \\
            5.856134254316347  0.21182745173697148  \\
            5.988929127090425  0.22407742253519308  \\
            6.121723999864503  0.2368952337021319  \\
        }
        ;
    \addlegendentry {Group 3}
\end{axis}
\end{tikzpicture}

%% file: figures/intro_figure_right.tikz
\begin{tikzpicture}[/tikz/background rectangle/.style={fill={rgb,1:red,1.0;green,1.0;blue,1.0}, draw opacity={1.0}}, show background rectangle]
\begin{axis}[point meta max={nan}, point meta min={nan}, legend cell align={left}, title={}, title style={at={{(0.5,1)}}, anchor={south}, font={{\fontsize{25.2 pt}{32.76 pt}\selectfont}}, color={rgb,1:red,0.0;green,0.0;blue,0.0}, draw opacity={1.0}, rotate={0.0}}, legend style={color={rgb,1:red,0.0;green,0.0;blue,0.0}, draw opacity={0.0}, line width={1.8}, solid, fill={rgb,1:red,0.0;green,0.0;blue,0.0}, fill opacity={0.0}, text opacity={1.0}, font={{\fontsize{14.4 pt}{18.720000000000002 pt}\selectfont}}, text={rgb,1:red,0.0;green,0.0;blue,0.0}, at={(1.02, 1)}, anchor={north west}}, axis background/.style={fill={rgb,1:red,1.0;green,1.0;blue,1.0}, opacity={1.0}}, anchor={north west}, xshift={1.0mm}, yshift={-1.0mm}, width={132.7mm}, height={99.6mm}, scaled x ticks={false}, xlabel={$\sigmacv$}, x tick style={color={rgb,1:red,0.0;green,0.0;blue,0.0}, opacity={1.0}}, x tick label style={color={rgb,1:red,0.0;green,0.0;blue,0.0}, opacity={1.0}, rotate={0}}, xlabel style={at={(ticklabel cs:0.5)}, anchor=near ticklabel, font={{\fontsize{19.8 pt}{25.740000000000002 pt}\selectfont}}, color={rgb,1:red,0.0;green,0.0;blue,0.0}, draw opacity={1.0}, rotate={0.0}}, xmajorgrids={false}, xmin={-0.1700970721713377}, xmax={6.304980924292732}, xtick={{0.0,1.0,2.0,3.0,4.0,5.0,6.0}}, xticklabels={{$0$,$1$,$2$,$3$,$4$,$5$,$6$}}, xtick align={inside}, xticklabel style={font={{\fontsize{14.4 pt}{18.720000000000002 pt}\selectfont}}, color={rgb,1:red,0.0;green,0.0;blue,0.0}, draw opacity={1.0}, rotate={0.0}}, x grid style={color={rgb,1:red,0.0;green,0.0;blue,0.0}, draw opacity={0.1}, line width={0.9}, solid}, x axis line style={color={rgb,1:red,0.0;green,0.0;blue,0.0}, draw opacity={1.0}, line width={1.8}, solid}, scaled y ticks={false}, ylabel={$\loo(\sigmacv)$}, y tick style={color={rgb,1:red,0.0;green,0.0;blue,0.0}, opacity={1.0}}, y tick label style={color={rgb,1:red,0.0;green,0.0;blue,0.0}, opacity={1.0}, rotate={0}}, ylabel style={at={(ticklabel cs:0.5)}, anchor=near ticklabel, font={{\fontsize{19.8 pt}{25.740000000000002 pt}\selectfont}}, color={rgb,1:red,0.0;green,0.0;blue,0.0}, draw opacity={1.0}, rotate={0.0}}, ymajorgrids={false}, ymin={18.2}, ymax={19.2}, ytick={{18.4,18.6,18.8,19.0}}, yticklabels={{$18.4$,$18.6$,$18.8$,$19.0$}}, ytick align={inside}, yticklabel style={font={{\fontsize{14.4 pt}{18.720000000000002 pt}\selectfont}}, color={rgb,1:red,0.0;green,0.0;blue,0.0}, draw opacity={1.0}, rotate={0.0}}, y grid style={color={rgb,1:red,0.0;green,0.0;blue,0.0}, draw opacity={0.1}, line width={0.9}, solid}, y axis line style={color={rgb,1:red,0.0;green,0.0;blue,0.0}, draw opacity={1.0}, line width={1.8}, solid}]
    \addplot[color={rgb,1:red,0.0;green,0.0;blue,0.5451}, name path={c8c96bdd-b321-4b36-bd9d-3be14c832a24}, draw opacity={1.0}, line width={1.8}, solid, forget plot]
        table[row sep={\\}]
        {
            \\
            0.013159852256890666  18.985366404378816  \\
            0.1459547250309692  18.983198101913946  \\
            0.27874959780504777  18.977429079580265  \\
            0.4115444705791263  18.968109840794117  \\
            0.5443393433532048  18.955322054349217  \\
            0.6771342161272833  18.939177966755498  \\
            0.809929088901362  18.91981960434526  \\
            0.9427239616754405  18.8974177800051  \\
            1.075518834449519  18.872170922845513  \\
            1.2083137072235974  18.844303752119803  \\
            1.341108579997676  18.81406581919019  \\
            1.4739034527717545  18.78172994327504  \\
            1.6066983255458331  18.74759056807823  \\
            1.7394931983199116  18.711962067203935  \\
            1.8722880710939902  18.675177026519272  \\
            2.005082943868069  18.637584531383943  \\
            2.1378778166421473  18.599548485974832  \\
            2.2706726894162257  18.561445990861348  \\
            2.4034675621903046  18.523665803609543  \\
            2.536262434964383  18.48660690559099  \\
            2.6690573077384614  18.450677196429794  \\
            2.80185218051254  18.41629233572118  \\
            2.9346470532866187  18.383874749879954  \\
            3.067441926060697  18.353852820303587  \\
            3.2002367988347755  18.326660267536116  \\
            3.333031671608854  18.302735744861526  \\
            3.465826544382933  18.282522653801603  \\
            3.5986214171570112  18.266469193399363  \\
            3.7314162899310896  18.255028654989427  \\
            3.864211162705168  18.248659974441708  \\
            3.997006035479247  18.247828554666803  \\
            4.129800908253325  18.25300737254505  \\
            4.262595781027404  18.264678386446185  \\
            4.395390653801482  18.28333426321337  \\
            4.528185526575561  18.30948044697714  \\
            4.66098039934964  18.343637596543466  \\
            4.793775272123718  18.386344423492154  \\
            4.926570144897797  18.43816096968496  \\
            5.059365017671875  18.49967237081402  \\
            5.192159890445954  18.57149316217041  \\
            5.324954763220032  18.654272194289494  \\
            5.4577496359941104  18.748698239929773  \\
            5.590544508768189  18.85550639045799  \\
            5.723339381542268  18.975485359771852  \\
            5.856134254316347  19.10948583818286  \\
            5.988929127090425  19.258430068210075  \\
            6.121723999864503  19.423322850278247  \\
        }
        ;
\end{axis}
\end{tikzpicture}

%% file: figures/christmas_tree_left.tikz
\begin{tikzpicture}[/tikz/background rectangle/.style={fill={rgb,1:red,1.0;green,1.0;blue,1.0}, draw opacity={1.0}}, show background rectangle]
\begin{axis}[point meta max={nan}, point meta min={nan}, legend cell align={left}, title={}, title style={at={{(0.5,1)}}, anchor={south}, font={{\fontsize{25.2 pt}{32.76 pt}\selectfont}}, color={rgb,1:red,0.0;green,0.0;blue,0.0}, draw opacity={1.0}, rotate={0.0}}, legend style={color={rgb,1:red,0.0;green,0.0;blue,0.0}, draw opacity={0.0}, line width={1.8}, solid, fill={rgb,1:red,0.0;green,0.0;blue,0.0}, fill opacity={0.0}, text opacity={1.0}, font={{\fontsize{14.4 pt}{18.720000000000002 pt}\selectfont}}, text={rgb,1:red,0.0;green,0.0;blue,0.0}, at={(0.02, 0.98)}, anchor={north west}}, axis background/.style={fill={rgb,1:red,1.0;green,1.0;blue,1.0}, opacity={1.0}}, anchor={north west}, xshift={1.0mm}, yshift={-1.0mm}, width={137.7mm}, height={99.6mm}, scaled x ticks={false}, xlabel={$\sigmacv$}, x tick style={color={rgb,1:red,0.0;green,0.0;blue,0.0}, opacity={1.0}}, x tick label style={color={rgb,1:red,0.0;green,0.0;blue,0.0}, opacity={1.0}, rotate={0}}, xlabel style={at={(ticklabel cs:0.5)}, anchor=near ticklabel, font={{\fontsize{19.8 pt}{25.740000000000002 pt}\selectfont}}, color={rgb,1:red,0.0;green,0.0;blue,0.0}, draw opacity={1.0}, rotate={0.0}}, xmajorgrids={false}, xmin={-0.3812409198821226}, xmax={13.55425302902968}, xtick={{0.0,2.0,4.0,6.0,8.0,10.0,12.0}}, xticklabels={{$0$,$2$,$4$,$6$,$8$,$10$,$12$}}, xtick align={inside}, xticklabel style={font={{\fontsize{14.4 pt}{18.720000000000002 pt}\selectfont}}, color={rgb,1:red,0.0;green,0.0;blue,0.0}, draw opacity={1.0}, rotate={0.0}}, x grid style={color={rgb,1:red,0.0;green,0.0;blue,0.0}, draw opacity={0.1}, line width={0.9}, solid}, x axis line style={color={rgb,1:red,0.0;green,0.0;blue,0.0}, draw opacity={1.0}, line width={1.8}, solid}, scaled y ticks={false}, ylabel={$\widehat{\lambda}_g(\sigmacv)$}, y tick style={color={rgb,1:red,0.0;green,0.0;blue,0.0}, opacity={1.0}}, y tick label style={color={rgb,1:red,0.0;green,0.0;blue,0.0}, opacity={1.0}, rotate={0}}, ylabel style={at={(ticklabel cs:0.5)}, anchor=near ticklabel, font={{\fontsize{19.8 pt}{25.740000000000002 pt}\selectfont}}, color={rgb,1:red,0.0;green,0.0;blue,0.0}, draw opacity={1.0}, rotate={0.0}}, ymajorgrids={false}, ymin={0}, ymax={15}, ytick={{0.0,5.0,10.0,15.0}}, yticklabels={{$0$,$5$,$10$,$15$}}, ytick align={inside}, yticklabel style={font={{\fontsize{14.4 pt}{18.720000000000002 pt}\selectfont}}, color={rgb,1:red,0.0;green,0.0;blue,0.0}, draw opacity={1.0}, rotate={0.0}}, y grid style={color={rgb,1:red,0.0;green,0.0;blue,0.0}, draw opacity={0.1}, line width={0.9}, solid}, y axis line style={color={rgb,1:red,0.0;green,0.0;blue,0.0}, draw opacity={1.0}, line width={1.8}, solid}]
    \addplot[color={rgb,1:red,0.502;green,0.0;blue,0.502}, name path={8ff3b116-278f-4c7f-b166-8703a4de2fb9}, draw opacity={1.0}, line width={1.8}, dotted]
        table[row sep={\\}]
        {
            \\
            0.013159852256890666  2.743639823400004e-6  \\
            0.1459547250309692  0.00033761975251406825  \\
            0.27874959780504777  0.0012327205795672012  \\
            0.4115444705791263  0.0026914934270527184  \\
            0.5443393433532048  0.004719579196816768  \\
            0.6771342161272833  0.007324867157792461  \\
            0.809929088901362  0.010517572208620807  \\
            0.9427239616754405  0.01431033605479869  \\
            1.075518834449519  0.018718354191550633  \\
            1.2083137072235974  0.023759531109061625  \\
            1.341108579997676  0.02945466673718111  \\
            1.4739034527717545  0.03582767784325298  \\
            1.6066983255458331  0.0429058589149553  \\
            1.7394931983199116  0.050720188031533206  \\
            1.8722880710939902  0.05930568439097097  \\
            2.005082943868069  0.06870182556703933  \\
            2.1378778166421473  0.07895303428166843  \\
            2.2706726894162257  0.0901092465751254  \\
            2.4034675621903046  0.1022265758425291  \\
            2.536262434964383  0.11536809041450527  \\
            2.6690573077384614  0.1296047263674751  \\
            2.80185218051254  0.14501636228555628  \\
            2.9346470532866187  0.16169308906649005  \\
            3.067441926060697  0.17973671597569632  \\
            3.2002367988347755  0.199262564553493  \\
            3.333031671608854  0.22040161541417777  \\
            3.465826544382933  0.24330309045973403  \\
            3.5986214171570112  0.26813757597141946  \\
            3.7314162899310896  0.2951008224013575  \\
            3.864211162705168  0.32441839723009647  \\
            3.997006035479247  0.3563514219301928  \\
            4.129800908253325  0.39120369857408094  \\
            4.262595781027404  0.4293306342702289  \\
            4.395390653801482  0.4711505147407071  \\
            4.528185526575561  0.5171588805209781  \\
            4.66098039934964  0.5679470488282781  \\
            4.793775272123718  0.624226245208503  \\
            4.926570144897797  0.686859431511303  \\
            5.059365017671875  0.7569038535765367  \\
            5.192159890445954  0.8356687701791603  \\
            5.324954763220032  0.924795081206935  \\
            5.4577496359941104  1.0263671999379433  \\
            5.590544508768189  1.1430735034260822  \\
            5.723339381542268  1.2784418892793514  \\
            5.856134254316347  1.43719492471534  \\
            5.988929127090425  1.6258019708720648  \\
            6.121723999864503  1.8533687180693506  \\
            6.254518872638582  2.133131973098582  \\
            6.38731374541266  2.4851016090627307  \\
            6.520108618186739  2.9410274286255222  \\
            6.652903490960818  3.554488094925424  \\
            6.785698363734896  4.4235436019963155  \\
            6.918493236508975  5.748984352427512  \\
            7.051288109283053  8.01659376470863  \\
            7.184082982057132  12.780872678145714  \\
            7.31687785483121  29.195973865602703  \\
        }
        ;
    \addlegendentry {Group 1}
    \addplot[color={rgb,1:red,0.0;green,0.502;blue,0.0}, name path={ccc5c3f9-686e-4532-b673-59da6ffd60df}, draw opacity={1.0}, line width={1.8}, dashdotted]
        table[row sep={\\}]
        {
            \\
            0.013159852256890666  1.2759585810281856e-6  \\
            0.1459547250309692  0.00015698167520514965  \\
            0.27874959780504777  0.0005728615159524226  \\
            0.4115444705791263  0.0012496655671031546  \\
            0.5443393433532048  0.0021886168326235903  \\
            0.6771342161272833  0.003391416770532719  \\
            0.809929088901362  0.004860253026257719  \\
            0.9427239616754405  0.006597809428825099  \\
            1.075518834449519  0.00860727833559284  \\
            1.2083137072235974  0.01089237543227447  \\
            1.341108579997676  0.013457357117546215  \\
            1.4739034527717545  0.016307040625900802  \\
            1.6066983255458331  0.019446827068993297  \\
            1.7394931983199116  0.022882727604938178  \\
            1.8722880710939902  0.026621392977344498  \\
            2.005082943868069  0.030670146701865984  \\
            2.1378778166421473  0.035037022218329605  \\
            2.2706726894162257  0.039730804371819445  \\
            2.4034675621903046  0.044761075637286624  \\
            2.536262434964383  0.05013826756032249  \\
            2.6690573077384614  0.055873717952840796  \\
            2.80185218051254  0.06197973445793407  \\
            2.9346470532866187  0.06846966518472612  \\
            3.067441926060697  0.07535797721355515  \\
            3.2002367988347755  0.08266034388657822  \\
            3.333031671608854  0.09039374193161381  \\
            3.465826544382933  0.09857655962099948  \\
            3.5986214171570112  0.10722871734635517  \\
            3.7314162899310896  0.11637180219913225  \\
            3.864211162705168  0.12602921839140027  \\
            3.997006035479247  0.13622635563840615  \\
            4.129800908253325  0.1469907779624591  \\
            4.262595781027404  0.15835243577695515  \\
            4.395390653801482  0.17034390458248258  \\
            4.528185526575561  0.1830006541695075  \\
            4.66098039934964  0.1963613528933502  \\
            4.793775272123718  0.21046821239090435  \\
            4.926570144897797  0.22536737907459597  \\
            5.059365017671875  0.24110937990470657  \\
            5.192159890445954  0.25774963135334605  \\
            5.324954763220032  0.27534902219139706  \\
            5.4577496359941104  0.293974582829149  \\
            5.590544508768189  0.31370025651855316  \\
            5.723339381542268  0.33460779090406856  \\
            5.856134254316347  0.35678777235026643  \\
            5.988929127090425  0.3803408303863391  \\
            6.121723999864503  0.40537904576360106  \\
            6.254518872638582  0.43202760338219187  \\
            6.38731374541266  0.46042674118582344  \\
            6.520108618186739  0.4907340586886723  \\
            6.652903490960818  0.5231272649497103  \\
            6.785698363734896  0.5578074667212748  \\
            6.918493236508975  0.5950031247822273  \\
            7.051288109283053  0.6349748423546095  \\
            7.184082982057132  0.6780211971213916  \\
            7.31687785483121  0.7244858921321864  \\
            7.4496727276052885  0.7747766102674493  \\
            7.582467600379367  0.8293986008104144  \\
            7.715262473153446  0.8888553649068226  \\
            7.848057345927525  0.9537878133657404  \\
            7.980852218701603  1.0249562873392755  \\
            8.11364709147568  1.1032697012503625  \\
            8.24644196424976  1.1898236588006839  \\
            8.37923683702384  1.2859509395953108  \\
            8.512031709797917  1.3932893082062405  \\
            8.644826582571996  1.5138740002261486  \\
            8.777621455346074  1.6502660388673374  \\
            8.910416328120153  1.8057336947291163  \\
            9.04321120089423  1.9845146703280105  \\
            9.17600607366831  2.192204261118006  \\
            9.308800946442389  2.4363462589140106  \\
            9.441595819216467  2.727361922613578  \\
            9.574390691990546  3.0800664511169282  \\
            9.707185564764623  3.5162575060870305  \\
            9.839980437538703  4.069378057686087  \\
            9.97277531031278  4.793491802743984  \\
            10.10557018308686  5.782072558063061  \\
            10.238365055860939  7.211916919739639  \\
            10.371159928635016  9.46260928738467  \\
            10.503954801409096  13.52416353467576  \\
            10.636749674183173  23.049122580710293  \\
            10.769544546957253  71.6701272772952  \\
        }
        ;
    \addlegendentry {Group 2}
    \addplot[color={rgb,1:red,0.502;green,0.502;blue,0.502}, name path={78938277-4ae6-426e-9a55-b31ff1b3c550}, draw opacity={1.0}, line width={1.8}, dashed]
        table[row sep={\\}]
        {
            \\
            0.013159852256890666  8.581160979195409e-7  \\
            0.1459547250309692  0.0001055682373995495  \\
            0.27874959780504777  0.0003851826907965093  \\
            0.4115444705791263  0.0008400431309457082  \\
            0.5443393433532048  0.0014707060545453962  \\
            0.6771342161272833  0.002277944504813442  \\
            0.809929088901362  0.0032627504456954597  \\
            0.9427239616754405  0.0044263378194472295  \\
            1.075518834449519  0.005770146305245945  \\
            1.2083137072235974  0.007295845800659396  \\
            1.341108579997676  0.009005341652181474  \\
            1.4739034527717545  0.010900780665669695  \\
            1.6066983255458331  0.012984557932442263  \\
            1.7394931983199116  0.015259324512059337  \\
            1.8722880710939902  0.01772799601848161  \\
            2.005082943868069  0.020393762162430135  \\
            2.1378778166421473  0.0232600973094334  \\
            2.2706726894162257  0.026330772120317047  \\
            2.4034675621903046  0.029609866348853833  \\
            2.536262434964383  0.0331017828800429  \\
            2.6690573077384614  0.036811263102137386  \\
            2.80185218051254  0.040743403716209915  \\
            2.9346470532866187  0.04490367509887877  \\
            3.067441926060697  0.049297941346970686  \\
            3.2002367988347755  0.053932482147555215  \\
            3.333031671608854  0.05881401663315512  \\
            3.465826544382933  0.06394972940026153  \\
            3.5986214171570112  0.06934729888983561  \\
            3.7314162899310896  0.07501492835157847  \\
            3.864211162705168  0.08096137963976636  \\
            3.997006035479247  0.08719601011780714  \\
            4.129800908253325  0.09372881298186721  \\
            4.262595781027404  0.10057046135152567  \\
            4.395390653801482  0.10773235651810247  \\
            4.528185526575561  0.11522668078986567  \\
            4.66098039934964  0.12306645542866929  \\
            4.793775272123718  0.13126560423578357  \\
            4.926570144897797  0.13983902341702875  \\
            5.059365017671875  0.14880265844030408  \\
            5.192159890445954  0.1581735886939952  \\
            5.324954763220032  0.1679701208646437  \\
            5.4577496359941104  0.17821189207917484  \\
            5.590544508768189  0.18891998400388918  \\
            5.723339381542268  0.20011704926289417  \\
            5.856134254316347  0.21182745173697148  \\
            5.988929127090425  0.22407742253519308  \\
            6.121723999864503  0.2368952337021319  \\
            6.254518872638582  0.2503113920408057  \\
            6.38731374541266  0.2643588558046961  \\
            6.520108618186739  0.279073277452469  \\
            6.652903490960818  0.2944932761800768  \\
            6.785698363734896  0.3106607445635271  \\
            6.918493236508975  0.3276211943825311  \\
            7.051288109283053  0.3454241475761353  \\
            7.184082982057132  0.36412357933829836  \\
            7.31687785483121  0.38377842163410714  \\
            7.4496727276052885  0.4044559254577677  \\
            7.582467600379367  0.42623797922714185  \\
            7.715262473153446  0.4491908869123634  \\
            7.848057345927525  0.4734004100466561  \\
            7.980852218701603  0.4989611646750881  \\
            8.11364709147568  0.5259777852952874  \\
            8.24644196424976  0.5545662777119348  \\
            8.37923683702384  0.5848555975554285  \\
            8.512031709797917  0.6169894996200842  \\
            8.644826582571996  0.6511287138083264  \\
            8.777621455346074  0.687453516999512  \\
            8.910416328120153  0.7261667875025253  \\
            9.04321120089423  0.7674976511292344  \\
            9.17600607366831  0.8117058570217145  \\
            9.308800946442389  0.8590870594977593  \\
            9.441595819216467  0.909979232576069  \\
            9.574390691990546  0.9647705110505448  \\
            9.707185564764623  1.0239088424731089  \\
            9.839980437538703  1.0879139574965595  \\
            9.97277531031278  1.1573923353183941  \\
            10.10557018308686  1.2330560765564975  \\
            10.238365055860939  1.315746927941446  \\
            10.371159928635016  1.4064671777064812  \\
            10.503954801409096  1.5064198287860944  \\
            10.636749674183173  1.6170614715058347  \\
            10.769544546957253  1.7401727998588816  \\
            10.90233941973133  1.878073296235325  \\
            11.03513429250541  2.0335613083468234  \\
            11.167929165279489  2.210045571192189  \\
            11.300724038053566  2.4120403433535587  \\
            11.433518910827646  2.645455049749713  \\
            11.566313783601723  2.9181783600889286  \\
            11.699108656375802  3.2409825072716383  \\
            11.83190352914988  3.6289732927126788  \\
            11.96469840192396  4.104012894091302  \\
            12.097493274698037  4.698972757073293  \\
            12.230288147472116  5.465661608238835  \\
            12.363083020246195  6.4907571587658826  \\
            12.495877893020273  7.9310752636006105  \\
            12.628672765794352  10.102503949869117  \\
            12.76146763856843  13.750233946707711  \\
            12.89426251134251  21.153348280245773  \\
            13.027057384116587  44.25555751033649  \\
        }
        ;
    \addlegendentry {Group 3}
\end{axis}
\end{tikzpicture}

%% file: figures/oracle_risks_identity/oracle_risk1.tikz
\begin{tikzpicture}[/tikz/background rectangle/.style={fill={rgb,1:red,1.0;green,1.0;blue,1.0}, draw opacity={1.0}}, show background rectangle]
\begin{axis}[point meta max={nan}, point meta min={nan}, legend cell align={left}, title={$\gamma_1 = \gamma_2 = \frac{1}{4},\;\; \alpha_1^2 + \alpha_2^2 = 1$}, title style={at={{(0.5,1)}}, anchor={south}, font={{\fontsize{30.800000000000004 pt}{40.040000000000006 pt}\selectfont}}, color={rgb,1:red,0.0;green,0.0;blue,0.0}, draw opacity={1.0}, rotate={0.0}}, legend style={color={rgb,1:red,0.0;green,0.0;blue,0.0}, draw opacity={0.0}, line width={2.2}, solid, fill={rgb,1:red,0.0;green,0.0;blue,0.0}, fill opacity={0.0}, text opacity={1.0}, font={{\fontsize{26.400000000000002 pt}{34.32000000000001 pt}\selectfont}}, text={rgb,1:red,0.0;green,0.0;blue,0.0}, at={(0.02, 0.98)}, anchor={north west}}, axis background/.style={fill={rgb,1:red,1.0;green,1.0;blue,1.0}, opacity={1.0}}, anchor={north west}, xshift={1.0mm}, yshift={-1.0mm}, width={163.1mm}, height={125.0mm}, scaled x ticks={false}, xlabel={$\alpha_1^2/(\alpha_1^2 + \alpha_2^2)$}, x tick style={color={rgb,1:red,0.0;green,0.0;blue,0.0}, opacity={1.0}}, x tick label style={color={rgb,1:red,0.0;green,0.0;blue,0.0}, opacity={1.0}, rotate={0}}, xlabel style={at={(ticklabel cs:0.5)}, anchor=near ticklabel, font={{\fontsize{24.200000000000003 pt}{31.460000000000004 pt}\selectfont}}, color={rgb,1:red,0.0;green,0.0;blue,0.0}, draw opacity={1.0}, rotate={0.0}}, xmajorgrids={false}, xmin={-0.03}, xmax={1.03}, xtick={{0.0,0.25,0.5,0.75,1.0}}, xticklabels={{$0.00$,$0.25$,$0.50$,$0.75$,$1.00$}}, xtick align={inside}, xticklabel style={font={{\fontsize{17.6 pt}{22.880000000000003 pt}\selectfont}}, color={rgb,1:red,0.0;green,0.0;blue,0.0}, draw opacity={1.0}, rotate={0.0}}, x grid style={color={rgb,1:red,0.0;green,0.0;blue,0.0}, draw opacity={0.1}, line width={1.1}, solid}, x axis line style={color={rgb,1:red,0.0;green,0.0;blue,0.0}, draw opacity={1.0}, line width={2.2}, solid}, scaled y ticks={false}, ylabel={$\risk{\blambda}- \sigma^2$}, y tick style={color={rgb,1:red,0.0;green,0.0;blue,0.0}, opacity={1.0}}, y tick label style={color={rgb,1:red,0.0;green,0.0;blue,0.0}, opacity={1.0}, rotate={0}}, ylabel style={at={(ticklabel cs:0.5)}, anchor=near ticklabel, font={{\fontsize{24.200000000000003 pt}{31.460000000000004 pt}\selectfont}}, color={rgb,1:red,0.0;green,0.0;blue,0.0}, draw opacity={1.0}, rotate={0.0}}, ymajorgrids={false}, ymin={0}, ymax={2.5}, ytick={{0.0,0.5,1.0,1.5,2.0,2.5}}, yticklabels={{$0.0$,$0.5$,$1.0$,$1.5$,$2.0$,$2.5$}}, ytick align={inside}, yticklabel style={font={{\fontsize{17.6 pt}{22.880000000000003 pt}\selectfont}}, color={rgb,1:red,0.0;green,0.0;blue,0.0}, draw opacity={1.0}, rotate={0.0}}, y grid style={color={rgb,1:red,0.0;green,0.0;blue,0.0}, draw opacity={0.1}, line width={1.1}, solid}, y axis line style={color={rgb,1:red,0.0;green,0.0;blue,0.0}, draw opacity={1.0}, line width={2.2}, solid}]
    \addplot[color={rgb,1:red,0.502;green,0.502;blue,0.502}, name path={e83658fd-ed76-4798-9e84-7c0a019e2d75}, draw opacity={1.0}, line width={2.2}, dotted]
        table[row sep={\\}]
        {
            \\
            0.0  0.9998000496736081  \\
            0.034482758620689655  0.9975826593509294  \\
            0.06896551724137931  0.9914281250833361  \\
            0.10344827586206896  0.9819607831101558  \\
            0.13793103448275862  0.9696545738253419  \\
            0.1724137931034483  0.9548809831783389  \\
            0.20689655172413793  0.9379394667307874  \\
            0.2413793103448276  0.9190663046873864  \\
            0.27586206896551724  0.8984672179334492  \\
            0.3103448275862069  0.876303554474744  \\
            0.3448275862068966  0.8527242577063086  \\
            0.3793103448275862  0.8278439032774907  \\
            0.41379310344827586  0.8017730919068531  \\
            0.4482758620689655  0.7746021732393586  \\
            0.4827586206896552  0.746412145175593  \\
            0.5172413793103449  0.7172743305143487  \\
            0.5517241379310345  0.6872509175243025  \\
            0.5862068965517241  0.6563974559160775  \\
            0.6206896551724138  0.6247655266748902  \\
            0.6551724137931034  0.5924016096769666  \\
            0.6896551724137931  0.5593478021598135  \\
            0.7241379310344828  0.5256405123968608  \\
            0.7586206896551724  0.4913166649260732  \\
            0.7931034482758621  0.45640329216951536  \\
            0.8275862068965517  0.42093280714261505  \\
            0.8620689655172413  0.38492961163143513  \\
            0.896551724137931  0.3484197299279894  \\
            0.9310344827586207  0.3114261168911332  \\
            0.9655172413793104  0.27396903466769773  \\
            1.0  0.23606797999813733  \\
        }
        ;
    \addlegendentry {$\;$Optimal $\blambda = (\lambda, \infty)$}
    \addplot[color={rgb,1:red,0.502;green,0.0;blue,0.502}, name path={60bc648f-87f7-4aad-8f6e-2b68e8f5726b}, draw opacity={1.0}, line width={2.2}, dashdotted]
        table[row sep={\\}]
        {
            \\
            0.0  0.41421356237309515  \\
            0.034482758620689655  0.41421356237309515  \\
            0.06896551724137931  0.41421356237309515  \\
            0.10344827586206896  0.41421356237309515  \\
            0.13793103448275862  0.41421356237309515  \\
            0.1724137931034483  0.41421356237309515  \\
            0.20689655172413793  0.41421356237309515  \\
            0.2413793103448276  0.41421356237309515  \\
            0.27586206896551724  0.41421356237309515  \\
            0.3103448275862069  0.41421356237309515  \\
            0.3448275862068966  0.41421356237309515  \\
            0.3793103448275862  0.41421356237309515  \\
            0.41379310344827586  0.41421356237309515  \\
            0.4482758620689655  0.41421356237309515  \\
            0.4827586206896552  0.41421356237309515  \\
            0.5172413793103449  0.41421356237309515  \\
            0.5517241379310345  0.41421356237309515  \\
            0.5862068965517241  0.41421356237309515  \\
            0.6206896551724138  0.41421356237309515  \\
            0.6551724137931034  0.41421356237309515  \\
            0.6896551724137931  0.41421356237309515  \\
            0.7241379310344828  0.41421356237309515  \\
            0.7586206896551724  0.41421356237309515  \\
            0.7931034482758621  0.41421356237309515  \\
            0.8275862068965517  0.41421356237309515  \\
            0.8620689655172413  0.41421356237309515  \\
            0.896551724137931  0.41421356237309515  \\
            0.9310344827586207  0.41421356237309515  \\
            0.9655172413793104  0.41421356237309515  \\
            1.0  0.41421356237309515  \\
        }
        ;
    \addlegendentry {$\;$Optimal $\blambda = (\lambda, \lambda)$}
    \addplot[color={rgb,1:red,0.0;green,0.502;blue,0.0}, name path={1e769706-2b19-45bb-a6c6-b58c866697cf}, draw opacity={1.0}, line width={2.2}, dashed]
        table[row sep={\\}]
        {
            \\
            0.0  0.23606797999813733  \\
            0.034482758620689655  0.2700407735673722  \\
            0.06896551724137931  0.29742511781089487  \\
            0.10344827586206896  0.3199688524327935  \\
            0.13793103448275862  0.3387673335294945  \\
            0.1724137931034483  0.35455403694140175  \\
            0.20689655172413793  0.36784340054865616  \\
            0.2413793103448276  0.37900829673322933  \\
            0.27586206896551724  0.388325199986169  \\
            0.3103448275862069  0.39600199591491214  \\
            0.3448275862068966  0.40219582035984014  \\
            0.3793103448275862  0.40702484366083036  \\
            0.41379310344827586  0.41057618970312193  \\
            0.4482758620689655  0.4129112646671653  \\
            0.4827586206896552  0.41406925612878154  \\
            0.5172413793103449  0.41406925612878154  \\
            0.5517241379310345  0.4129112646671653  \\
            0.5862068965517241  0.41057618970312193  \\
            0.6206896551724138  0.40702484366083036  \\
            0.6551724137931034  0.40219582035984014  \\
            0.6896551724137931  0.39600199591491214  \\
            0.7241379310344828  0.388325199986169  \\
            0.7586206896551724  0.37900829673322933  \\
            0.7931034482758621  0.36784340054865616  \\
            0.8275862068965517  0.35455403694140175  \\
            0.8620689655172413  0.3387673335294945  \\
            0.896551724137931  0.3199688524327935  \\
            0.9310344827586207  0.29742511781089487  \\
            0.9655172413793104  0.2700407735673722  \\
            1.0  0.23606797999813733  \\
        }
        ;
    \addlegendentry {$\;$Optimal $\blambda = (\lambda_1, \lambda_2)$}
    \addplot[color={rgb,1:red,0.502;green,0.502;blue,0.502}, name path={70e612cb-95e6-4e5e-bff5-dce13d9e1da9}, only marks, draw opacity={1.0}, line width={0.0}, solid, mark={triangle*}, mark size={6.6000000000000005 pt}, mark repeat={1}, mark options={color={rgb,1:red,0.0;green,0.0;blue,0.0}, draw opacity={0.0}, fill={rgb,1:red,0.502;green,0.502;blue,0.502}, fill opacity={1.0}, line width={1.6500000000000001}, rotate={0}, solid}, forget plot]
        table[row sep={\\}]
        {
            \\
            0.0  0.9750025152643609  \\
            0.034482758620689655  0.9605445403608222  \\
            0.06896551724137931  0.8939709480683145  \\
            0.10344827586206896  0.9427634352942129  \\
            0.13793103448275862  0.970832104082977  \\
            0.1724137931034483  0.9160730101972516  \\
            0.20689655172413793  0.9567137210462697  \\
            0.2413793103448276  0.9834136871024355  \\
            0.27586206896551724  0.9697344492271405  \\
            0.3103448275862069  1.0415527936941245  \\
            0.3448275862068966  0.7734614068922667  \\
            0.3793103448275862  0.7825119361279969  \\
            0.41379310344827586  0.7174074633025656  \\
            0.4482758620689655  0.7713622569366212  \\
            0.4827586206896552  0.7531332484286453  \\
            0.5172413793103449  0.6445930503315274  \\
            0.5517241379310345  0.5921323922572319  \\
            0.5862068965517241  0.7116620387319279  \\
            0.6206896551724138  0.6753522465875805  \\
            0.6551724137931034  0.5133570042828968  \\
            0.6896551724137931  0.5446993685313501  \\
            0.7241379310344828  0.48015777980432106  \\
            0.7586206896551724  0.5010034728486734  \\
            0.7931034482758621  0.5011330664347817  \\
            0.8275862068965517  0.41937489294513464  \\
            0.8620689655172413  0.3730664708424487  \\
            0.896551724137931  0.38470191828976885  \\
            0.9310344827586207  0.27556983876824925  \\
            0.9655172413793104  0.2522342184698201  \\
            1.0  0.2194140826655886  \\
        }
        ;
    \addplot[color={rgb,1:red,0.502;green,0.0;blue,0.502}, name path={b5e2eb0c-a814-43eb-8ba9-deee0bb812df}, only marks, draw opacity={1.0}, line width={0.0}, solid, mark={triangle*}, mark size={6.6000000000000005 pt}, mark repeat={1}, mark options={color={rgb,1:red,0.0;green,0.0;blue,0.0}, draw opacity={0.0}, fill={rgb,1:red,0.502;green,0.0;blue,0.502}, fill opacity={1.0}, line width={1.6500000000000001}, rotate={0}, solid}, forget plot]
        table[row sep={\\}]
        {
            \\
            0.0  0.3513963854347859  \\
            0.034482758620689655  0.3826055820798806  \\
            0.06896551724137931  0.43108377622194594  \\
            0.10344827586206896  0.3980391473147604  \\
            0.13793103448275862  0.39570244317183745  \\
            0.1724137931034483  0.4495059132510799  \\
            0.20689655172413793  0.43003189538625164  \\
            0.2413793103448276  0.46114951198505216  \\
            0.27586206896551724  0.42762585936901343  \\
            0.3103448275862069  0.49186924686821243  \\
            0.3448275862068966  0.4138982844066845  \\
            0.3793103448275862  0.4027672682567165  \\
            0.41379310344827586  0.363604076605174  \\
            0.4482758620689655  0.41261948679455407  \\
            0.4827586206896552  0.3932962818318051  \\
            0.5172413793103449  0.3500910820583283  \\
            0.5517241379310345  0.38335137154319554  \\
            0.5862068965517241  0.45763060220347906  \\
            0.6206896551724138  0.4324927291560874  \\
            0.6551724137931034  0.39448553901823713  \\
            0.6896551724137931  0.43490303774138384  \\
            0.7241379310344828  0.4037519837227672  \\
            0.7586206896551724  0.43674423060837664  \\
            0.7931034482758621  0.4326472170361031  \\
            0.8275862068965517  0.3485167751256055  \\
            0.8620689655172413  0.416251173235618  \\
            0.896551724137931  0.4455477212084975  \\
            0.9310344827586207  0.3974013964949976  \\
            0.9655172413793104  0.42968715703266436  \\
            1.0  0.40982390257385526  \\
        }
        ;
    \addplot[color={rgb,1:red,0.0;green,0.502;blue,0.0}, name path={ae7b1ed6-8123-4150-8774-24e162f826be}, only marks, draw opacity={1.0}, line width={0.0}, solid, mark={triangle*}, mark size={6.6000000000000005 pt}, mark repeat={1}, mark options={color={rgb,1:red,0.0;green,0.0;blue,0.0}, draw opacity={0.0}, fill={rgb,1:red,0.0;green,0.502;blue,0.0}, fill opacity={1.0}, line width={1.6500000000000001}, rotate={0}, solid}, forget plot]
        table[row sep={\\}]
        {
            \\
            0.0  0.1928217537202066  \\
            0.034482758620689655  0.25074579928958185  \\
            0.06896551724137931  0.30543842508684516  \\
            0.10344827586206896  0.33095366719385666  \\
            0.13793103448275862  0.3379951412318516  \\
            0.1724137931034483  0.39224679433857723  \\
            0.20689655172413793  0.3790065355872638  \\
            0.2413793103448276  0.420878765202223  \\
            0.27586206896551724  0.39887418902064975  \\
            0.3103448275862069  0.45613180627726213  \\
            0.3448275862068966  0.40001834980282025  \\
            0.3793103448275862  0.38874031531109465  \\
            0.41379310344827586  0.3580344868196228  \\
            0.4482758620689655  0.4116165599160073  \\
            0.4827586206896552  0.39245293957913896  \\
            0.5172413793103449  0.3492824228264908  \\
            0.5517241379310345  0.3790166218851423  \\
            0.5862068965517241  0.45742607147732817  \\
            0.6206896551724138  0.42934554478263265  \\
            0.6551724137931034  0.3640315420096494  \\
            0.6896551724137931  0.42725442630068056  \\
            0.7241379310344828  0.369641530747701  \\
            0.7586206896551724  0.38990541771327436  \\
            0.7931034482758621  0.3827371101192609  \\
            0.8275862068965517  0.3160897942762022  \\
            0.8620689655172413  0.3353256605011361  \\
            0.896551724137931  0.3429972035129991  \\
            0.9310344827586207  0.2642501044173746  \\
            0.9655172413793104  0.2504313521087722  \\
            1.0  0.2194140826655886  \\
        }
        ;
\end{axis}
\end{tikzpicture}

%% file: figures/oracle_risks_identity/oracle_risk2.tikz
\begin{tikzpicture}[/tikz/background rectangle/.style={fill={rgb,1:red,1.0;green,1.0;blue,1.0}, draw opacity={1.0}}, show background rectangle]
\begin{axis}[point meta max={nan}, point meta min={nan}, legend cell align={left}, title={$\gamma_1 = \frac{1}{10},\; \gamma_2 = \frac{4}{10},\;\; \alpha_1^2 + \alpha_2^2 = 1$}, title style={at={{(0.5,1)}}, anchor={south}, font={{\fontsize{30.800000000000004 pt}{40.040000000000006 pt}\selectfont}}, color={rgb,1:red,0.0;green,0.0;blue,0.0}, draw opacity={1.0}, rotate={0.0}}, legend style={color={rgb,1:red,0.0;green,0.0;blue,0.0}, draw opacity={0.0}, line width={2.2}, solid, fill={rgb,1:red,0.0;green,0.0;blue,0.0}, fill opacity={0.0}, text opacity={1.0}, font={{\fontsize{26.400000000000002 pt}{34.32000000000001 pt}\selectfont}}, text={rgb,1:red,0.0;green,0.0;blue,0.0}, at={(1.02, 1)}, anchor={north west}}, axis background/.style={fill={rgb,1:red,1.0;green,1.0;blue,1.0}, opacity={1.0}}, anchor={north west}, xshift={1.0mm}, yshift={-1.0mm}, width={163.1mm}, height={125.0mm}, scaled x ticks={false}, xlabel={$\alpha_1^2/(\alpha_1^2 + \alpha_2^2)$}, x tick style={color={rgb,1:red,0.0;green,0.0;blue,0.0}, opacity={1.0}}, x tick label style={color={rgb,1:red,0.0;green,0.0;blue,0.0}, opacity={1.0}, rotate={0}}, xlabel style={at={(ticklabel cs:0.5)}, anchor=near ticklabel, font={{\fontsize{24.200000000000003 pt}{31.460000000000004 pt}\selectfont}}, color={rgb,1:red,0.0;green,0.0;blue,0.0}, draw opacity={1.0}, rotate={0.0}}, xmajorgrids={false}, xmin={-0.03}, xmax={1.03}, xtick={{0.0,0.25,0.5,0.75,1.0}}, xticklabels={{$0.00$,$0.25$,$0.50$,$0.75$,$1.00$}}, xtick align={inside}, xticklabel style={font={{\fontsize{17.6 pt}{22.880000000000003 pt}\selectfont}}, color={rgb,1:red,0.0;green,0.0;blue,0.0}, draw opacity={1.0}, rotate={0.0}}, x grid style={color={rgb,1:red,0.0;green,0.0;blue,0.0}, draw opacity={0.1}, line width={1.1}, solid}, x axis line style={color={rgb,1:red,0.0;green,0.0;blue,0.0}, draw opacity={1.0}, line width={2.2}, solid}, scaled y ticks={false}, ylabel={$\risk{\blambda}- \sigma^2$}, y tick style={color={rgb,1:red,0.0;green,0.0;blue,0.0}, opacity={1.0}}, y tick label style={color={rgb,1:red,0.0;green,0.0;blue,0.0}, opacity={1.0}, rotate={0}}, ylabel style={at={(ticklabel cs:0.5)}, anchor=near ticklabel, font={{\fontsize{24.200000000000003 pt}{31.460000000000004 pt}\selectfont}}, color={rgb,1:red,0.0;green,0.0;blue,0.0}, draw opacity={1.0}, rotate={0.0}}, ymajorgrids={false}, ymin={0}, ymax={2.5}, ytick={{0.0,0.5,1.0,1.5,2.0,2.5}}, yticklabels={{$0.0$,$0.5$,$1.0$,$1.5$,$2.0$,$2.5$}}, ytick align={inside}, yticklabel style={font={{\fontsize{17.6 pt}{22.880000000000003 pt}\selectfont}}, color={rgb,1:red,0.0;green,0.0;blue,0.0}, draw opacity={1.0}, rotate={0.0}}, y grid style={color={rgb,1:red,0.0;green,0.0;blue,0.0}, draw opacity={0.1}, line width={1.1}, solid}, y axis line style={color={rgb,1:red,0.0;green,0.0;blue,0.0}, draw opacity={1.0}, line width={2.2}, solid}]
    \addplot[color={rgb,1:red,0.502;green,0.502;blue,0.502}, name path={a101c4e1-c26f-4a61-b9f1-4d0e26be3805}, draw opacity={1.0}, line width={2.2}, dotted]
        table[row sep={\\}]
        {
            \\
            0.0  0.9998000498868627  \\
            0.034482758620689655  0.9947275900588419  \\
            0.06896551724137931  0.9820160395078577  \\
            0.10344827586206896  0.9641395763668039  \\
            0.13793103448275862  0.9425637636515225  \\
            0.1724137931034483  0.9182297936154189  \\
            0.20689655172413793  0.8917735063201415  \\
            0.2413793103448276  0.8636444128418339  \\
            0.27586206896551724  0.8341695354707235  \\
            0.3103448275862069  0.8035931285682034  \\
            0.3448275862068966  0.7721045191933671  \\
            0.3793103448275862  0.7398486580858983  \\
            0.41379310344827586  0.7069430338721108  \\
            0.4482758620689655  0.6734789975530355  \\
            0.4827586206896552  0.6395342760566849  \\
            0.5172413793103449  0.6051707500263865  \\
            0.5517241379310345  0.5704400580894964  \\
            0.5862068965517241  0.5353876695751929  \\
            0.6206896551724138  0.5000487866206886  \\
            0.6551724137931034  0.4644561607615314  \\
            0.6896551724137931  0.42863527355678155  \\
            0.7241379310344828  0.3926106681467847  \\
            0.7586206896551724  0.35640180590277915  \\
            0.7931034482758621  0.3200260292276882  \\
            0.8275862068965517  0.28349914932560516  \\
            0.8620689655172413  0.24683452695668096  \\
            0.896551724137931  0.21004442923225053  \\
            0.9310344827586207  0.17313896909456328  \\
            0.9655172413793104  0.13612793067154416  \\
            1.0  0.09901951758307614  \\
        }
        ;
    \addplot[color={rgb,1:red,0.502;green,0.0;blue,0.502}, name path={2f5adae5-22c6-4619-aadf-6d5c73e50b82}, draw opacity={1.0}, line width={2.2}, dashdotted]
        table[row sep={\\}]
        {
            \\
            0.0  0.4142135623730856  \\
            0.034482758620689655  0.41421356237308293  \\
            0.06896551724137931  0.4142135623730887  \\
            0.10344827586206896  0.41421356237308804  \\
            0.13793103448275862  0.41421356237309204  \\
            0.1724137931034483  0.41421356237309315  \\
            0.20689655172413793  0.41421356237309537  \\
            0.2413793103448276  0.41421356237309714  \\
            0.27586206896551724  0.4142135623730987  \\
            0.3103448275862069  0.4142135623731005  \\
            0.3448275862068966  0.41421356237310203  \\
            0.3793103448275862  0.4142135623731038  \\
            0.41379310344827586  0.41421356237310536  \\
            0.4482758620689655  0.41421356237310714  \\
            0.4827586206896552  0.4142135623731087  \\
            0.5172413793103449  0.41421356237311047  \\
            0.5517241379310345  0.414213562373112  \\
            0.5862068965517241  0.4142135623731138  \\
            0.6206896551724138  0.41421356237311535  \\
            0.6551724137931034  0.41421356237311713  \\
            0.6896551724137931  0.4142135623731187  \\
            0.7241379310344828  0.41421356237312046  \\
            0.7586206896551724  0.414213562373122  \\
            0.7931034482758621  0.4142135623731238  \\
            0.8275862068965517  0.4142135623731411  \\
            0.8620689655172413  0.4142135623731271  \\
            0.896551724137931  0.4142135623731462  \\
            0.9310344827586207  0.41421356237313045  \\
            0.9655172413793104  0.4142135623731513  \\
            1.0  0.4142135623731338  \\
        }
        ;
    \addplot[color={rgb,1:red,0.0;green,0.502;blue,0.0}, name path={74939c41-3bca-4542-bf9b-324cebf3f9f1}, draw opacity={1.0}, line width={2.2}, dashed]
        table[row sep={\\}]
        {
            \\
            0.0  0.35078106024941436  \\
            0.034482758620689655  0.3785585795123474  \\
            0.06896551724137931  0.3950545253151272  \\
            0.10344827586206896  0.405040761536003  \\
            0.13793103448275862  0.41080343447694845  \\
            0.1724137931034483  0.41359841380963003  \\
            0.20689655172413793  0.41417803330447467  \\
            0.2413793103448276  0.4130201173574335  \\
            0.27586206896551724  0.41044041659531816  \\
            0.3103448275862069  0.4066528106015368  \\
            0.3448275862068966  0.4018036950335455  \\
            0.3793103448275862  0.39599260054805674  \\
            0.41379310344827586  0.38928499718482823  \\
            0.4482758620689655  0.38172041621095176  \\
            0.4827586206896552  0.37331761778456896  \\
            0.5172413793103449  0.3640777898105221  \\
            0.5517241379310345  0.3539863459220953  \\
            0.5862068965517241  0.3430136395562162  \\
            0.6206896551724138  0.3311147454645289  \\
            0.6551724137931034  0.31822833417361673  \\
            0.6896551724137931  0.30427454872718585  \\
            0.7241379310344828  0.2891516602499882  \\
            0.7586206896551724  0.272731097683355  \\
            0.7931034482758621  0.2548501678347932  \\
            0.8275862068965517  0.23530131474529314  \\
            0.8620689655172413  0.21381593382356412  \\
            0.896551724137931  0.19003917481466215  \\
            0.9310344827586207  0.16348895882523085  \\
            0.9655172413793104  0.13348538021573964  \\
            1.0  0.09901951758307614  \\
        }
        ;
    \addplot[color={rgb,1:red,0.502;green,0.502;blue,0.502}, name path={51eda290-e999-4954-9acf-8779e2c165ac}, only marks, draw opacity={1.0}, line width={0.0}, solid, mark={triangle*}, mark size={6.6000000000000005 pt}, mark repeat={1}, mark options={color={rgb,1:red,0.0;green,0.0;blue,0.0}, draw opacity={0.0}, fill={rgb,1:red,0.502;green,0.502;blue,0.502}, fill opacity={1.0}, line width={1.6500000000000001}, rotate={0}, solid}]
        table[row sep={\\}]
        {
            \\
            0.0  0.9940926142593831  \\
            0.034482758620689655  0.9777677038246504  \\
            0.06896551724137931  1.0781628554192073  \\
            0.10344827586206896  0.8866092492240589  \\
            0.13793103448275862  0.8713382603073216  \\
            0.1724137931034483  0.8883752736306705  \\
            0.20689655172413793  0.9547702679107839  \\
            0.2413793103448276  0.8979272606954707  \\
            0.27586206896551724  0.841521910119637  \\
            0.3103448275862069  0.8527478709417629  \\
            0.3448275862068966  0.7931048326693038  \\
            0.3793103448275862  0.6967654491099899  \\
            0.41379310344827586  0.5968803666773659  \\
            0.4482758620689655  0.6772737148303696  \\
            0.4827586206896552  0.5904770547609752  \\
            0.5172413793103449  0.5598074282147183  \\
            0.5517241379310345  0.5081760399130133  \\
            0.5862068965517241  0.5070220603592639  \\
            0.6206896551724138  0.5664178786755194  \\
            0.6551724137931034  0.44708971757779437  \\
            0.6896551724137931  0.45215459906941513  \\
            0.7241379310344828  0.36814139315368855  \\
            0.7586206896551724  0.3237102282441937  \\
            0.7931034482758621  0.3159793666523434  \\
            0.8275862068965517  0.3049483639335  \\
            0.8620689655172413  0.2453763674240952  \\
            0.896551724137931  0.18462227577358759  \\
            0.9310344827586207  0.14162156312551044  \\
            0.9655172413793104  0.15683049490109013  \\
            1.0  0.07478960336052398  \\
        }
        ;
    \addplot[color={rgb,1:red,0.502;green,0.0;blue,0.502}, name path={275d663b-53c9-4b66-8894-cd8f9a7eff6a}, only marks, draw opacity={1.0}, line width={0.0}, solid, mark={triangle*}, mark size={6.6000000000000005 pt}, mark repeat={1}, mark options={color={rgb,1:red,0.0;green,0.0;blue,0.0}, draw opacity={0.0}, fill={rgb,1:red,0.502;green,0.0;blue,0.502}, fill opacity={1.0}, line width={1.6500000000000001}, rotate={0}, solid}]
        table[row sep={\\}]
        {
            \\
            0.0  0.4256872403196965  \\
            0.034482758620689655  0.42594807064379125  \\
            0.06896551724137931  0.4256999214305124  \\
            0.10344827586206896  0.444679517706666  \\
            0.13793103448275862  0.34914636503436536  \\
            0.1724137931034483  0.38751972442101623  \\
            0.20689655172413793  0.4623332702488261  \\
            0.2413793103448276  0.3827177929082668  \\
            0.27586206896551724  0.40339217928775106  \\
            0.3103448275862069  0.4201640837334151  \\
            0.3448275862068966  0.4131141521776507  \\
            0.3793103448275862  0.4076464735326739  \\
            0.41379310344827586  0.4034163867378633  \\
            0.4482758620689655  0.48576932242602844  \\
            0.4827586206896552  0.406974019281078  \\
            0.5172413793103449  0.3819724654880643  \\
            0.5517241379310345  0.3378312434633455  \\
            0.5862068965517241  0.40044827241309155  \\
            0.6206896551724138  0.41592545240598855  \\
            0.6551724137931034  0.3765161881688541  \\
            0.6896551724137931  0.45857871409014805  \\
            0.7241379310344828  0.42027429245652703  \\
            0.7586206896551724  0.3952280206277652  \\
            0.7931034482758621  0.4239422106294892  \\
            0.8275862068965517  0.4307870565064369  \\
            0.8620689655172413  0.415798104068168  \\
            0.896551724137931  0.40185417633990994  \\
            0.9310344827586207  0.4156253498965077  \\
            0.9655172413793104  0.4686182329719668  \\
            1.0  0.4038517167044626  \\
        }
        ;
    \addplot[color={rgb,1:red,0.0;green,0.502;blue,0.0}, name path={a615e250-b2df-4f94-81e1-9cabb1ecc00c}, only marks, draw opacity={1.0}, line width={0.0}, solid, mark={triangle*}, mark size={6.6000000000000005 pt}, mark repeat={1}, mark options={color={rgb,1:red,0.0;green,0.0;blue,0.0}, draw opacity={0.0}, fill={rgb,1:red,0.0;green,0.502;blue,0.0}, fill opacity={1.0}, line width={1.6500000000000001}, rotate={0}, solid}]
        table[row sep={\\}]
        {
            \\
            0.0  0.3668943351745486  \\
            0.034482758620689655  0.393554297323748  \\
            0.06896551724137931  0.39955750648629884  \\
            0.10344827586206896  0.4336938653545399  \\
            0.13793103448275862  0.3491286976685486  \\
            0.1724137931034483  0.3860319275231274  \\
            0.20689655172413793  0.46272586962081275  \\
            0.2413793103448276  0.3849646943819254  \\
            0.27586206896551724  0.39946061813630873  \\
            0.3103448275862069  0.4134795059404337  \\
            0.3448275862068966  0.40777464954414544  \\
            0.3793103448275862  0.3865083139807133  \\
            0.41379310344827586  0.35904352966578235  \\
            0.4482758620689655  0.44890681756155515  \\
            0.4827586206896552  0.3398364735560637  \\
            0.5172413793103449  0.3249988594977229  \\
            0.5517241379310345  0.3092836199826534  \\
            0.5862068965517241  0.3270017289246985  \\
            0.6206896551724138  0.35978926945772627  \\
            0.6551724137931034  0.28425360583093795  \\
            0.6896551724137931  0.31736444450160883  \\
            0.7241379310344828  0.2851210483079758  \\
            0.7586206896551724  0.2572271152624812  \\
            0.7931034482758621  0.2592643383749167  \\
            0.8275862068965517  0.2509851407858521  \\
            0.8620689655172413  0.20935744035596482  \\
            0.896551724137931  0.16927255360907578  \\
            0.9310344827586207  0.1352712553474451  \\
            0.9655172413793104  0.15273472070960925  \\
            1.0  0.07478960336052398  \\
        }
        ;
\end{axis}
\end{tikzpicture}

%% file: figures/oracle_risks_identity/oracle_risk3.tikz
\begin{tikzpicture}[/tikz/background rectangle/.style={fill={rgb,1:red,1.0;green,1.0;blue,1.0}, draw opacity={1.0}}, show background rectangle]
\begin{axis}[point meta max={nan}, point meta min={nan}, legend cell align={left}, title={$\gamma_1 = \gamma_2 = 1,\;\; \alpha_1^2 + \alpha_2^2 = 1$}, title style={at={{(0.5,1)}}, anchor={south}, font={{\fontsize{30.800000000000004 pt}{40.040000000000006 pt}\selectfont}}, color={rgb,1:red,0.0;green,0.0;blue,0.0}, draw opacity={1.0}, rotate={0.0}}, legend style={color={rgb,1:red,0.0;green,0.0;blue,0.0}, draw opacity={0.0}, line width={2.2}, solid, fill={rgb,1:red,0.0;green,0.0;blue,0.0}, fill opacity={0.0}, text opacity={1.0}, font={{\fontsize{26.400000000000002 pt}{34.32000000000001 pt}\selectfont}}, text={rgb,1:red,0.0;green,0.0;blue,0.0}, at={(1.02, 1)}, anchor={north west}}, axis background/.style={fill={rgb,1:red,1.0;green,1.0;blue,1.0}, opacity={1.0}}, anchor={north west}, xshift={1.0mm}, yshift={-1.0mm}, width={163.1mm}, height={125.0mm}, scaled x ticks={false}, xlabel={$\alpha_1^2/(\alpha_1^2 + \alpha_2^2)$}, x tick style={color={rgb,1:red,0.0;green,0.0;blue,0.0}, opacity={1.0}}, x tick label style={color={rgb,1:red,0.0;green,0.0;blue,0.0}, opacity={1.0}, rotate={0}}, xlabel style={at={(ticklabel cs:0.5)}, anchor=near ticklabel, font={{\fontsize{24.200000000000003 pt}{31.460000000000004 pt}\selectfont}}, color={rgb,1:red,0.0;green,0.0;blue,0.0}, draw opacity={1.0}, rotate={0.0}}, xmajorgrids={false}, xmin={-0.03}, xmax={1.03}, xtick={{0.0,0.25,0.5,0.75,1.0}}, xticklabels={{$0.00$,$0.25$,$0.50$,$0.75$,$1.00$}}, xtick align={inside}, xticklabel style={font={{\fontsize{17.6 pt}{22.880000000000003 pt}\selectfont}}, color={rgb,1:red,0.0;green,0.0;blue,0.0}, draw opacity={1.0}, rotate={0.0}}, x grid style={color={rgb,1:red,0.0;green,0.0;blue,0.0}, draw opacity={0.1}, line width={1.1}, solid}, x axis line style={color={rgb,1:red,0.0;green,0.0;blue,0.0}, draw opacity={1.0}, line width={2.2}, solid}, scaled y ticks={false}, ylabel={$\risk{\blambda}- \sigma^2$}, y tick style={color={rgb,1:red,0.0;green,0.0;blue,0.0}, opacity={1.0}}, y tick label style={color={rgb,1:red,0.0;green,0.0;blue,0.0}, opacity={1.0}, rotate={0}}, ylabel style={at={(ticklabel cs:0.5)}, anchor=near ticklabel, font={{\fontsize{24.200000000000003 pt}{31.460000000000004 pt}\selectfont}}, color={rgb,1:red,0.0;green,0.0;blue,0.0}, draw opacity={1.0}, rotate={0.0}}, ymajorgrids={false}, ymin={0}, ymax={2.5}, ytick={{0.0,0.5,1.0,1.5,2.0,2.5}}, yticklabels={{$0.0$,$0.5$,$1.0$,$1.5$,$2.0$,$2.5$}}, ytick align={inside}, yticklabel style={font={{\fontsize{17.6 pt}{22.880000000000003 pt}\selectfont}}, color={rgb,1:red,0.0;green,0.0;blue,0.0}, draw opacity={1.0}, rotate={0.0}}, y grid style={color={rgb,1:red,0.0;green,0.0;blue,0.0}, draw opacity={0.1}, line width={1.1}, solid}, y axis line style={color={rgb,1:red,0.0;green,0.0;blue,0.0}, draw opacity={1.0}, line width={2.2}, solid}]
    \addplot[color={rgb,1:red,0.502;green,0.502;blue,0.502}, name path={b149c564-a725-4702-877b-a4e9f94d577c}, draw opacity={1.0}, line width={2.2}, dotted]
        table[row sep={\\}]
        {
            \\
            0.0  0.9998001100228961  \\
            0.034482758620689655  0.9992256641070205  \\
            0.06896551724137931  0.9975184153691004  \\
            0.10344827586206896  0.9947295265800038  \\
            0.13793103448275862  0.9909020605652135  \\
            0.1724137931034483  0.9860756430836075  \\
            0.20689655172413793  0.98028509821226  \\
            0.2413793103448276  0.9735599686552165  \\
            0.27586206896551724  0.9659272926862552  \\
            0.3103448275862069  0.9574103910919876  \\
            0.3448275862068966  0.9480296549142753  \\
            0.3793103448275862  0.9378033454375461  \\
            0.41379310344827586  0.9267476054992834  \\
            0.4482758620689655  0.9148752874251147  \\
            0.4827586206896552  0.9021978569031233  \\
            0.5172413793103449  0.8887258653941219  \\
            0.5517241379310345  0.874466951624745  \\
            0.5862068965517241  0.8594256744896063  \\
            0.6206896551724138  0.8436089387550487  \\
            0.6551724137931034  0.8270167784575448  \\
            0.6896551724137931  0.8096529316699383  \\
            0.7241379310344828  0.7915150206002735  \\
            0.7586206896551724  0.7726038711815226  \\
            0.7931034482758621  0.7529134670290984  \\
            0.8275862068965517  0.7324413651747774  \\
            0.8620689655172413  0.7111781996155686  \\
            0.896551724137931  0.6891174553682133  \\
            0.9310344827586207  0.6662479254283771  \\
            0.9655172413793104  0.6425584404256584  \\
            1.0  0.6180339971274516  \\
        }
        ;
    \addplot[color={rgb,1:red,0.502;green,0.0;blue,0.502}, name path={207e8409-336b-4fef-b118-034d4b2fb537}, draw opacity={1.0}, line width={2.2}, dashdotted]
        table[row sep={\\}]
        {
            \\
            0.0  0.7807764064044151  \\
            0.034482758620689655  0.7807764064044149  \\
            0.06896551724137931  0.7807764064044151  \\
            0.10344827586206896  0.7807764064044149  \\
            0.13793103448275862  0.7807764064044151  \\
            0.1724137931034483  0.7807764064044149  \\
            0.20689655172413793  0.7807764064044151  \\
            0.2413793103448276  0.7807764064044151  \\
            0.27586206896551724  0.7807764064044151  \\
            0.3103448275862069  0.7807764064044151  \\
            0.3448275862068966  0.7807764064044151  \\
            0.3793103448275862  0.7807764064044151  \\
            0.41379310344827586  0.7807764064044151  \\
            0.4482758620689655  0.7807764064044151  \\
            0.4827586206896552  0.7807764064044151  \\
            0.5172413793103449  0.7807764064044151  \\
            0.5517241379310345  0.7807764064044151  \\
            0.5862068965517241  0.7807764064044151  \\
            0.6206896551724138  0.7807764064044151  \\
            0.6551724137931034  0.7807764064044151  \\
            0.6896551724137931  0.7807764064044151  \\
            0.7241379310344828  0.7807764064044151  \\
            0.7586206896551724  0.7807764064044151  \\
            0.7931034482758621  0.7807764064044151  \\
            0.8275862068965517  0.7807764064044149  \\
            0.8620689655172413  0.7807764064044151  \\
            0.896551724137931  0.7807764064044149  \\
            0.9310344827586207  0.7807764064044151  \\
            0.9655172413793104  0.7807764064044149  \\
            1.0  0.7807764064044151  \\
        }
        ;
    \addplot[color={rgb,1:red,0.0;green,0.502;blue,0.0}, name path={1a4116f9-fe08-4ec3-abcc-ede16f37ca3a}, draw opacity={1.0}, line width={2.2}, dashed]
        table[row sep={\\}]
        {
            \\
            0.0  0.6180339971274516  \\
            0.034482758620689655  0.6417413022093053  \\
            0.06896551724137931  0.6631067467660756  \\
            0.10344827586206896  0.6823184993719913  \\
            0.13793103448275862  0.6995291517072184  \\
            0.1724137931034483  0.7148635623056023  \\
            0.20689655172413793  0.7284244993706976  \\
            0.2413793103448276  0.7402967761120787  \\
            0.27586206896551724  0.7505503283442794  \\
            0.3103448275862069  0.7592425305598138  \\
            0.3448275862068966  0.7664199500393809  \\
            0.3793103448275862  0.7721196758552611  \\
            0.41379310344827586  0.7763703177458681  \\
            0.4482758620689655  0.7791927410343595  \\
            0.4827586206896552  0.7806005832904612  \\
            0.5172413793103449  0.7806005832904612  \\
            0.5517241379310345  0.7791927410343595  \\
            0.5862068965517241  0.7763703177458681  \\
            0.6206896551724138  0.7721196758552611  \\
            0.6551724137931034  0.7664199500393809  \\
            0.6896551724137931  0.7592425305598138  \\
            0.7241379310344828  0.7505503283442794  \\
            0.7586206896551724  0.7402967761120787  \\
            0.7931034482758621  0.7284244993706976  \\
            0.8275862068965517  0.7148635623056023  \\
            0.8620689655172413  0.6995291517072184  \\
            0.896551724137931  0.6823184993719913  \\
            0.9310344827586207  0.6631067467660756  \\
            0.9655172413793104  0.6417413022093053  \\
            1.0  0.6180339971274516  \\
        }
        ;
    \addplot[color={rgb,1:red,0.502;green,0.502;blue,0.502}, name path={8b9eeaed-5558-4253-bb7b-df146b490587}, only marks, draw opacity={1.0}, line width={0.0}, solid, mark={triangle*}, mark size={6.6000000000000005 pt}, mark repeat={1}, mark options={color={rgb,1:red,0.0;green,0.0;blue,0.0}, draw opacity={0.0}, fill={rgb,1:red,0.502;green,0.502;blue,0.502}, fill opacity={1.0}, line width={1.6500000000000001}, rotate={0}, solid}]
        table[row sep={\\}]
        {
            \\
            0.0  1.0199370022187315  \\
            0.034482758620689655  1.0220899245920232  \\
            0.06896551724137931  1.0315515324865991  \\
            0.10344827586206896  0.9921994946112915  \\
            0.13793103448275862  0.9484361322095494  \\
            0.1724137931034483  0.9337700232401722  \\
            0.20689655172413793  1.0324346983306243  \\
            0.2413793103448276  0.9537956138176051  \\
            0.27586206896551724  0.9809026161163708  \\
            0.3103448275862069  0.9898066401807954  \\
            0.3448275862068966  0.8936222165727565  \\
            0.3793103448275862  0.9970334196517778  \\
            0.41379310344827586  0.8699003431414749  \\
            0.4482758620689655  0.8721850061902496  \\
            0.4827586206896552  0.8853134646817324  \\
            0.5172413793103449  0.9219289889948581  \\
            0.5517241379310345  0.8889744947929143  \\
            0.5862068965517241  0.9028212114645575  \\
            0.6206896551724138  0.8245142227572535  \\
            0.6551724137931034  0.8617674049066737  \\
            0.6896551724137931  0.8375437280300562  \\
            0.7241379310344828  0.7934886647627537  \\
            0.7586206896551724  0.7194981678126897  \\
            0.7931034482758621  0.7808104847556734  \\
            0.8275862068965517  0.7090197924951287  \\
            0.8620689655172413  0.7278116940912529  \\
            0.896551724137931  0.6621205870073561  \\
            0.9310344827586207  0.6947785839497538  \\
            0.9655172413793104  0.5820239999286609  \\
            1.0  0.6235321830324168  \\
        }
        ;
    \addplot[color={rgb,1:red,0.502;green,0.0;blue,0.502}, name path={1ce36b85-293e-4a08-82e1-dcb01a4859fc}, only marks, draw opacity={1.0}, line width={0.0}, solid, mark={triangle*}, mark size={6.6000000000000005 pt}, mark repeat={1}, mark options={color={rgb,1:red,0.0;green,0.0;blue,0.0}, draw opacity={0.0}, fill={rgb,1:red,0.502;green,0.0;blue,0.502}, fill opacity={1.0}, line width={1.6500000000000001}, rotate={0}, solid}]
        table[row sep={\\}]
        {
            \\
            0.0  0.8283232486413195  \\
            0.034482758620689655  0.7971446535467905  \\
            0.06896551724137931  0.8559868894781302  \\
            0.10344827586206896  0.7918982815199325  \\
            0.13793103448275862  0.7496114142890655  \\
            0.1724137931034483  0.7260352813547288  \\
            0.20689655172413793  0.8012750571411831  \\
            0.2413793103448276  0.7414078458389128  \\
            0.27586206896551724  0.7609151806416434  \\
            0.3103448275862069  0.8110956457039855  \\
            0.3448275862068966  0.7164390258602242  \\
            0.3793103448275862  0.8526194932788771  \\
            0.41379310344827586  0.7526774450014633  \\
            0.4482758620689655  0.7164035826416337  \\
            0.4827586206896552  0.7670475603541953  \\
            0.5172413793103449  0.8136632273851101  \\
            0.5517241379310345  0.8017827279780165  \\
            0.5862068965517241  0.8183888976261147  \\
            0.6206896551724138  0.7800745911870564  \\
            0.6551724137931034  0.8486454569496753  \\
            0.6896551724137931  0.7737813027314893  \\
            0.7241379310344828  0.7688830558399273  \\
            0.7586206896551724  0.7043468707837732  \\
            0.7931034482758621  0.8099742516303348  \\
            0.8275862068965517  0.7557120215990929  \\
            0.8620689655172413  0.7866333737702023  \\
            0.896551724137931  0.7547209817741236  \\
            0.9310344827586207  0.8134257524832742  \\
            0.9655172413793104  0.7360596066658784  \\
            1.0  0.7723602391692965  \\
        }
        ;
    \addplot[color={rgb,1:red,0.0;green,0.502;blue,0.0}, name path={69e05703-91b9-43d1-a0da-378d3dad43b6}, only marks, draw opacity={1.0}, line width={0.0}, solid, mark={triangle*}, mark size={6.6000000000000005 pt}, mark repeat={1}, mark options={color={rgb,1:red,0.0;green,0.0;blue,0.0}, draw opacity={0.0}, fill={rgb,1:red,0.0;green,0.502;blue,0.0}, fill opacity={1.0}, line width={1.6500000000000001}, rotate={0}, solid}]
        table[row sep={\\}]
        {
            \\
            0.0  0.665778591511313  \\
            0.034482758620689655  0.6670693105194245  \\
            0.06896551724137931  0.7251354326725599  \\
            0.10344827586206896  0.6853278513501915  \\
            0.13793103448275862  0.6721668104683944  \\
            0.1724137931034483  0.6739414580545082  \\
            0.20689655172413793  0.7419754270331804  \\
            0.2413793103448276  0.6990276793390868  \\
            0.27586206896551724  0.7178830378767915  \\
            0.3103448275862069  0.7858880847963814  \\
            0.3448275862068966  0.7063109231434304  \\
            0.3793103448275862  0.845337778606551  \\
            0.41379310344827586  0.7527479596185414  \\
            0.4482758620689655  0.716604381255737  \\
            0.4827586206896552  0.7671436523472737  \\
            0.5172413793103449  0.8139448074811426  \\
            0.5517241379310345  0.7993363960078852  \\
            0.5862068965517241  0.8155541482064232  \\
            0.6206896551724138  0.7652352462243037  \\
            0.6551724137931034  0.8244216376774414  \\
            0.6896551724137931  0.7653974166404653  \\
            0.7241379310344828  0.74345556290857  \\
            0.7586206896551724  0.6838557456933061  \\
            0.7931034482758621  0.7538691949223044  \\
            0.8275862068965517  0.6879449992544424  \\
            0.8620689655172413  0.711833653512095  \\
            0.896551724137931  0.6554285684566108  \\
            0.9310344827586207  0.6925960139028096  \\
            0.9655172413793104  0.5822378834996944  \\
            1.0  0.6235321830324168  \\
        }
        ;
\end{axis}
\end{tikzpicture}

%% file: figures/oracle_risks_identity/oracle_risk4.tikz
\begin{tikzpicture}[/tikz/background rectangle/.style={fill={rgb,1:red,1.0;green,1.0;blue,1.0}, draw opacity={1.0}}, show background rectangle]
\begin{axis}[point meta max={nan}, point meta min={nan}, legend cell align={left}, title={$\gamma_1 = \gamma_2 = \frac{1}{4},\;\; \alpha_1^2 + \alpha_2^2 = 2$}, title style={at={{(0.5,1)}}, anchor={south}, font={{\fontsize{30.800000000000004 pt}{40.040000000000006 pt}\selectfont}}, color={rgb,1:red,0.0;green,0.0;blue,0.0}, draw opacity={1.0}, rotate={0.0}}, legend style={color={rgb,1:red,0.0;green,0.0;blue,0.0}, draw opacity={0.0}, line width={2.2}, solid, fill={rgb,1:red,0.0;green,0.0;blue,0.0}, fill opacity={0.0}, text opacity={1.0}, font={{\fontsize{26.400000000000002 pt}{34.32000000000001 pt}\selectfont}}, text={rgb,1:red,0.0;green,0.0;blue,0.0}, at={(1.02, 1)}, anchor={north west}}, axis background/.style={fill={rgb,1:red,1.0;green,1.0;blue,1.0}, opacity={1.0}}, anchor={north west}, xshift={1.0mm}, yshift={-1.0mm}, width={163.1mm}, height={125.0mm}, scaled x ticks={false}, xlabel={$\alpha_1^2/(\alpha_1^2 + \alpha_2^2)$}, x tick style={color={rgb,1:red,0.0;green,0.0;blue,0.0}, opacity={1.0}}, x tick label style={color={rgb,1:red,0.0;green,0.0;blue,0.0}, opacity={1.0}, rotate={0}}, xlabel style={at={(ticklabel cs:0.5)}, anchor=near ticklabel, font={{\fontsize{24.200000000000003 pt}{31.460000000000004 pt}\selectfont}}, color={rgb,1:red,0.0;green,0.0;blue,0.0}, draw opacity={1.0}, rotate={0.0}}, xmajorgrids={false}, xmin={-0.03}, xmax={1.03}, xtick={{0.0,0.25,0.5,0.75,1.0}}, xticklabels={{$0.00$,$0.25$,$0.50$,$0.75$,$1.00$}}, xtick align={inside}, xticklabel style={font={{\fontsize{17.6 pt}{22.880000000000003 pt}\selectfont}}, color={rgb,1:red,0.0;green,0.0;blue,0.0}, draw opacity={1.0}, rotate={0.0}}, x grid style={color={rgb,1:red,0.0;green,0.0;blue,0.0}, draw opacity={0.1}, line width={1.1}, solid}, x axis line style={color={rgb,1:red,0.0;green,0.0;blue,0.0}, draw opacity={1.0}, line width={2.2}, solid}, scaled y ticks={false}, ylabel={$\risk{\blambda}- \sigma^2$}, y tick style={color={rgb,1:red,0.0;green,0.0;blue,0.0}, opacity={1.0}}, y tick label style={color={rgb,1:red,0.0;green,0.0;blue,0.0}, opacity={1.0}, rotate={0}}, ylabel style={at={(ticklabel cs:0.5)}, anchor=near ticklabel, font={{\fontsize{24.200000000000003 pt}{31.460000000000004 pt}\selectfont}}, color={rgb,1:red,0.0;green,0.0;blue,0.0}, draw opacity={1.0}, rotate={0.0}}, ymajorgrids={false}, ymin={0}, ymax={2.5}, ytick={{0.0,0.5,1.0,1.5,2.0,2.5}}, yticklabels={{$0.0$,$0.5$,$1.0$,$1.5$,$2.0$,$2.5$}}, ytick align={inside}, yticklabel style={font={{\fontsize{17.6 pt}{22.880000000000003 pt}\selectfont}}, color={rgb,1:red,0.0;green,0.0;blue,0.0}, draw opacity={1.0}, rotate={0.0}}, y grid style={color={rgb,1:red,0.0;green,0.0;blue,0.0}, draw opacity={0.1}, line width={1.1}, solid}, y axis line style={color={rgb,1:red,0.0;green,0.0;blue,0.0}, draw opacity={1.0}, line width={2.2}, solid}]
    \addplot[color={rgb,1:red,0.502;green,0.502;blue,0.502}, name path={38156b4e-adb8-4cc0-97a3-fcb9d074e7bb}, draw opacity={1.0}, line width={2.2}, dotted]
        table[row sep={\\}]
        {
            \\
            0.0  1.9996000943487005  \\
            0.034482758620689655  1.9938030442849293  \\
            0.06896551724137931  1.9780841040073973  \\
            0.10344827586206896  1.9543858362206654  \\
            0.13793103448275862  1.9240989566835451  \\
            0.1724137931034483  1.8882440853747626  \\
            0.20689655172413793  1.8476106259305425  \\
            0.2413793103448276  1.8027984229004241  \\
            0.27586206896551724  1.7543186679772345  \\
            0.3103448275862069  1.7025652305601873  \\
            0.3448275862068966  1.647883488364391  \\
            0.3793103448275862  1.590535039568083  \\
            0.41379310344827586  1.5307744130896377  \\
            0.4482758620689655  1.4688073771229924  \\
            0.4827586206896552  1.4048010081455402  \\
            0.5172413793103449  1.3389280913044006  \\
            0.5517241379310345  1.2713040613120565  \\
            0.5862068965517241  1.2020669465363776  \\
            0.6206896551724138  1.1313076857008544  \\
            0.6551724137931034  1.0591274606026806  \\
            0.6896551724137931  0.9856161884235317  \\
            0.7241379310344828  0.9108542371240467  \\
            0.7586206896551724  0.8349115470176094  \\
            0.7931034482758621  0.757847736852753  \\
            0.8275862068965517  0.6797248853340159  \\
            0.8620689655172413  0.600590156771521  \\
            0.896551724137931  0.5205011192661246  \\
            0.9310344827586207  0.4394966118840342  \\
            0.9655172413793104  0.35762319097927087  \\
            1.0  0.27491721989550433  \\
        }
        ;
    \addplot[color={rgb,1:red,0.502;green,0.0;blue,0.502}, name path={c50f0a36-2a0a-4ab1-9c92-de844b6731b8}, draw opacity={1.0}, line width={2.2}, dashdotted]
        table[row sep={\\}]
        {
            \\
            0.0  0.5615528128088303  \\
            0.034482758620689655  0.5615528128088303  \\
            0.06896551724137931  0.5615528128088303  \\
            0.10344827586206896  0.5615528128088303  \\
            0.13793103448275862  0.5615528128088303  \\
            0.1724137931034483  0.5615528128088303  \\
            0.20689655172413793  0.5615528128088303  \\
            0.2413793103448276  0.5615528128088303  \\
            0.27586206896551724  0.5615528128088303  \\
            0.3103448275862069  0.5615528128088303  \\
            0.3448275862068966  0.5615528128088303  \\
            0.3793103448275862  0.5615528128088303  \\
            0.41379310344827586  0.5615528128088303  \\
            0.4482758620689655  0.5615528128088303  \\
            0.4827586206896552  0.5615528128088303  \\
            0.5172413793103449  0.5615528128088303  \\
            0.5517241379310345  0.5615528128088303  \\
            0.5862068965517241  0.5615528128088303  \\
            0.6206896551724138  0.5615528128088303  \\
            0.6551724137931034  0.5615528128088303  \\
            0.6896551724137931  0.5615528128088303  \\
            0.7241379310344828  0.5615528128088303  \\
            0.7586206896551724  0.5615528128088303  \\
            0.7931034482758621  0.5615528128088303  \\
            0.8275862068965517  0.5615528128088303  \\
            0.8620689655172413  0.5615528128088303  \\
            0.896551724137931  0.5615528128088303  \\
            0.9310344827586207  0.5615528128088303  \\
            0.9655172413793104  0.5615528128088303  \\
            1.0  0.5615528128088303  \\
        }
        ;
    \addplot[color={rgb,1:red,0.0;green,0.502;blue,0.0}, name path={bc745465-78c7-43c8-9ea2-2fc4f47dfb62}, draw opacity={1.0}, line width={2.2}, dashed]
        table[row sep={\\}]
        {
            \\
            0.0  0.27491721989550433  \\
            0.034482758620689655  0.3432872715400339  \\
            0.06896551724137931  0.3919787749709025  \\
            0.10344827586206896  0.428895122390615  \\
            0.13793103448275862  0.457888007404317  \\
            0.1724137931034483  0.4811435425932382  \\
            0.20689655172413793  0.5000234669253398  \\
            0.2413793103448276  0.5154301211937984  \\
            0.27586206896551724  0.5279873351457043  \\
            0.3103448275862069  0.5381388020662525  \\
            0.3448275862068966  0.5462053139442489  \\
            0.3793103448275862  0.552419731461737  \\
            0.41379310344827586  0.5569490601945657  \\
            0.4482758620689655  0.5599085906258088  \\
            0.4827586206896552  0.5613708409036002  \\
            0.5172413793103449  0.5613708409036002  \\
            0.5517241379310345  0.5599085906258088  \\
            0.5862068965517241  0.5569490601945657  \\
            0.6206896551724138  0.552419731461737  \\
            0.6551724137931034  0.5462053139442489  \\
            0.6896551724137931  0.5381388020662525  \\
            0.7241379310344828  0.5279873351457043  \\
            0.7586206896551724  0.5154301211937984  \\
            0.7931034482758621  0.5000234669253398  \\
            0.8275862068965517  0.4811435425932382  \\
            0.8620689655172413  0.457888007404317  \\
            0.896551724137931  0.428895122390615  \\
            0.9310344827586207  0.3919787749709025  \\
            0.9655172413793104  0.3432872715400339  \\
            1.0  0.27491721989550433  \\
        }
        ;
    \addplot[color={rgb,1:red,0.502;green,0.502;blue,0.502}, name path={733320c7-aed0-470b-8b1e-ceb5574f739a}, only marks, draw opacity={1.0}, line width={0.0}, solid, mark={triangle*}, mark size={6.6000000000000005 pt}, mark repeat={1}, mark options={color={rgb,1:red,0.0;green,0.0;blue,0.0}, draw opacity={0.0}, fill={rgb,1:red,0.502;green,0.502;blue,0.502}, fill opacity={1.0}, line width={1.6500000000000001}, rotate={0}, solid}]
        table[row sep={\\}]
        {
            \\
            0.0  1.923114277127934  \\
            0.034482758620689655  2.05219974899184  \\
            0.06896551724137931  2.0388269205506524  \\
            0.10344827586206896  1.7717541793639264  \\
            0.13793103448275862  2.183777937900382  \\
            0.1724137931034483  1.9742763495281062  \\
            0.20689655172413793  1.916111508411189  \\
            0.2413793103448276  1.9699482777410582  \\
            0.27586206896551724  1.8038449759029804  \\
            0.3103448275862069  1.5313312562257764  \\
            0.3448275862068966  1.8760050101798567  \\
            0.3793103448275862  1.6636686639044251  \\
            0.41379310344827586  1.5451420145648735  \\
            0.4482758620689655  1.3512179427333888  \\
            0.4827586206896552  1.355991089197527  \\
            0.5172413793103449  1.4034264626475337  \\
            0.5517241379310345  1.320061473023892  \\
            0.5862068965517241  1.3403364675346556  \\
            0.6206896551724138  1.043308545729249  \\
            0.6551724137931034  0.9854345808284102  \\
            0.6896551724137931  0.9859065223934977  \\
            0.7241379310344828  0.968158868246281  \\
            0.7586206896551724  0.739716962409825  \\
            0.7931034482758621  0.730595613181874  \\
            0.8275862068965517  0.6073778867441781  \\
            0.8620689655172413  0.6194948538929577  \\
            0.896551724137931  0.4919084158737128  \\
            0.9310344827586207  0.4406019979721987  \\
            0.9655172413793104  0.3326662048863387  \\
            1.0  0.2438008314641138  \\
        }
        ;
    \addplot[color={rgb,1:red,0.502;green,0.0;blue,0.502}, name path={954d8a96-92ee-4f0c-aa9f-e26e7a2fe1c8}, only marks, draw opacity={1.0}, line width={0.0}, solid, mark={triangle*}, mark size={6.6000000000000005 pt}, mark repeat={1}, mark options={color={rgb,1:red,0.0;green,0.0;blue,0.0}, draw opacity={0.0}, fill={rgb,1:red,0.502;green,0.0;blue,0.502}, fill opacity={1.0}, line width={1.6500000000000001}, rotate={0}, solid}]
        table[row sep={\\}]
        {
            \\
            0.0  0.5795268065853014  \\
            0.034482758620689655  0.5919599269774445  \\
            0.06896551724137931  0.5063504198097002  \\
            0.10344827586206896  0.5634372336849978  \\
            0.13793103448275862  0.5988998558614906  \\
            0.1724137931034483  0.5944220147745263  \\
            0.20689655172413793  0.6104257641347606  \\
            0.2413793103448276  0.4963167194088012  \\
            0.27586206896551724  0.4668249707944019  \\
            0.3103448275862069  0.4817377670609593  \\
            0.3448275862068966  0.6331201413281025  \\
            0.3793103448275862  0.6039558832423757  \\
            0.41379310344827586  0.5767631763365766  \\
            0.4482758620689655  0.5691388623261835  \\
            0.4827586206896552  0.5698940453390657  \\
            0.5172413793103449  0.6015592161171543  \\
            0.5517241379310345  0.5770318544832509  \\
            0.5862068965517241  0.5480534035394986  \\
            0.6206896551724138  0.5194870046047266  \\
            0.6551724137931034  0.5793354820187637  \\
            0.6896551724137931  0.5823621833925163  \\
            0.7241379310344828  0.5228224044748602  \\
            0.7586206896551724  0.5355162079024292  \\
            0.7931034482758621  0.5803221736074837  \\
            0.8275862068965517  0.5107724121493691  \\
            0.8620689655172413  0.636291777876556  \\
            0.896551724137931  0.4814236651895065  \\
            0.9310344827586207  0.5714993219604543  \\
            0.9655172413793104  0.5606193045576813  \\
            1.0  0.4873840278183499  \\
        }
        ;
    \addplot[color={rgb,1:red,0.0;green,0.502;blue,0.0}, name path={2e193594-7a75-4e03-a7c8-6dde1b941b32}, only marks, draw opacity={1.0}, line width={0.0}, solid, mark={triangle*}, mark size={6.6000000000000005 pt}, mark repeat={1}, mark options={color={rgb,1:red,0.0;green,0.0;blue,0.0}, draw opacity={0.0}, fill={rgb,1:red,0.0;green,0.502;blue,0.0}, fill opacity={1.0}, line width={1.6500000000000001}, rotate={0}, solid}]
        table[row sep={\\}]
        {
            \\
            0.0  0.27730164095308174  \\
            0.034482758620689655  0.36607547608698776  \\
            0.06896551724137931  0.38240107180730143  \\
            0.10344827586206896  0.45587191020715645  \\
            0.13793103448275862  0.4445167139895354  \\
            0.1724137931034483  0.5171625828674382  \\
            0.20689655172413793  0.5128457097826709  \\
            0.2413793103448276  0.4610201297918932  \\
            0.27586206896551724  0.45067819075812077  \\
            0.3103448275862069  0.47163091308952243  \\
            0.3448275862068966  0.611869865596504  \\
            0.3793103448275862  0.5816133032636264  \\
            0.41379310344827586  0.5698550373554809  \\
            0.4482758620689655  0.5666671963526384  \\
            0.4827586206896552  0.5673716043842261  \\
            0.5172413793103449  0.6017379353922174  \\
            0.5517241379310345  0.5700506067277853  \\
            0.5862068965517241  0.5487919743739877  \\
            0.6206896551724138  0.5118910485630239  \\
            0.6551724137931034  0.564203850966644  \\
            0.6896551724137931  0.5328558779361945  \\
            0.7241379310344828  0.48878077329031155  \\
            0.7586206896551724  0.49117401118627346  \\
            0.7931034482758621  0.49560283421476115  \\
            0.8275862068965517  0.4333442172070585  \\
            0.8620689655172413  0.5164242933652043  \\
            0.896551724137931  0.3905396399515466  \\
            0.9310344827586207  0.3990864091586641  \\
            0.9655172413793104  0.32416407904248556  \\
            1.0  0.2438008314641138  \\
        }
        ;
\end{axis}
\end{tikzpicture}

%% file: figures/oracle_risks_identity/oracle_risk5.tikz
\begin{tikzpicture}[/tikz/background rectangle/.style={fill={rgb,1:red,1.0;green,1.0;blue,1.0}, draw opacity={1.0}}, show background rectangle]
\begin{axis}[point meta max={nan}, point meta min={nan}, legend cell align={left}, title={$\gamma_1 = \frac{1}{10},\; \gamma_2 = \frac{4}{10},\;\; \alpha_1^2 + \alpha_2^2 = 2$}, title style={at={{(0.5,1)}}, anchor={south}, font={{\fontsize{30.800000000000004 pt}{40.040000000000006 pt}\selectfont}}, color={rgb,1:red,0.0;green,0.0;blue,0.0}, draw opacity={1.0}, rotate={0.0}}, legend style={color={rgb,1:red,0.0;green,0.0;blue,0.0}, draw opacity={0.0}, line width={2.2}, solid, fill={rgb,1:red,0.0;green,0.0;blue,0.0}, fill opacity={0.0}, text opacity={1.0}, font={{\fontsize{26.400000000000002 pt}{34.32000000000001 pt}\selectfont}}, text={rgb,1:red,0.0;green,0.0;blue,0.0}, at={(1.02, 1)}, anchor={north west}}, axis background/.style={fill={rgb,1:red,1.0;green,1.0;blue,1.0}, opacity={1.0}}, anchor={north west}, xshift={1.0mm}, yshift={-1.0mm}, width={163.1mm}, height={125.0mm}, scaled x ticks={false}, xlabel={$\alpha_1^2/(\alpha_1^2 + \alpha_2^2)$}, x tick style={color={rgb,1:red,0.0;green,0.0;blue,0.0}, opacity={1.0}}, x tick label style={color={rgb,1:red,0.0;green,0.0;blue,0.0}, opacity={1.0}, rotate={0}}, xlabel style={at={(ticklabel cs:0.5)}, anchor=near ticklabel, font={{\fontsize{24.200000000000003 pt}{31.460000000000004 pt}\selectfont}}, color={rgb,1:red,0.0;green,0.0;blue,0.0}, draw opacity={1.0}, rotate={0.0}}, xmajorgrids={false}, xmin={-0.03}, xmax={1.03}, xtick={{0.0,0.25,0.5,0.75,1.0}}, xticklabels={{$0.00$,$0.25$,$0.50$,$0.75$,$1.00$}}, xtick align={inside}, xticklabel style={font={{\fontsize{17.6 pt}{22.880000000000003 pt}\selectfont}}, color={rgb,1:red,0.0;green,0.0;blue,0.0}, draw opacity={1.0}, rotate={0.0}}, x grid style={color={rgb,1:red,0.0;green,0.0;blue,0.0}, draw opacity={0.1}, line width={1.1}, solid}, x axis line style={color={rgb,1:red,0.0;green,0.0;blue,0.0}, draw opacity={1.0}, line width={2.2}, solid}, scaled y ticks={false}, ylabel={$\risk{\blambda}- \sigma^2$}, y tick style={color={rgb,1:red,0.0;green,0.0;blue,0.0}, opacity={1.0}}, y tick label style={color={rgb,1:red,0.0;green,0.0;blue,0.0}, opacity={1.0}, rotate={0}}, ylabel style={at={(ticklabel cs:0.5)}, anchor=near ticklabel, font={{\fontsize{24.200000000000003 pt}{31.460000000000004 pt}\selectfont}}, color={rgb,1:red,0.0;green,0.0;blue,0.0}, draw opacity={1.0}, rotate={0.0}}, ymajorgrids={false}, ymin={0}, ymax={2.5}, ytick={{0.0,0.5,1.0,1.5,2.0,2.5}}, yticklabels={{$0.0$,$0.5$,$1.0$,$1.5$,$2.0$,$2.5$}}, ytick align={inside}, yticklabel style={font={{\fontsize{17.6 pt}{22.880000000000003 pt}\selectfont}}, color={rgb,1:red,0.0;green,0.0;blue,0.0}, draw opacity={1.0}, rotate={0.0}}, y grid style={color={rgb,1:red,0.0;green,0.0;blue,0.0}, draw opacity={0.1}, line width={1.1}, solid}, y axis line style={color={rgb,1:red,0.0;green,0.0;blue,0.0}, draw opacity={1.0}, line width={2.2}, solid}]
    \addplot[color={rgb,1:red,0.502;green,0.502;blue,0.502}, name path={4049b2cc-e908-4f9e-95cb-e2ff092a7254}, draw opacity={1.0}, line width={2.2}, dotted]
        table[row sep={\\}]
        {
            \\
            0.0  1.9996000947752384  \\
            0.034482758620689655  1.9866838159622597  \\
            0.06896551724137931  1.955754519815347  \\
            0.10344827586206896  1.9137474773179104  \\
            0.13793103448275862  1.8643346016641114  \\
            0.1724137931034483  1.8096806449618557  \\
            0.20689655172413793  1.7511574683717486  \\
            0.2413793103448276  1.689687578280358  \\
            0.27586206896551724  1.6259200779437126  \\
            0.3103448275862069  1.560317184704612  \\
            0.3448275862068966  1.4932211070951134  \\
            0.3793103448275862  1.4249011542276273  \\
            0.41379310344827586  1.355562523791503  \\
            0.4482758620689655  1.2853598032860951  \\
            0.4827586206896552  1.2144216764872025  \\
            0.5172413793103449  1.142854076370849  \\
            0.5517241379310345  1.0707420760667725  \\
            0.5862068965517241  0.9981554329076119  \\
            0.6206896551724138  0.9251499504667189  \\
            0.6551724137931034  0.8517786362967525  \\
            0.6896551724137931  0.7780801183433601  \\
            0.7241379310344828  0.7040932093379677  \\
            0.7586206896551724  0.6298455948906292  \\
            0.7931034482758621  0.5553643702741684  \\
            0.8275862068965517  0.480673714366068  \\
            0.8620689655172413  0.40579348120392544  \\
            0.896551724137931  0.33073970213233905  \\
            0.9310344827586207  0.25552788583927777  \\
            0.9655172413793104  0.18017294849199939  \\
            1.0  0.10468636006794108  \\
        }
        ;
    \addplot[color={rgb,1:red,0.502;green,0.0;blue,0.502}, name path={a6198627-b3f6-4959-ae10-eeef6767794b}, draw opacity={1.0}, line width={2.2}, dashdotted]
        table[row sep={\\}]
        {
            \\
            0.0  0.5615528128088081  \\
            0.034482758620689655  0.5615528128087748  \\
            0.06896551724137931  0.5615528128087863  \\
            0.10344827586206896  0.5615528128087912  \\
            0.13793103448275862  0.5615528128088094  \\
            0.1724137931034483  0.5615528128088192  \\
            0.20689655172413793  0.5615528128088325  \\
            0.2413793103448276  0.561552812808847  \\
            0.27586206896551724  0.5615528128088609  \\
            0.3103448275862069  0.5615528128088749  \\
            0.3448275862068966  0.5615528128088889  \\
            0.3793103448275862  0.5615528128089027  \\
            0.41379310344827586  0.561552812808902  \\
            0.4482758620689655  0.5615528128089138  \\
            0.4827586206896552  0.5615528128089446  \\
            0.5172413793103449  0.5615528128089586  \\
            0.5517241379310345  0.5615528128089484  \\
            0.5862068965517241  0.56155281280896  \\
            0.6206896551724138  0.5615528128090004  \\
            0.6551724137931034  0.5615528128090144  \\
            0.6896551724137931  0.5615528128090284  \\
            0.7241379310344828  0.5615528128090421  \\
            0.7586206896551724  0.5615528128090561  \\
            0.7931034482758621  0.5615528128090295  \\
            0.8275862068965517  0.5615528128090841  \\
            0.8620689655172413  0.5615528128090526  \\
            0.896551724137931  0.5615528128091118  \\
            0.9310344827586207  0.5615528128090759  \\
            0.9655172413793104  0.5615528128090874  \\
            1.0  0.5615528128089193  \\
        }
        ;
    \addplot[color={rgb,1:red,0.0;green,0.502;blue,0.0}, name path={6c5a1cd6-80f6-4f57-9cf0-fc6b3774a496}, draw opacity={1.0}, line width={2.2}, dashed]
        table[row sep={\\}]
        {
            \\
            0.0  0.44948974324983104  \\
            0.034482758620689655  0.5066133702944691  \\
            0.06896551724137931  0.5341934075141777  \\
            0.10344827586206896  0.5490845332088043  \\
            0.13793103448275862  0.5570747776949136  \\
            0.1724137931034483  0.5607651188517573  \\
            0.20689655172413793  0.5615081671562248  \\
            0.2413793103448276  0.5600743687441234  \\
            0.27586206896551724  0.5569288512779367  \\
            0.3103448275862069  0.5523618921896449  \\
            0.3448275862068966  0.5465561437307136  \\
            0.3793103448275862  0.5396235742974143  \\
            0.41379310344827586  0.531626672393319  \\
            0.4482758620689655  0.5225908715528642  \\
            0.4827586206896552  0.5125117264231123  \\
            0.5172413793103449  0.5013586936501992  \\
            0.5517241379310345  0.4890764834093386  \\
            0.5862068965517241  0.4755844273434957  \\
            0.6206896551724138  0.4607739527925414  \\
            0.6551724137931034  0.4445039458232085  \\
            0.6896551724137931  0.4265934380117873  \\
            0.7241379310344828  0.40681056164801754  \\
            0.7586206896551724  0.3848559197927117  \\
            0.7931034482758621  0.36033709127203917  \\
            0.8275862068965517  0.3327282453386293  \\
            0.8620689655172413  0.3013031459772788  \\
            0.896551724137931  0.26501693909412816  \\
            0.9310344827586207  0.22227958033556572  \\
            0.9655172413793104  0.17046867217652162  \\
            1.0  0.10468636006794108  \\
        }
        ;
    \addplot[color={rgb,1:red,0.502;green,0.502;blue,0.502}, name path={0a97a5fe-3097-490e-8889-d21989a73000}, only marks, draw opacity={1.0}, line width={0.0}, solid, mark={triangle*}, mark size={6.6000000000000005 pt}, mark repeat={1}, mark options={color={rgb,1:red,0.0;green,0.0;blue,0.0}, draw opacity={0.0}, fill={rgb,1:red,0.502;green,0.502;blue,0.502}, fill opacity={1.0}, line width={1.6500000000000001}, rotate={0}, solid}]
        table[row sep={\\}]
        {
            \\
            0.0  2.1544400238773793  \\
            0.034482758620689655  1.8071906113167127  \\
            0.06896551724137931  2.0623786638541732  \\
            0.10344827586206896  1.7947827190469483  \\
            0.13793103448275862  1.9617918153263325  \\
            0.1724137931034483  1.681188680994675  \\
            0.20689655172413793  1.699688076954097  \\
            0.2413793103448276  1.7059249931901266  \\
            0.27586206896551724  1.70268339278882  \\
            0.3103448275862069  1.6095387678674022  \\
            0.3448275862068966  1.6235873593221357  \\
            0.3793103448275862  1.3910082625241107  \\
            0.41379310344827586  1.4176839736877178  \\
            0.4482758620689655  1.362559229501838  \\
            0.4827586206896552  1.2242271747068387  \\
            0.5172413793103449  1.1279977435369561  \\
            0.5517241379310345  0.979458460107923  \\
            0.5862068965517241  0.9738302147190763  \\
            0.6206896551724138  0.8598593382176831  \\
            0.6551724137931034  0.8126926181220995  \\
            0.6896551724137931  0.7039001832605034  \\
            0.7241379310344828  0.7033615039638483  \\
            0.7586206896551724  0.6581985396157746  \\
            0.7931034482758621  0.616982015650269  \\
            0.8275862068965517  0.4979386528477925  \\
            0.8620689655172413  0.3897414352670372  \\
            0.896551724137931  0.3329403049585131  \\
            0.9310344827586207  0.23765249593684912  \\
            0.9655172413793104  0.17425364353291584  \\
            1.0  0.13383213795561266  \\
        }
        ;
    \addplot[color={rgb,1:red,0.502;green,0.0;blue,0.502}, name path={07d597e0-9b50-4b2a-bdff-43d653928f25}, only marks, draw opacity={1.0}, line width={0.0}, solid, mark={triangle*}, mark size={6.6000000000000005 pt}, mark repeat={1}, mark options={color={rgb,1:red,0.0;green,0.0;blue,0.0}, draw opacity={0.0}, fill={rgb,1:red,0.502;green,0.0;blue,0.502}, fill opacity={1.0}, line width={1.6500000000000001}, rotate={0}, solid}]
        table[row sep={\\}]
        {
            \\
            0.0  0.59678898635228  \\
            0.034482758620689655  0.5417968816913235  \\
            0.06896551724137931  0.5588213069002421  \\
            0.10344827586206896  0.5106795693908806  \\
            0.13793103448275862  0.5893939932484791  \\
            0.1724137931034483  0.5273808788964005  \\
            0.20689655172413793  0.5454598159207027  \\
            0.2413793103448276  0.5250426960007877  \\
            0.27586206896551724  0.5325469656115143  \\
            0.3103448275862069  0.6016470239622984  \\
            0.3448275862068966  0.5267624863192206  \\
            0.3793103448275862  0.6116223425696197  \\
            0.41379310344827586  0.5847078536627446  \\
            0.4482758620689655  0.5518052287835651  \\
            0.4827586206896552  0.5514460074185912  \\
            0.5172413793103449  0.5479887104518641  \\
            0.5517241379310345  0.5182203265464524  \\
            0.5862068965517241  0.5598070045018431  \\
            0.6206896551724138  0.5578273034422625  \\
            0.6551724137931034  0.5591859121282219  \\
            0.6896551724137931  0.5350381434771199  \\
            0.7241379310344828  0.5454222343693296  \\
            0.7586206896551724  0.5687260017072462  \\
            0.7931034482758621  0.5465730745543533  \\
            0.8275862068965517  0.5840461497835701  \\
            0.8620689655172413  0.6429927544027538  \\
            0.896551724137931  0.5364493890820912  \\
            0.9310344827586207  0.45130285494779265  \\
            0.9655172413793104  0.5014438590089765  \\
            1.0  0.5214505887872076  \\
        }
        ;
    \addplot[color={rgb,1:red,0.0;green,0.502;blue,0.0}, name path={5627daf4-3e96-49fc-a40c-27387cbca69e}, only marks, draw opacity={1.0}, line width={0.0}, solid, mark={triangle*}, mark size={6.6000000000000005 pt}, mark repeat={1}, mark options={color={rgb,1:red,0.0;green,0.0;blue,0.0}, draw opacity={0.0}, fill={rgb,1:red,0.0;green,0.502;blue,0.0}, fill opacity={1.0}, line width={1.6500000000000001}, rotate={0}, solid}]
        table[row sep={\\}]
        {
            \\
            0.0  0.4731652581027854  \\
            0.034482758620689655  0.5069171829587327  \\
            0.06896551724137931  0.5251653306002488  \\
            0.10344827586206896  0.4949588007736052  \\
            0.13793103448275862  0.5865080493434016  \\
            0.1724137931034483  0.5274509225642738  \\
            0.20689655172413793  0.5454049037798905  \\
            0.2413793103448276  0.5237617851788312  \\
            0.27586206896551724  0.5413536734698408  \\
            0.3103448275862069  0.5973292461364925  \\
            0.3448275862068966  0.5162030350821236  \\
            0.3793103448275862  0.584664501333586  \\
            0.41379310344827586  0.5546937945630301  \\
            0.4482758620689655  0.4976869473432104  \\
            0.4827586206896552  0.5037194556423505  \\
            0.5172413793103449  0.48396291958843385  \\
            0.5517241379310345  0.4616391720834849  \\
            0.5862068965517241  0.4593808049560937  \\
            0.6206896551724138  0.41925333710498225  \\
            0.6551724137931034  0.4473852205346516  \\
            0.6896551724137931  0.3991020844755928  \\
            0.7241379310344828  0.38646305298134687  \\
            0.7586206896551724  0.3888415711637936  \\
            0.7931034482758621  0.38560476988715053  \\
            0.8275862068965517  0.32708185793755873  \\
            0.8620689655172413  0.3251419193695575  \\
            0.896551724137931  0.2729301478134467  \\
            0.9310344827586207  0.1960190866168421  \\
            0.9655172413793104  0.16292808684275095  \\
            1.0  0.13383213795561266  \\
        }
        ;
\end{axis}
\end{tikzpicture}

%% file: figures/oracle_risks_identity/oracle_risk6.tikz
\begin{tikzpicture}[/tikz/background rectangle/.style={fill={rgb,1:red,1.0;green,1.0;blue,1.0}, draw opacity={1.0}}, show background rectangle]
\begin{axis}[point meta max={nan}, point meta min={nan}, legend cell align={left}, title={$\gamma_1 = \gamma_2 = 1,\;\; \alpha_1^2 + \alpha_2^2 = 2$}, title style={at={{(0.5,1)}}, anchor={south}, font={{\fontsize{30.800000000000004 pt}{40.040000000000006 pt}\selectfont}}, color={rgb,1:red,0.0;green,0.0;blue,0.0}, draw opacity={1.0}, rotate={0.0}}, legend style={color={rgb,1:red,0.0;green,0.0;blue,0.0}, draw opacity={0.0}, line width={2.2}, solid, fill={rgb,1:red,0.0;green,0.0;blue,0.0}, fill opacity={0.0}, text opacity={1.0}, font={{\fontsize{26.400000000000002 pt}{34.32000000000001 pt}\selectfont}}, text={rgb,1:red,0.0;green,0.0;blue,0.0}, at={(1.02, 1)}, anchor={north west}}, axis background/.style={fill={rgb,1:red,1.0;green,1.0;blue,1.0}, opacity={1.0}}, anchor={north west}, xshift={1.0mm}, yshift={-1.0mm}, width={163.1mm}, height={125.0mm}, scaled x ticks={false}, xlabel={$\alpha_1^2/(\alpha_1^2 + \alpha_2^2)$}, x tick style={color={rgb,1:red,0.0;green,0.0;blue,0.0}, opacity={1.0}}, x tick label style={color={rgb,1:red,0.0;green,0.0;blue,0.0}, opacity={1.0}, rotate={0}}, xlabel style={at={(ticklabel cs:0.5)}, anchor=near ticklabel, font={{\fontsize{24.200000000000003 pt}{31.460000000000004 pt}\selectfont}}, color={rgb,1:red,0.0;green,0.0;blue,0.0}, draw opacity={1.0}, rotate={0.0}}, xmajorgrids={false}, xmin={-0.03}, xmax={1.03}, xtick={{0.0,0.25,0.5,0.75,1.0}}, xticklabels={{$0.00$,$0.25$,$0.50$,$0.75$,$1.00$}}, xtick align={inside}, xticklabel style={font={{\fontsize{17.6 pt}{22.880000000000003 pt}\selectfont}}, color={rgb,1:red,0.0;green,0.0;blue,0.0}, draw opacity={1.0}, rotate={0.0}}, x grid style={color={rgb,1:red,0.0;green,0.0;blue,0.0}, draw opacity={0.1}, line width={1.1}, solid}, x axis line style={color={rgb,1:red,0.0;green,0.0;blue,0.0}, draw opacity={1.0}, line width={2.2}, solid}, scaled y ticks={false}, ylabel={$\risk{\blambda}- \sigma^2$}, y tick style={color={rgb,1:red,0.0;green,0.0;blue,0.0}, opacity={1.0}}, y tick label style={color={rgb,1:red,0.0;green,0.0;blue,0.0}, opacity={1.0}, rotate={0}}, ylabel style={at={(ticklabel cs:0.5)}, anchor=near ticklabel, font={{\fontsize{24.200000000000003 pt}{31.460000000000004 pt}\selectfont}}, color={rgb,1:red,0.0;green,0.0;blue,0.0}, draw opacity={1.0}, rotate={0.0}}, ymajorgrids={false}, ymin={0}, ymax={2.5}, ytick={{0.0,0.5,1.0,1.5,2.0,2.5}}, yticklabels={{$0.0$,$0.5$,$1.0$,$1.5$,$2.0$,$2.5$}}, ytick align={inside}, yticklabel style={font={{\fontsize{17.6 pt}{22.880000000000003 pt}\selectfont}}, color={rgb,1:red,0.0;green,0.0;blue,0.0}, draw opacity={1.0}, rotate={0.0}}, y grid style={color={rgb,1:red,0.0;green,0.0;blue,0.0}, draw opacity={0.1}, line width={1.1}, solid}, y axis line style={color={rgb,1:red,0.0;green,0.0;blue,0.0}, draw opacity={1.0}, line width={2.2}, solid}]
    \addplot[color={rgb,1:red,0.502;green,0.502;blue,0.502}, name path={4ee925e3-ead4-47b5-8395-42c8a3d824ed}, draw opacity={1.0}, line width={2.2}, dotted]
        table[row sep={\\}]
        {
            \\
            0.0  1.9996002000578015  \\
            0.034482758620689655  1.9980716655547912  \\
            0.06896551724137931  1.9935684769182385  \\
            0.10344827586206896  1.9862587948883896  \\
            0.13793103448275862  1.976284053517165  \\
            0.1724137931034483  1.9637644750898353  \\
            0.20689655172413793  1.9488018758184626  \\
            0.2413793103448276  1.93147810904677  \\
            0.27586206896551724  1.9118694181459155  \\
            0.3103448275862069  1.8900365762967506  \\
            0.3448275862068966  1.8660239304605089  \\
            0.3793103448275862  1.8398734846849614  \\
            0.41379310344827586  1.8116185905435924  \\
            0.4482758620689655  1.781279688406375  \\
            0.4827586206896552  1.7488682444450712  \\
            0.5172413793103449  1.7144011141592124  \\
            0.5517241379310345  1.6778682368487843  \\
            0.5862068965517241  1.6392704343134703  \\
            0.6206896551724138  1.5985857671397974  \\
            0.6551724137931034  1.5557958519921207  \\
            0.6896551724137931  1.5108681961572343  \\
            0.7241379310344828  1.4637629845970075  \\
            0.7586206896551724  1.4144330513001973  \\
            0.7931034482758621  1.3628188944675292  \\
            0.8275862068965517  1.3088443698437318  \\
            0.8620689655172413  1.2524254389267844  \\
            0.896551724137931  1.1934631654581889  \\
            0.9310344827586207  1.1318370172833525  \\
            0.9655172413793104  1.067405574527501  \\
            1.0  1.0000000077067681  \\
        }
        ;
    \addplot[color={rgb,1:red,0.502;green,0.0;blue,0.502}, name path={3d43a4dc-c068-424e-9e44-87f4dd91e98c}, draw opacity={1.0}, line width={2.2}, dashdotted]
        table[row sep={\\}]
        {
            \\
            0.0  1.4142135623730954  \\
            0.034482758620689655  1.414213562373095  \\
            0.06896551724137931  1.414213562373095  \\
            0.10344827586206896  1.414213562373095  \\
            0.13793103448275862  1.414213562373095  \\
            0.1724137931034483  1.414213562373095  \\
            0.20689655172413793  1.414213562373095  \\
            0.2413793103448276  1.414213562373095  \\
            0.27586206896551724  1.414213562373095  \\
            0.3103448275862069  1.414213562373095  \\
            0.3448275862068966  1.414213562373095  \\
            0.3793103448275862  1.414213562373095  \\
            0.41379310344827586  1.414213562373095  \\
            0.4482758620689655  1.414213562373095  \\
            0.4827586206896552  1.414213562373095  \\
            0.5172413793103449  1.414213562373095  \\
            0.5517241379310345  1.414213562373095  \\
            0.5862068965517241  1.414213562373095  \\
            0.6206896551724138  1.414213562373095  \\
            0.6551724137931034  1.414213562373095  \\
            0.6896551724137931  1.414213562373095  \\
            0.7241379310344828  1.414213562373095  \\
            0.7586206896551724  1.414213562373095  \\
            0.7931034482758621  1.414213562373095  \\
            0.8275862068965517  1.414213562373095  \\
            0.8620689655172413  1.414213562373095  \\
            0.896551724137931  1.414213562373095  \\
            0.9310344827586207  1.414213562373095  \\
            0.9655172413793104  1.414213562373095  \\
            1.0  1.4142135623730954  \\
        }
        ;
    \addplot[color={rgb,1:red,0.0;green,0.502;blue,0.0}, name path={d502acb9-f823-4dd5-9ffe-0f95641fe71f}, draw opacity={1.0}, line width={2.2}, dashed]
        table[row sep={\\}]
        {
            \\
            0.0  1.0000000077067681  \\
            0.034482758620689655  1.0645068303746985  \\
            0.06896551724137931  1.1210874625523353  \\
            0.10344827586206896  1.1708915066448031  \\
            0.13793103448275862  1.2147527554350717  \\
            0.1724137931034483  1.25329529968181  \\
            0.20689655172413793  1.2869970763118292  \\
            0.2413793103448276  1.3162299782063664  \\
            0.27586206896551724  1.3412862302604638  \\
            0.3103448275862069  1.3623962941474765  \\
            0.3448275862068966  1.3797413119822455  \\
            0.3793103448275862  1.3934618862880517  \\
            0.41379310344827586  1.403664306003284  \\
            0.4482758620689655  1.410424919739142  \\
            0.4827586206896552  1.4137931034482758  \\
            0.5172413793103449  1.4137931034482758  \\
            0.5517241379310345  1.410424919739142  \\
            0.5862068965517241  1.403664306003284  \\
            0.6206896551724138  1.3934618862880517  \\
            0.6551724137931034  1.3797413119822455  \\
            0.6896551724137931  1.3623962941474765  \\
            0.7241379310344828  1.3412862302604638  \\
            0.7586206896551724  1.3162299782063664  \\
            0.7931034482758621  1.2869970763118292  \\
            0.8275862068965517  1.25329529968181  \\
            0.8620689655172413  1.2147527554350717  \\
            0.896551724137931  1.1708915066448031  \\
            0.9310344827586207  1.1210874625523353  \\
            0.9655172413793104  1.0645068303746985  \\
            1.0  1.0000000077067681  \\
        }
        ;
    \addplot[color={rgb,1:red,0.502;green,0.502;blue,0.502}, name path={4f7228dc-a75d-4c5a-af1c-557256a78fae}, only marks, draw opacity={1.0}, line width={0.0}, solid, mark={triangle*}, mark size={6.6000000000000005 pt}, mark repeat={1}, mark options={color={rgb,1:red,0.0;green,0.0;blue,0.0}, draw opacity={0.0}, fill={rgb,1:red,0.502;green,0.502;blue,0.502}, fill opacity={1.0}, line width={1.6500000000000001}, rotate={0}, solid}]
        table[row sep={\\}]
        {
            \\
            0.0  2.0721009541718884  \\
            0.034482758620689655  1.7870583824526496  \\
            0.06896551724137931  2.0282666611012012  \\
            0.10344827586206896  1.9541549926054382  \\
            0.13793103448275862  2.055704989355616  \\
            0.1724137931034483  1.8807484552748543  \\
            0.20689655172413793  1.8986929092076075  \\
            0.2413793103448276  2.0039519283392213  \\
            0.27586206896551724  1.8356580111637628  \\
            0.3103448275862069  1.7217768523681238  \\
            0.3448275862068966  1.8283286022047696  \\
            0.3793103448275862  1.7278664673939654  \\
            0.41379310344827586  1.837040717054395  \\
            0.4482758620689655  1.7417805596641136  \\
            0.4827586206896552  1.8412817914610593  \\
            0.5172413793103449  1.6881511525166037  \\
            0.5517241379310345  1.715052158921691  \\
            0.5862068965517241  1.730633660048189  \\
            0.6206896551724138  1.6491522811876922  \\
            0.6551724137931034  1.574000080889531  \\
            0.6896551724137931  1.4967680698139207  \\
            0.7241379310344828  1.4121064163424637  \\
            0.7586206896551724  1.4580317444444155  \\
            0.7931034482758621  1.3813601401875828  \\
            0.8275862068965517  1.29859035530201  \\
            0.8620689655172413  1.202570978732909  \\
            0.896551724137931  1.2304381609254453  \\
            0.9310344827586207  1.1441506411357825  \\
            0.9655172413793104  1.0746719003901286  \\
            1.0  0.9958597868426915  \\
        }
        ;
    \addplot[color={rgb,1:red,0.502;green,0.0;blue,0.502}, name path={af863d8b-7854-413c-95f3-7acce11452ff}, only marks, draw opacity={1.0}, line width={0.0}, solid, mark={triangle*}, mark size={6.6000000000000005 pt}, mark repeat={1}, mark options={color={rgb,1:red,0.0;green,0.0;blue,0.0}, draw opacity={0.0}, fill={rgb,1:red,0.502;green,0.0;blue,0.502}, fill opacity={1.0}, line width={1.6500000000000001}, rotate={0}, solid}]
        table[row sep={\\}]
        {
            \\
            0.0  1.408181482805702  \\
            0.034482758620689655  1.2512113043236508  \\
            0.06896551724137931  1.4574456567096248  \\
            0.10344827586206896  1.4082222983257058  \\
            0.13793103448275862  1.4356152983702377  \\
            0.1724137931034483  1.3675353018087306  \\
            0.20689655172413793  1.3576143077038134  \\
            0.2413793103448276  1.4820601624236804  \\
            0.27586206896551724  1.3847727462372577  \\
            0.3103448275862069  1.2725267381700234  \\
            0.3448275862068966  1.4154338650783735  \\
            0.3793103448275862  1.3426710271092461  \\
            0.41379310344827586  1.4455638711311583  \\
            0.4482758620689655  1.4063399513018378  \\
            0.4827586206896552  1.4574379480128483  \\
            0.5172413793103449  1.4166290229647935  \\
            0.5517241379310345  1.449538480769336  \\
            0.5862068965517241  1.4638100903234608  \\
            0.6206896551724138  1.4959070745325591  \\
            0.6551724137931034  1.419136249906865  \\
            0.6896551724137931  1.3960858258483553  \\
            0.7241379310344828  1.3783306722161957  \\
            0.7586206896551724  1.4187866625223453  \\
            0.7931034482758621  1.4766988296909496  \\
            0.8275862068965517  1.4220105950156778  \\
            0.8620689655172413  1.3175951456371848  \\
            0.896551724137931  1.4523620642156128  \\
            0.9310344827586207  1.423398999758326  \\
            0.9655172413793104  1.4195926290328997  \\
            1.0  1.4356593207536212  \\
        }
        ;
    \addplot[color={rgb,1:red,0.0;green,0.502;blue,0.0}, name path={f2590328-41e5-426a-a64e-d8b731c95397}, only marks, draw opacity={1.0}, line width={0.0}, solid, mark={triangle*}, mark size={6.6000000000000005 pt}, mark repeat={1}, mark options={color={rgb,1:red,0.0;green,0.0;blue,0.0}, draw opacity={0.0}, fill={rgb,1:red,0.0;green,0.502;blue,0.0}, fill opacity={1.0}, line width={1.6500000000000001}, rotate={0}, solid}]
        table[row sep={\\}]
        {
            \\
            0.0  1.0272158854243645  \\
            0.034482758620689655  0.9508767134166323  \\
            0.06896551724137931  1.1190007666426598  \\
            0.10344827586206896  1.1863434630789476  \\
            0.13793103448275862  1.2017806060474379  \\
            0.1724137931034483  1.2002023856793298  \\
            0.20689655172413793  1.219808037905457  \\
            0.2413793103448276  1.3592103822058221  \\
            0.27586206896551724  1.3150490548825844  \\
            0.3103448275862069  1.2308765096223788  \\
            0.3448275862068966  1.3947170109332894  \\
            0.3793103448275862  1.3238963811541518  \\
            0.41379310344827586  1.438583142325517  \\
            0.4482758620689655  1.4036183794305646  \\
            0.4827586206896552  1.4570284417477022  \\
            0.5172413793103449  1.4132262197099719  \\
            0.5517241379310345  1.4486761019149967  \\
            0.5862068965517241  1.4602905388789988  \\
            0.6206896551724138  1.4642017565931047  \\
            0.6551724137931034  1.3914794818535303  \\
            0.6896551724137931  1.3423047087687738  \\
            0.7241379310344828  1.2977432921634144  \\
            0.7586206896551724  1.3423277191354517  \\
            0.7931034482758621  1.3153028888297968  \\
            0.8275862068965517  1.245098523078417  \\
            0.8620689655172413  1.1523271102402664  \\
            0.896551724137931  1.2198377469595485  \\
            0.9310344827586207  1.1344097099411115  \\
            0.9655172413793104  1.0736030525566616  \\
            1.0  0.9958597868426915  \\
        }
        ;
\end{axis}
\end{tikzpicture}

%% file: figures/simulations/simulations_ar_phalf.tikz
\begin{tikzpicture}[/tikz/background rectangle/.style={fill={rgb,1:red,1.0;green,1.0;blue,1.0}, draw opacity={1.0}}, show background rectangle]
\begin{axis}[point meta max={nan}, point meta min={nan}, legend cell align={left}, title={$\bSigma=\textrm{AR}(0.8),\;n=p/2$}, title style={at={{(0.5,1)}}, anchor={south}, font={{\fontsize{32.199999999999996 pt}{41.86 pt}\selectfont}}, color={rgb,1:red,0.0;green,0.0;blue,0.0}, draw opacity={1.0}, rotate={0.0}}, legend style={color={rgb,1:red,0.0;green,0.0;blue,0.0}, draw opacity={0.0}, line width={2.3}, solid, fill={rgb,1:red,0.0;green,0.0;blue,0.0}, fill opacity={0.0}, text opacity={1.0}, font={{\fontsize{18.4 pt}{23.919999999999998 pt}\selectfont}}, text={rgb,1:red,0.0;green,0.0;blue,0.0}, at={(0.02, 0.98)}, anchor={north west}}, axis background/.style={fill={rgb,1:red,1.0;green,1.0;blue,1.0}, opacity={1.0}}, anchor={north west}, xshift={1.0mm}, yshift={-1.0mm}, width={137.7mm}, height={109.76mm}, scaled x ticks={false}, xlabel={$K$}, x tick style={color={rgb,1:red,0.0;green,0.0;blue,0.0}, opacity={1.0}}, x tick label style={color={rgb,1:red,0.0;green,0.0;blue,0.0}, opacity={1.0}, rotate={0}}, xlabel style={at={(ticklabel cs:0.5)}, anchor=near ticklabel, font={{\fontsize{25.299999999999997 pt}{32.89 pt}\selectfont}}, color={rgb,1:red,0.0;green,0.0;blue,0.0}, draw opacity={1.0}, rotate={0.0}}, xmode={log}, log basis x={2}, xmajorgrids={false}, xmin={1.8403753012497501}, xmax={34.77551560083386}, xtick={{2.0,4.0,8.0,16.0,32.0}}, xticklabels={{$2^{1}$,$2^{2}$,$2^{3}$,$2^{4}$,$2^{5}$}}, xtick align={inside}, xticklabel style={font={{\fontsize{18.4 pt}{23.919999999999998 pt}\selectfont}}, color={rgb,1:red,0.0;green,0.0;blue,0.0}, draw opacity={1.0}, rotate={0.0}}, x grid style={color={rgb,1:red,0.0;green,0.0;blue,0.0}, draw opacity={0.1}, line width={1.15}, solid}, x axis line style={color={rgb,1:red,0.0;green,0.0;blue,0.0}, draw opacity={1.0}, line width={2.3}, solid}, scaled y ticks={false}, ylabel={$\mathbb{E}[\risk{\widehat{\bw}}] - \sigma^2$}, y tick style={color={rgb,1:red,0.0;green,0.0;blue,0.0}, opacity={1.0}}, y tick label style={color={rgb,1:red,0.0;green,0.0;blue,0.0}, opacity={1.0}, rotate={0}}, ylabel style={at={(ticklabel cs:0.5)}, anchor=near ticklabel, font={{\fontsize{25.299999999999997 pt}{32.89 pt}\selectfont}}, color={rgb,1:red,0.0;green,0.0;blue,0.0}, draw opacity={1.0}, rotate={0.0}}, ymajorgrids={false}, ymin={0}, ymax={795}, ytick={{0.0,200.0,400.0,600.0}}, yticklabels={{$0$,$200$,$400$,$600$}}, ytick align={inside}, yticklabel style={font={{\fontsize{18.4 pt}{23.919999999999998 pt}\selectfont}}, color={rgb,1:red,0.0;green,0.0;blue,0.0}, draw opacity={1.0}, rotate={0.0}}, y grid style={color={rgb,1:red,0.0;green,0.0;blue,0.0}, draw opacity={0.1}, line width={1.15}, solid}, y axis line style={color={rgb,1:red,0.0;green,0.0;blue,0.0}, draw opacity={1.0}, line width={2.3}, solid}]
    \addplot[color={rgb,1:red,0.2745;green,0.5098;blue,0.7059}, name path={0966a3c2-29d5-459a-af17-f005d83ac3fc}, draw opacity={1.0}, line width={2.3}, dotted, mark={triangle*}, mark size={6.8999999999999995 pt}, mark repeat={1}, mark options={color={rgb,1:red,0.0;green,0.0;blue,0.0}, draw opacity={0.6}, fill={rgb,1:red,0.2745;green,0.5098;blue,0.7059}, fill opacity={0.6}, line width={1.7249999999999999}, rotate={0}, solid}]
        table[row sep={\\}]
        {
            \\
            2.0  207.65534619251667  \\
            4.0  207.65534619251667  \\
            8.0  207.65534619251667  \\
            16.0  207.65534619251667  \\
            32.0  207.65534619251667  \\
        }
        ;
    \addlegendentry {Single Ridge}
    \addplot[color={rgb,1:red,0.0;green,0.502;blue,0.0}, name path={8bdbee3d-a639-4fe9-842d-7ffc3ce36f79}, draw opacity={1.0}, line width={2.3}, dashdotted, mark={triangle*}, mark size={6.8999999999999995 pt}, mark repeat={1}, mark options={color={rgb,1:red,0.0;green,0.0;blue,0.0}, draw opacity={0.6}, fill={rgb,1:red,0.0;green,0.502;blue,0.0}, fill opacity={0.6}, line width={1.7249999999999999}, rotate={180}, solid}]
        table[row sep={\\}]
        {
            \\
            2.0  172.12962242694547  \\
            4.0  162.3317539875045  \\
            8.0  160.4452474967526  \\
            16.0  161.07444227789054  \\
            32.0  163.51244360935289  \\
        }
        ;
    \addlegendentry {Group Lasso}
    \addplot[color={rgb,1:red,1.0;green,0.6471;blue,0.0}, name path={ed23a7a4-503a-414f-b928-a293e643f13a}, draw opacity={1.0}, line width={2.3}, dashdotdotted, mark={diamond*}, mark size={6.8999999999999995 pt}, mark repeat={1}, mark options={color={rgb,1:red,0.0;green,0.0;blue,0.0}, draw opacity={0.6}, fill={rgb,1:red,1.0;green,0.6471;blue,0.0}, fill opacity={0.6}, line width={1.7249999999999999}, rotate={0}, solid}]
        table[row sep={\\}]
        {
            \\
            2.0  156.49462264053886  \\
            4.0  154.57418298174736  \\
            8.0  177.2574272549391  \\
            16.0  311.3284233856493  \\
            32.0  400.60601842142387  \\
        }
        ;
    \addlegendentry {Multi Ridge}
    \addplot[color={rgb,1:red,0.502;green,0.0;blue,0.502}, name path={e4fd59cb-03f5-4b25-896e-59535422e56a}, draw opacity={1.0}, line width={2.3}, solid, mark={pentagon*}, mark size={6.8999999999999995 pt}, mark repeat={1}, mark options={color={rgb,1:red,0.0;green,0.0;blue,0.0}, draw opacity={0.6}, fill={rgb,1:red,0.502;green,0.0;blue,0.502}, fill opacity={0.6}, line width={1.7249999999999999}, rotate={0}, solid}]
        table[row sep={\\}]
        {
            \\
            2.0  157.2190944709665  \\
            4.0  143.82354874310235  \\
            8.0  143.47236134151922  \\
            16.0  151.11929493806997  \\
            32.0  166.4308677732943  \\
        }
        ;
    \addlegendentry {$\sigmacv\textrm{-Ridge}$}
    \addplot[color={rgb,1:red,0.502;green,0.502;blue,0.502}, name path={a2793773-c800-4a61-be7c-1dda48fb44de}, draw opacity={1.0}, line width={2.3}, dashed, mark={*}, mark size={6.8999999999999995 pt}, mark repeat={1}, mark options={color={rgb,1:red,0.0;green,0.0;blue,0.0}, draw opacity={0.6}, fill={rgb,1:red,0.502;green,0.502;blue,0.502}, fill opacity={0.6}, line width={1.7249999999999999}, rotate={0}, solid}]
        table[row sep={\\}]
        {
            \\
            2.0  130.97316246922074  \\
            4.0  130.97316246922074  \\
            8.0  130.97316246922074  \\
            16.0  130.97316246922074  \\
            32.0  130.97316246922074  \\
        }
        ;
    \addlegendentry {Bayes}
\end{axis}
\end{tikzpicture}

%% file: figures/simulations/simulations_ar_p.tikz
\begin{tikzpicture}[/tikz/background rectangle/.style={fill={rgb,1:red,1.0;green,1.0;blue,1.0}, draw opacity={1.0}}, show background rectangle]
\begin{axis}[point meta max={nan}, point meta min={nan}, legend cell align={left}, title={$\bSigma=\textrm{AR}(0.8),\;n=p$}, title style={at={{(0.5,1)}}, anchor={south}, font={{\fontsize{32.199999999999996 pt}{41.86 pt}\selectfont}}, color={rgb,1:red,0.0;green,0.0;blue,0.0}, draw opacity={1.0}, rotate={0.0}}, legend style={color={rgb,1:red,0.0;green,0.0;blue,0.0}, draw opacity={0.0}, line width={2.3}, solid, fill={rgb,1:red,0.0;green,0.0;blue,0.0}, fill opacity={0.0}, text opacity={1.0}, font={{\fontsize{18.4 pt}{23.919999999999998 pt}\selectfont}}, text={rgb,1:red,0.0;green,0.0;blue,0.0}, at={(1.02, 1)}, anchor={north west}}, axis background/.style={fill={rgb,1:red,1.0;green,1.0;blue,1.0}, opacity={1.0}}, anchor={north west}, xshift={1.0mm}, yshift={-1.0mm}, width={132.7mm}, height={109.76mm}, scaled x ticks={false}, xlabel={$K$}, x tick style={color={rgb,1:red,0.0;green,0.0;blue,0.0}, opacity={1.0}}, x tick label style={color={rgb,1:red,0.0;green,0.0;blue,0.0}, opacity={1.0}, rotate={0}}, xlabel style={at={(ticklabel cs:0.5)}, anchor=near ticklabel, font={{\fontsize{25.299999999999997 pt}{32.89 pt}\selectfont}}, color={rgb,1:red,0.0;green,0.0;blue,0.0}, draw opacity={1.0}, rotate={0.0}}, xmode={log}, log basis x={2}, xmajorgrids={false}, xmin={1.8403753012497501}, xmax={34.77551560083386}, xtick={{2.0,4.0,8.0,16.0,32.0}}, xticklabels={{$2^{1}$,$2^{2}$,$2^{3}$,$2^{4}$,$2^{5}$}}, xtick align={inside}, xticklabel style={font={{\fontsize{18.4 pt}{23.919999999999998 pt}\selectfont}}, color={rgb,1:red,0.0;green,0.0;blue,0.0}, draw opacity={1.0}, rotate={0.0}}, x grid style={color={rgb,1:red,0.0;green,0.0;blue,0.0}, draw opacity={0.1}, line width={1.15}, solid}, x axis line style={color={rgb,1:red,0.0;green,0.0;blue,0.0}, draw opacity={1.0}, line width={2.3}, solid}, scaled y ticks={false}, ylabel={$\mathbb{E}[\risk{\widehat{\bw}}] - \sigma^2$}, y tick style={color={rgb,1:red,0.0;green,0.0;blue,0.0}, opacity={1.0}}, y tick label style={color={rgb,1:red,0.0;green,0.0;blue,0.0}, opacity={1.0}, rotate={0}}, ylabel style={at={(ticklabel cs:0.5)}, anchor=near ticklabel, font={{\fontsize{25.299999999999997 pt}{32.89 pt}\selectfont}}, color={rgb,1:red,0.0;green,0.0;blue,0.0}, draw opacity={1.0}, rotate={0.0}}, ymajorgrids={false}, ymin={0}, ymax={285}, ytick={{0.0,50.0,100.0,150.0,200.0,250.0}}, yticklabels={{$0$,$50$,$100$,$150$,$200$,$250$}}, ytick align={inside}, yticklabel style={font={{\fontsize{18.4 pt}{23.919999999999998 pt}\selectfont}}, color={rgb,1:red,0.0;green,0.0;blue,0.0}, draw opacity={1.0}, rotate={0.0}}, y grid style={color={rgb,1:red,0.0;green,0.0;blue,0.0}, draw opacity={0.1}, line width={1.15}, solid}, y axis line style={color={rgb,1:red,0.0;green,0.0;blue,0.0}, draw opacity={1.0}, line width={2.3}, solid}]
    \addplot[color={rgb,1:red,0.2745;green,0.5098;blue,0.7059}, name path={f70758a9-e090-4597-9051-ae18742f2ed2}, draw opacity={1.0}, line width={2.3}, dotted, mark={triangle*}, mark size={6.8999999999999995 pt}, mark repeat={1}, mark options={color={rgb,1:red,0.0;green,0.0;blue,0.0}, draw opacity={0.6}, fill={rgb,1:red,0.2745;green,0.5098;blue,0.7059}, fill opacity={0.6}, line width={1.7249999999999999}, rotate={0}, solid}, forget plot]
        table[row sep={\\}]
        {
            \\
            2.0  72.44689148830057  \\
            4.0  72.44689148830057  \\
            8.0  72.44689148830057  \\
            16.0  72.44689148830057  \\
            32.0  72.44689148830057  \\
        }
        ;
    \addplot[color={rgb,1:red,0.0;green,0.502;blue,0.0}, name path={2253d09b-d673-41f5-ae6f-b141cfc3b906}, draw opacity={1.0}, line width={2.3}, dashdotted, mark={triangle*}, mark size={6.8999999999999995 pt}, mark repeat={1}, mark options={color={rgb,1:red,0.0;green,0.0;blue,0.0}, draw opacity={0.6}, fill={rgb,1:red,0.0;green,0.502;blue,0.0}, fill opacity={0.6}, line width={1.7249999999999999}, rotate={180}, solid}, forget plot]
        table[row sep={\\}]
        {
            \\
            2.0  59.88052184412956  \\
            4.0  55.76981763708534  \\
            8.0  54.815402166175744  \\
            16.0  54.75940790601871  \\
            32.0  55.104421299071404  \\
        }
        ;
    \addplot[color={rgb,1:red,1.0;green,0.6471;blue,0.0}, name path={bd5a6e77-c85a-490a-b07a-4a2d30812202}, draw opacity={1.0}, line width={2.3}, dashdotdotted, mark={diamond*}, mark size={6.8999999999999995 pt}, mark repeat={1}, mark options={color={rgb,1:red,0.0;green,0.0;blue,0.0}, draw opacity={0.6}, fill={rgb,1:red,1.0;green,0.6471;blue,0.0}, fill opacity={0.6}, line width={1.7249999999999999}, rotate={0}, solid}, forget plot]
        table[row sep={\\}]
        {
            \\
            2.0  54.00858681200842  \\
            4.0  56.05589203922405  \\
            8.0  84.47588120875069  \\
            16.0  176.27278860656796  \\
            32.0  252.42931066095764  \\
        }
        ;
    \addplot[color={rgb,1:red,0.502;green,0.0;blue,0.502}, name path={639303bf-8215-4edf-be5a-239d17212a9f}, draw opacity={1.0}, line width={2.3}, solid, mark={pentagon*}, mark size={6.8999999999999995 pt}, mark repeat={1}, mark options={color={rgb,1:red,0.0;green,0.0;blue,0.0}, draw opacity={0.6}, fill={rgb,1:red,0.502;green,0.0;blue,0.502}, fill opacity={0.6}, line width={1.7249999999999999}, rotate={0}, solid}, forget plot]
        table[row sep={\\}]
        {
            \\
            2.0  53.953355978612635  \\
            4.0  47.88746700004553  \\
            8.0  46.811826946959954  \\
            16.0  47.94532809962344  \\
            32.0  50.63563902240429  \\
        }
        ;
    \addplot[color={rgb,1:red,0.502;green,0.502;blue,0.502}, name path={2e9741aa-75a6-442b-9169-fb4140ff2e46}, draw opacity={1.0}, line width={2.3}, dashed, mark={*}, mark size={6.8999999999999995 pt}, mark repeat={1}, mark options={color={rgb,1:red,0.0;green,0.0;blue,0.0}, draw opacity={0.6}, fill={rgb,1:red,0.502;green,0.502;blue,0.502}, fill opacity={0.6}, line width={1.7249999999999999}, rotate={0}, solid}, forget plot]
        table[row sep={\\}]
        {
            \\
            2.0  44.30917778528904  \\
            4.0  44.30917778528904  \\
            8.0  44.30917778528904  \\
            16.0  44.30917778528904  \\
            32.0  44.30917778528904  \\
        }
        ;
\end{axis}
\end{tikzpicture}

%% file: figures/simulations/simulations_ar_ptwice.tikz
\begin{tikzpicture}[/tikz/background rectangle/.style={fill={rgb,1:red,1.0;green,1.0;blue,1.0}, draw opacity={1.0}}, show background rectangle]
\begin{axis}[point meta max={nan}, point meta min={nan}, legend cell align={left}, title={$\bSigma=\textrm{AR}(0.8),\;n=2p$}, title style={at={{(0.5,1)}}, anchor={south}, font={{\fontsize{32.199999999999996 pt}{41.86 pt}\selectfont}}, color={rgb,1:red,0.0;green,0.0;blue,0.0}, draw opacity={1.0}, rotate={0.0}}, legend style={color={rgb,1:red,0.0;green,0.0;blue,0.0}, draw opacity={0.0}, line width={2.3}, solid, fill={rgb,1:red,0.0;green,0.0;blue,0.0}, fill opacity={0.0}, text opacity={1.0}, font={{\fontsize{18.4 pt}{23.919999999999998 pt}\selectfont}}, text={rgb,1:red,0.0;green,0.0;blue,0.0}, at={(1.02, 1)}, anchor={north west}}, axis background/.style={fill={rgb,1:red,1.0;green,1.0;blue,1.0}, opacity={1.0}}, anchor={north west}, xshift={1.0mm}, yshift={-1.0mm}, width={132.7mm}, height={109.76mm}, scaled x ticks={false}, xlabel={$K$}, x tick style={color={rgb,1:red,0.0;green,0.0;blue,0.0}, opacity={1.0}}, x tick label style={color={rgb,1:red,0.0;green,0.0;blue,0.0}, opacity={1.0}, rotate={0}}, xlabel style={at={(ticklabel cs:0.5)}, anchor=near ticklabel, font={{\fontsize{25.299999999999997 pt}{32.89 pt}\selectfont}}, color={rgb,1:red,0.0;green,0.0;blue,0.0}, draw opacity={1.0}, rotate={0.0}}, xmode={log}, log basis x={2}, xmajorgrids={false}, xmin={1.8403753012497501}, xmax={34.77551560083386}, xtick={{2.0,4.0,8.0,16.0,32.0}}, xticklabels={{$2^{1}$,$2^{2}$,$2^{3}$,$2^{4}$,$2^{5}$}}, xtick align={inside}, xticklabel style={font={{\fontsize{18.4 pt}{23.919999999999998 pt}\selectfont}}, color={rgb,1:red,0.0;green,0.0;blue,0.0}, draw opacity={1.0}, rotate={0.0}}, x grid style={color={rgb,1:red,0.0;green,0.0;blue,0.0}, draw opacity={0.1}, line width={1.15}, solid}, x axis line style={color={rgb,1:red,0.0;green,0.0;blue,0.0}, draw opacity={1.0}, line width={2.3}, solid}, scaled y ticks={false}, ylabel={$\mathbb{E}[\risk{\widehat{\bw}}] - \sigma^2$}, y tick style={color={rgb,1:red,0.0;green,0.0;blue,0.0}, opacity={1.0}}, y tick label style={color={rgb,1:red,0.0;green,0.0;blue,0.0}, opacity={1.0}, rotate={0}}, ylabel style={at={(ticklabel cs:0.5)}, anchor=near ticklabel, font={{\fontsize{25.299999999999997 pt}{32.89 pt}\selectfont}}, color={rgb,1:red,0.0;green,0.0;blue,0.0}, draw opacity={1.0}, rotate={0.0}}, ymajorgrids={false}, ymin={0}, ymax={55}, ytick={{0.0,10.0,20.0,30.0,40.0,50.0}}, yticklabels={{$0$,$10$,$20$,$30$,$40$,$50$}}, ytick align={inside}, yticklabel style={font={{\fontsize{18.4 pt}{23.919999999999998 pt}\selectfont}}, color={rgb,1:red,0.0;green,0.0;blue,0.0}, draw opacity={1.0}, rotate={0.0}}, y grid style={color={rgb,1:red,0.0;green,0.0;blue,0.0}, draw opacity={0.1}, line width={1.15}, solid}, y axis line style={color={rgb,1:red,0.0;green,0.0;blue,0.0}, draw opacity={1.0}, line width={2.3}, solid}]
    \addplot[color={rgb,1:red,0.2745;green,0.5098;blue,0.7059}, name path={866ad952-7f01-45d9-aafc-77a33a9a7e69}, draw opacity={1.0}, line width={2.3}, dotted, mark={triangle*}, mark size={6.8999999999999995 pt}, mark repeat={1}, mark options={color={rgb,1:red,0.0;green,0.0;blue,0.0}, draw opacity={0.6}, fill={rgb,1:red,0.2745;green,0.5098;blue,0.7059}, fill opacity={0.6}, line width={1.7249999999999999}, rotate={0}, solid}, forget plot]
        table[row sep={\\}]
        {
            \\
            2.0  21.145740075114105  \\
            4.0  21.145740075114105  \\
            8.0  21.145740075114105  \\
            16.0  21.145740075114105  \\
            32.0  21.145740075114105  \\
        }
        ;
    \addplot[color={rgb,1:red,0.0;green,0.502;blue,0.0}, name path={3c033d5a-b5f4-4e6c-9a91-5b9c7a663acd}, draw opacity={1.0}, line width={2.3}, dashdotted, mark={triangle*}, mark size={6.8999999999999995 pt}, mark repeat={1}, mark options={color={rgb,1:red,0.0;green,0.0;blue,0.0}, draw opacity={0.6}, fill={rgb,1:red,0.0;green,0.502;blue,0.0}, fill opacity={0.6}, line width={1.7249999999999999}, rotate={180}, solid}, forget plot]
        table[row sep={\\}]
        {
            \\
            2.0  19.637606238395044  \\
            4.0  18.672389933228544  \\
            8.0  18.385020081857782  \\
            16.0  18.335416716882385  \\
            32.0  18.358106061658006  \\
        }
        ;
    \addplot[color={rgb,1:red,1.0;green,0.6471;blue,0.0}, name path={68219ce8-b8da-450f-a8ff-60b425139573}, draw opacity={1.0}, line width={2.3}, dashdotdotted, mark={diamond*}, mark size={6.8999999999999995 pt}, mark repeat={1}, mark options={color={rgb,1:red,0.0;green,0.0;blue,0.0}, draw opacity={0.6}, fill={rgb,1:red,1.0;green,0.6471;blue,0.0}, fill opacity={0.6}, line width={1.7249999999999999}, rotate={0}, solid}, forget plot]
        table[row sep={\\}]
        {
            \\
            2.0  18.779838074254783  \\
            4.0  20.548165844403037  \\
            8.0  48.75621587384589  \\
            16.0  117.05546416132847  \\
            32.0  184.88264343577944  \\
        }
        ;
    \addplot[color={rgb,1:red,0.502;green,0.0;blue,0.502}, name path={06068c2d-9396-4922-9d8a-32b1c5dd316d}, draw opacity={1.0}, line width={2.3}, solid, mark={pentagon*}, mark size={6.8999999999999995 pt}, mark repeat={1}, mark options={color={rgb,1:red,0.0;green,0.0;blue,0.0}, draw opacity={0.6}, fill={rgb,1:red,0.502;green,0.0;blue,0.502}, fill opacity={0.6}, line width={1.7249999999999999}, rotate={0}, solid}, forget plot]
        table[row sep={\\}]
        {
            \\
            2.0  18.598086899914904  \\
            4.0  16.86196893750028  \\
            8.0  16.3123448060759  \\
            16.0  16.308906475585246  \\
            32.0  16.63105741455042  \\
        }
        ;
    \addplot[color={rgb,1:red,0.502;green,0.502;blue,0.502}, name path={33a15078-a891-4ab4-900b-f3212ab4297b}, draw opacity={1.0}, line width={2.3}, dashed, mark={*}, mark size={6.8999999999999995 pt}, mark repeat={1}, mark options={color={rgb,1:red,0.0;green,0.0;blue,0.0}, draw opacity={0.6}, fill={rgb,1:red,0.502;green,0.502;blue,0.502}, fill opacity={0.6}, line width={1.7249999999999999}, rotate={0}, solid}, forget plot]
        table[row sep={\\}]
        {
            \\
            2.0  15.827837127726468  \\
            4.0  15.827837127726468  \\
            8.0  15.827837127726468  \\
            16.0  15.827837127726468  \\
            32.0  15.827837127726468  \\
        }
        ;
\end{axis}
\end{tikzpicture}

%% file: figures/simulations/simulations_id_phalf.tikz
\begin{tikzpicture}[/tikz/background rectangle/.style={fill={rgb,1:red,1.0;green,1.0;blue,1.0}, draw opacity={1.0}}, show background rectangle]
\begin{axis}[point meta max={nan}, point meta min={nan}, legend cell align={left}, title={$\bSigma=\bI,\; n=p/2$}, title style={at={{(0.5,1)}}, anchor={south}, font={{\fontsize{32.199999999999996 pt}{41.86 pt}\selectfont}}, color={rgb,1:red,0.0;green,0.0;blue,0.0}, draw opacity={1.0}, rotate={0.0}}, legend style={color={rgb,1:red,0.0;green,0.0;blue,0.0}, draw opacity={0.0}, line width={2.3}, solid, fill={rgb,1:red,0.0;green,0.0;blue,0.0}, fill opacity={0.0}, text opacity={1.0}, font={{\fontsize{18.4 pt}{23.919999999999998 pt}\selectfont}}, text={rgb,1:red,0.0;green,0.0;blue,0.0}, at={(0.02, 0.98)}, anchor={north west}}, axis background/.style={fill={rgb,1:red,1.0;green,1.0;blue,1.0}, opacity={1.0}}, anchor={north west}, xshift={1.0mm}, yshift={-1.0mm}, width={137.7mm}, height={109.76mm}, scaled x ticks={false}, xlabel={$K$}, x tick style={color={rgb,1:red,0.0;green,0.0;blue,0.0}, opacity={1.0}}, x tick label style={color={rgb,1:red,0.0;green,0.0;blue,0.0}, opacity={1.0}, rotate={0}}, xlabel style={at={(ticklabel cs:0.5)}, anchor=near ticklabel, font={{\fontsize{25.299999999999997 pt}{32.89 pt}\selectfont}}, color={rgb,1:red,0.0;green,0.0;blue,0.0}, draw opacity={1.0}, rotate={0.0}}, xmode={log}, log basis x={2}, xmajorgrids={false}, xmin={1.8403753012497501}, xmax={34.77551560083386}, xtick={{2.0,4.0,8.0,16.0,32.0}}, xticklabels={{$2^{1}$,$2^{2}$,$2^{3}$,$2^{4}$,$2^{5}$}}, xtick align={inside}, xticklabel style={font={{\fontsize{18.4 pt}{23.919999999999998 pt}\selectfont}}, color={rgb,1:red,0.0;green,0.0;blue,0.0}, draw opacity={1.0}, rotate={0.0}}, x grid style={color={rgb,1:red,0.0;green,0.0;blue,0.0}, draw opacity={0.1}, line width={1.15}, solid}, x axis line style={color={rgb,1:red,0.0;green,0.0;blue,0.0}, draw opacity={1.0}, line width={2.3}, solid}, scaled y ticks={false}, ylabel={$\mathbb{E}[\risk{\widehat{\bw}}] - \sigma^2$}, y tick style={color={rgb,1:red,0.0;green,0.0;blue,0.0}, opacity={1.0}}, y tick label style={color={rgb,1:red,0.0;green,0.0;blue,0.0}, opacity={1.0}, rotate={0}}, ylabel style={at={(ticklabel cs:0.5)}, anchor=near ticklabel, font={{\fontsize{25.299999999999997 pt}{32.89 pt}\selectfont}}, color={rgb,1:red,0.0;green,0.0;blue,0.0}, draw opacity={1.0}, rotate={0.0}}, ymajorgrids={false}, ymin={0}, ymax={795}, ytick={{0.0,200.0,400.0,600.0}}, yticklabels={{$0$,$200$,$400$,$600$}}, ytick align={inside}, yticklabel style={font={{\fontsize{18.4 pt}{23.919999999999998 pt}\selectfont}}, color={rgb,1:red,0.0;green,0.0;blue,0.0}, draw opacity={1.0}, rotate={0.0}}, y grid style={color={rgb,1:red,0.0;green,0.0;blue,0.0}, draw opacity={0.1}, line width={1.15}, solid}, y axis line style={color={rgb,1:red,0.0;green,0.0;blue,0.0}, draw opacity={1.0}, line width={2.3}, solid}]
    \addplot[color={rgb,1:red,0.2745;green,0.5098;blue,0.7059}, name path={cf0aed43-a683-4c50-95d7-5c22b7891209}, draw opacity={1.0}, line width={2.3}, dotted, mark={triangle*}, mark size={6.8999999999999995 pt}, mark repeat={1}, mark options={color={rgb,1:red,0.0;green,0.0;blue,0.0}, draw opacity={0.6}, fill={rgb,1:red,0.2745;green,0.5098;blue,0.7059}, fill opacity={0.6}, line width={1.7249999999999999}, rotate={0}, solid}, forget plot]
        table[row sep={\\}]
        {
            \\
            2.0  565.7779563379962  \\
            4.0  565.7779563379962  \\
            8.0  565.7779563379962  \\
            16.0  565.7779563379962  \\
            32.0  565.7779563379962  \\
        }
        ;
    \addplot[color={rgb,1:red,0.0;green,0.502;blue,0.0}, name path={c294bcf8-1746-4a20-bb8b-b20d4a3ada70}, draw opacity={1.0}, line width={2.3}, dashdotted, mark={triangle*}, mark size={6.8999999999999995 pt}, mark repeat={1}, mark options={color={rgb,1:red,0.0;green,0.0;blue,0.0}, draw opacity={0.6}, fill={rgb,1:red,0.0;green,0.502;blue,0.0}, fill opacity={0.6}, line width={1.7249999999999999}, rotate={180}, solid}, forget plot]
        table[row sep={\\}]
        {
            \\
            2.0  437.2349134284431  \\
            4.0  408.7014953514028  \\
            8.0  403.6561532327714  \\
            16.0  407.1407588687848  \\
            32.0  417.3230105315693  \\
        }
        ;
    \addplot[color={rgb,1:red,1.0;green,0.6471;blue,0.0}, name path={d62551dd-f54e-4d8a-bd7f-20a2e40ba8b7}, draw opacity={1.0}, line width={2.3}, dashdotdotted, mark={diamond*}, mark size={6.8999999999999995 pt}, mark repeat={1}, mark options={color={rgb,1:red,0.0;green,0.0;blue,0.0}, draw opacity={0.6}, fill={rgb,1:red,1.0;green,0.6471;blue,0.0}, fill opacity={0.6}, line width={1.7249999999999999}, rotate={0}, solid}, forget plot]
        table[row sep={\\}]
        {
            \\
            2.0  391.6922256613011  \\
            4.0  361.45977308722365  \\
            8.0  446.37577449035695  \\
            16.0  618.474333516128  \\
            32.0  776.950888504575  \\
        }
        ;
    \addplot[color={rgb,1:red,0.502;green,0.0;blue,0.502}, name path={dad7fa0f-8f35-4c10-82dd-bc1bb1d0123f}, draw opacity={1.0}, line width={2.3}, solid, mark={pentagon*}, mark size={6.8999999999999995 pt}, mark repeat={1}, mark options={color={rgb,1:red,0.0;green,0.0;blue,0.0}, draw opacity={0.6}, fill={rgb,1:red,0.502;green,0.0;blue,0.502}, fill opacity={0.6}, line width={1.7249999999999999}, rotate={0}, solid}, forget plot]
        table[row sep={\\}]
        {
            \\
            2.0  394.9766234192815  \\
            4.0  359.84926641685564  \\
            8.0  363.0034670447661  \\
            16.0  383.43989427996604  \\
            32.0  424.0934683481697  \\
        }
        ;
    \addplot[color={rgb,1:red,0.502;green,0.502;blue,0.502}, name path={f18c71b9-6969-4acd-a9df-01e99085bac0}, draw opacity={1.0}, line width={2.3}, dashed, mark={*}, mark size={6.8999999999999995 pt}, mark repeat={1}, mark options={color={rgb,1:red,0.0;green,0.0;blue,0.0}, draw opacity={0.6}, fill={rgb,1:red,0.502;green,0.502;blue,0.502}, fill opacity={0.6}, line width={1.7249999999999999}, rotate={0}, solid}, forget plot]
        table[row sep={\\}]
        {
            \\
            2.0  331.57426083467385  \\
            4.0  331.57426083467385  \\
            8.0  331.57426083467385  \\
            16.0  331.57426083467385  \\
            32.0  331.57426083467385  \\
        }
        ;
\end{axis}
\end{tikzpicture}

%% file: figures/simulations/simulations_id_p.tikz
\begin{tikzpicture}[/tikz/background rectangle/.style={fill={rgb,1:red,1.0;green,1.0;blue,1.0}, draw opacity={1.0}}, show background rectangle]
\begin{axis}[point meta max={nan}, point meta min={nan}, legend cell align={left}, title={$\bSigma=\bI,\; n=p$}, title style={at={{(0.5,1)}}, anchor={south}, font={{\fontsize{32.199999999999996 pt}{41.86 pt}\selectfont}}, color={rgb,1:red,0.0;green,0.0;blue,0.0}, draw opacity={1.0}, rotate={0.0}}, legend style={color={rgb,1:red,0.0;green,0.0;blue,0.0}, draw opacity={0.0}, line width={2.3}, solid, fill={rgb,1:red,0.0;green,0.0;blue,0.0}, fill opacity={0.0}, text opacity={1.0}, font={{\fontsize{18.4 pt}{23.919999999999998 pt}\selectfont}}, text={rgb,1:red,0.0;green,0.0;blue,0.0}, at={(1.02, 1)}, anchor={north west}}, axis background/.style={fill={rgb,1:red,1.0;green,1.0;blue,1.0}, opacity={1.0}}, anchor={north west}, xshift={1.0mm}, yshift={-1.0mm}, width={132.7mm}, height={109.76mm}, scaled x ticks={false}, xlabel={$K$}, x tick style={color={rgb,1:red,0.0;green,0.0;blue,0.0}, opacity={1.0}}, x tick label style={color={rgb,1:red,0.0;green,0.0;blue,0.0}, opacity={1.0}, rotate={0}}, xlabel style={at={(ticklabel cs:0.5)}, anchor=near ticklabel, font={{\fontsize{25.299999999999997 pt}{32.89 pt}\selectfont}}, color={rgb,1:red,0.0;green,0.0;blue,0.0}, draw opacity={1.0}, rotate={0.0}}, xmode={log}, log basis x={2}, xmajorgrids={false}, xmin={1.8403753012497501}, xmax={34.77551560083386}, xtick={{2.0,4.0,8.0,16.0,32.0}}, xticklabels={{$2^{1}$,$2^{2}$,$2^{3}$,$2^{4}$,$2^{5}$}}, xtick align={inside}, xticklabel style={font={{\fontsize{18.4 pt}{23.919999999999998 pt}\selectfont}}, color={rgb,1:red,0.0;green,0.0;blue,0.0}, draw opacity={1.0}, rotate={0.0}}, x grid style={color={rgb,1:red,0.0;green,0.0;blue,0.0}, draw opacity={0.1}, line width={1.15}, solid}, x axis line style={color={rgb,1:red,0.0;green,0.0;blue,0.0}, draw opacity={1.0}, line width={2.3}, solid}, scaled y ticks={false}, ylabel={$\mathbb{E}[\risk{\widehat{\bw}}] - \sigma^2$}, y tick style={color={rgb,1:red,0.0;green,0.0;blue,0.0}, opacity={1.0}}, y tick label style={color={rgb,1:red,0.0;green,0.0;blue,0.0}, opacity={1.0}, rotate={0}}, ylabel style={at={(ticklabel cs:0.5)}, anchor=near ticklabel, font={{\fontsize{25.299999999999997 pt}{32.89 pt}\selectfont}}, color={rgb,1:red,0.0;green,0.0;blue,0.0}, draw opacity={1.0}, rotate={0.0}}, ymajorgrids={false}, ymin={0}, ymax={285}, ytick={{0.0,50.0,100.0,150.0,200.0,250.0}}, yticklabels={{$0$,$50$,$100$,$150$,$200$,$250$}}, ytick align={inside}, yticklabel style={font={{\fontsize{18.4 pt}{23.919999999999998 pt}\selectfont}}, color={rgb,1:red,0.0;green,0.0;blue,0.0}, draw opacity={1.0}, rotate={0.0}}, y grid style={color={rgb,1:red,0.0;green,0.0;blue,0.0}, draw opacity={0.1}, line width={1.15}, solid}, y axis line style={color={rgb,1:red,0.0;green,0.0;blue,0.0}, draw opacity={1.0}, line width={2.3}, solid}]
    \addplot[color={rgb,1:red,0.2745;green,0.5098;blue,0.7059}, name path={2c25cef4-77b6-4c0b-956c-a7dc36a6803d}, draw opacity={1.0}, line width={2.3}, dotted, mark={triangle*}, mark size={6.8999999999999995 pt}, mark repeat={1}, mark options={color={rgb,1:red,0.0;green,0.0;blue,0.0}, draw opacity={0.6}, fill={rgb,1:red,0.2745;green,0.5098;blue,0.7059}, fill opacity={0.6}, line width={1.7249999999999999}, rotate={0}, solid}, forget plot]
        table[row sep={\\}]
        {
            \\
            2.0  155.1337108704425  \\
            4.0  155.1337108704425  \\
            8.0  155.1337108704425  \\
            16.0  155.1337108704425  \\
            32.0  155.1337108704425  \\
        }
        ;
    \addplot[color={rgb,1:red,0.0;green,0.502;blue,0.0}, name path={fb45ff00-5709-4e79-b7b2-db72f47019cf}, draw opacity={1.0}, line width={2.3}, dashdotted, mark={triangle*}, mark size={6.8999999999999995 pt}, mark repeat={1}, mark options={color={rgb,1:red,0.0;green,0.0;blue,0.0}, draw opacity={0.6}, fill={rgb,1:red,0.0;green,0.502;blue,0.0}, fill opacity={0.6}, line width={1.7249999999999999}, rotate={180}, solid}, forget plot]
        table[row sep={\\}]
        {
            \\
            2.0  115.03354718172338  \\
            4.0  100.26625482943545  \\
            8.0  97.09515316904364  \\
            16.0  96.662834817969  \\
            32.0  97.36415564689493  \\
        }
        ;
    \addplot[color={rgb,1:red,1.0;green,0.6471;blue,0.0}, name path={a9dfbf7f-2bfa-407b-826b-6a96cba26c61}, draw opacity={1.0}, line width={2.3}, dashdotdotted, mark={diamond*}, mark size={6.8999999999999995 pt}, mark repeat={1}, mark options={color={rgb,1:red,0.0;green,0.0;blue,0.0}, draw opacity={0.6}, fill={rgb,1:red,1.0;green,0.6471;blue,0.0}, fill opacity={0.6}, line width={1.7249999999999999}, rotate={0}, solid}, forget plot]
        table[row sep={\\}]
        {
            \\
            2.0  103.96792921565716  \\
            4.0  95.06683731471394  \\
            8.0  113.68892335950846  \\
            16.0  264.6433629123305  \\
            32.0  427.2564133230057  \\
        }
        ;
    \addplot[color={rgb,1:red,0.502;green,0.0;blue,0.502}, name path={3cf0e0d2-7424-4086-b2b1-5ec9baf637df}, draw opacity={1.0}, line width={2.3}, solid, mark={pentagon*}, mark size={6.8999999999999995 pt}, mark repeat={1}, mark options={color={rgb,1:red,0.0;green,0.0;blue,0.0}, draw opacity={0.6}, fill={rgb,1:red,0.502;green,0.0;blue,0.502}, fill opacity={0.6}, line width={1.7249999999999999}, rotate={0}, solid}, forget plot]
        table[row sep={\\}]
        {
            \\
            2.0  109.2009835080031  \\
            4.0  87.69286629422615  \\
            8.0  80.78406134148543  \\
            16.0  81.54586860617736  \\
            32.0  83.66076519947866  \\
        }
        ;
    \addplot[color={rgb,1:red,0.502;green,0.502;blue,0.502}, name path={36f91f14-62c1-4a52-b427-2159cc861d2a}, draw opacity={1.0}, line width={2.3}, dashed, mark={*}, mark size={6.8999999999999995 pt}, mark repeat={1}, mark options={color={rgb,1:red,0.0;green,0.0;blue,0.0}, draw opacity={0.6}, fill={rgb,1:red,0.502;green,0.502;blue,0.502}, fill opacity={0.6}, line width={1.7249999999999999}, rotate={0}, solid}, forget plot]
        table[row sep={\\}]
        {
            \\
            2.0  73.94182624744589  \\
            4.0  73.94182624744589  \\
            8.0  73.94182624744589  \\
            16.0  73.94182624744589  \\
            32.0  73.94182624744589  \\
        }
        ;
\end{axis}
\end{tikzpicture}

%% file: figures/simulations/simulations_id_ptwice.tikz
\begin{tikzpicture}[/tikz/background rectangle/.style={fill={rgb,1:red,1.0;green,1.0;blue,1.0}, draw opacity={1.0}}, show background rectangle]
\begin{axis}[point meta max={nan}, point meta min={nan}, legend cell align={left}, title={$\bSigma=\bI,\; n=2p$}, title style={at={{(0.5,1)}}, anchor={south}, font={{\fontsize{32.199999999999996 pt}{41.86 pt}\selectfont}}, color={rgb,1:red,0.0;green,0.0;blue,0.0}, draw opacity={1.0}, rotate={0.0}}, legend style={color={rgb,1:red,0.0;green,0.0;blue,0.0}, draw opacity={0.0}, line width={2.3}, solid, fill={rgb,1:red,0.0;green,0.0;blue,0.0}, fill opacity={0.0}, text opacity={1.0}, font={{\fontsize{18.4 pt}{23.919999999999998 pt}\selectfont}}, text={rgb,1:red,0.0;green,0.0;blue,0.0}, at={(1.02, 1)}, anchor={north west}}, axis background/.style={fill={rgb,1:red,1.0;green,1.0;blue,1.0}, opacity={1.0}}, anchor={north west}, xshift={1.0mm}, yshift={-1.0mm}, width={132.7mm}, height={109.76mm}, scaled x ticks={false}, xlabel={$K$}, x tick style={color={rgb,1:red,0.0;green,0.0;blue,0.0}, opacity={1.0}}, x tick label style={color={rgb,1:red,0.0;green,0.0;blue,0.0}, opacity={1.0}, rotate={0}}, xlabel style={at={(ticklabel cs:0.5)}, anchor=near ticklabel, font={{\fontsize{25.299999999999997 pt}{32.89 pt}\selectfont}}, color={rgb,1:red,0.0;green,0.0;blue,0.0}, draw opacity={1.0}, rotate={0.0}}, xmode={log}, log basis x={2}, xmajorgrids={false}, xmin={1.8403753012497501}, xmax={34.77551560083386}, xtick={{2.0,4.0,8.0,16.0,32.0}}, xticklabels={{$2^{1}$,$2^{2}$,$2^{3}$,$2^{4}$,$2^{5}$}}, xtick align={inside}, xticklabel style={font={{\fontsize{18.4 pt}{23.919999999999998 pt}\selectfont}}, color={rgb,1:red,0.0;green,0.0;blue,0.0}, draw opacity={1.0}, rotate={0.0}}, x grid style={color={rgb,1:red,0.0;green,0.0;blue,0.0}, draw opacity={0.1}, line width={1.15}, solid}, x axis line style={color={rgb,1:red,0.0;green,0.0;blue,0.0}, draw opacity={1.0}, line width={2.3}, solid}, scaled y ticks={false}, ylabel={$\mathbb{E}[\risk{\widehat{\bw}}] - \sigma^2$}, y tick style={color={rgb,1:red,0.0;green,0.0;blue,0.0}, opacity={1.0}}, y tick label style={color={rgb,1:red,0.0;green,0.0;blue,0.0}, opacity={1.0}, rotate={0}}, ylabel style={at={(ticklabel cs:0.5)}, anchor=near ticklabel, font={{\fontsize{25.299999999999997 pt}{32.89 pt}\selectfont}}, color={rgb,1:red,0.0;green,0.0;blue,0.0}, draw opacity={1.0}, rotate={0.0}}, ymajorgrids={false}, ymin={0}, ymax={55}, ytick={{0.0,10.0,20.0,30.0,40.0,50.0}}, yticklabels={{$0$,$10$,$20$,$30$,$40$,$50$}}, ytick align={inside}, yticklabel style={font={{\fontsize{18.4 pt}{23.919999999999998 pt}\selectfont}}, color={rgb,1:red,0.0;green,0.0;blue,0.0}, draw opacity={1.0}, rotate={0.0}}, y grid style={color={rgb,1:red,0.0;green,0.0;blue,0.0}, draw opacity={0.1}, line width={1.15}, solid}, y axis line style={color={rgb,1:red,0.0;green,0.0;blue,0.0}, draw opacity={1.0}, line width={2.3}, solid}]
    \addplot[color={rgb,1:red,0.2745;green,0.5098;blue,0.7059}, name path={162b4530-c6b4-466b-8364-89af74059dca}, draw opacity={1.0}, line width={2.3}, dotted, mark={triangle*}, mark size={6.8999999999999995 pt}, mark repeat={1}, mark options={color={rgb,1:red,0.0;green,0.0;blue,0.0}, draw opacity={0.6}, fill={rgb,1:red,0.2745;green,0.5098;blue,0.7059}, fill opacity={0.6}, line width={1.7249999999999999}, rotate={0}, solid}, forget plot]
        table[row sep={\\}]
        {
            \\
            2.0  23.952750323698737  \\
            4.0  23.952750323698737  \\
            8.0  23.952750323698737  \\
            16.0  23.952750323698737  \\
            32.0  23.952750323698737  \\
        }
        ;
    \addplot[color={rgb,1:red,0.0;green,0.502;blue,0.0}, name path={12043297-efb8-4b69-80f1-b5903ecd091b}, draw opacity={1.0}, line width={2.3}, dashdotted, mark={triangle*}, mark size={6.8999999999999995 pt}, mark repeat={1}, mark options={color={rgb,1:red,0.0;green,0.0;blue,0.0}, draw opacity={0.6}, fill={rgb,1:red,0.0;green,0.502;blue,0.0}, fill opacity={0.6}, line width={1.7249999999999999}, rotate={180}, solid}, forget plot]
        table[row sep={\\}]
        {
            \\
            2.0  23.169333786971592  \\
            4.0  22.250868405621716  \\
            8.0  21.724522266528567  \\
            16.0  21.599205058210337  \\
            32.0  21.60527559799018  \\
        }
        ;
    \addplot[color={rgb,1:red,1.0;green,0.6471;blue,0.0}, name path={fc15c38f-66db-4edc-bf69-b42cdaf72c54}, draw opacity={1.0}, line width={2.3}, dashdotdotted, mark={diamond*}, mark size={6.8999999999999995 pt}, mark repeat={1}, mark options={color={rgb,1:red,0.0;green,0.0;blue,0.0}, draw opacity={0.6}, fill={rgb,1:red,1.0;green,0.6471;blue,0.0}, fill opacity={0.6}, line width={1.7249999999999999}, rotate={0}, solid}, forget plot]
        table[row sep={\\}]
        {
            \\
            2.0  22.733005757111158  \\
            4.0  22.259085781576033  \\
            8.0  43.018507135346354  \\
            16.0  145.23386407321527  \\
            32.0  263.3424113350597  \\
        }
        ;
    \addplot[color={rgb,1:red,0.502;green,0.0;blue,0.502}, name path={ad293829-dd34-4412-9c15-7aabd13f85b6}, draw opacity={1.0}, line width={2.3}, solid, mark={pentagon*}, mark size={6.8999999999999995 pt}, mark repeat={1}, mark options={color={rgb,1:red,0.0;green,0.0;blue,0.0}, draw opacity={0.6}, fill={rgb,1:red,0.502;green,0.0;blue,0.502}, fill opacity={0.6}, line width={1.7249999999999999}, rotate={0}, solid}, forget plot]
        table[row sep={\\}]
        {
            \\
            2.0  22.725548362072807  \\
            4.0  20.952700258941675  \\
            8.0  19.657783772037455  \\
            16.0  19.32758563786711  \\
            32.0  19.426945255485087  \\
        }
        ;
    \addplot[color={rgb,1:red,0.502;green,0.502;blue,0.502}, name path={f2444df6-7d87-4724-bcf5-319340524e12}, draw opacity={1.0}, line width={2.3}, dashed, mark={*}, mark size={6.8999999999999995 pt}, mark repeat={1}, mark options={color={rgb,1:red,0.0;green,0.0;blue,0.0}, draw opacity={0.6}, fill={rgb,1:red,0.502;green,0.502;blue,0.502}, fill opacity={0.6}, line width={1.7249999999999999}, rotate={0}, solid}, forget plot]
        table[row sep={\\}]
        {
            \\
            2.0  19.065373889708063  \\
            4.0  19.065373889708063  \\
            8.0  19.065373889708063  \\
            16.0  19.065373889708063  \\
            32.0  19.065373889708063  \\
        }
        ;
\end{axis}
\end{tikzpicture}

%% file: tables/cll_analysis.tex
\begin{tabular}{lllllll}
 & Tuning & $\widehat{\lambda}_{\text{Drugs}}$ & $\widehat{\lambda}_{\text{Methyl}}$ & $\widehat{\lambda}_{\text{RNA}}$ & Time (s) & RMSE \\
\hline 
$\sigmacv$\textbf{-Ridge} & $\widehat{\sigmacv}$ = 0.00195 & 1.44e-5 & $\infty$ & $\infty$ & 12.1 & 0.0510 \\
	\textbf{Single Ridge} &  & 0.00082 & 0.00082 & 0.00082 & 5.88 & 0.0785 \\
	\textbf{Multi Ridge} &  & 0.00381 & 0.00082 & 0.00082 & 256.0 & 0.0859 \\
\textbf{Group Lasso} & $\widehat{\lambda}^{gl}$ = 2.39 & 2.33 & $\infty$ & $\infty$ & 50.3 & 0.0514 \\
\end{tabular}

%% file: tables/cll_analysis_noise.tex
\begin{tabular}{lllllllll}
 & Tuning & $\widehat{\lambda}_{\text{Drugs}}$ & $\widehat{\lambda}_{\text{Methyl}}$ & $\widehat{\lambda}_{\text{RNA}}$ & $\widehat{\lambda}_{\text{Noise}_1}$ & $\widehat{\lambda}_{\text{Noise}_2}$ & Time (s) &  RMSE \\
\hline 
$\sigmacv$\textbf{-Ridge} & $\widehat{\sigmacv}$ = 0.00201 & 1.5e-5 & $\infty$ & $\infty$ & $\infty$ & $\infty$ & 13.5 & 0.0510 \\
	\textbf{Single Ridge} &  & 0.00082 & 0.00082 & 0.00082 & 0.00082 & 0.00082 & 6.1 & 0.0799 \\
	\textbf{Multi Ridge} &  & 0.00108 & 0.00125 & 0.00143 & 470.0 & 203.0 & 296.0 & 0.0961 \\
\textbf{Group Lasso} & $\widehat{\lambda}^{gl}$ = 2.81 & 2.87 & $\infty$ & $\infty$ & $\infty$ & $\infty$ & 54.1 & 0.0515 \\
\end{tabular}

%% file: figures/oracle_risks_exponential/oracle_risk1.tikz
\begin{tikzpicture}[/tikz/background rectangle/.style={fill={rgb,1:red,1.0;green,1.0;blue,1.0}, draw opacity={1.0}}, show background rectangle]
\begin{axis}[point meta max={nan}, point meta min={nan}, legend cell align={left}, title={$\gamma_1 = \gamma_2 = \frac{1}{4},\;\; \alpha_1^2 + \alpha_2^2 = 1$}, title style={at={{(0.5,1)}}, anchor={south}, font={{\fontsize{30.800000000000004 pt}{40.040000000000006 pt}\selectfont}}, color={rgb,1:red,0.0;green,0.0;blue,0.0}, draw opacity={1.0}, rotate={0.0}}, legend style={color={rgb,1:red,0.0;green,0.0;blue,0.0}, draw opacity={0.0}, line width={2.2}, solid, fill={rgb,1:red,0.0;green,0.0;blue,0.0}, fill opacity={0.0}, text opacity={1.0}, font={{\fontsize{26.400000000000002 pt}{34.32000000000001 pt}\selectfont}}, text={rgb,1:red,0.0;green,0.0;blue,0.0}, at={(0.02, 0.98)}, anchor={north west}}, axis background/.style={fill={rgb,1:red,1.0;green,1.0;blue,1.0}, opacity={1.0}}, anchor={north west}, xshift={1.0mm}, yshift={-1.0mm}, width={163.1mm}, height={125.0mm}, scaled x ticks={false}, xlabel={$\alpha_1^2/(\alpha_1^2 + \alpha_2^2)$}, x tick style={color={rgb,1:red,0.0;green,0.0;blue,0.0}, opacity={1.0}}, x tick label style={color={rgb,1:red,0.0;green,0.0;blue,0.0}, opacity={1.0}, rotate={0}}, xlabel style={at={(ticklabel cs:0.5)}, anchor=near ticklabel, font={{\fontsize{24.200000000000003 pt}{31.460000000000004 pt}\selectfont}}, color={rgb,1:red,0.0;green,0.0;blue,0.0}, draw opacity={1.0}, rotate={0.0}}, xmajorgrids={false}, xmin={-0.03}, xmax={1.03}, xtick={{0.0,0.25,0.5,0.75,1.0}}, xticklabels={{$0.00$,$0.25$,$0.50$,$0.75$,$1.00$}}, xtick align={inside}, xticklabel style={font={{\fontsize{17.6 pt}{22.880000000000003 pt}\selectfont}}, color={rgb,1:red,0.0;green,0.0;blue,0.0}, draw opacity={1.0}, rotate={0.0}}, x grid style={color={rgb,1:red,0.0;green,0.0;blue,0.0}, draw opacity={0.1}, line width={1.1}, solid}, x axis line style={color={rgb,1:red,0.0;green,0.0;blue,0.0}, draw opacity={1.0}, line width={2.2}, solid}, scaled y ticks={false}, ylabel={$\risk{\blambda}- \sigma^2$}, y tick style={color={rgb,1:red,0.0;green,0.0;blue,0.0}, opacity={1.0}}, y tick label style={color={rgb,1:red,0.0;green,0.0;blue,0.0}, opacity={1.0}, rotate={0}}, ylabel style={at={(ticklabel cs:0.5)}, anchor=near ticklabel, font={{\fontsize{24.200000000000003 pt}{31.460000000000004 pt}\selectfont}}, color={rgb,1:red,0.0;green,0.0;blue,0.0}, draw opacity={1.0}, rotate={0.0}}, ymajorgrids={false}, ymin={0}, ymax={6}, ytick={{0.0,1.0,2.0,3.0,4.0,5.0,6.0}}, yticklabels={{$0$,$1$,$2$,$3$,$4$,$5$,$6$}}, ytick align={inside}, yticklabel style={font={{\fontsize{17.6 pt}{22.880000000000003 pt}\selectfont}}, color={rgb,1:red,0.0;green,0.0;blue,0.0}, draw opacity={1.0}, rotate={0.0}}, y grid style={color={rgb,1:red,0.0;green,0.0;blue,0.0}, draw opacity={0.1}, line width={1.1}, solid}, y axis line style={color={rgb,1:red,0.0;green,0.0;blue,0.0}, draw opacity={1.0}, line width={2.2}, solid}]
    \addplot[color={rgb,1:red,0.502;green,0.502;blue,0.502}, name path={910d0821-fcb2-46e3-81b3-ecfe203cf050}, draw opacity={1.0}, line width={2.2}, dotted]
        table[row sep={\\}]
        {
            \\
            0.0  1.9956474898983485  \\
            0.034482758620689655  1.9869797416532653  \\
            0.06896551724137931  1.9646316465913864  \\
            0.10344827586206896  1.9329663141705087  \\
            0.13793103448275862  1.8945214334600182  \\
            0.1724137931034483  1.8508623660804533  \\
            0.20689655172413793  1.8030310404701515  \\
            0.2413793103448276  1.7517577772407655  \\
            0.27586206896551724  1.69758887287699  \\
            0.3103448275862069  1.6409181812035678  \\
            0.3448275862068966  1.5820775192393302  \\
            0.3793103448275862  1.5212753765396734  \\
            0.41379310344827586  1.4587464556127143  \\
            0.4482758620689655  1.394671246957806  \\
            0.4827586206896552  1.3291518089541805  \\
            0.5172413793103449  1.2623382127864904  \\
            0.5517241379310345  1.1942877201534294  \\
            0.5862068965517241  1.125160428657758  \\
            0.6206896551724138  1.0549596321133432  \\
            0.6551724137931034  0.9837762247562494  \\
            0.6896551724137931  0.9116575329403307  \\
            0.7241379310344828  0.8386856187988105  \\
            0.7586206896551724  0.7648631797745757  \\
            0.7931034482758621  0.690256917493264  \\
            0.8275862068965517  0.6148841117653931  \\
            0.8620689655172413  0.5387828689658813  \\
            0.896551724137931  0.4619735264802338  \\
            0.9310344827586207  0.3844877901142394  \\
            0.9655172413793104  0.30634238386201695  \\
            1.0  0.22756214520030182  \\
        }
        ;
    \addlegendentry {$\;$Optimal $\blambda = (\lambda, \infty)$}
    \addplot[color={rgb,1:red,0.502;green,0.0;blue,0.502}, name path={9ea69add-603e-4d00-ba03-15a2ff5afa3f}, draw opacity={1.0}, line width={2.2}, dashdotted]
        table[row sep={\\}]
        {
            \\
            0.0  0.43022777585493444  \\
            0.034482758620689655  0.43022777585493444  \\
            0.06896551724137931  0.43022777585493444  \\
            0.10344827586206896  0.43022777585493444  \\
            0.13793103448275862  0.43022777585493444  \\
            0.1724137931034483  0.43022777585493444  \\
            0.20689655172413793  0.43022777585493444  \\
            0.2413793103448276  0.43022777585493444  \\
            0.27586206896551724  0.43022777585493444  \\
            0.3103448275862069  0.43022777585493444  \\
            0.3448275862068966  0.43022777585493444  \\
            0.3793103448275862  0.43022777585493444  \\
            0.41379310344827586  0.43022777585493444  \\
            0.4482758620689655  0.43022777585493444  \\
            0.4827586206896552  0.43022777585493444  \\
            0.5172413793103449  0.43022777585493444  \\
            0.5517241379310345  0.43022777585493444  \\
            0.5862068965517241  0.43022777585493444  \\
            0.6206896551724138  0.43022777585493444  \\
            0.6551724137931034  0.43022777585493444  \\
            0.6896551724137931  0.43022777585493444  \\
            0.7241379310344828  0.43022777585493444  \\
            0.7586206896551724  0.43022777585493444  \\
            0.7931034482758621  0.43022777585493444  \\
            0.8275862068965517  0.43022777585493444  \\
            0.8620689655172413  0.43022777585493444  \\
            0.896551724137931  0.43022777585493444  \\
            0.9310344827586207  0.43022777585493444  \\
            0.9655172413793104  0.43022777585493444  \\
            1.0  0.43022777585493444  \\
        }
        ;
    \addlegendentry {$\;$Optimal $\blambda = (\lambda, \lambda)$}
    \addplot[color={rgb,1:red,0.0;green,0.502;blue,0.0}, name path={7ec9fc23-f605-4a39-9057-8691f4208a5b}, draw opacity={1.0}, line width={2.2}, dashed]
        table[row sep={\\}]
        {
            \\
            0.0  0.22756214520030182  \\
            0.034482758620689655  0.28441701051115675  \\
            0.06896551724137931  0.3193977510984123  \\
            0.10344827586206896  0.34443751994257177  \\
            0.13793103448275862  0.36355301368660964  \\
            0.1724137931034483  0.37865591089874995  \\
            0.20689655172413793  0.39081768867629085  \\
            0.2413793103448276  0.4007003621030574  \\
            0.27586206896551724  0.4087397301172777  \\
            0.3103448275862069  0.41523497062158166  \\
            0.3448275862068966  0.42039683785065396  \\
            0.3793103448275862  0.4243754091315961  \\
            0.41379310344827586  0.42727681814048846  \\
            0.4482758620689655  0.4291736047085217  \\
            0.4827586206896552  0.43011109296859296  \\
            0.5172413793103449  0.43011109296859296  \\
            0.5517241379310345  0.4291736047085217  \\
            0.5862068965517241  0.42727681814048846  \\
            0.6206896551724138  0.4243754091315961  \\
            0.6551724137931034  0.42039683785065396  \\
            0.6896551724137931  0.41523497062158166  \\
            0.7241379310344828  0.4087397301172777  \\
            0.7586206896551724  0.4007003621030574  \\
            0.7931034482758621  0.39081768867629085  \\
            0.8275862068965517  0.37865591089874995  \\
            0.8620689655172413  0.36355301368660964  \\
            0.896551724137931  0.34443751994257177  \\
            0.9310344827586207  0.3193977510984123  \\
            0.9655172413793104  0.28441701051115675  \\
            1.0  0.22756214520030182  \\
        }
        ;
    \addlegendentry {$\;$Optimal $\blambda = (\lambda_1, \lambda_2)$}
    \addplot[color={rgb,1:red,0.502;green,0.502;blue,0.502}, name path={e65f564f-fe71-4501-8a01-a9a57a39f237}, only marks, draw opacity={1.0}, line width={0.0}, solid, mark={triangle*}, mark size={6.6000000000000005 pt}, mark repeat={1}, mark options={color={rgb,1:red,0.0;green,0.0;blue,0.0}, draw opacity={0.0}, fill={rgb,1:red,0.502;green,0.502;blue,0.502}, fill opacity={1.0}, line width={1.6500000000000001}, rotate={0}, solid}, forget plot]
        table[row sep={\\}]
        {
            \\
            0.0  2.16071865906206  \\
            0.034482758620689655  2.2151813627433703  \\
            0.06896551724137931  1.8686219450108612  \\
            0.10344827586206896  1.8307512862234807  \\
            0.13793103448275862  2.0302246517053177  \\
            0.1724137931034483  1.4099313107428184  \\
            0.20689655172413793  1.6008827463595514  \\
            0.2413793103448276  1.5426110096762669  \\
            0.27586206896551724  1.7530881269087613  \\
            0.3103448275862069  1.4793475513922698  \\
            0.3448275862068966  1.4887223431031558  \\
            0.3793103448275862  1.6073821150771082  \\
            0.41379310344827586  1.2317994642786654  \\
            0.4482758620689655  1.2181039564892204  \\
            0.4827586206896552  1.543873487002947  \\
            0.5172413793103449  1.179663999434018  \\
            0.5517241379310345  1.225881107929344  \\
            0.5862068965517241  1.1292078996328385  \\
            0.6206896551724138  1.0255651465166578  \\
            0.6551724137931034  1.0367513878193115  \\
            0.6896551724137931  1.0028311556095386  \\
            0.7241379310344828  0.8032769836766107  \\
            0.7586206896551724  0.8156377784506779  \\
            0.7931034482758621  0.7795745753798762  \\
            0.8275862068965517  0.6068620374352909  \\
            0.8620689655172413  0.5492845276968434  \\
            0.896551724137931  0.44380014080334074  \\
            0.9310344827586207  0.45166420089117865  \\
            0.9655172413793104  0.32625751966787075  \\
            1.0  0.26502726035628377  \\
        }
        ;
    \addplot[color={rgb,1:red,0.502;green,0.0;blue,0.502}, name path={271c8a92-12ee-438e-bcca-1b35b5dc22fe}, only marks, draw opacity={1.0}, line width={0.0}, solid, mark={triangle*}, mark size={6.6000000000000005 pt}, mark repeat={1}, mark options={color={rgb,1:red,0.0;green,0.0;blue,0.0}, draw opacity={0.0}, fill={rgb,1:red,0.502;green,0.0;blue,0.502}, fill opacity={1.0}, line width={1.6500000000000001}, rotate={0}, solid}, forget plot]
        table[row sep={\\}]
        {
            \\
            0.0  0.4237311322045467  \\
            0.034482758620689655  0.46940619148643226  \\
            0.06896551724137931  0.4170539058979126  \\
            0.10344827586206896  0.43287587013325823  \\
            0.13793103448275862  0.38422477876975636  \\
            0.1724137931034483  0.39609867717304614  \\
            0.20689655172413793  0.41123540355529387  \\
            0.2413793103448276  0.40814713338800135  \\
            0.27586206896551724  0.4941687894911708  \\
            0.3103448275862069  0.38936062236810076  \\
            0.3448275862068966  0.3729326844875227  \\
            0.3793103448275862  0.41833096329232133  \\
            0.41379310344827586  0.3738042332793783  \\
            0.4482758620689655  0.38269554340848266  \\
            0.4827586206896552  0.5046642771137195  \\
            0.5172413793103449  0.390631918976144  \\
            0.5517241379310345  0.3993961596646862  \\
            0.5862068965517241  0.49644858990970997  \\
            0.6206896551724138  0.4468740702678484  \\
            0.6551724137931034  0.416586757103544  \\
            0.6896551724137931  0.4100410793346041  \\
            0.7241379310344828  0.5106945119890003  \\
            0.7586206896551724  0.42164006726207925  \\
            0.7931034482758621  0.4854139061372451  \\
            0.8275862068965517  0.40121765268762855  \\
            0.8620689655172413  0.40543740523286487  \\
            0.896551724137931  0.41810700514009835  \\
            0.9310344827586207  0.4952649819983803  \\
            0.9655172413793104  0.47445353347931385  \\
            1.0  0.46081589597259565  \\
        }
        ;
    \addplot[color={rgb,1:red,0.0;green,0.502;blue,0.0}, name path={3d28e336-7b63-466b-861d-d28b467d4dbe}, only marks, draw opacity={1.0}, line width={0.0}, solid, mark={triangle*}, mark size={6.6000000000000005 pt}, mark repeat={1}, mark options={color={rgb,1:red,0.0;green,0.0;blue,0.0}, draw opacity={0.0}, fill={rgb,1:red,0.0;green,0.502;blue,0.0}, fill opacity={1.0}, line width={1.6500000000000001}, rotate={0}, solid}, forget plot]
        table[row sep={\\}]
        {
            \\
            0.0  0.21813644568257318  \\
            0.034482758620689655  0.3007782551898879  \\
            0.06896551724137931  0.3032000287961527  \\
            0.10344827586206896  0.3301805737684549  \\
            0.13793103448275862  0.29726039810264737  \\
            0.1724137931034483  0.3454995577141302  \\
            0.20689655172413793  0.3810865248198658  \\
            0.2413793103448276  0.39383506468303575  \\
            0.27586206896551724  0.46269085341673555  \\
            0.3103448275862069  0.3731927235498791  \\
            0.3448275862068966  0.3628039373645662  \\
            0.3793103448275862  0.4134473674969923  \\
            0.41379310344827586  0.3737958469027034  \\
            0.4482758620689655  0.3843390872383976  \\
            0.4827586206896552  0.5052587494984717  \\
            0.5172413793103449  0.3910073035589561  \\
            0.5517241379310345  0.39894488892044033  \\
            0.5862068965517241  0.4910388143406499  \\
            0.6206896551724138  0.4459341428620356  \\
            0.6551724137931034  0.39660569966458215  \\
            0.6896551724137931  0.4109471929920052  \\
            0.7241379310344828  0.4764256050941247  \\
            0.7586206896551724  0.40492193899652595  \\
            0.7931034482758621  0.4380828528246261  \\
            0.8275862068965517  0.35758870430247436  \\
            0.8620689655172413  0.3431818670970608  \\
            0.896551724137931  0.3306996970122007  \\
            0.9310344827586207  0.3849971150851972  \\
            0.9655172413793104  0.3028574687027161  \\
            1.0  0.26502726035628377  \\
        }
        ;
\end{axis}
\end{tikzpicture}

%% file: figures/oracle_risks_exponential/oracle_risk2.tikz
\begin{tikzpicture}[/tikz/background rectangle/.style={fill={rgb,1:red,1.0;green,1.0;blue,1.0}, draw opacity={1.0}}, show background rectangle]
\begin{axis}[point meta max={nan}, point meta min={nan}, legend cell align={left}, title={$\gamma_1 = \frac{1}{10},\; \gamma_2 = \frac{4}{10},\;\; \alpha_1^2 + \alpha_2^2 = 1$}, title style={at={{(0.5,1)}}, anchor={south}, font={{\fontsize{30.800000000000004 pt}{40.040000000000006 pt}\selectfont}}, color={rgb,1:red,0.0;green,0.0;blue,0.0}, draw opacity={1.0}, rotate={0.0}}, legend style={color={rgb,1:red,0.0;green,0.0;blue,0.0}, draw opacity={0.0}, line width={2.2}, solid, fill={rgb,1:red,0.0;green,0.0;blue,0.0}, fill opacity={0.0}, text opacity={1.0}, font={{\fontsize{26.400000000000002 pt}{34.32000000000001 pt}\selectfont}}, text={rgb,1:red,0.0;green,0.0;blue,0.0}, at={(1.02, 1)}, anchor={north west}}, axis background/.style={fill={rgb,1:red,1.0;green,1.0;blue,1.0}, opacity={1.0}}, anchor={north west}, xshift={1.0mm}, yshift={-1.0mm}, width={163.1mm}, height={125.0mm}, scaled x ticks={false}, xlabel={$\alpha_1^2/(\alpha_1^2 + \alpha_2^2)$}, x tick style={color={rgb,1:red,0.0;green,0.0;blue,0.0}, opacity={1.0}}, x tick label style={color={rgb,1:red,0.0;green,0.0;blue,0.0}, opacity={1.0}, rotate={0}}, xlabel style={at={(ticklabel cs:0.5)}, anchor=near ticklabel, font={{\fontsize{24.200000000000003 pt}{31.460000000000004 pt}\selectfont}}, color={rgb,1:red,0.0;green,0.0;blue,0.0}, draw opacity={1.0}, rotate={0.0}}, xmajorgrids={false}, xmin={-0.03}, xmax={1.03}, xtick={{0.0,0.25,0.5,0.75,1.0}}, xticklabels={{$0.00$,$0.25$,$0.50$,$0.75$,$1.00$}}, xtick align={inside}, xticklabel style={font={{\fontsize{17.6 pt}{22.880000000000003 pt}\selectfont}}, color={rgb,1:red,0.0;green,0.0;blue,0.0}, draw opacity={1.0}, rotate={0.0}}, x grid style={color={rgb,1:red,0.0;green,0.0;blue,0.0}, draw opacity={0.1}, line width={1.1}, solid}, x axis line style={color={rgb,1:red,0.0;green,0.0;blue,0.0}, draw opacity={1.0}, line width={2.2}, solid}, scaled y ticks={false}, ylabel={$\risk{\blambda}- \sigma^2$}, y tick style={color={rgb,1:red,0.0;green,0.0;blue,0.0}, opacity={1.0}}, y tick label style={color={rgb,1:red,0.0;green,0.0;blue,0.0}, opacity={1.0}, rotate={0}}, ylabel style={at={(ticklabel cs:0.5)}, anchor=near ticklabel, font={{\fontsize{24.200000000000003 pt}{31.460000000000004 pt}\selectfont}}, color={rgb,1:red,0.0;green,0.0;blue,0.0}, draw opacity={1.0}, rotate={0.0}}, ymajorgrids={false}, ymin={0}, ymax={6}, ytick={{0.0,1.0,2.0,3.0,4.0,5.0,6.0}}, yticklabels={{$0$,$1$,$2$,$3$,$4$,$5$,$6$}}, ytick align={inside}, yticklabel style={font={{\fontsize{17.6 pt}{22.880000000000003 pt}\selectfont}}, color={rgb,1:red,0.0;green,0.0;blue,0.0}, draw opacity={1.0}, rotate={0.0}}, y grid style={color={rgb,1:red,0.0;green,0.0;blue,0.0}, draw opacity={0.1}, line width={1.1}, solid}, y axis line style={color={rgb,1:red,0.0;green,0.0;blue,0.0}, draw opacity={1.0}, line width={2.2}, solid}]
    \addplot[color={rgb,1:red,0.502;green,0.502;blue,0.502}, name path={17780092-a4cf-4356-b821-8ab07b2cf5fd}, draw opacity={1.0}, line width={2.2}, dotted]
        table[row sep={\\}]
        {
            \\
            0.0  1.9966806612966073  \\
            0.034482758620689655  1.978536573006398  \\
            0.06896551724137931  1.9400129324259243  \\
            0.10344827586206896  1.8917237507578446  \\
            0.13793103448275862  1.8376302182400535  \\
            0.1724137931034483  1.7796620038118975  \\
            0.20689655172413793  1.7189135854305757  \\
            0.2413793103448276  1.6560656702945167  \\
            0.27586206896551724  1.5915743234521473  \\
            0.3103448275862069  1.5257639068661302  \\
            0.3448275862068966  1.4588560196075506  \\
            0.3793103448275862  1.391038624790922  \\
            0.41379310344827586  1.3224444879494532  \\
            0.4482758620689655  1.2531785149115606  \\
            0.4827586206896552  1.1833236159645404  \\
            0.5172413793103449  1.1129551898466863  \\
            0.5517241379310345  1.0421238822125183  \\
            0.5862068965517241  0.9708885506433966  \\
            0.6206896551724138  0.8992793967727697  \\
            0.6551724137931034  0.8273356221687729  \\
            0.6896551724137931  0.7550857133144042  \\
            0.7241379310344828  0.6825539985323756  \\
            0.7586206896551724  0.6097656311947763  \\
            0.7931034482758621  0.5367374497497368  \\
            0.8275862068965517  0.4634881673350282  \\
            0.8620689655172413  0.3900306490405605  \\
            0.896551724137931  0.3163797984842973  \\
            0.9310344827586207  0.2425472073592141  \\
            0.9655172413793104  0.16854319005965746  \\
            1.0  0.09437720536203775  \\
        }
        ;
    \addplot[color={rgb,1:red,0.502;green,0.0;blue,0.502}, name path={4d18bda9-2459-470b-8e7b-cff55b9d743d}, draw opacity={1.0}, line width={2.2}, dashdotted]
        table[row sep={\\}]
        {
            \\
            0.0  0.43022284982851455  \\
            0.034482758620689655  0.43022376086899117  \\
            0.06896551724137931  0.43022467190975067  \\
            0.10344827586206896  0.43022558295028634  \\
            0.13793103448275862  0.4302264939909868  \\
            0.1724137931034483  0.4302274050315813  \\
            0.20689655172413793  0.4302283160722231  \\
            0.2413793103448276  0.4302292271128412  \\
            0.27586206896551724  0.43023013815345923  \\
            0.3103448275862069  0.4302310491940773  \\
            0.3448275862068966  0.43023196023469534  \\
            0.3793103448275862  0.4302328712753134  \\
            0.41379310344827586  0.43023378231593146  \\
            0.4482758620689655  0.4302346933565495  \\
            0.4827586206896552  0.43023560439716757  \\
            0.5172413793103449  0.4302365154377856  \\
            0.5517241379310345  0.4302374264784037  \\
            0.5862068965517241  0.43023833751902174  \\
            0.6206896551724138  0.4302392485596398  \\
            0.6551724137931034  0.43024015960025785  \\
            0.6896551724137931  0.4302410706408759  \\
            0.7241379310344828  0.43024198168149397  \\
            0.7586206896551724  0.430242892722112  \\
            0.7931034482758621  0.4302438037627303  \\
            0.8275862068965517  0.4302447148038846  \\
            0.8620689655172413  0.4302456258439664  \\
            0.896551724137931  0.43024653688517955  \\
            0.9310344827586207  0.43024744792520253  \\
            0.9655172413793104  0.43024835896647473  \\
            1.0  0.43024927000643864  \\
        }
        ;
    \addplot[color={rgb,1:red,0.0;green,0.502;blue,0.0}, name path={77cd3832-4b56-4acb-ba49-081f5579841e}, draw opacity={1.0}, line width={2.2}, dashed]
        table[row sep={\\}]
        {
            \\
            0.0  0.35233003914808503  \\
            0.034482758620689655  0.39554740759999385  \\
            0.06896551724137931  0.4131325966439199  \\
            0.10344827586206896  0.4224049937021832  \\
            0.13793103448275862  0.4273963363672286  \\
            0.1724137931034483  0.42972547046470866  \\
            0.20689655172413793  0.43019964165396773  \\
            0.2413793103448276  0.4292727682624564  \\
            0.27586206896551724  0.42721917434875234  \\
            0.3103448275862069  0.4242114331685545  \\
            0.3448275862068966  0.42035934918217555  \\
            0.3793103448275862  0.4157310259892122  \\
            0.41379310344827586  0.41036481619346277  \\
            0.4482758620689655  0.4042762188685214  \\
            0.4827586206896552  0.39746173146104624  \\
            0.5172413793103449  0.3899006761687662  \\
            0.5517241379310345  0.3815554970251285  \\
            0.5862068965517241  0.3723707037697679  \\
            0.6206896551724138  0.3622703947984869  \\
            0.6551724137931034  0.35115404333271694  \\
            0.6896551724137931  0.33888990011724696  \\
            0.7241379310344828  0.3253048364531508  \\
            0.7586206896551724  0.3101685021240539  \\
            0.7931034482758621  0.293167821843614  \\
            0.8275862068965517  0.2738639399106355  \\
            0.8620689655172413  0.2516146349926687  \\
            0.896551724137931  0.22542146238020844  \\
            0.9310344827586207  0.19358776454789361  \\
            0.9655172413793104  0.1527864472781697  \\
            1.0  0.09437720536203775  \\
        }
        ;
    \addplot[color={rgb,1:red,0.502;green,0.502;blue,0.502}, name path={676d28ea-5116-4c38-9c2d-c32785b57894}, only marks, draw opacity={1.0}, line width={0.0}, solid, mark={triangle*}, mark size={6.6000000000000005 pt}, mark repeat={1}, mark options={color={rgb,1:red,0.0;green,0.0;blue,0.0}, draw opacity={0.0}, fill={rgb,1:red,0.502;green,0.502;blue,0.502}, fill opacity={1.0}, line width={1.6500000000000001}, rotate={0}, solid}]
        table[row sep={\\}]
        {
            \\
            0.0  1.7594263357917361  \\
            0.034482758620689655  2.173170141975562  \\
            0.06896551724137931  1.917437028735304  \\
            0.10344827586206896  2.062874932439652  \\
            0.13793103448275862  1.7634917605013944  \\
            0.1724137931034483  1.9804991639144194  \\
            0.20689655172413793  1.7913588745155806  \\
            0.2413793103448276  1.6171299342557175  \\
            0.27586206896551724  1.7135446281850562  \\
            0.3103448275862069  1.435437644961778  \\
            0.3448275862068966  1.4587327450970986  \\
            0.3793103448275862  1.3988947778487497  \\
            0.41379310344827586  1.3528356439361522  \\
            0.4482758620689655  1.166221736320637  \\
            0.4827586206896552  1.2689459665371818  \\
            0.5172413793103449  1.1302207352280633  \\
            0.5517241379310345  1.1997130414021928  \\
            0.5862068965517241  0.9211123037576738  \\
            0.6206896551724138  0.9921251724623492  \\
            0.6551724137931034  0.767000175284136  \\
            0.6896551724137931  0.6448231788158063  \\
            0.7241379310344828  0.6710788700141566  \\
            0.7586206896551724  0.5892226852264288  \\
            0.7931034482758621  0.5680287516325742  \\
            0.8275862068965517  0.44809263869551486  \\
            0.8620689655172413  0.3442297377844463  \\
            0.896551724137931  0.3312236073944297  \\
            0.9310344827586207  0.23139236527695428  \\
            0.9655172413793104  0.18942901249538902  \\
            1.0  0.057067722785991926  \\
        }
        ;
    \addplot[color={rgb,1:red,0.502;green,0.0;blue,0.502}, name path={5cc7866c-9b9d-44fd-b211-6dba8a46eed4}, only marks, draw opacity={1.0}, line width={0.0}, solid, mark={triangle*}, mark size={6.6000000000000005 pt}, mark repeat={1}, mark options={color={rgb,1:red,0.0;green,0.0;blue,0.0}, draw opacity={0.0}, fill={rgb,1:red,0.502;green,0.0;blue,0.502}, fill opacity={1.0}, line width={1.6500000000000001}, rotate={0}, solid}]
        table[row sep={\\}]
        {
            \\
            0.0  0.4119496333870696  \\
            0.034482758620689655  0.35771512941524186  \\
            0.06896551724137931  0.4902491512494187  \\
            0.10344827586206896  0.47460020058017194  \\
            0.13793103448275862  0.4330687866630756  \\
            0.1724137931034483  0.41363227639121214  \\
            0.20689655172413793  0.4813476482021015  \\
            0.2413793103448276  0.4192632266124159  \\
            0.27586206896551724  0.3510916148683081  \\
            0.3103448275862069  0.3609602004444554  \\
            0.3448275862068966  0.39174834191785934  \\
            0.3793103448275862  0.46856961633129646  \\
            0.41379310344827586  0.3605069298168735  \\
            0.4482758620689655  0.40861115976960094  \\
            0.4827586206896552  0.42037247533359356  \\
            0.5172413793103449  0.4619380385434937  \\
            0.5517241379310345  0.42168685999126865  \\
            0.5862068965517241  0.38131098152488563  \\
            0.6206896551724138  0.45566810938499214  \\
            0.6551724137931034  0.3934436938758943  \\
            0.6896551724137931  0.3840456261388083  \\
            0.7241379310344828  0.4442108121362134  \\
            0.7586206896551724  0.3984533185065502  \\
            0.7931034482758621  0.3970581450767956  \\
            0.8275862068965517  0.4106584313405939  \\
            0.8620689655172413  0.41542878104375514  \\
            0.896551724137931  0.46925740837065866  \\
            0.9310344827586207  0.410768273458914  \\
            0.9655172413793104  0.395359754264891  \\
            1.0  0.43611503288393827  \\
        }
        ;
    \addplot[color={rgb,1:red,0.0;green,0.502;blue,0.0}, name path={242c41d9-7d8d-41d7-ae89-6eb49182a943}, only marks, draw opacity={1.0}, line width={0.0}, solid, mark={triangle*}, mark size={6.6000000000000005 pt}, mark repeat={1}, mark options={color={rgb,1:red,0.0;green,0.0;blue,0.0}, draw opacity={0.0}, fill={rgb,1:red,0.0;green,0.502;blue,0.0}, fill opacity={1.0}, line width={1.6500000000000001}, rotate={0}, solid}]
        table[row sep={\\}]
        {
            \\
            0.0  0.33323773800218626  \\
            0.034482758620689655  0.322534508931994  \\
            0.06896551724137931  0.4762703203911174  \\
            0.10344827586206896  0.47865072364863415  \\
            0.13793103448275862  0.42940399738942925  \\
            0.1724137931034483  0.4141447323442875  \\
            0.20689655172413793  0.48163046358458494  \\
            0.2413793103448276  0.4172201006475196  \\
            0.27586206896551724  0.3529069820225399  \\
            0.3103448275862069  0.3513043832604854  \\
            0.3448275862068966  0.3895729667444885  \\
            0.3793103448275862  0.4233460910376503  \\
            0.41379310344827586  0.36854876447420626  \\
            0.4482758620689655  0.38501853204782566  \\
            0.4827586206896552  0.4171604478274149  \\
            0.5172413793103449  0.4322076116183551  \\
            0.5517241379310345  0.366600234045402  \\
            0.5862068965517241  0.3226558667316386  \\
            0.6206896551724138  0.40632223164676606  \\
            0.6551724137931034  0.30480592431599485  \\
            0.6896551724137931  0.30651220371727894  \\
            0.7241379310344828  0.3363566259082884  \\
            0.7586206896551724  0.2861689322019185  \\
            0.7931034482758621  0.25712778508462475  \\
            0.8275862068965517  0.23958907979761368  \\
            0.8620689655172413  0.24142479786423432  \\
            0.896551724137931  0.28469756923966205  \\
            0.9310344827586207  0.1735611496375511  \\
            0.9655172413793104  0.16379206167111815  \\
            1.0  0.057067722785991926  \\
        }
        ;
\end{axis}
\end{tikzpicture}

%% file: figures/oracle_risks_exponential/oracle_risk3.tikz
\begin{tikzpicture}[/tikz/background rectangle/.style={fill={rgb,1:red,1.0;green,1.0;blue,1.0}, draw opacity={1.0}}, show background rectangle]
\begin{axis}[point meta max={nan}, point meta min={nan}, legend cell align={left}, title={$\gamma_1 = \gamma_2 = 1,\;\; \alpha_1^2 + \alpha_2^2 = 1$}, title style={at={{(0.5,1)}}, anchor={south}, font={{\fontsize{30.800000000000004 pt}{40.040000000000006 pt}\selectfont}}, color={rgb,1:red,0.0;green,0.0;blue,0.0}, draw opacity={1.0}, rotate={0.0}}, legend style={color={rgb,1:red,0.0;green,0.0;blue,0.0}, draw opacity={0.0}, line width={2.2}, solid, fill={rgb,1:red,0.0;green,0.0;blue,0.0}, fill opacity={0.0}, text opacity={1.0}, font={{\fontsize{26.400000000000002 pt}{34.32000000000001 pt}\selectfont}}, text={rgb,1:red,0.0;green,0.0;blue,0.0}, at={(1.02, 1)}, anchor={north west}}, axis background/.style={fill={rgb,1:red,1.0;green,1.0;blue,1.0}, opacity={1.0}}, anchor={north west}, xshift={1.0mm}, yshift={-1.0mm}, width={163.1mm}, height={125.0mm}, scaled x ticks={false}, xlabel={$\alpha_1^2/(\alpha_1^2 + \alpha_2^2)$}, x tick style={color={rgb,1:red,0.0;green,0.0;blue,0.0}, opacity={1.0}}, x tick label style={color={rgb,1:red,0.0;green,0.0;blue,0.0}, opacity={1.0}, rotate={0}}, xlabel style={at={(ticklabel cs:0.5)}, anchor=near ticklabel, font={{\fontsize{24.200000000000003 pt}{31.460000000000004 pt}\selectfont}}, color={rgb,1:red,0.0;green,0.0;blue,0.0}, draw opacity={1.0}, rotate={0.0}}, xmajorgrids={false}, xmin={-0.03}, xmax={1.03}, xtick={{0.0,0.25,0.5,0.75,1.0}}, xticklabels={{$0.00$,$0.25$,$0.50$,$0.75$,$1.00$}}, xtick align={inside}, xticklabel style={font={{\fontsize{17.6 pt}{22.880000000000003 pt}\selectfont}}, color={rgb,1:red,0.0;green,0.0;blue,0.0}, draw opacity={1.0}, rotate={0.0}}, x grid style={color={rgb,1:red,0.0;green,0.0;blue,0.0}, draw opacity={0.1}, line width={1.1}, solid}, x axis line style={color={rgb,1:red,0.0;green,0.0;blue,0.0}, draw opacity={1.0}, line width={2.2}, solid}, scaled y ticks={false}, ylabel={$\risk{\blambda}- \sigma^2$}, y tick style={color={rgb,1:red,0.0;green,0.0;blue,0.0}, opacity={1.0}}, y tick label style={color={rgb,1:red,0.0;green,0.0;blue,0.0}, opacity={1.0}, rotate={0}}, ylabel style={at={(ticklabel cs:0.5)}, anchor=near ticklabel, font={{\fontsize{24.200000000000003 pt}{31.460000000000004 pt}\selectfont}}, color={rgb,1:red,0.0;green,0.0;blue,0.0}, draw opacity={1.0}, rotate={0.0}}, ymajorgrids={false}, ymin={0}, ymax={6}, ytick={{0.0,1.0,2.0,3.0,4.0,5.0,6.0}}, yticklabels={{$0$,$1$,$2$,$3$,$4$,$5$,$6$}}, ytick align={inside}, yticklabel style={font={{\fontsize{17.6 pt}{22.880000000000003 pt}\selectfont}}, color={rgb,1:red,0.0;green,0.0;blue,0.0}, draw opacity={1.0}, rotate={0.0}}, y grid style={color={rgb,1:red,0.0;green,0.0;blue,0.0}, draw opacity={0.1}, line width={1.1}, solid}, y axis line style={color={rgb,1:red,0.0;green,0.0;blue,0.0}, draw opacity={1.0}, line width={2.2}, solid}]
    \addplot[color={rgb,1:red,0.502;green,0.502;blue,0.502}, name path={63f7f38e-4e72-4c48-8efa-e9c841801eb3}, draw opacity={1.0}, line width={2.2}, dotted]
        table[row sep={\\}]
        {
            \\
            0.0  1.9977145038247501  \\
            0.034482758620689655  1.9954096852296068  \\
            0.06896551724137931  1.9883711356238862  \\
            0.10344827586206896  1.9769324211945025  \\
            0.13793103448275862  1.9614890168762162  \\
            0.1724137931034483  1.942368550512076  \\
            0.20689655172413793  1.9198743762673405  \\
            0.2413793103448276  1.8942889750119023  \\
            0.27586206896551724  1.865771055152491  \\
            0.3103448275862069  1.8345205490122591  \\
            0.3448275862068966  1.8007015226878664  \\
            0.3793103448275862  1.7644178354401001  \\
            0.41379310344827586  1.7257749394368354  \\
            0.4482758620689655  1.6848839837793674  \\
            0.4827586206896552  1.6417852318211241  \\
            0.5172413793103449  1.5965386717676964  \\
            0.5517241379310345  1.5492187885361899  \\
            0.5862068965517241  1.4998628669085639  \\
            0.6206896551724138  1.4485027321676278  \\
            0.6551724137931034  1.3951599838032682  \\
            0.6896551724137931  1.3398372248795578  \\
            0.7241379310344828  1.282553025184023  \\
            0.7586206896551724  1.2232951853350595  \\
            0.7931034482758621  1.1620830765766295  \\
            0.8275862068965517  1.0988900056175916  \\
            0.8620689655172413  1.0336702864330642  \\
            0.896551724137931  0.9664492500326851  \\
            0.9310344827586207  0.8971548469955453  \\
            0.9655172413793104  0.8257393590020956  \\
            1.0  0.752172427366026  \\
        }
        ;
    \addplot[color={rgb,1:red,0.502;green,0.0;blue,0.502}, name path={ac7e7a1c-6b1c-4569-92cc-60d968444248}, draw opacity={1.0}, line width={2.2}, dashdotted]
        table[row sep={\\}]
        {
            \\
            0.0  1.1385760270139569  \\
            0.034482758620689655  1.1385760270139569  \\
            0.06896551724137931  1.1385760270139569  \\
            0.10344827586206896  1.1385760270139569  \\
            0.13793103448275862  1.1385760270139569  \\
            0.1724137931034483  1.1385760270139569  \\
            0.20689655172413793  1.1385760270139569  \\
            0.2413793103448276  1.1385760270139569  \\
            0.27586206896551724  1.1385760270139569  \\
            0.3103448275862069  1.1385760270139569  \\
            0.3448275862068966  1.1385760270139569  \\
            0.3793103448275862  1.1385760270139569  \\
            0.41379310344827586  1.1385760270139569  \\
            0.4482758620689655  1.1385760270139569  \\
            0.4827586206896552  1.1385760270139569  \\
            0.5172413793103449  1.1385760270139569  \\
            0.5517241379310345  1.1385760270139569  \\
            0.5862068965517241  1.1385760270139569  \\
            0.6206896551724138  1.1385760270139569  \\
            0.6551724137931034  1.1385760270139569  \\
            0.6896551724137931  1.1385760270139569  \\
            0.7241379310344828  1.1385760270139569  \\
            0.7586206896551724  1.1385760270139569  \\
            0.7931034482758621  1.1385760270139569  \\
            0.8275862068965517  1.1385760270139569  \\
            0.8620689655172413  1.1385760270139569  \\
            0.896551724137931  1.1385760270139569  \\
            0.9310344827586207  1.1385760270139569  \\
            0.9655172413793104  1.1385760270139569  \\
            1.0  1.1385760270139569  \\
        }
        ;
    \addplot[color={rgb,1:red,0.0;green,0.502;blue,0.0}, name path={8ad58bf4-e8b2-4f9f-a671-32a14f6c027c}, draw opacity={1.0}, line width={2.2}, dashed]
        table[row sep={\\}]
        {
            \\
            0.0  0.752172427366026  \\
            0.034482758620689655  0.8196250310149906  \\
            0.06896551724137931  0.8755942898942788  \\
            0.10344827586206896  0.9230095897182382  \\
            0.13793103448275862  0.963614399109102  \\
            0.1724137931034483  0.9985452660601299  \\
            0.20689655172413793  1.0285883674507534  \\
            0.2413793103448276  1.0543098644132223  \\
            0.27586206896551724  1.076128661445039  \\
            0.3103448275862069  1.0943598732613076  \\
            0.3448275862068966  1.109242122107088  \\
            0.3793103448275862  1.1209552999806358  \\
            0.41379310344827586  1.129632378020614  \\
            0.4482758620689655  1.1353672939586343  \\
            0.4827586206896552  1.138220105177406  \\
            0.5172413793103449  1.138220105177406  \\
            0.5517241379310345  1.1353672939586343  \\
            0.5862068965517241  1.129632378020614  \\
            0.6206896551724138  1.1209552999806358  \\
            0.6551724137931034  1.109242122107088  \\
            0.6896551724137931  1.0943598732613076  \\
            0.7241379310344828  1.076128661445039  \\
            0.7586206896551724  1.0543098644132223  \\
            0.7931034482758621  1.0285883674507534  \\
            0.8275862068965517  0.9985452660601299  \\
            0.8620689655172413  0.963614399109102  \\
            0.896551724137931  0.9230095897182382  \\
            0.9310344827586207  0.8755942898942788  \\
            0.9655172413793104  0.8196250310149906  \\
            1.0  0.752172427366026  \\
        }
        ;
    \addplot[color={rgb,1:red,0.502;green,0.502;blue,0.502}, name path={84564398-89f7-4967-806a-1accd8b759e0}, only marks, draw opacity={1.0}, line width={0.0}, solid, mark={triangle*}, mark size={6.6000000000000005 pt}, mark repeat={1}, mark options={color={rgb,1:red,0.0;green,0.0;blue,0.0}, draw opacity={0.0}, fill={rgb,1:red,0.502;green,0.502;blue,0.502}, fill opacity={1.0}, line width={1.6500000000000001}, rotate={0}, solid}]
        table[row sep={\\}]
        {
            \\
            0.0  1.9708983766041137  \\
            0.034482758620689655  2.085190573914419  \\
            0.06896551724137931  1.9466360256611739  \\
            0.10344827586206896  1.917446875902015  \\
            0.13793103448275862  2.126375005726164  \\
            0.1724137931034483  2.1039621942381004  \\
            0.20689655172413793  1.893256004086548  \\
            0.2413793103448276  1.8375408205864674  \\
            0.27586206896551724  1.7662742654073469  \\
            0.3103448275862069  1.943302451366125  \\
            0.3448275862068966  1.8467203379860928  \\
            0.3793103448275862  1.7385818072043748  \\
            0.41379310344827586  1.8511687417705267  \\
            0.4482758620689655  1.725640069242599  \\
            0.4827586206896552  1.8238817972674064  \\
            0.5172413793103449  1.7013482914992637  \\
            0.5517241379310345  1.482341616821624  \\
            0.5862068965517241  1.4910525355874151  \\
            0.6206896551724138  1.6190489033711257  \\
            0.6551724137931034  1.3312441934367758  \\
            0.6896551724137931  1.3695861538691414  \\
            0.7241379310344828  1.3161041372447477  \\
            0.7586206896551724  1.195725171985957  \\
            0.7931034482758621  1.3007857621196832  \\
            0.8275862068965517  0.9613775452109956  \\
            0.8620689655172413  1.0917800752085478  \\
            0.896551724137931  1.0413637769115032  \\
            0.9310344827586207  0.9916493133401219  \\
            0.9655172413793104  0.7970593921890918  \\
            1.0  0.814212803352323  \\
        }
        ;
    \addplot[color={rgb,1:red,0.502;green,0.0;blue,0.502}, name path={4824fc4e-58b3-410b-a0fe-b60ca1ea0c0f}, only marks, draw opacity={1.0}, line width={0.0}, solid, mark={triangle*}, mark size={6.6000000000000005 pt}, mark repeat={1}, mark options={color={rgb,1:red,0.0;green,0.0;blue,0.0}, draw opacity={0.0}, fill={rgb,1:red,0.502;green,0.0;blue,0.502}, fill opacity={1.0}, line width={1.6500000000000001}, rotate={0}, solid}]
        table[row sep={\\}]
        {
            \\
            0.0  1.0917808310788968  \\
            0.034482758620689655  1.2190819272738849  \\
            0.06896551724137931  1.1188335133525733  \\
            0.10344827586206896  1.1563164063191143  \\
            0.13793103448275862  1.1228817317763138  \\
            0.1724137931034483  1.1343049516872838  \\
            0.20689655172413793  1.1525740355257734  \\
            0.2413793103448276  1.110287167791793  \\
            0.27586206896551724  1.092575297583866  \\
            0.3103448275862069  1.1547441084457692  \\
            0.3448275862068966  1.1539049964210681  \\
            0.3793103448275862  1.1302045122405233  \\
            0.41379310344827586  1.1278335266109125  \\
            0.4482758620689655  1.1830833195617005  \\
            0.4827586206896552  1.2108711620463168  \\
            0.5172413793103449  1.139028663201342  \\
            0.5517241379310345  1.0852674892597585  \\
            0.5862068965517241  1.1700800130288975  \\
            0.6206896551724138  1.1903914489976581  \\
            0.6551724137931034  1.105780882393224  \\
            0.6896551724137931  1.1609332914176864  \\
            0.7241379310344828  1.1514906914624974  \\
            0.7586206896551724  1.1316239633615472  \\
            0.7931034482758621  1.2329055218547467  \\
            0.8275862068965517  1.0245695855180927  \\
            0.8620689655172413  1.2446064215913748  \\
            0.896551724137931  1.201600482917053  \\
            0.9310344827586207  1.220148986869849  \\
            0.9655172413793104  1.102470349598991  \\
            1.0  1.1864746051146438  \\
        }
        ;
    \addplot[color={rgb,1:red,0.0;green,0.502;blue,0.0}, name path={7833f99a-04b7-44ed-a255-d01e1223263f}, only marks, draw opacity={1.0}, line width={0.0}, solid, mark={triangle*}, mark size={6.6000000000000005 pt}, mark repeat={1}, mark options={color={rgb,1:red,0.0;green,0.0;blue,0.0}, draw opacity={0.0}, fill={rgb,1:red,0.0;green,0.502;blue,0.0}, fill opacity={1.0}, line width={1.6500000000000001}, rotate={0}, solid}]
        table[row sep={\\}]
        {
            \\
            0.0  0.7531482626190786  \\
            0.034482758620689655  0.8933695180707786  \\
            0.06896551724137931  0.8735898664004751  \\
            0.10344827586206896  0.9612272960712165  \\
            0.13793103448275862  0.9776306156857939  \\
            0.1724137931034483  0.9438748837762105  \\
            0.20689655172413793  1.0395597189567907  \\
            0.2413793103448276  1.0107100449288717  \\
            0.27586206896551724  1.049209567625656  \\
            0.3103448275862069  1.0947783836575917  \\
            0.3448275862068966  1.1232738527053012  \\
            0.3793103448275862  1.121712703159663  \\
            0.41379310344827586  1.1243203755771174  \\
            0.4482758620689655  1.1790275917002107  \\
            0.4827586206896552  1.2090023143027935  \\
            0.5172413793103449  1.1402829602384381  \\
            0.5517241379310345  1.0804762902750915  \\
            0.5862068965517241  1.1543184284178962  \\
            0.6206896551724138  1.1807887923125944  \\
            0.6551724137931034  1.0790453939789133  \\
            0.6896551724137931  1.115540794396153  \\
            0.7241379310344828  1.0966387625907008  \\
            0.7586206896551724  1.0236190024085037  \\
            0.7931034482758621  1.1370315272723732  \\
            0.8275862068965517  0.8815983337764788  \\
            0.8620689655172413  1.0359182504126814  \\
            0.896551724137931  0.9952321979880976  \\
            0.9310344827586207  0.9637617543842281  \\
            0.9655172413793104  0.7910085877195698  \\
            1.0  0.814212803352323  \\
        }
        ;
\end{axis}
\end{tikzpicture}

%% file: figures/oracle_risks_exponential/oracle_risk4.tikz
\begin{tikzpicture}[/tikz/background rectangle/.style={fill={rgb,1:red,1.0;green,1.0;blue,1.0}, draw opacity={1.0}}, show background rectangle]
\begin{axis}[point meta max={nan}, point meta min={nan}, legend cell align={left}, title={$\gamma_1 = \gamma_2 = \frac{1}{4},\;\; \alpha_1^2 + \alpha_2^2 = 2$}, title style={at={{(0.5,1)}}, anchor={south}, font={{\fontsize{30.800000000000004 pt}{40.040000000000006 pt}\selectfont}}, color={rgb,1:red,0.0;green,0.0;blue,0.0}, draw opacity={1.0}, rotate={0.0}}, legend style={color={rgb,1:red,0.0;green,0.0;blue,0.0}, draw opacity={0.0}, line width={2.2}, solid, fill={rgb,1:red,0.0;green,0.0;blue,0.0}, fill opacity={0.0}, text opacity={1.0}, font={{\fontsize{26.400000000000002 pt}{34.32000000000001 pt}\selectfont}}, text={rgb,1:red,0.0;green,0.0;blue,0.0}, at={(1.02, 1)}, anchor={north west}}, axis background/.style={fill={rgb,1:red,1.0;green,1.0;blue,1.0}, opacity={1.0}}, anchor={north west}, xshift={1.0mm}, yshift={-1.0mm}, width={163.1mm}, height={125.0mm}, scaled x ticks={false}, xlabel={$\alpha_1^2/(\alpha_1^2 + \alpha_2^2)$}, x tick style={color={rgb,1:red,0.0;green,0.0;blue,0.0}, opacity={1.0}}, x tick label style={color={rgb,1:red,0.0;green,0.0;blue,0.0}, opacity={1.0}, rotate={0}}, xlabel style={at={(ticklabel cs:0.5)}, anchor=near ticklabel, font={{\fontsize{24.200000000000003 pt}{31.460000000000004 pt}\selectfont}}, color={rgb,1:red,0.0;green,0.0;blue,0.0}, draw opacity={1.0}, rotate={0.0}}, xmajorgrids={false}, xmin={-0.03}, xmax={1.03}, xtick={{0.0,0.25,0.5,0.75,1.0}}, xticklabels={{$0.00$,$0.25$,$0.50$,$0.75$,$1.00$}}, xtick align={inside}, xticklabel style={font={{\fontsize{17.6 pt}{22.880000000000003 pt}\selectfont}}, color={rgb,1:red,0.0;green,0.0;blue,0.0}, draw opacity={1.0}, rotate={0.0}}, x grid style={color={rgb,1:red,0.0;green,0.0;blue,0.0}, draw opacity={0.1}, line width={1.1}, solid}, x axis line style={color={rgb,1:red,0.0;green,0.0;blue,0.0}, draw opacity={1.0}, line width={2.2}, solid}, scaled y ticks={false}, ylabel={$\risk{\blambda}- \sigma^2$}, y tick style={color={rgb,1:red,0.0;green,0.0;blue,0.0}, opacity={1.0}}, y tick label style={color={rgb,1:red,0.0;green,0.0;blue,0.0}, opacity={1.0}, rotate={0}}, ylabel style={at={(ticklabel cs:0.5)}, anchor=near ticklabel, font={{\fontsize{24.200000000000003 pt}{31.460000000000004 pt}\selectfont}}, color={rgb,1:red,0.0;green,0.0;blue,0.0}, draw opacity={1.0}, rotate={0.0}}, ymajorgrids={false}, ymin={0}, ymax={6}, ytick={{0.0,1.0,2.0,3.0,4.0,5.0,6.0}}, yticklabels={{$0$,$1$,$2$,$3$,$4$,$5$,$6$}}, ytick align={inside}, yticklabel style={font={{\fontsize{17.6 pt}{22.880000000000003 pt}\selectfont}}, color={rgb,1:red,0.0;green,0.0;blue,0.0}, draw opacity={1.0}, rotate={0.0}}, y grid style={color={rgb,1:red,0.0;green,0.0;blue,0.0}, draw opacity={0.1}, line width={1.1}, solid}, y axis line style={color={rgb,1:red,0.0;green,0.0;blue,0.0}, draw opacity={1.0}, line width={2.2}, solid}]
    \addplot[color={rgb,1:red,0.502;green,0.502;blue,0.502}, name path={343bc0f5-de44-4eec-8380-8e3613af96b7}, draw opacity={1.0}, line width={2.2}, dotted]
        table[row sep={\\}]
        {
            \\
            0.0  3.9912949402777507  \\
            0.034482758620689655  3.9728795646894683  \\
            0.06896551724137931  3.9241118202679495  \\
            0.10344827586206896  3.8551552086577408  \\
            0.13793103448275862  3.771898195512274  \\
            0.1724137931034483  3.6778893660602128  \\
            0.20689655172413793  3.5753542098762656  \\
            0.2413793103448276  3.4658863760213032  \\
            0.27586206896551724  3.3505406026989615  \\
            0.3103448275862069  3.2302366128057276  \\
            0.3448275862068966  3.1055639404026394  \\
            0.3793103448275862  2.9770494711891917  \\
            0.41379310344827586  2.845107805292002  \\
            0.4482758620689655  2.7100476030956013  \\
            0.4827586206896552  2.5721202773827114  \\
            0.5172413793103449  2.431564936679373  \\
            0.5517241379310345  2.2886681380170684  \\
            0.5862068965517241  2.1434890847536217  \\
            0.6206896551724138  1.9961373690165383  \\
            0.6551724137931034  1.8468252173862885  \\
            0.6896551724137931  1.695589608623468  \\
            0.7241379310344828  1.5425391561757422  \\
            0.7586206896551724  1.38777697593313  \\
            0.7931034482758621  1.2313396236339682  \\
            0.8275862068965517  1.0733201621703277  \\
            0.8620689655172413  0.9136982538892422  \\
            0.896551724137931  0.7525486193714113  \\
            0.9310344827586207  0.589921090934518  \\
            0.9655172413793104  0.42581754519516646  \\
            1.0  0.2602573203321421  \\
        }
        ;
    \addplot[color={rgb,1:red,0.502;green,0.0;blue,0.502}, name path={d4b4e54b-2eb6-4958-a448-045fd9c43e9a}, draw opacity={1.0}, line width={2.2}, dashdotted]
        table[row sep={\\}]
        {
            \\
            0.0  0.5464257821337792  \\
            0.034482758620689655  0.5464257821337792  \\
            0.06896551724137931  0.5464257821337792  \\
            0.10344827586206896  0.5464257821337792  \\
            0.13793103448275862  0.5464257821337792  \\
            0.1724137931034483  0.5464257821337792  \\
            0.20689655172413793  0.5464257821337792  \\
            0.2413793103448276  0.5464257821337792  \\
            0.27586206896551724  0.5464257821337792  \\
            0.3103448275862069  0.5464257821337792  \\
            0.3448275862068966  0.5464257821337792  \\
            0.3793103448275862  0.5464257821337792  \\
            0.41379310344827586  0.5464257821337792  \\
            0.4482758620689655  0.5464257821337792  \\
            0.4827586206896552  0.5464257821337792  \\
            0.5172413793103449  0.5464257821337792  \\
            0.5517241379310345  0.5464257821337792  \\
            0.5862068965517241  0.5464257821337792  \\
            0.6206896551724138  0.5464257821337792  \\
            0.6551724137931034  0.5464257821337792  \\
            0.6896551724137931  0.5464257821337792  \\
            0.7241379310344828  0.5464257821337792  \\
            0.7586206896551724  0.5464257821337792  \\
            0.7931034482758621  0.5464257821337792  \\
            0.8275862068965517  0.5464257821337792  \\
            0.8620689655172413  0.5464257821337792  \\
            0.896551724137931  0.5464257821337792  \\
            0.9310344827586207  0.5464257821337792  \\
            0.9655172413793104  0.5464257821337792  \\
            1.0  0.5464257821337792  \\
        }
        ;
    \addplot[color={rgb,1:red,0.0;green,0.502;blue,0.0}, name path={1773c26c-6012-402b-80b5-9117cb7af8a5}, draw opacity={1.0}, line width={2.2}, dashed]
        table[row sep={\\}]
        {
            \\
            0.0  0.2602573203321421  \\
            0.034482758620689655  0.3592102360956091  \\
            0.06896551724137931  0.40959402755776964  \\
            0.10344827586206896  0.44301907533397444  \\
            0.13793103448275862  0.46739479700523257  \\
            0.1724137931034483  0.4860551359201386  \\
            0.20689655172413793  0.5007350698005182  \\
            0.2413793103448276  0.5124523635380784  \\
            0.27586206896551724  0.5218515253466509  \\
            0.3103448275862069  0.5293619284627527  \\
            0.3448275862068966  0.5352789815942822  \\
            0.3793103448275862  0.539809167273956  \\
            0.41379310344827586  0.5430964162246164  \\
            0.4482758620689655  0.5452380448540359  \\
            0.4827586206896552  0.5462944031197754  \\
            0.5172413793103449  0.5462944031197754  \\
            0.5517241379310345  0.5452380448540359  \\
            0.5862068965517241  0.5430964162246164  \\
            0.6206896551724138  0.539809167273956  \\
            0.6551724137931034  0.5352789815942822  \\
            0.6896551724137931  0.5293619284627527  \\
            0.7241379310344828  0.5218515253466509  \\
            0.7586206896551724  0.5124523635380784  \\
            0.7931034482758621  0.5007350698005182  \\
            0.8275862068965517  0.4860551359201386  \\
            0.8620689655172413  0.46739479700523257  \\
            0.896551724137931  0.44301907533397444  \\
            0.9310344827586207  0.40959402755776964  \\
            0.9655172413793104  0.3592102360956091  \\
            1.0  0.2602573203321421  \\
        }
        ;
    \addplot[color={rgb,1:red,0.502;green,0.502;blue,0.502}, name path={522e6744-ae23-4cc7-b290-f4cadb7d9b73}, only marks, draw opacity={1.0}, line width={0.0}, solid, mark={triangle*}, mark size={6.6000000000000005 pt}, mark repeat={1}, mark options={color={rgb,1:red,0.0;green,0.0;blue,0.0}, draw opacity={0.0}, fill={rgb,1:red,0.502;green,0.502;blue,0.502}, fill opacity={1.0}, line width={1.6500000000000001}, rotate={0}, solid}]
        table[row sep={\\}]
        {
            \\
            0.0  3.8288322364023273  \\
            0.034482758620689655  3.7098558983113294  \\
            0.06896551724137931  3.7658275143942825  \\
            0.10344827586206896  3.6209215623421445  \\
            0.13793103448275862  4.618666515354004  \\
            0.1724137931034483  3.125992241410022  \\
            0.20689655172413793  3.504621319799381  \\
            0.2413793103448276  3.3984594921231244  \\
            0.27586206896551724  3.2457033028504467  \\
            0.3103448275862069  3.3275171619670507  \\
            0.3448275862068966  2.911546791092845  \\
            0.3793103448275862  2.593525412525097  \\
            0.41379310344827586  2.7543595157537646  \\
            0.4482758620689655  2.3664335961521514  \\
            0.4827586206896552  1.9057451993809469  \\
            0.5172413793103449  2.2663066852281504  \\
            0.5517241379310345  2.020840329185838  \\
            0.5862068965517241  1.8721385151384524  \\
            0.6206896551724138  1.6367183167008696  \\
            0.6551724137931034  1.8988595028114155  \\
            0.6896551724137931  1.5676371578467245  \\
            0.7241379310344828  1.384228938598313  \\
            0.7586206896551724  1.2595097755443074  \\
            0.7931034482758621  1.0352507488603493  \\
            0.8275862068965517  1.0946412544890105  \\
            0.8620689655172413  0.8141646258004487  \\
            0.896551724137931  0.6389565491265283  \\
            0.9310344827586207  0.6060444747626939  \\
            0.9655172413793104  0.37743985157283566  \\
            1.0  0.3063947282998811  \\
        }
        ;
    \addplot[color={rgb,1:red,0.502;green,0.0;blue,0.502}, name path={73b9848d-1ced-4b44-8420-b1cb8e54c68a}, only marks, draw opacity={1.0}, line width={0.0}, solid, mark={triangle*}, mark size={6.6000000000000005 pt}, mark repeat={1}, mark options={color={rgb,1:red,0.0;green,0.0;blue,0.0}, draw opacity={0.0}, fill={rgb,1:red,0.502;green,0.0;blue,0.502}, fill opacity={1.0}, line width={1.6500000000000001}, rotate={0}, solid}]
        table[row sep={\\}]
        {
            \\
            0.0  0.5104120847310394  \\
            0.034482758620689655  0.5595923747195364  \\
            0.06896551724137931  0.548187310719704  \\
            0.10344827586206896  0.579995025154342  \\
            0.13793103448275862  0.5805957293405464  \\
            0.1724137931034483  0.5264685935291027  \\
            0.20689655172413793  0.6358076730301694  \\
            0.2413793103448276  0.5329764965161838  \\
            0.27586206896551724  0.5036894942243249  \\
            0.3103448275862069  0.6036354844784528  \\
            0.3448275862068966  0.5233820114388577  \\
            0.3793103448275862  0.49237925037113905  \\
            0.41379310344827586  0.5804466911543462  \\
            0.4482758620689655  0.5222720923527377  \\
            0.4827586206896552  0.548889035783654  \\
            0.5172413793103449  0.4687273987927074  \\
            0.5517241379310345  0.5161182479384789  \\
            0.5862068965517241  0.5504169671734875  \\
            0.6206896551724138  0.5805463672384428  \\
            0.6551724137931034  0.5262503289595202  \\
            0.6896551724137931  0.49203432810109216  \\
            0.7241379310344828  0.5242850433725428  \\
            0.7586206896551724  0.5164369305972818  \\
            0.7931034482758621  0.5328482192440471  \\
            0.8275862068965517  0.6032858025929666  \\
            0.8620689655172413  0.4857612117509209  \\
            0.896551724137931  0.5266301950375538  \\
            0.9310344827586207  0.5502651996834644  \\
            0.9655172413793104  0.46524879831777044  \\
            1.0  0.5895650537206854  \\
        }
        ;
    \addplot[color={rgb,1:red,0.0;green,0.502;blue,0.0}, name path={f30eebfc-cd23-406c-981d-92d57f015591}, only marks, draw opacity={1.0}, line width={0.0}, solid, mark={triangle*}, mark size={6.6000000000000005 pt}, mark repeat={1}, mark options={color={rgb,1:red,0.0;green,0.0;blue,0.0}, draw opacity={0.0}, fill={rgb,1:red,0.0;green,0.502;blue,0.0}, fill opacity={1.0}, line width={1.6500000000000001}, rotate={0}, solid}]
        table[row sep={\\}]
        {
            \\
            0.0  0.25258713719992953  \\
            0.034482758620689655  0.3614845928356438  \\
            0.06896551724137931  0.43745951637991176  \\
            0.10344827586206896  0.42664369208409303  \\
            0.13793103448275862  0.4784870005048927  \\
            0.1724137931034483  0.46089980868300917  \\
            0.20689655172413793  0.5751125383574618  \\
            0.2413793103448276  0.5037161278387918  \\
            0.27586206896551724  0.4578796891075143  \\
            0.3103448275862069  0.5804537287779858  \\
            0.3448275862068966  0.518433797318604  \\
            0.3793103448275862  0.4992644740412695  \\
            0.41379310344827586  0.569444427602041  \\
            0.4482758620689655  0.5237359265503838  \\
            0.4827586206896552  0.5486522420244833  \\
            0.5172413793103449  0.46838996859480275  \\
            0.5517241379310345  0.515283156558811  \\
            0.5862068965517241  0.5448781688414468  \\
            0.6206896551724138  0.5935903493599606  \\
            0.6551724137931034  0.5044872448432349  \\
            0.6896551724137931  0.49555408795610245  \\
            0.7241379310344828  0.495622762957753  \\
            0.7586206896551724  0.45347356801265315  \\
            0.7931034482758621  0.46475799686645725  \\
            0.8275862068965517  0.5334853383311899  \\
            0.8620689655172413  0.4058480194720355  \\
            0.896551724137931  0.41961051312178976  \\
            0.9310344827586207  0.3852726871375711  \\
            0.9655172413793104  0.27970234017470585  \\
            1.0  0.3063947282998811  \\
        }
        ;
\end{axis}
\end{tikzpicture}

%% file: figures/oracle_risks_exponential/oracle_risk5.tikz
\begin{tikzpicture}[/tikz/background rectangle/.style={fill={rgb,1:red,1.0;green,1.0;blue,1.0}, draw opacity={1.0}}, show background rectangle]
\begin{axis}[point meta max={nan}, point meta min={nan}, legend cell align={left}, title={$\gamma_1 = \frac{1}{10},\; \gamma_2 = \frac{4}{10},\;\; \alpha_1^2 + \alpha_2^2 = 2$}, title style={at={{(0.5,1)}}, anchor={south}, font={{\fontsize{30.800000000000004 pt}{40.040000000000006 pt}\selectfont}}, color={rgb,1:red,0.0;green,0.0;blue,0.0}, draw opacity={1.0}, rotate={0.0}}, legend style={color={rgb,1:red,0.0;green,0.0;blue,0.0}, draw opacity={0.0}, line width={2.2}, solid, fill={rgb,1:red,0.0;green,0.0;blue,0.0}, fill opacity={0.0}, text opacity={1.0}, font={{\fontsize{26.400000000000002 pt}{34.32000000000001 pt}\selectfont}}, text={rgb,1:red,0.0;green,0.0;blue,0.0}, at={(1.02, 1)}, anchor={north west}}, axis background/.style={fill={rgb,1:red,1.0;green,1.0;blue,1.0}, opacity={1.0}}, anchor={north west}, xshift={1.0mm}, yshift={-1.0mm}, width={163.1mm}, height={125.0mm}, scaled x ticks={false}, xlabel={$\alpha_1^2/(\alpha_1^2 + \alpha_2^2)$}, x tick style={color={rgb,1:red,0.0;green,0.0;blue,0.0}, opacity={1.0}}, x tick label style={color={rgb,1:red,0.0;green,0.0;blue,0.0}, opacity={1.0}, rotate={0}}, xlabel style={at={(ticklabel cs:0.5)}, anchor=near ticklabel, font={{\fontsize{24.200000000000003 pt}{31.460000000000004 pt}\selectfont}}, color={rgb,1:red,0.0;green,0.0;blue,0.0}, draw opacity={1.0}, rotate={0.0}}, xmajorgrids={false}, xmin={-0.03}, xmax={1.03}, xtick={{0.0,0.25,0.5,0.75,1.0}}, xticklabels={{$0.00$,$0.25$,$0.50$,$0.75$,$1.00$}}, xtick align={inside}, xticklabel style={font={{\fontsize{17.6 pt}{22.880000000000003 pt}\selectfont}}, color={rgb,1:red,0.0;green,0.0;blue,0.0}, draw opacity={1.0}, rotate={0.0}}, x grid style={color={rgb,1:red,0.0;green,0.0;blue,0.0}, draw opacity={0.1}, line width={1.1}, solid}, x axis line style={color={rgb,1:red,0.0;green,0.0;blue,0.0}, draw opacity={1.0}, line width={2.2}, solid}, scaled y ticks={false}, ylabel={$\risk{\blambda}- \sigma^2$}, y tick style={color={rgb,1:red,0.0;green,0.0;blue,0.0}, opacity={1.0}}, y tick label style={color={rgb,1:red,0.0;green,0.0;blue,0.0}, opacity={1.0}, rotate={0}}, ylabel style={at={(ticklabel cs:0.5)}, anchor=near ticklabel, font={{\fontsize{24.200000000000003 pt}{31.460000000000004 pt}\selectfont}}, color={rgb,1:red,0.0;green,0.0;blue,0.0}, draw opacity={1.0}, rotate={0.0}}, ymajorgrids={false}, ymin={0}, ymax={6}, ytick={{0.0,1.0,2.0,3.0,4.0,5.0,6.0}}, yticklabels={{$0$,$1$,$2$,$3$,$4$,$5$,$6$}}, ytick align={inside}, yticklabel style={font={{\fontsize{17.6 pt}{22.880000000000003 pt}\selectfont}}, color={rgb,1:red,0.0;green,0.0;blue,0.0}, draw opacity={1.0}, rotate={0.0}}, y grid style={color={rgb,1:red,0.0;green,0.0;blue,0.0}, draw opacity={0.1}, line width={1.1}, solid}, y axis line style={color={rgb,1:red,0.0;green,0.0;blue,0.0}, draw opacity={1.0}, line width={2.2}, solid}]
    \addplot[color={rgb,1:red,0.502;green,0.502;blue,0.502}, name path={30a528e3-8249-45a5-a3a3-a235c1059ea0}, draw opacity={1.0}, line width={2.2}, dotted]
        table[row sep={\\}]
        {
            \\
            0.0  3.993361283062117  \\
            0.034482758620689655  3.954149591304083  \\
            0.06896551724137931  3.8713234049146728  \\
            0.10344827586206896  3.7689101476797475  \\
            0.13793103448275862  3.6553099696083997  \\
            0.1724137931034483  3.5344031588690017  \\
            0.20689655172413793  3.4083505194176436  \\
            0.2413793103448276  3.2784462002093635  \\
            0.27586206896551724  3.1455796799043156  \\
            0.3103448275862069  3.010328690548948  \\
            0.3448275862068966  2.87312319406736  \\
            0.3793103448275862  2.734296658803478  \\
            0.41379310344827586  2.5940808286267356  \\
            0.4482758620689655  2.4526728987863917  \\
            0.4827586206896552  2.310252691370628  \\
            0.5172413793103449  2.1668994894945124  \\
            0.5517241379310345  2.0227438413200485  \\
            0.5862068965517241  1.877856757357503  \\
            0.6206896551724138  1.7323242143578632  \\
            0.6551724137931034  1.5861884923249416  \\
            0.6896551724137931  1.4395176123273026  \\
            0.7241379310344828  1.2923415097922053  \\
            0.7586206896551724  1.1447074246282751  \\
            0.7931034482758621  0.9966411724835611  \\
            0.8275862068965517  0.8481751105253388  \\
            0.8620689655172413  0.6993354366532603  \\
            0.896551724137931  0.5501370840153434  \\
            0.9310344827586207  0.40060629332493214  \\
            0.9655172413793104  0.25075705159500905  \\
            1.0  0.1006038234211819  \\
        }
        ;
    \addplot[color={rgb,1:red,0.502;green,0.0;blue,0.502}, name path={e12e9fc9-e6e7-4a08-9473-7760d58e976e}, draw opacity={1.0}, line width={2.2}, dashdotted]
        table[row sep={\\}]
        {
            \\
            0.0  0.5464208103773449  \\
            0.034482758620689655  0.5464219544301405  \\
            0.06896551724137931  0.5464230984830163  \\
            0.10344827586206896  0.5464242425359087  \\
            0.13793103448275862  0.5464253865887674  \\
            0.1724137931034483  0.5464265306416476  \\
            0.20689655172413793  0.5464276746945185  \\
            0.2413793103448276  0.5464288187473867  \\
            0.27586206896551724  0.5464299628002562  \\
            0.3103448275862069  0.5464311068531258  \\
            0.3448275862068966  0.5464322509059953  \\
            0.3793103448275862  0.5464333949588649  \\
            0.41379310344827586  0.5464345390117717  \\
            0.4482758620689655  0.5464356830646473  \\
            0.4827586206896552  0.5464368271174733  \\
            0.5172413793103449  0.5464379711703429  \\
            0.5517241379310345  0.5464391152232739  \\
            0.5862068965517241  0.5464402592761495  \\
            0.6206896551724138  0.5464414033289513  \\
            0.6551724137931034  0.5464425473818209  \\
            0.6896551724137931  0.5464436914346904  \\
            0.7241379310344828  0.54644483548756  \\
            0.7586206896551724  0.5464459795404295  \\
            0.7931034482758621  0.546447123593403  \\
            0.8275862068965517  0.5464482676461684  \\
            0.8620689655172413  0.5464494116991541  \\
            0.896551724137931  0.5464505557519075  \\
            0.9310344827586207  0.5464516998049052  \\
            0.9655172413793104  0.5464528438577807  \\
            1.0  0.5464539879103374  \\
        }
        ;
    \addplot[color={rgb,1:red,0.0;green,0.502;blue,0.0}, name path={0c3efddf-8aa5-415a-9a3d-f5b032c7db1c}, draw opacity={1.0}, line width={2.2}, dashed]
        table[row sep={\\}]
        {
            \\
            0.0  0.42987099736670165  \\
            0.034482758620689655  0.5026286602705097  \\
            0.06896551724137931  0.5258937161923432  \\
            0.10344827586206896  0.5372848397543302  \\
            0.13793103448275862  0.5431758847081731  \\
            0.1724137931034483  0.545857272369104  \\
            0.20689655172413793  0.5463954431721842  \\
            0.2413793103448276  0.5453607790939903  \\
            0.27586206896551724  0.5430841328142364  \\
            0.3103448275862069  0.5397658368172542  \\
            0.3448275862068966  0.5355280556884141  \\
            0.3793103448275862  0.5304420893932393  \\
            0.41379310344827586  0.5245433426470609  \\
            0.4482758620689655  0.5178396294009198  \\
            0.4827586206896552  0.5103155183758288  \\
            0.5172413793103449  0.5019340418297289  \\
            0.5517241379310345  0.49263636438677083  \\
            0.5862068965517241  0.4823395588619137  \\
            0.6206896551724138  0.47093227337877797  \\
            0.6551724137931034  0.45826767199777896  \\
            0.6896551724137931  0.4441524515915358  \\
            0.7241379310344828  0.4283297548446552  \\
            0.7586206896551724  0.410451952930873  \\
            0.7931034482758621  0.3900355043814936  \\
            0.8275862068965517  0.3663817178076354  \\
            0.8620689655172413  0.33842656136552973  \\
            0.896551724137931  0.30442412190108326  \\
            0.9310344827586207  0.26116735394820156  \\
            0.9655172413793104  0.20151236650953663  \\
            1.0  0.1006038234211819  \\
        }
        ;
    \addplot[color={rgb,1:red,0.502;green,0.502;blue,0.502}, name path={a33a87b2-2243-4922-8053-89f642e17d77}, only marks, draw opacity={1.0}, line width={0.0}, solid, mark={triangle*}, mark size={6.6000000000000005 pt}, mark repeat={1}, mark options={color={rgb,1:red,0.0;green,0.0;blue,0.0}, draw opacity={0.0}, fill={rgb,1:red,0.502;green,0.502;blue,0.502}, fill opacity={1.0}, line width={1.6500000000000001}, rotate={0}, solid}]
        table[row sep={\\}]
        {
            \\
            0.0  3.8579648345785387  \\
            0.034482758620689655  3.2948768547310934  \\
            0.06896551724137931  4.295873218439311  \\
            0.10344827586206896  4.303290293928124  \\
            0.13793103448275862  3.3653193922137676  \\
            0.1724137931034483  3.335301436313718  \\
            0.20689655172413793  3.521311873731852  \\
            0.2413793103448276  3.2883625935287997  \\
            0.27586206896551724  3.835444544685722  \\
            0.3103448275862069  2.889824609696034  \\
            0.3448275862068966  2.816844712882101  \\
            0.3793103448275862  2.861336480511568  \\
            0.41379310344827586  2.806092234954168  \\
            0.4482758620689655  2.3245740637910655  \\
            0.4827586206896552  2.323382245263342  \\
            0.5172413793103449  2.224898960551111  \\
            0.5517241379310345  2.225065970863364  \\
            0.5862068965517241  1.6330357757848506  \\
            0.6206896551724138  1.4904451467664748  \\
            0.6551724137931034  1.5853007200525155  \\
            0.6896551724137931  1.612571998139574  \\
            0.7241379310344828  1.3992121915440658  \\
            0.7586206896551724  1.0747917137364773  \\
            0.7931034482758621  1.1669342351162975  \\
            0.8275862068965517  0.8275047608662902  \\
            0.8620689655172413  0.7291780959104752  \\
            0.896551724137931  0.5937390557564792  \\
            0.9310344827586207  0.41020705308457805  \\
            0.9655172413793104  0.22342925529516533  \\
            1.0  0.10496207510705902  \\
        }
        ;
    \addplot[color={rgb,1:red,0.502;green,0.0;blue,0.502}, name path={8854c1c7-db08-4c41-926e-8dd9f03e9c87}, only marks, draw opacity={1.0}, line width={0.0}, solid, mark={triangle*}, mark size={6.6000000000000005 pt}, mark repeat={1}, mark options={color={rgb,1:red,0.0;green,0.0;blue,0.0}, draw opacity={0.0}, fill={rgb,1:red,0.502;green,0.0;blue,0.502}, fill opacity={1.0}, line width={1.6500000000000001}, rotate={0}, solid}]
        table[row sep={\\}]
        {
            \\
            0.0  0.57542506737789  \\
            0.034482758620689655  0.47309221129313817  \\
            0.06896551724137931  0.575342653301514  \\
            0.10344827586206896  0.5864595885972568  \\
            0.13793103448275862  0.5234407888859627  \\
            0.1724137931034483  0.5414503173862761  \\
            0.20689655172413793  0.5978877305095196  \\
            0.2413793103448276  0.5756863048511929  \\
            0.27586206896551724  0.6059640616573254  \\
            0.3103448275862069  0.3873748365800618  \\
            0.3448275862068966  0.504384743385726  \\
            0.3793103448275862  0.5739239412851806  \\
            0.41379310344827586  0.5534996148899913  \\
            0.4482758620689655  0.6141141492102771  \\
            0.4827586206896552  0.5755699432227459  \\
            0.5172413793103449  0.5419512984889272  \\
            0.5517241379310345  0.5925177806520856  \\
            0.5862068965517241  0.5566899809601062  \\
            0.6206896551724138  0.5051313918754039  \\
            0.6551724137931034  0.5087734985258272  \\
            0.6896551724137931  0.5404453195966481  \\
            0.7241379310344828  0.5869358880586246  \\
            0.7586206896551724  0.5539326948410999  \\
            0.7931034482758621  0.6255858783890436  \\
            0.8275862068965517  0.4980466923679019  \\
            0.8620689655172413  0.48335760200815203  \\
            0.896551724137931  0.5993170774022669  \\
            0.9310344827586207  0.5066102224190083  \\
            0.9655172413793104  0.46145727371042855  \\
            1.0  0.49369853039577216  \\
        }
        ;
    \addplot[color={rgb,1:red,0.0;green,0.502;blue,0.0}, name path={7094c1e5-6b7a-4e63-a6bb-8895ca89913c}, only marks, draw opacity={1.0}, line width={0.0}, solid, mark={triangle*}, mark size={6.6000000000000005 pt}, mark repeat={1}, mark options={color={rgb,1:red,0.0;green,0.0;blue,0.0}, draw opacity={0.0}, fill={rgb,1:red,0.0;green,0.502;blue,0.0}, fill opacity={1.0}, line width={1.6500000000000001}, rotate={0}, solid}]
        table[row sep={\\}]
        {
            \\
            0.0  0.4528885239780496  \\
            0.034482758620689655  0.44621035609886195  \\
            0.06896551724137931  0.5601388761635913  \\
            0.10344827586206896  0.5721445614080489  \\
            0.13793103448275862  0.5185817045248267  \\
            0.1724137931034483  0.5415724295190707  \\
            0.20689655172413793  0.5982410498553774  \\
            0.2413793103448276  0.5717833897525195  \\
            0.27586206896551724  0.6101572673197069  \\
            0.3103448275862069  0.3865784987832306  \\
            0.3448275862068966  0.4879681199695085  \\
            0.3793103448275862  0.5530371846822086  \\
            0.41379310344827586  0.5381670505800158  \\
            0.4482758620689655  0.5596482624673076  \\
            0.4827586206896552  0.5431809215610752  \\
            0.5172413793103449  0.49802215592277155  \\
            0.5517241379310345  0.5678945577229844  \\
            0.5862068965517241  0.5057582296239096  \\
            0.6206896551724138  0.4326086353760239  \\
            0.6551724137931034  0.4306092406983655  \\
            0.6896551724137931  0.42368868339182963  \\
            0.7241379310344828  0.4513055040813576  \\
            0.7586206896551724  0.43829194994941134  \\
            0.7931034482758621  0.4544661812032429  \\
            0.8275862068965517  0.3654842726348311  \\
            0.8620689655172413  0.297313292418109  \\
            0.896551724137931  0.31864941965123617  \\
            0.9310344827586207  0.24316600987059034  \\
            0.9655172413793104  0.1789475951441979  \\
            1.0  0.10496207510705902  \\
        }
        ;
\end{axis}
\end{tikzpicture}

%% file: figures/oracle_risks_exponential/oracle_risk6.tikz
\begin{tikzpicture}[/tikz/background rectangle/.style={fill={rgb,1:red,1.0;green,1.0;blue,1.0}, draw opacity={1.0}}, show background rectangle]
\begin{axis}[point meta max={nan}, point meta min={nan}, legend cell align={left}, title={$\gamma_1 = \gamma_2 = 1,\;\; \alpha_1^2 + \alpha_2^2 = 2$}, title style={at={{(0.5,1)}}, anchor={south}, font={{\fontsize{30.800000000000004 pt}{40.040000000000006 pt}\selectfont}}, color={rgb,1:red,0.0;green,0.0;blue,0.0}, draw opacity={1.0}, rotate={0.0}}, legend style={color={rgb,1:red,0.0;green,0.0;blue,0.0}, draw opacity={0.0}, line width={2.2}, solid, fill={rgb,1:red,0.0;green,0.0;blue,0.0}, fill opacity={0.0}, text opacity={1.0}, font={{\fontsize{26.400000000000002 pt}{34.32000000000001 pt}\selectfont}}, text={rgb,1:red,0.0;green,0.0;blue,0.0}, at={(1.02, 1)}, anchor={north west}}, axis background/.style={fill={rgb,1:red,1.0;green,1.0;blue,1.0}, opacity={1.0}}, anchor={north west}, xshift={1.0mm}, yshift={-1.0mm}, width={163.1mm}, height={125.0mm}, scaled x ticks={false}, xlabel={$\alpha_1^2/(\alpha_1^2 + \alpha_2^2)$}, x tick style={color={rgb,1:red,0.0;green,0.0;blue,0.0}, opacity={1.0}}, x tick label style={color={rgb,1:red,0.0;green,0.0;blue,0.0}, opacity={1.0}, rotate={0}}, xlabel style={at={(ticklabel cs:0.5)}, anchor=near ticklabel, font={{\fontsize{24.200000000000003 pt}{31.460000000000004 pt}\selectfont}}, color={rgb,1:red,0.0;green,0.0;blue,0.0}, draw opacity={1.0}, rotate={0.0}}, xmajorgrids={false}, xmin={-0.03}, xmax={1.03}, xtick={{0.0,0.25,0.5,0.75,1.0}}, xticklabels={{$0.00$,$0.25$,$0.50$,$0.75$,$1.00$}}, xtick align={inside}, xticklabel style={font={{\fontsize{17.6 pt}{22.880000000000003 pt}\selectfont}}, color={rgb,1:red,0.0;green,0.0;blue,0.0}, draw opacity={1.0}, rotate={0.0}}, x grid style={color={rgb,1:red,0.0;green,0.0;blue,0.0}, draw opacity={0.1}, line width={1.1}, solid}, x axis line style={color={rgb,1:red,0.0;green,0.0;blue,0.0}, draw opacity={1.0}, line width={2.2}, solid}, scaled y ticks={false}, ylabel={$\risk{\blambda}- \sigma^2$}, y tick style={color={rgb,1:red,0.0;green,0.0;blue,0.0}, opacity={1.0}}, y tick label style={color={rgb,1:red,0.0;green,0.0;blue,0.0}, opacity={1.0}, rotate={0}}, ylabel style={at={(ticklabel cs:0.5)}, anchor=near ticklabel, font={{\fontsize{24.200000000000003 pt}{31.460000000000004 pt}\selectfont}}, color={rgb,1:red,0.0;green,0.0;blue,0.0}, draw opacity={1.0}, rotate={0.0}}, ymajorgrids={false}, ymin={0}, ymax={6}, ytick={{0.0,1.0,2.0,3.0,4.0,5.0,6.0}}, yticklabels={{$0$,$1$,$2$,$3$,$4$,$5$,$6$}}, ytick align={inside}, yticklabel style={font={{\fontsize{17.6 pt}{22.880000000000003 pt}\selectfont}}, color={rgb,1:red,0.0;green,0.0;blue,0.0}, draw opacity={1.0}, rotate={0.0}}, y grid style={color={rgb,1:red,0.0;green,0.0;blue,0.0}, draw opacity={0.1}, line width={1.1}, solid}, y axis line style={color={rgb,1:red,0.0;green,0.0;blue,0.0}, draw opacity={1.0}, line width={2.2}, solid}]
    \addplot[color={rgb,1:red,0.502;green,0.502;blue,0.502}, name path={9f22af14-a38f-459b-ad74-9267ad359e55}, draw opacity={1.0}, line width={2.2}, dotted]
        table[row sep={\\}]
        {
            \\
            0.0  3.99542884846883  \\
            0.034482758620689655  3.990852415083304  \\
            0.06896551724137931  3.9759007138982447  \\
            0.10344827586206896  3.950953216154094  \\
            0.13793103448275862  3.9168504640616675  \\
            0.1724137931034483  3.8744022294902587  \\
            0.20689655172413793  3.824339282518425  \\
            0.2413793103448276  3.767256989570239  \\
            0.27586206896551724  3.7035997121436504  \\
            0.3103448275862069  3.6337523965881013  \\
            0.3448275862068966  3.558074198832105  \\
            0.3793103448275862  3.4768430109360215  \\
            0.41379310344827586  3.390155080341767  \\
            0.4482758620689655  3.2983076400959197  \\
            0.4827586206896552  3.201298355068107  \\
            0.5172413793103449  3.0993000737335787  \\
            0.5517241379310345  2.9923470428607253  \\
            0.5862068965517241  2.880474682854018  \\
            0.6206896551724138  2.7636120957682664  \\
            0.6551724137931034  2.6417604616523134  \\
            0.6896551724137931  2.514900673648736  \\
            0.7241379310344828  2.3829504864033817  \\
            0.7586206896551724  2.2456299228436984  \\
            0.7931034482758621  2.102982576849634  \\
            0.8275862068965517  1.9546414079288015  \\
            0.8620689655172413  1.8003946756336422  \\
            0.896551724137931  1.6399172698315483  \\
            0.9310344827586207  1.4728527603776564  \\
            0.9655172413793104  1.2986863520603351  \\
            1.0  1.1168536169101158  \\
        }
        ;
    \addplot[color={rgb,1:red,0.502;green,0.0;blue,0.502}, name path={e0d1eb8a-e04f-4353-aa74-073a7a4ac953}, draw opacity={1.0}, line width={2.2}, dashdotted]
        table[row sep={\\}]
        {
            \\
            0.0  1.9512510559932261  \\
            0.034482758620689655  1.9512510559932261  \\
            0.06896551724137931  1.9512510559932261  \\
            0.10344827586206896  1.9512510559932261  \\
            0.13793103448275862  1.9512510559932261  \\
            0.1724137931034483  1.9512510559932261  \\
            0.20689655172413793  1.9512510559932261  \\
            0.2413793103448276  1.9512510559932261  \\
            0.27586206896551724  1.9512510559932261  \\
            0.3103448275862069  1.9512510559932261  \\
            0.3448275862068966  1.9512510559932261  \\
            0.3793103448275862  1.9512510559932261  \\
            0.41379310344827586  1.9512510559932261  \\
            0.4482758620689655  1.9512510559932261  \\
            0.4827586206896552  1.9512510559932261  \\
            0.5172413793103449  1.9512510559932261  \\
            0.5517241379310345  1.9512510559932261  \\
            0.5862068965517241  1.9512510559932261  \\
            0.6206896551724138  1.9512510559932261  \\
            0.6551724137931034  1.9512510559932261  \\
            0.6896551724137931  1.9512510559932261  \\
            0.7241379310344828  1.9512510559932261  \\
            0.7586206896551724  1.9512510559932261  \\
            0.7931034482758621  1.9512510559932261  \\
            0.8275862068965517  1.9512510559932261  \\
            0.8620689655172413  1.9512510559932261  \\
            0.896551724137931  1.9512510559932261  \\
            0.9310344827586207  1.9512510559932261  \\
            0.9655172413793104  1.9512510559932261  \\
            1.0  1.9512510559932261  \\
        }
        ;
    \addplot[color={rgb,1:red,0.0;green,0.502;blue,0.0}, name path={d25800b0-851f-450f-9a7e-2d20023189ba}, draw opacity={1.0}, line width={2.2}, dashed]
        table[row sep={\\}]
        {
            \\
            0.0  1.1168536169101158  \\
            0.034482758620689655  1.2774496389031262  \\
            0.06896551724137931  1.402972046047541  \\
            0.10344827586206896  1.505814598574731  \\
            0.13793103448275862  1.5920331836036516  \\
            0.1724137931034483  1.6651324158673821  \\
            0.20689655172413793  1.7273498968739758  \\
            0.2413793103448276  1.7802073618934053  \\
            0.27586206896551724  1.8247839533439887  \\
            0.3103448275862069  1.86186573768381  \\
            0.3448275862068966  1.8920334760364872  \\
            0.3793103448275862  1.91571680463107  \\
            0.41379310344827586  1.9332288002299824  \\
            0.4482758620689655  1.944788340203477  \\
            0.4827586206896552  1.9505343657753325  \\
            0.5172413793103449  1.9505343657753325  \\
            0.5517241379310345  1.944788340203477  \\
            0.5862068965517241  1.9332288002299824  \\
            0.6206896551724138  1.91571680463107  \\
            0.6551724137931034  1.8920334760364872  \\
            0.6896551724137931  1.86186573768381  \\
            0.7241379310344828  1.8247839533439887  \\
            0.7586206896551724  1.7802073618934053  \\
            0.7931034482758621  1.7273498968739758  \\
            0.8275862068965517  1.6651324158673821  \\
            0.8620689655172413  1.5920331836036516  \\
            0.896551724137931  1.505814598574731  \\
            0.9310344827586207  1.402972046047541  \\
            0.9655172413793104  1.2774496389031262  \\
            1.0  1.1168536169101158  \\
        }
        ;
    \addplot[color={rgb,1:red,0.502;green,0.502;blue,0.502}, name path={9c15a403-a1f8-46d2-910e-4ad4ce505347}, only marks, draw opacity={1.0}, line width={0.0}, solid, mark={triangle*}, mark size={6.6000000000000005 pt}, mark repeat={1}, mark options={color={rgb,1:red,0.0;green,0.0;blue,0.0}, draw opacity={0.0}, fill={rgb,1:red,0.502;green,0.502;blue,0.502}, fill opacity={1.0}, line width={1.6500000000000001}, rotate={0}, solid}]
        table[row sep={\\}]
        {
            \\
            0.0  3.933927282397053  \\
            0.034482758620689655  3.9341257485309713  \\
            0.06896551724137931  3.743949831466769  \\
            0.10344827586206896  3.8021857452280656  \\
            0.13793103448275862  3.5294857415493057  \\
            0.1724137931034483  3.5811386969967893  \\
            0.20689655172413793  3.7958046154051877  \\
            0.2413793103448276  3.6223698737720245  \\
            0.27586206896551724  3.4565974541506543  \\
            0.3103448275862069  3.8398150683124843  \\
            0.3448275862068966  3.8939785321527483  \\
            0.3793103448275862  3.621347892522004  \\
            0.41379310344827586  3.2025707513292447  \\
            0.4482758620689655  3.4076808864612396  \\
            0.4827586206896552  3.1754713235757226  \\
            0.5172413793103449  3.009843639804928  \\
            0.5517241379310345  2.988921593648816  \\
            0.5862068965517241  3.0348290732676197  \\
            0.6206896551724138  2.9685923997949475  \\
            0.6551724137931034  2.659858025934342  \\
            0.6896551724137931  2.7574945420212935  \\
            0.7241379310344828  2.246376664661464  \\
            0.7586206896551724  2.200300304943451  \\
            0.7931034482758621  2.0773792350770304  \\
            0.8275862068965517  1.951796273811825  \\
            0.8620689655172413  1.7930610616451208  \\
            0.896551724137931  1.5561979473768752  \\
            0.9310344827586207  1.312457950279585  \\
            0.9655172413793104  1.344921785294066  \\
            1.0  1.121445294828273  \\
        }
        ;
    \addplot[color={rgb,1:red,0.502;green,0.0;blue,0.502}, name path={89375816-748f-49ff-9e63-7e32b2c05dec}, only marks, draw opacity={1.0}, line width={0.0}, solid, mark={triangle*}, mark size={6.6000000000000005 pt}, mark repeat={1}, mark options={color={rgb,1:red,0.0;green,0.0;blue,0.0}, draw opacity={0.0}, fill={rgb,1:red,0.502;green,0.0;blue,0.502}, fill opacity={1.0}, line width={1.6500000000000001}, rotate={0}, solid}]
        table[row sep={\\}]
        {
            \\
            0.0  1.8958251476808878  \\
            0.034482758620689655  2.0277851559204945  \\
            0.06896551724137931  1.9529197571664283  \\
            0.10344827586206896  1.9423760569042527  \\
            0.13793103448275862  1.809799493959805  \\
            0.1724137931034483  2.034922247192794  \\
            0.20689655172413793  2.0041222694721976  \\
            0.2413793103448276  1.8821719201018596  \\
            0.27586206896551724  1.8250689864876612  \\
            0.3103448275862069  2.0248076853694896  \\
            0.3448275862068966  2.0269880829434697  \\
            0.3793103448275862  1.8864295031868896  \\
            0.41379310344827586  1.8587100866600799  \\
            0.4482758620689655  2.0344657776438924  \\
            0.4827586206896552  2.0040999274516262  \\
            0.5172413793103449  2.044765560656113  \\
            0.5517241379310345  1.9290325595510143  \\
            0.5862068965517241  1.9650414743376143  \\
            0.6206896551724138  1.934321715120082  \\
            0.6551724137931034  2.0313740134747764  \\
            0.6896551724137931  1.9731301329232496  \\
            0.7241379310344828  1.8991668676949716  \\
            0.7586206896551724  1.9608940769303715  \\
            0.7931034482758621  1.8676407997324191  \\
            0.8275862068965517  2.0159322075102626  \\
            0.8620689655172413  1.9303013583289976  \\
            0.896551724137931  1.8500458616614233  \\
            0.9310344827586207  1.8732039911710512  \\
            0.9655172413793104  1.9156328154944018  \\
            1.0  2.0006367188651093  \\
        }
        ;
    \addplot[color={rgb,1:red,0.0;green,0.502;blue,0.0}, name path={ae57b662-94ea-4ac9-afbf-78d493531518}, only marks, draw opacity={1.0}, line width={0.0}, solid, mark={triangle*}, mark size={6.6000000000000005 pt}, mark repeat={1}, mark options={color={rgb,1:red,0.0;green,0.0;blue,0.0}, draw opacity={0.0}, fill={rgb,1:red,0.0;green,0.502;blue,0.0}, fill opacity={1.0}, line width={1.6500000000000001}, rotate={0}, solid}]
        table[row sep={\\}]
        {
            \\
            0.0  1.1192932284070825  \\
            0.034482758620689655  1.3838023365326109  \\
            0.06896551724137931  1.4394362256621438  \\
            0.10344827586206896  1.5342476610166353  \\
            0.13793103448275862  1.502993350889959  \\
            0.1724137931034483  1.8436319671205679  \\
            0.20689655172413793  1.7890514585001274  \\
            0.2413793103448276  1.7557782292846755  \\
            0.27586206896551724  1.6808551097338271  \\
            0.3103448275862069  1.9216586175314583  \\
            0.3448275862068966  1.8793114170968641  \\
            0.3793103448275862  1.8229703238516004  \\
            0.41379310344827586  1.8488601667594917  \\
            0.4482758620689655  2.0286251714413224  \\
            0.4827586206896552  1.9994824652415568  \\
            0.5172413793103449  2.0420543818393893  \\
            0.5517241379310345  1.927229281971516  \\
            0.5862068965517241  1.9481401613278488  \\
            0.6206896551724138  1.919748856109626  \\
            0.6551724137931034  1.9762618171349562  \\
            0.6896551724137931  1.9361174652732633  \\
            0.7241379310344828  1.7564529086410823  \\
            0.7586206896551724  1.7923550891317235  \\
            0.7931034482758621  1.6997808023378806  \\
            0.8275862068965517  1.650372247439662  \\
            0.8620689655172413  1.5795030502647367  \\
            0.896551724137931  1.430225892538934  \\
            0.9310344827586207  1.2689218405587588  \\
            0.9655172413793104  1.3212915275094348  \\
            1.0  1.121445294828273  \\
        }
        ;
\end{axis}
\end{tikzpicture}